\documentclass[a4paper, 11pt, oneside]{Thesis}  
\graphicspath{{./Figures/}}  

\usepackage[square, numbers, comma, sort&compress]{natbib}  
\usepackage{hypernat}
\usepackage{verbatim}  
\usepackage{vector}  
\usepackage{dsfont}
\hypersetup{urlcolor=MidnightBlue, colorlinks=true}  

\begin{document}
\frontmatter	  

\title  {Noise Correlations and Coherent \\Coupling in Solid-State Qubits}
\authors  {\texorpdfstring
            {\href{http://www.icmm.csic.es/dmarcos}{David Marcos}}
            {David Marcos}
            }
\addresses  {\groupname\\\deptname\\\univname}  
\date {Madrid, January 2011}
\subject    {}

\maketitle

\newpage{\
\thispagestyle{empty}}
\newpage{\
\thispagestyle{empty}}

\setstretch{1.3}  

\fancyhead{}  
\rhead{\thepage}  
\lhead{}  

\pagestyle{fancy}  

\pagestyle{empty}  

\null\vfill

\begin{flushright}

\textit{Cuando encuentre el silencio \\
y la palabra \\
callar\'e para siempre. \\
(Quiz\'as entonces hable \\
lo que hoy calla) \\
He de callar, \\
si encuentro una palabra \\
que baste no decir.}

Jos\'e Corredor-Matheos
\end{flushright}

\vfill\vfill\vfill\vfill\vfill\vfill\null
\clearpage  

\newpage{\
\thispagestyle{empty}}
\newpage{\
\thispagestyle{empty}}


\begin{flushright}

Thesis Advisor: \href{http://www.icmm.csic.es/raguado/} {\bf{Ram\'on Aguado}}

\end{flushright}

\Committee{

\centerline{Prof. Markus B\"uttiker} 
\vskip 12mm
\centerline{Prof. Yuli Nazarov} 
\vskip 12mm
\centerline{Prof. Tobias Brandes} 
\vskip 12mm
\centerline{Prof. Juan Carlos Cuevas} 
\vskip 12mm
\centerline{Prof. Mar\'ia Jos\'e Calder\'on} 

}

\clearpage  

\pagestyle{empty}  

\null\vfill

\begin{flushleft}

\textit{``And if you take one from three hundred and sixty-five, what remains?''\\
``Three hundred and sixty-four, of course.''\\
Humpty Dumpty looked doubtful. ``I'd rather see that done on paper,'' he said.\\
\ldots\\
Humpty Dumpty took the book, and looked at it carefully. ``That seems to be done right -- '' he began.\\
``You're holding it upside down!'' Alice interrupted.\\
``To be sure I was!'' Humpty Dumpty said gaily, as she turned it round for him. ``I thought it looked a little queer. As I was saying, that seems to be done right -- though I haven't time to look it over thoroughly just now\ldots''\\
\ldots\\
``When I use a word,'' Humpty Dumpty said in rather a scornful tone, ``it means just what I choose it to mean -- neither more nor less.''\\
``The question is,'' said Alice, ``whether you can make words mean so many different things.''\\
``The question is,' said Humpty Dumpty, ``which is to be master -- that's all.''}
\end{flushleft}

\begin{flushleft}
Lewis Carroll -- \textit{Alice Through the Looking-Glass.}
\end{flushleft}

\vfill\vfill\vfill\vfill\vfill\vfill\null
\clearpage  


\acknowledgements{
\addtocontents{toc}{\vspace{0em}}  

\begin{flushright}

\textit{``Freedom is not worth having if it does not\\ include the freedom to make mistakes.''}

Mahatma Gandhi







\end{flushright}

A PhD thesis is certainly something that changes someone's life. At least that was my case. When I look back over the last five years of my life, I find a mixture of feelings that range from extreme desperation to absolute joy. I find incredible stories to tell, and most importantly, I find many people to whom I am grateful and that I will never forget. Without them my life would not be the same, and my PhD thesis would simply not exist. Growing up with them was really standing on giants' shoulders, and I am aware that I will not properly comprehend their role on my scientific and personal life until many years from now.

First, I want to acknowledge my advisor Ram\'on Aguado. It was back in December 2004 when we first met, and had a conversation on spontaneous symmetry breaking. At that time my eye was focusing on a quantum optics group in Paris, and on a quantum information group in Geneva. But seeing Ramon's interest in understanding every conceptual detail of what we discussed, I finally stayed in Madrid. Thank you, Ram\'on, for having shown me the way towards critical thinking. Also, for teaching me that smooth seas do not make skillful sailors, and for insisting on the fact that every small piece of physics can contain a great deal of information (or not, but one should check!). Above all, thank you for your extraordinary support.

On September 2nd. 2006, I took a flight to stay some months in Tobias Brandes' group at the Technical University of Berlin. This period cemented in me the scientific grounds from which any research conducted in the next years could grow. Thank you, Tobias, for the excitement for physics that you always transmit, and for teaching us how to run a group in the most efficient way.

In August 2007, I went to Charlie Marcus' group at Harvard, where I would stay until the end of the year. Without question, this was the experience that most affected me over the course of my PhD. Although finding my way as an experimentalist was not easy, I now remember my time at Harvard as one of the happiest periods of my life. Charlie always found the time for a memorable group meeting, or for a night conversation where we would set up the experiment with one hand, while holding coffee with the other. Thank you, Charlie, for the great time in your lab.

During my time in Boston I used to watch some of the ITAMP online talks, and one of them was particularly fascinating to me. It was about hybrid quantum systems, and given by Anders S\o rensen. The topic, and the way of doing physics appeared so interesting that I contacted Anders to spend some time in his group in Copenhagen. That finally happened from September to January 2008. Anders was my mentor in quantum optics. Thank you for your patience during my learning process. From Anders I learned the art of making approximations in physics, and how a good intuition can be used to do great science. Also, that theoretical physics can be easy up to factors of two, pi, or minus signs. Anders, and the science he does, is for me the example to follow in physics.

Reaching the end of my PhD, I decided to take a last adventure, and visit Leo Kowenhoven's group, where I stayed the last four months of 2009. Just when one thinks that he has learned everything in life, a new experience proves otherwise. Delft was a challenging and fascinating experience from all perspectives. Leo, and the time in QT, showed me the way towards scientific independence, and how small ideas can be turned into an outstanding project. Thank you, Leo, for your support and making this possible.

Over these years I have collaborated with a number of people. Without their insight, I would still be sitting at my desk, trying to finish some calculations to complete my PhD. Some of them are: Clive Emary, from whom I learned how to work efficiently. Thank you for your priceless contribution to my PhD. Hugh Churchill, my mentor in experimental physics. Thanks for your patience and great sense of humor, particularly in those moments when things did not work out in the lab, and you showed me the picture of the first transistor. Pol Forn-D\'iaz, thanks to whom I finally found my way in Delft. Thank you for having taught me all I know about superconducting qubits. I wish you all the best in your American adventure with Roser. I hope we continue working together for many years, and that those discussions on the gmail chat are repeated quite soon. Martijn Wubs, who supervised every detail of the notes I sent. Thank you for always having encouraging words for me. Ferdinand Kuemmeth, one of my best friends in Boston. Your intuition in physics was an example to follow for all of us. Thank you for the best Gl\"uhwein I have ever drunk. I am also indebted to some of the most cultivated people I know: Jake Taylor, Enrique Solano, Juanjo Garc\'ia-Ripoll, Kees Harmans, Hans Mooij, and Misha Lukin for many discussions. 

Over the last years I have found many colleagues in different parts of the world. First of all I want to thank Christian Flindt and Alessandro Braggio. Their impact on my work has been tremendous, and they were always willing to discuss and give me a hand. Thank you for having referred with pleasure this manuscript, and for your nice words in the evaluation. Here I would like to thank my thesis committee: Markus B\"uttiker, Yuli Nazarov, Tobias Brandes, Mar\'ia Jos\'e Calder\'on, Juan Carlos Cuevas, Juanjo Garc\'ia-Ripoll and Fernando Sols. 
Also many thanks to Peter Zoller, Aash Clerk, Frank Wilhelm, Markus B\"uttiker, Sankar Das Sarma, and Alexandre Blais for having kindly invited me to visit their groups. 

Many other colleagues have left an important footprint on my PhD. One example is Leo Di Carlo. Thank you for sharing with me your happiness and knowledge, fundamental ingredients to make an experiment work. Also many thanks to Edward Laird and the rest of the group in Charlie's lab. In Yacoby's lab, I wish all the best to Sandra Foletti and Gilad Barak. Thank you to my colleagues during my time in Berlin, Ying-Tsan Tang, one of the funniest people I've known, Philipp Zedler, Jin-Jun Liang, Hannes H\"ubener, Neil Lambert, and Anna Grodecka-Grad. I am grateful to  Michael Kastoryano for the good times spent in Copenhagen. Also, to Jonatan Bohr Brask and Dirk Witthaut. In Delft I left many friends: Katja Nowack, Gary Steele, Lyosha Feofanov, Stijn Goosens, Moira Hocevar, Lan Liu, Toeno van der Sar, Wolfgang Pfaff, Ronald Hanson, Lieven Vandersypen, and a long etcetera. In particular, I wish all the best in his new life (with money) to my office-mate and friend Wei Tang.

In ICMM and UAM I have spent one third of my life, and there I have probably learned more than 90\% of the physics I know. Wherever I am, that campus will come with me. The wonder years of my life were my first years as an undergraduate. We were a group of friends to which today I am grateful, because thanks to them I am now writing these lines. These are Paula P\'erez Rubio, Jos\'e Rodr\'iguez Lozano, Ana Mar\'ia Rodr\'iguez S\'anchez, Eva Gallardo, and Javier Sabio. I am specially grateful to Eva, for all the beautiful moments that we spent together. I am sure you will find the best job in the world because you truly deserve it. Also, I am particularly grateful to Javi, with whom I've shared moments of enthusiasm for physics and life that I will never forget. You have been one of the most influential people in my life. I really hope that our adventures together don't stop here. 

In the last years of undergraduate and first years of graduate school, life on campus brought many other friends. Miriam del Valle would always help me in difficult times, Bel\'en Lasanta is the kindest person I've known, Daniel Garc\'ia Figueroa an example of physicist to follow, and Elena del Valle someone I admire for her determination to make the physics community a more human environment. I am also grateful to Juanjo L\'opez Villarejo, Diego Alonso, C\'esar Seo\'anez, and many other friends that came to my life during that time. Inside ICMM I specially thank Carlos L\'opez-Mon\'is. Thank you for combining moments of joy and deep thinking that gave rise to the most pleasant conversations. Fernando de Juan, with whom I enjoyed very much every discussion. Gladys Le\'on, whose presence greatly improved the atmosphere in the scientific community. Lunch was so much better with you guys. Also, special thanks to Mar\'ia \'Angeles Vozmediano for her (quite successful) attempts to improve the Spanish scientific and political system. Many thanks to the theory department, including in particular Eduardo Castro, Elsa Prada, Pablo San-Jos\'e, Mathias Lunde, Adolfo Grushin, Maria Busl, Virginia Est\'evez, Sigmund Kohler, Leni Bascones, Bel\'en Valenzuela, Gloria Platero, Luis Brey, and Carlos Tejedor (in UAM).

My closest friends, although many of them not physicists, have played a fundamental role in my thesis. First, I would like to thank Alejandro Garc\'ia Alfonso. I am aware that I will find very few friends in life like you. Probably none. Thanks to Sara S\'aiz Bailador. You are the friend that everyone dreams to have by his side. Thanks for being by mine. Jos\'e Carlos Martinez Lozoya, your loyalty as a friend should be taken to a museum. Thanks to Encarnaci\'on Aguado for all those extremely funny situations and your patience with me. Thanks to Axel Alonso, for being there in the important moments. I know I can always count on you. Thanks to Justin Draft. I have learned a lot from you. Thanks for those beautiful trips and for teaching me how to get things done quickly! 
Thanks to Salvo Bagiante. With you as a neighbour, life was close to perfect in Delft. Your help and the adventures we shared are something that I will never forget. Thanks to my friends in Arcos de la Sierra. Mar\'ia and Patricia Clares, Sara Recuenco, Nuria Illana, Sergio Arribas, Jos\'e Garc\'ia, Paula Sanz, etc.
Friends from my institute, such as David Jim\'enez, were also very important in the context of this thesis.
Thanks to my friends around the world: Malika Mourot, Amanda Eyer, Yolanda Garc\'ia, Ryan Draft, Anita N\o rby, and many others.

Finally I would like to thank my family. This thesis probably started when my aunt brought me some physics books about home-made experiments, or when my parents included an extra physics class in my schedule. Probably, it started when I used to study at my grandmother's house, or probably it didn't start until I knew the love of my life. I don't know when my thesis began, but it certainly exists thanks to my family. All of them have contributed in one way or another. Thank you to my grandmothers, Julia and Inmaculada. Because you devoted your life to your family. 
Thank you to Marta. Because you have made my dreams come true.
Thank you to my parents, Valent\'in and Araceli. Because you valued my education above all things. Thank you for all the hard work dedicated to me. Now I dedicate this thesis to you.

Mark Twain said ``being politically correct means always having to say you're sorry''. Not for the same reason, I here want to send an apology to the people whose names should have been in these pages, but aren't. They contributed to this work in many ways. Thank you. I haven't forgotten you.

\textit{Scientiam do menti Cordi Virtutem.}

}
\clearpage  

\newpage{\
\thispagestyle{empty}}


\addtotoc{Abstract}  
\abstract{
\addtocontents{toc}{\vspace{0em}}  

\begin{flushright}

\textit{``A new idea is first condemned as ridiculous and then dismissed\\ as trivial, until finally, it becomes what everybody knows."}

William James

\end{flushright}

This thesis is devoted to the study of quantum mechanical effects that arise in systems of reduced dimensionality. Specifically, we investigate coherence and correlation effects in quantum transport models. In the first part, we present a theory of Markovian and non-Markovian current correlations in nanoscopic conductors. The theory is applied to obtain the spectrum of quantum noise and high-order current correlations at finite frequencies in quantum-dot systems. One of the main conclusions is that only the non-Markovian approach contains the physics of vacuum fluctuations. In the second part, we study the coupling of superconducting qubits to optical atomic systems and to cavity resonators. We propose a hybrid quantum system consisting of a flux qubit coupled to NV centers in diamond. We also demonstrate the existence of the so-called Bloch-Siegert shift in the ultra-strong coupling regime between a flux qubit and a $LC$ resonator. Throughout the thesis, we make special emphasis on the study of decoherence effects produced by the distinct dissipative baths to which the various types of qubits presented in this thesis are inevitably coupled.

}

\newpage{\
\thispagestyle{empty}}

\clearpage  


\resumen{





El objeto de esta tesis es el estudio de efectos mecano-cu\'anticos emergentes en sistemas de dimensionalidad reducida. En particular, investigamos efectos de coherencia y correlaci\'on cu\'antica en conductores el\'ectricos nanosc\'opicos. En la primera parte, presentamos una teor\'ia de correlaciones de corriente en sistemas Markovianos y no Markovianos. El formalismo es utilizado para derivar el espectro de ruido cu\'antico y funciones de correlaci\'on de corriente a frecuencia finita en puntos cu\'anticos. Una de las conclusiones principales de la primera parte es que s\'olo la aproximaci\'on no Markoviana contiene la f\'isica de fluctuaciones de vac\'io. En la segunda parte, estudiamos el acoplo de qubits superconductores a sistemas \'opticos at\'omicos y a cavidades resonantes. Proponemos un sistema h\'ibrido formado por un qubit de flujo acoplado a centros NV en diamante. Tambi\'en, demostramos la existencia de la variaci\'on energ\'etica ``Bloch-Siegert'' en el r\'egimen de acoplo ultra-fuerte entre un qubit de flujo y un resonador $LC$. A trav\'es de la tesis, prestamos especial atenci\'on al estudio de la decoherencia producida por los distintos ba\~nos disipativos a los que los qubits estudiados se encuentran inevitablemente acoplados. 

}

\newpage{\
\thispagestyle{empty}}

\clearpage  


\setstretch{1.3}  

\pagestyle{fancy}  

\lhead{\emph{Contents}}  
\tableofcontents  

\newpage{\
\thispagestyle{empty}}

\clearpage

\lhead{\emph{List of Figures}}  
\listoffigures  


\setstretch{1.5}  
\clearpage  
\lhead{\emph{Abbreviations}}  
\listofsymbols{ll}  
{
\textbf{Qubit} & \textbf{Q}uantum \textbf{B}it \\
\textbf{QD} & \textbf{Q}uantum \textbf{D}ot \\
\textbf{2DEG} & Two-\textbf{D}imensional \textbf{E}lectron \textbf{G}as \\
\textbf{LED} & \textbf{L}ight-\textbf{E}mitting \textbf{D}iode \\
\textbf{SET} & \textbf{S}ingle \textbf{E}lectron \textbf{T}ransistor \\
\textbf{SEM} & \textbf{S}canning \textbf{E}lectron \textbf{M}icrograph \\
\textbf{CB} & \textbf{C}oulomb \textbf{B}lockade \\
\textbf{CNT} & \textbf{C}arbon \textbf{N}ano\textbf{T}ube \\
\textbf{CNTQD} & \textbf{C}arbon-\textbf{N}ano\textbf{T}ube \textbf{Q}uantum \textbf{D}ot \\
\textbf{SC} & \textbf{S}uper\textbf{C}onducting \\
\textbf{EOM} & \textbf{E}quation \textbf{O}f \textbf{M}otion \\
\textbf{JJ} & \textbf{J}osephson \textbf{J}unction \\
\textbf{CPB} & \textbf{C}ooper \textbf{P}air \textbf{B}ox \\
\textbf{FQ} & \textbf{F}lux \textbf{Q}ubit \\
\textbf{PQ} & \textbf{P}hase \textbf{Q}ubit \\
\textbf{QPC} & \textbf{Q}uantum \textbf{P}oint \textbf{C}ontact\\
\textbf{FDT} & \textbf{F}luctuation \textbf{D}issipation \textbf{T}heorem\\
\textbf{NEFDT} & \textbf{N}on-\textbf{E}quilibrium \textbf{F}luctuation \textbf{D}issipation \textbf{T}heorem\\
\textbf{H}.\textbf{c}. & \textbf{H}ermitian \textbf{C}onjugate\\
\textbf{2LS} & Two-\textbf{L}evel \textbf{S}ystem\\
\textbf{JC} & \textbf{J}aynes-\textbf{C}ummings \\
\textbf{RWA} & \textbf{R}otating \textbf{W}ave \textbf{A}pproximation \\
\textbf{GF} & \textbf{G}reen's \textbf{F}unction \\
\textbf{FCS} & \textbf{F}ull \textbf{C}ounting \textbf{S}tatistics \\
\textbf{QME} & \textbf{Q}uantum \textbf{M}aster \textbf{E}quation \\
\textbf{ME} & \textbf{M}aster \textbf{E}quation \\
\textbf{CGF} & \textbf{C}umulant \textbf{G}enerating \textbf{F}unction \\
\textbf{MGF} & \textbf{M}oment \textbf{G}enerating \textbf{F}unction \\
\textbf{SRL} & \textbf{S}ingle \textbf{R}esonant \textbf{L}evel \\
\textbf{DQD} & \textbf{D}ouble \textbf{Q}uantum \textbf{D}ot \\
\textbf{CNTDQD} & \textbf{C}arbon-\textbf{N}ano\textbf{T}ube \textbf{D}ouble \textbf{Q}uantum \textbf{D}ot \\
\textbf{DM} & \textbf{D}ensity \textbf{M}atrix \\
\textbf{QNR} & \textbf{Q}uantum \textbf{N}oise \textbf{R}egime \\
\textbf{MA} & \textbf{M}arkovian \textbf{A}pproximation \\
\textbf{DO} & \textbf{D}ensity \textbf{O}perator \\
\textbf{NM} & \textbf{N}on-\textbf{M}arkovian \\
\textbf{NV} & \textbf{N}itrogen-\textbf{V}acancy \\
\textbf{QED} & \textbf{Q}uantum \textbf{E}lectro\textbf{D}ynamics \\
\textbf{CRT} & \textbf{C}ounter-\textbf{R}otating \textbf{T}erms \\ 
\textbf{USC} & \textbf{U}ltra-\textbf{S}trong \textbf{C}oupling 
}

\clearpage  
\lhead{\emph{Physical Constants}}  
\listofconstants{lrcl}  
{
Boltzmann constant & $k_B/k$ & $=$ & $1.380\ 6504\times10^{-23}\ \mbox{JK}^{-1}$ \\
Planck constant & $h$ & $=$ & $6.626\ 068\ 96\times10^{-34}\ \mbox{Js}$ \\
			 & $\hbar$ & $=$ & $1.054\ 571\ 63\times10^{-34}\ \mbox{Js}$ \\
Conductance quantum & $G_0$ & $=$ & $7.748\ 091\ 70\times10^{-5}\ \mbox{S}$ \\
Elementary charge & $e$ & $=$ & $1.602\ 176\ 49\times10^{-19}\ \mbox{C}$ \\
Electron g-factor & $g_e$ & $=$ & $2.002\ 319\ 304\ 3622 $ \\
Bohr magneton & $\mu_B$ & $=$ & $13.996\ 246\ 04\times10^{9}\ \mbox{HzT}^{-1}$ \\
Magnetic flux quantum & $\Phi_0$ & $=$ & $2.067\ 833\ 667\times10^{-15}\ \mbox{Wb}$ \\
Euler number & $\mathrm{e}$ & $=$ & $2.718\ 281\ 828\ ...$ \\
Speed of light & $c$ & $=$ & $299\ 792\ 458\ \mbox{ms}^{-1}$ \\
Magnetic permeability & $\mu_0$ & $=$ & $4\pi \times 10^{-7}\ \mbox{NA}^{-2}$ \\
Bohr radius & $a_0$ & $=$ & $0.529\ 177\ 208\ 59\times10^{-10}\ \mbox{m}$ \\
Electron g-factor & $g_e$ & $=$ & $2.002\ 319\ 304\ 36$ \\
Electron gyromagnetic ratio & $\gamma_e$ & $=$ & $1.760\ 859\ 770\ \times10^{11}\ \mbox{s}^{-1}\mbox{T}^{-1}$
}

\clearpage  
\lhead{\emph{Symbols}}  
\listofnomenclature{lll}  
{
$a$ & distance & m \\
$L$ & length & m \\
$m$ & mass & kg \\
$T$ & temperature & K \\
$G$ & conductance & S \\
$C$ & capacitance & F \\
$L$ & inductance & H \\
$\Phi$ & flux & Wb \\
$Q$ & charge & C \\
$I$ & current & A \\
$V$ & voltage & V \\
$\omega$ & frequency & (2$\pi$)Hz
}

\setstretch{1.3}  

\pagestyle{empty}  
\dedicatory{A mis padres, Valent\'in y Araceli, \\que lo dieron todo para que esta tesis fuera posible}

\addtocontents{toc}{\vspace{1em}}  

\newpage{\
\thispagestyle{empty}}

\mainmatter	  
\pagestyle{fancy}  



\chapter{Introduction} 
\label{Chapter1}
\lhead{Chapter 1. \emph{Introduction}} 

\begin{flushright}

\textit{``Don't listen to what I say; listen to what I mean!"}

R. P. Feynman

\end{flushright}

\begin{small}
In this chapter we motivate our work through an overview of the different systems and concepts that will be of interest in the thesis. We first present various phenomena that emerge when the dimensionality of a physical system is reduced, and establish the main goal of the thesis. In the first section, important concepts such as quantum coherence, entanglement or the density operator are introduced. In the second section we overview some examples of low dimensional systems, namely quantum dots, carbon nanotubes, NV centers in diamond, and superconducting qubits. These four will be subject of study in the next chapters. Next, we turn to briefly comment on the field of quantum transport, with emphasis in the area of quantum noise in mesoscopic systems. In particular, we discuss the various types of noise and related relevant experiments. The systems and concepts considered are of great importance in condensed matter, quantum optics, and in the field of quantum computation. We therefore discuss the more general context in which they take place in the final section.
\end{small}

\newpage

\section{Towards a quantum world}

Matter is made of atoms. This nowadays `trivial fact' is probably the discovery that has most affected the history of science, together with the emergence of the scientific method itself. The determination among the scientists to understand these `elementary' particles gave birth to different disciplines and to a technological revolution. Among them is condensed matter physics. The understanding of materials at a mesoscopic scale has, on one hand, brought great advances in technology, such as transistors, integrated circuits, light-emitting diodes, solid-state lasers, magnetic resonance imaging, magnetic recording disks, and liquid crystal displays. On the other hand, it has shown that the interplay between hundreds or thousands of particles needs to be understood through collective models capturing the essential degrees of freedom, since the emergent properties that arise when such a large number of particles interact, cannot be predicted from a microscopic theory accounting for all fundamental interactions and constraints in the system \citep{Anderson, LaughlinPines}.

The science of the atomic-size objects, or nanoscience, investigates, among others, the quantum effects that appear when we reduce dimensionality. The typical dimensions at which quantum phenomena may be observed are determined by the de Broglie wavelength $\lambda$ associated to the object under study. If it is to be observed or confined with a precision comparable to $\lambda$, then the emergence of quantum effects are expected. In condensed matter systems, when we deal with  a large number of particles, these effects may become visible if the average inter-particle spacing is of the same magnitude as the single particle wavelength. This criterium can be translated to a threshold temperature if we assume the ensemble to be in thermal equilibrium; then the equipartition theorem holds and we find that quantum effects come into sight if
\beq \label{quantumcriterium}
k_B T \lesssim \frac{h^2}{3ma^2},
\eeq
where $a$ refers to the mean inter-particle separation, $m$ to the individual mass, $T$ to the temperature, and $k_B$ and $h$ are the Boltzmann and Planck constants respectively. Notice the coincidence of this energy scale with the one required by energy quantization of a quantum particle confined in a box of side $L$, namely $h^2/(8mL^2)$; or with that of a traveling wave subject to periodic boundary conditions: $h^2/(2mL^2)$. The condition (\ref{quantumcriterium}) can be thus taken (with the corresponding length scale $a$ in the problem) as a rule of thumb to approach the quantum regime in different low dimensional systems. 

The search for quantum effects represents in itself an active topic of research, even nearly a century after the theory of quantum mechanics has been established. It has lead to some of the most spectacular experiments in the history of physics \cite{doubleslit}, and different scenarios of these are still pursued today (see e.g. \cite{Sinha10}). Materials science offers nowadays the possibility to realize experiments showing this type of phenomena in nanoscopic devices embedded in electrical circuits. In this way, \textit{quantum tranport} has emerged as a very relevant field in physics. This area of research has permitted us to answer fundamental questions, and has brought important technological applications, such as single electron transistors or ultra-precise magnetometers. 

In this thesis we will study the quantum effects that appear in quantum-transport and quantum-optical systems. More specifically, `quantum effects' mainly refer to \textit{quantum coherence} and \textit{entanglement}. These two concepts are genuine from a world in which there is a particle-wave duality. For the former, we will adopt the definition given by A. Albrecht \cite{Albrecht94}:
``To the extent that one needs to know the initial probability amplitudes (rather than just the probabilities) in order to do the right calculation, I will say that the system exhibits quantum coherence''. As for entanglement, we will take the notion introduced by Schr\"odinger in 1935 \cite{Schrodinger35}: ``When two systems, of which we know the states by their respective representatives, enter into temporary physical interaction due to known forces between them, and when after a time of mutual influence the systems separate again, then they can no longer be described in the same way as before, viz. by endowing each of them with a representative of its own. I would not call that \textit{one} but rather \textit{the} characteristic trait of quantum mechanics, the one that enforces its entire departure from classical lines of thought. By the interaction the two representatives (or $\psi$-functions) have become entangled. To disentangle them we must gather further information by experiment, although we knew as much as anybody could possibly know about all that happened. Of either system, taken separately, all previous knowledge may be entirely lost, leaving us but one privilege: to restrict the experiments to one only of the two systems. After reestablishing one representative by observation, the other one can be inferred simultaneously''.

In the last decade, an outstanding progress has been made in the direction of observing, controlling, and even designing at will, the quantum world. State-of-the-art experiments are bringing the mentioned quantum effects to a macroscopic level (see e.g. \cite{Connell10}), and this, of course, raises the question of whether there is a limiting size to observe quantum superpositions of two or more objects.
Other than the race in scaling up the quantum world, there is the race in trying to make the quantum effects to last as long as possible, more specifically, the fight against  \textit{decoherence}. This, as defined by Zurek \cite{Zurek03}, is ``the destruction of quantum coherence between preferred states associated with the observables monitored by the environment". Even at the level of a single particle (e.g. single spin), decoherence is difficult to avoid, because a sufficient degree of decoupling from the environment is in conflict with a good degree of control (needed to measure the particle quantum state), and these two are very hard to accomplish simultaneously. This, is the major limitation in the technological race towards a quantum computer; the building pieces of such, called qubits (or quantum bits), are two-level quantum systems subject to decoherence. 
At this point, the main questions that motivate our work have become apparent, and the major goal of this thesis can be summarized as:
\textit{To study the coupling between different quantum systems and with their environments.}

\subsection{More about quantum coherence} \label{moreCoherence}

In this subsection we define more precisely the notion of quantum coherence. To this end we introduce the density operator. Let a system be such that it can be in one of the possible states $|\psi_1\rangle,\ldots,|\psi_N\rangle$ with respective probabilities (or `statistical weights') $W_1,\ldots,W_N$. Then, the density operator for the system is defined as
\beq
\hat{\rho} := \sum_{\alpha=1}^N W_{\alpha} |\psi_{\alpha}\rangle\langle\psi_{\alpha}|.
\eeq
If there exists a basis in which the density operator is simply $\hat{\rho}=|\psi\rangle\langle\psi|$ for a particular state $|\psi\rangle$, then the system is said to be in a \textit{pure state}. Otherwise we say that it is in a \textit{mixed state}. In a concrete basis $\{|n\rangle\}$ the density operator reads
\beq
\hat{\rho} = \sum_{nm} \rho_{nm} |n\rangle\langle m|,
\eeq
with $\rho_{nm}\equiv \sum_{\alpha=1}^N W_{\alpha} c_n^{(\alpha)}c_m^{*(\alpha)}$, and where $c_n^{(\alpha)}$ is defined through the decomposition $|\psi_{\alpha}\rangle = \sum_n c_n^{(\alpha)} |n\rangle$. The diagonal elements $\rho_{nn}$, or \textit{populations}, correspond to the probability of occupying the state $\ket{n}$. The off-diagonal elements, $\rho_{nm}$ with $n\neq m$, describe the effect of quantum interference between the states $\ket{m}$ and $\ket{n}$. They are the so-called \textit{coherences}. The existence of these is thus a purely quantum mechanical effect, that appears as a consequence of the wave character of the particles and the superposition principle. Therefore, any phenomenon that reveals this kind of interference between particles is said to exhibit quantum coherence.
This concept must not to be confused with that of optical coherence of a field. This, introduced by Glauber \cite{Glauber63}, refers to the degree of correlation of the field at different space-time points. If the radiation field is in a so-called \textit{coherent state}, this correlation function can be factorized to all orders (perfect correlation), and the probability of finding $n$ photons in then given by a Poissonian distribution. This will be interesting for us when we study correlations in the context of quantum transport. In the same spirit that in quantum optics, if we encounter a Poissonian current distribution, we will be able to say that the electrons flow `coherently' through the conductor.

The density operator is a fundamental tool to study quantum coherence. To illustrate this point let us imagine that by repetitive measurements, we resolve that a quantum system can be in one of two possible states, $\ket{0}$ and $\ket{1}$, of a particular observable $\hat{\cal O}$, with equal probability. Yet, it is not straightforward that the system is in a coherent superposition of both. As mentioned, it can happen that the possible states are occupied randomly according to a certain probability distribution, and that the system is in what is called a \textit{statistical mixture} of states. In our example, this case would have as density operator $\hat{\rho}=\frac{1}{2}(\ket{0}\bra{0}+\ket{1}\bra{1})$. However, if there is some degree of coherence between $\ket{0}$ and $\ket{1}$, the density operator would then be $\hat{\rho}=\frac{1}{2}(\ket{0}\bra{0}+\ket{1}\bra{1} + x \ket{0}\bra{1} + x \ket{1}\bra{0})$, with $x=1$ for a pure state. In an experiment, to distinguish between these two possible scenarios, one would need to measure the same variable $\hat{\cal O}$ along a different degree of freedom. For example, if $\hat{\cal O}$ is the spin of a single electron, this would mean measuring along a different projection axis. Under this operation, the density operator describing the statistical mixture remains invariant, but the one corresponding to a system presenting quantum coherence does not. This, is the spirit of \textit{quantum state tomography}, by which the density matrix can be fully determined. The density operator is therefore a very powerful tool to describe quantum systems, particularly when they consist on statistical ensembles. Some general properties that the density operator fulfills are:
\begin{itemize}
\item $\hat{\rho}$ is Hermitian.
\item $\mathrm{Tr}\{\hat{\rho}\} = 1$.
\item $\rho_{nn}\geqslant 0$.
\item $\ew{\hat{\cal O}(t)}= \mathrm{Tr}\{\hat{\rho}(t) \hat{\cal O}\}$.
\item $ \mathrm{Tr}\{\hat{\rho}^2\}
 \left\{ \begin{array}{lcl} =1 & \Leftrightarrow & \mbox{pure state} \\ <1 & \Leftrightarrow & \mbox{mixed state} \end{array} \right.$
\item If a system is described by the Hamiltonian $\hat{{\cal H}}$, then the evolution of the system density operator is given by the von Neumann's equation: $\dot{\hat{\rho}}(t)=-\frac{i}{\hbar}[\hat{{\cal H}},\hat{\rho}(t)]$.
\end{itemize}

The phenomenon of quantum coherence has been shown in a huge range of experiments, from electronics to optics \cite{QuantumCoherence08}. For example, in the field of quantum transport, observations like the Aharonov-Bohm effect \cite{AharonovBohm59}, where the wave character of the electron becomes apparent, have been carried out in a variety of systems, such as metallic rings \cite{Webb85}, carbon nanotubes \cite{Bachtold99}, and recently in topological insulators \cite{Peng10}. Here we will show how through transport measurements of a classical variable, namely the current, one can infer coherence and other interesting properties of different quantum systems. In particular, it will be shown how correlation functions of this variable display much more information about the microscopic behaviour than the current itself. The area of research behind this idea is known as \textit{full counting statistics}.

\section{Examples of low-dimensional systems}

In this thesis we will deal with different solid-state systems which will help us to learn about the quantum behaviour and the interactions that govern the nanoscale. From a perspective of applications, we will be interested in systems with satisfactory properties in the context of quantum computation. In this section we give an overview of the systems studied in this thesis. The typical scales involved are summarized in table \ref{tabScales}. Generally speaking, the nanostructures we consider here are man-made atomic systems which can be accessed electronically or/and optically with a high degree of control. This permits us not only to observe their quantum properties, but also to modify them.

\begin{table}
\setcounter{table}{0}
\begin{center}
\begin{tabular}{| c || c | c |}
	\hline
\textbf{Variable}  & \textbf{Range} & \textbf{Typical} \\ \hline \hline
Temperature   & 10 mK - 300 K & 50 mK \\ \hline
Frequency   & 10 Hz - 10 GHz & 1 GHz \\ \hline
Voltage   & 10 $\mu$V - 1 V & 1 mV \\ \hline
Current   & 1 pA - 1 $\mu$A & 1 nA \\ \hline
Conductance   & (1 G$\Omega$)$^{-1}$ - (1 k$\Omega$)$^{-1}$ & (12.9 k$\Omega$)$^{-1}$ \\ \hline
Capacitance & 1 aF - 1 pF  & 1 fF \\ \hline
Inductance & 100 pH - 10 nH  & 1 nH \\ \hline
Magnetic field & 1 $\mu$T - 1 T  & 1 mT \\ \hline
Length & 1 nm - 10 $\mu$m & 100 nm \\ \hline
Time & 10 ns - 100 s  & 1 $\mu$s \\ \hline
\end{tabular}
\caption[Typical scales encountered throughout the text]{Typical scales encountered throughout the text.}
\label{tabScales}
\end{center}
\end{table}

\subsection{Quantum dots}

\begin{figure}
  \begin{center}
    \includegraphics[width=0.8\textwidth]{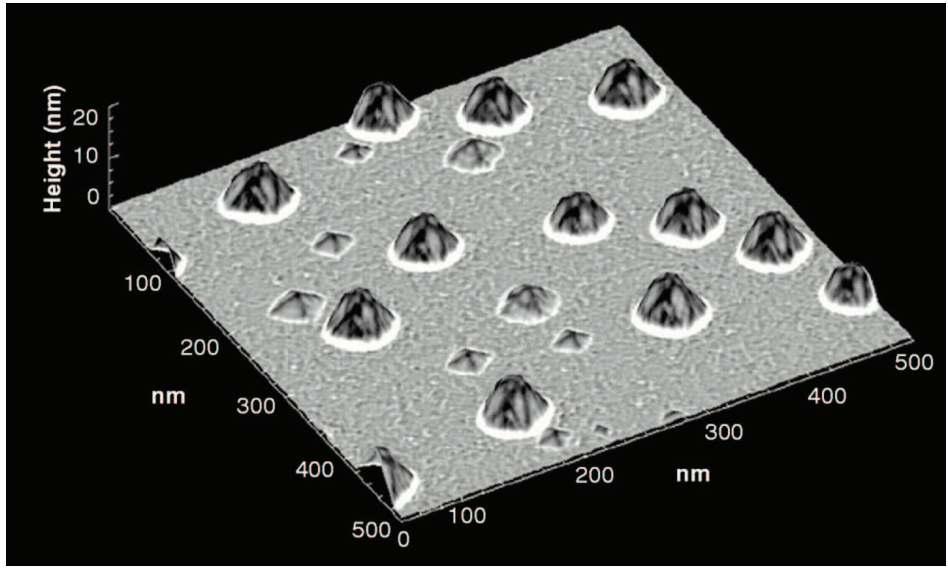}
  \end{center}  
  \caption[Self-assembled quantum dots]{
Germanium pyramids and domes formed on a silicon substrate. Small size nanocrystrals have a discrete energy spectrum (self-assembled quantum dots). Image taken from \cite{Medeiros-Ribeiro98}.
}
  \label{SAQDsFig}
\end{figure}

Although the term has been distorted, quantum dots (QDs) can be defined as semiconductor nanostructures in which a small number (1$\sim$1000) of conduction band electrons, valence band holes or excitons are confined in all three spatial dimensions. This confinement is usually performed either spontaneously, due to the high bandgap in the surrounding material, when a few monolayers of a semiconductor are deposited on a substrate with a lattice mismatch of about 5\% (\textit{self-assembled QDs}), or by lithographically-patterned gate electrodes, which etch typically two-dimensional electron gases (2DEGs)\footnote{These 2DEGs are usually formed in the interface between GaAs and AlGaAs.} by electrostatically-induced potential barriers (\textit{lithographically-defined QDs}). The most important remark about QDs is that they resemble artificial atoms, with a discrete energy spectrum, strong Coulomb interactions, and whose level filling follows Hund rules \cite{Kastner93,Ashoori96,Tarucha96,Kouw98,Imamoglu01,Kouw01}. As mentioned, QDs are zero-dimensional structures that, although can be fabricated in a number of different ways; however, for the most part, can be categorized into two families:
\begin{itemize}
\item \textbf{Self-assembled QDs:} These are formed due to the mismatch when two materials are highly strained \cite{Leonard93}. In Fig.~\ref{SAQDsFig} we show an image of Germanium nanocrystals form on a silicon surface. As these become smaller in size, their energy spectrum becomes quantized (quantum dots), the typical sizes of these being $\sim$ 20 nm of side (and $\sim$ 5 nm for QDs formed from colloidal semiconductor nanocristals). Typically, their charging energy is on the order of a few eV. This means that they emit in the visible or infrared optical regime. They are therefore manipulated with optical photons, usually at temperatures around 4 K, and up to room temperature.
\item \textbf{Lithographically-defined QDs:} These are usually defined by metallic contacts that confine high-mobility electrons or holes by means of applied voltages, which act as electrostatic potentials \cite{Meirav90}. The system (quantum well), with a typical size of $\sim$ 200 nm of side, contains a small number of particles that can be controlled by a gate voltage, and is in contact with electron reservoirs (leads). If these are at different chemical potentials, (a source-drain voltage is applied) a flow of particles through the system becomes possible. The charging energy of the dots is generally in the meV scale, and therefore work in the microwave regime.
\end{itemize}

QDs present a great variety of applications, derived from the possibility of integrating tunable atoms in solid-state systems. For example, they can be used as light-emitting diodes (LEDs), single electron transistors (SETs), charge sensors, amplifiers, and lasers.
Here we will focus on the fundamental properties of lithographically-defined QDs, and on the possibility of using them as qubits. We will show different theoretical techniques to study quantum transport through these systems. In particular, it is our goal to study their quantum mechanical behaviour when different energy scales (temperature, voltage and frequency) are of comparable magnitude. 

\begin{figure}
  \begin{center}
    \includegraphics[width=0.8\textwidth]{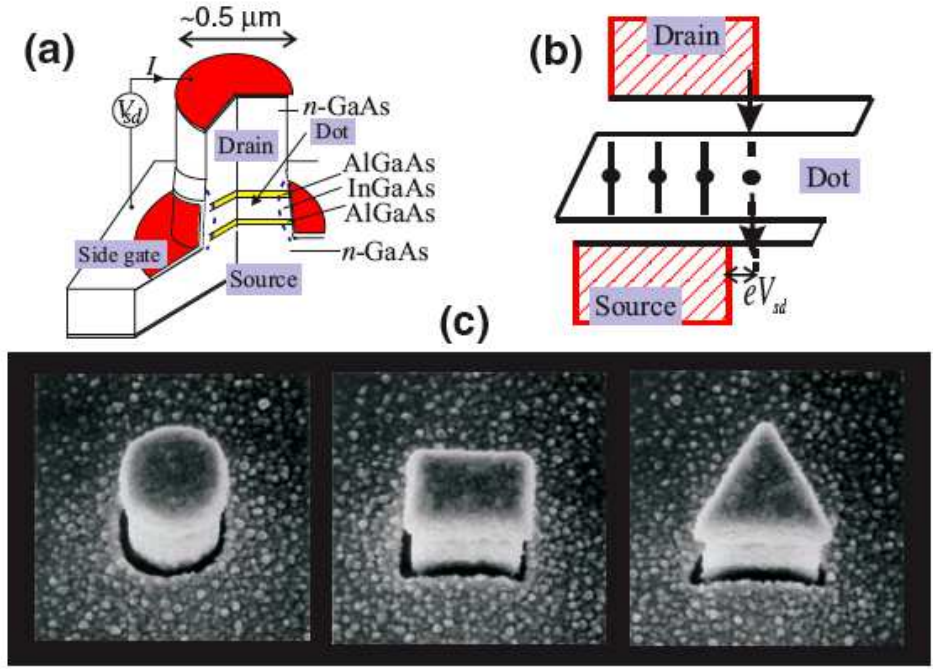}
  \end{center}
  \caption[Vertical quantum dots]{a) Schematic diagram of a vertical quantum dot formed in a layer of InGaAs embedded into two n-doped regions of GaAs. Two layers of AlGaAs act as potential barriers that isolate contacts and dot. b) Schematic energy diagram of the heterostructure. The application of a source-drain voltage permits electrons to tunnel through the quantum dot. However, the current is blocked if the bias (source-drain voltage) window lies in-between two states of the dot, and the device acts as a SET due to the so-called Coulomb Blockade effect. c) Scanning electron micrographs (SEMs) of vertical quantum dots. The shapes vary and the widths are about 500 nm. Image taken from \cite{Kouw01}.}
  \label{verticalQDsFig}
\end{figure}

\begin{figure}
  \begin{center}
    \includegraphics[width=0.6\textwidth]{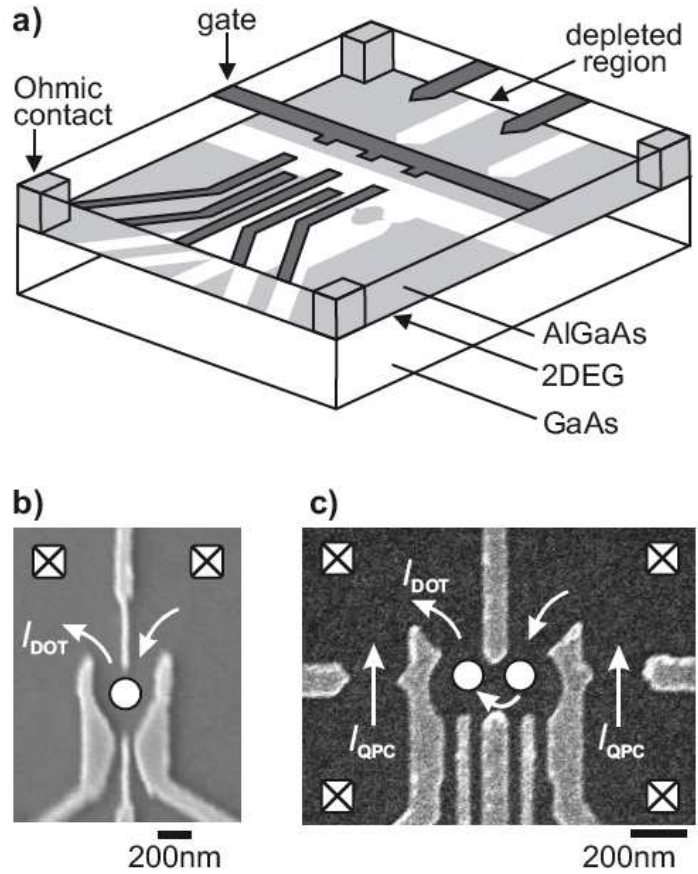}
  \end{center}
  \caption[Lithographically-defined lateral quantum dots]{Lithographically-defined lateral quantum dots. a) Schematic diagram. A 2DEG (light gray) is confined in the interface of GaAs and AlGaAs. Metalic contacts (dark gray) placed on top allow us to confine electrons in regions of $\sim$ 200 nm with a peaked density of states and discrete energy spectrum (quantum dots). Applying different voltages to the contacts enables a configuration of transport through the system or of Coulomb Blockade. b) SEM of a single quantum dot. Electrons can tunnel in our out the QD from two Fermi leads. The electrodes are shown in light gray, the surface in dark gray and the position of the dot is marked with a white spot. c) SEM of a double quantum dot, consisting on two dots coupled internally with each other and in contact with Fermi leads. This system acts as the paradigmatic quantum two-level system coupled to a fermionic bath. Notice the presence of constrictions at both sides defined by metallic contacts. These are quantum point contacts, which are used as electrometers to measure the charge state of the system. Image taken from \cite{Hanson07}.}
  \label{lateralQDsFig}
\end{figure}

\subsubsection{Coulomb Blockade} \label{CBsection}

\begin{figure}
  \begin{center}
    \includegraphics[width=0.8\textwidth]{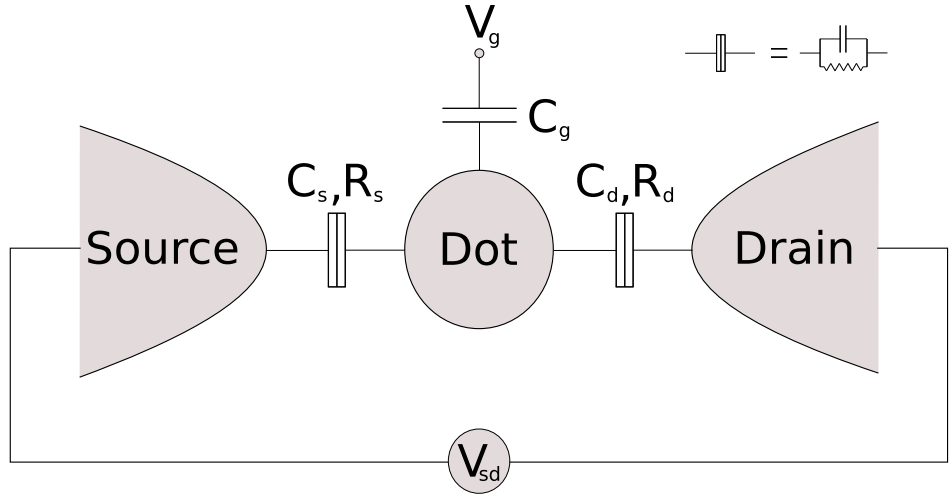}
  \end{center}
  \caption[Single electron transistor]{Schematic diagram of a SET. An island with discrete energy levels (quantum dot) is connected to Fermi leads to which a source-drain voltage $V_{sd}$ is applied. The QD is also connected to a gate electrode, whose potential $V_g$ controls the flow of electrons according to the CB effect. The strong Coulomb interaction between electrons in the QD can be treated with a classical model, named \textit{constant interaction model} \cite{Beenakker91}, in which this is captured through the total capacitance $C=C_s+C_d+C_g$ of the dot.}
  \label{CoulombCircuitFig}
\end{figure}

\begin{figure}
  \begin{center}
    \includegraphics[width=0.5\textwidth]{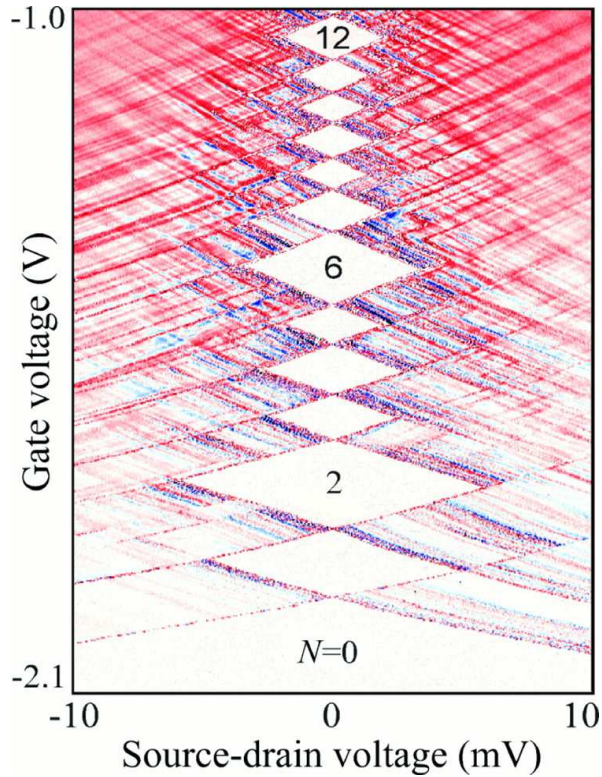}
  \end{center}
  \caption[Coulomb Diamonds]{Differential conductance $\partial I/\partial V_{sd}$ through a quantum dot plotted in color scale in the $V_g - V_{sd}$ plane. In the white diamond-shaped regions the current is suppressed and the number of electrons in the dot is well defined due to CB. Notice the different size of some diamonds, which is due to different charging energies as dictated by the shell structure. The lines outside the diamonds correspond to excited states of the dot. Sweeping the gate voltage for $V_{sd}=0$ gives a series of current oscillations called \textit{Coulomb peaks}. Image taken from \cite{Kouw97}.}
  \label{CoulombDiamondsFig}
\end{figure}

The phenomenon of Coulomb Blockade (CB) is based on the charge quantization and Coulomb interaction between electrons, and it is the basic principle that permits quantum dots to have a well defined number of electrons. CB was already suggested sixty years ago as a possible explanation for the increase of the resistance of metallic films observed at low temperatures \cite{Lambeir50}. But it was not until the development of the first SET \cite{Fulton87}, entirely made out of metals, and its semiconducting counterpart \cite{Scott89}, that a large number of experiments and theoretical studies followed on the topic. Coulomb blockade is the trapping of electrons in a nanostructure (and consequently a suppression of the current) due to Coulomb interactions. The general rule of thumb to observe CB is that the conductance through a nanostructure, as well as the temperature, are sufficiently low. More specifically:
\begin{equation}
\begin{array}{c}
G \ll G_0, \\
k_B T \ll E_C.
\end{array}
\end{equation}
Here $G$ is the conductance, $G_0:=\frac{2e^2}{h}$ (with $e$ the electron charge) is the quantum of conductance, and $E_C:= \frac{e^2}{2C}$ is the charging energy (where $C$ denotes the total capacitance of the dot). A simple but complete model describing CB was developed by Beenakker \cite{Beenakker91}. In particular, it explains the so-called \textit{Coulomb diamonds} (see Fig.~\ref{CoulombDiamondsFig}). To understand these let us consider the circuit depicted in Fig.~\ref{CoulombCircuitFig}. Due to the finite number of electrons (say N) in the dot, there can be an energy imbalance between dot and gates, which can be described by the electrostatic potential $\phi(Q)=Q/C+\phi_{\mathrm{ext}}$, where $Q=Ne$ is the charge of the dot and $C$ the total capacitance to its suroundings. Now, the mentioned energy will be given by $U(N)=\int_0^{Ne} \phi(q)dq$, that is
\beq \label{UchargeQD}
U(N) = \frac{\left(Ne - Q_{\mathrm{ext}}\right)^2}{2C} - \frac{Q_{\mathrm{ext}}^2}{2C}.
\eeq
Here $Q_{\mathrm{ext}}\equiv C\phi_{\mathrm{ext}}$ is the externally induced charge. Now, the condition to have current through the device is simply that the energy associated with the transfer of electrons is positive. If $E_F$ is the Fermi energy of the leads, $E_N$ the sum of the single-particle energies of the $N$ electrons in the dot, and defining $\mu_N:=U(N)-U(N-1)$, we find that this condition reads
\begin{equation}
\begin{array}{c}
\left.
\begin{array}{c}
\mu_{N+1}>eV_{sd}/2 + E_F - E_N \\
\mu_N<-eV_{sd}/2 + E_F - E_N
\end{array}
\right\} \mbox{If} \; V_{sd} > 0. \\
\\
\left.
\begin{array}{c}
\mu_{N+1}>-eV_{sd}/2 + E_F - E_N \\
\mu_N<eV_{sd}/2 + E_F - E_N
\end{array}
\right\} \mbox{If} \; V_{sd} < 0.
\end{array}
\end{equation}
This is what determines the the form of the Coulomb diamonds shown in Fig.~\ref{CoulombDiamondsFig}.

\subsection{Carbon nanotubes}

Carbon is a very versatile element that presents itself in a variety of forms in nature. These have attracted enormous attention as their mechanical, optical, and electronic properties allow us to explore fundamental questions, as well as to develop great advances in technology \cite{Avouris07}. Here we will focus on one of this forms, the carbon nanotubes (CNTs), in the context of electron transport and spintronics. Since they were synthesized by Sumio Iijima in 1991 \cite{Iijima91}, CNTs have been used for example to achieve novel integrated circuits (see e.g. \cite{Chen06}) or optical devices (see e.g. \cite{Gabor09}), and currently they are drawing a lot of attention also because of their mechanical properties \cite{Sazonova04, Steele09, Lassagne09}. 

\begin{figure}
  \begin{center}
    \includegraphics[width=0.8\textwidth]{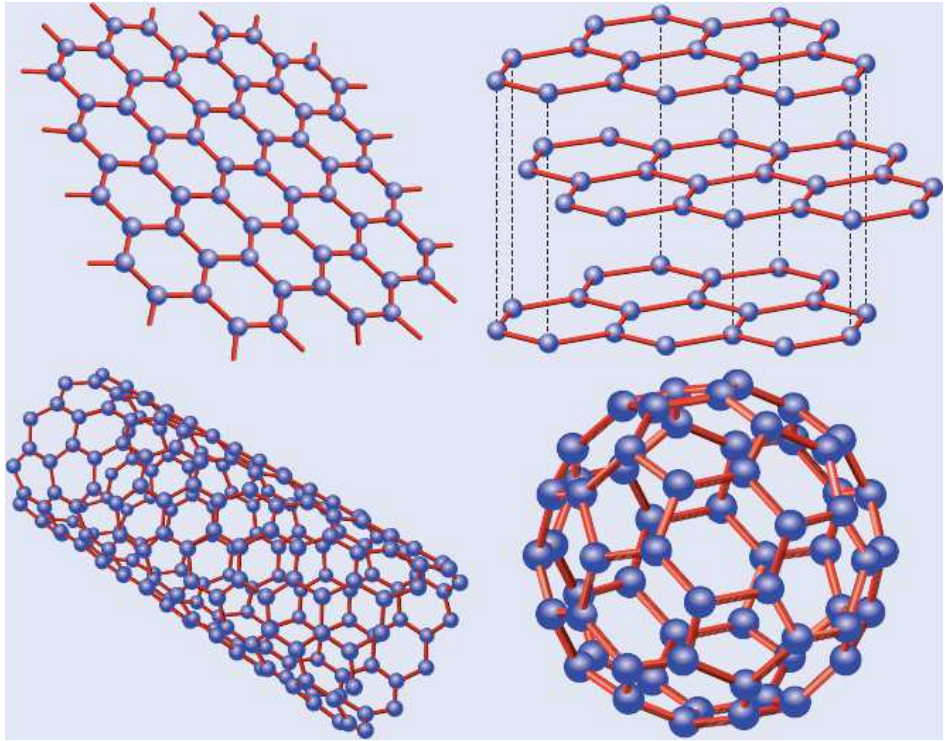}
  \end{center}
  \caption[Some allotropic forms of carbon]{Some allotropic forms of carbon. On the top left we find graphene, the two-dimensional allotrope of carbon, which consists of a hexagonal lattice of carbon atoms. On the top right we see the graphite lattice, composed of a stack of graphene layers. Carbon nanotubes (bottom left) are rolled-up graphene cilinders and the buckyballs (bottom right) are spheres of carbon atoms. Taken from \cite{CastroNeto06}.}
  \label{CarbonAllotropesFig}
\end{figure}

\begin{figure}
  \begin{center}
    \includegraphics[width=\textwidth]{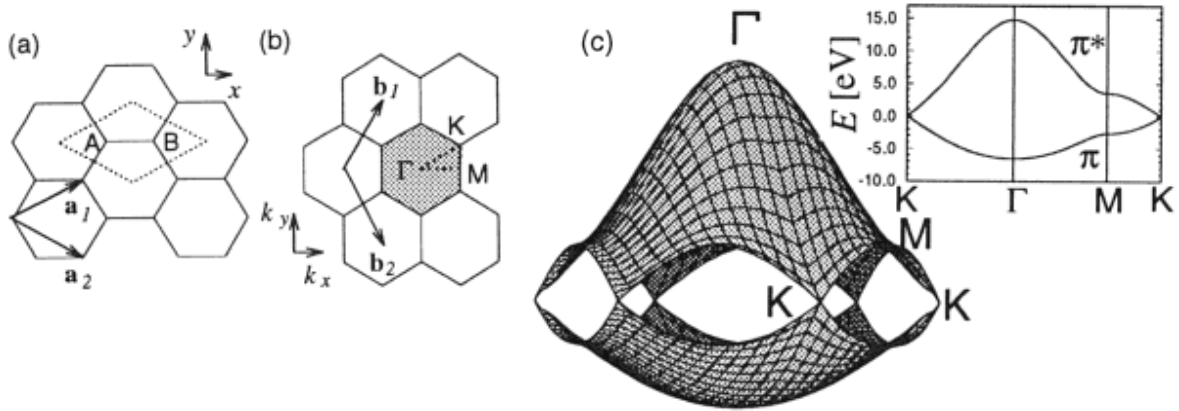}
  \end{center}
  \caption[Exact graphene band structure]{a) Graphene lattice. The first Brillouin zone (dotted rhombus) and the unit vectors (${\bf a}_1$, ${\bf a}_2$) are shown. b) Reciprocal lattice. The Brillouin zone is here the shaded hexagon and the unit vectors ${\bf b}_1$ and ${\bf b}_2$. c) Energy bands within the Brillouin zone for the $\pi$ electrons of graphene as given by equation (\ref{exactGrapheneEnergy}). Notice the electron-hole asymmetry that exists for the exact solution to the tight-binding model. The points $K$, $\Gamma$ and $M$ correspond to those of figure (b). Taken from \cite{Saito98}.}
  \label{GrapheneExactSpectrumFig}
\end{figure}

\begin{figure}
  \begin{center}
    \includegraphics[width=0.8\textwidth]{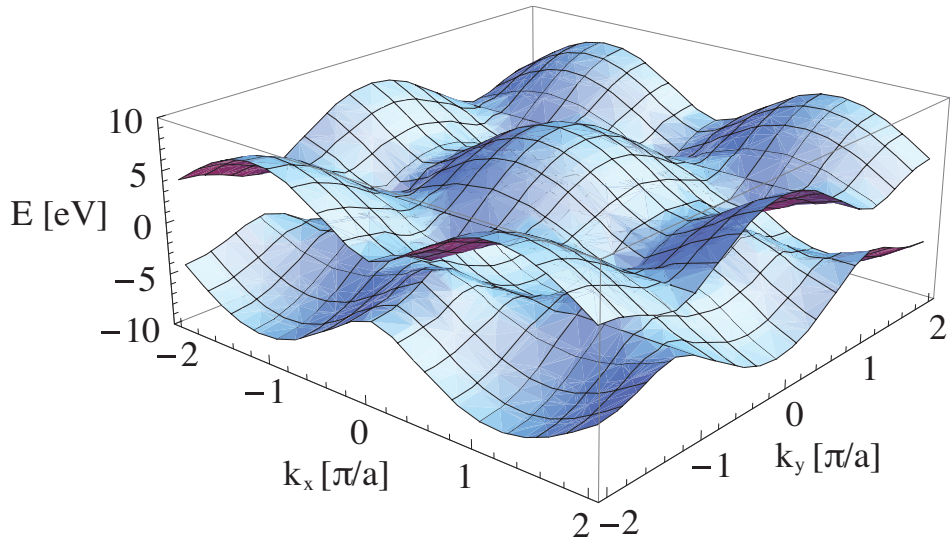}
  \end{center}
  \caption[Approximated graphene band structure]{Band structure of the graphene $\pi$ electrons as given by the approximated equation (\ref{approxGrapheneEnergy}) for $t=2.7$ eV. The value of the lattice constant is estimated to be $a=1.44\;\mathrm{\AA}\times\sqrt{3}=2.49\;\mathrm{\AA}$. Notice the electron-hole symmetry. The $\pi$ bands for a CNT are obtained from vertical cuts to this plot (given by the momentum quantization due to the boundary conditions). In graphene, close to the K points, and if the energy excitations are low, the electrons move according to a Dirac equation, with Fermi velocity $v_F\sim 10^6\mathrm{ms}^{-1}$, giving rise to many interesting phenomena such as Klein tunneling \cite{Katsnelson06}.}
  \label{GrapheneSpectrumFig}
\end{figure}

Carbon nanotubes are cylindrical graphene sheets (see Fig.~\ref{CarbonAllotropesFig}), with diameters in the range $\sim 0.7 - 10$ mm and lengths from hundreds of nm to a few cm. The electronic structure of CNTs can be thus understood from that of graphene with the proper boundary conditions. In this material, each carbon atom has four electrons in the second shell, three of which reside in the hybridized $s$, $p_x$ and $p_y$ atomic orbitals into three molecular orbitals $\sigma$. These are responsible for the bonding that forms the two-dimensional structure of graphene. The remaining electron from each carbon resides in the orbital $\pi$, and contributes to the transport properties here discussed. The motion of electrons therefore occurs in the honeycomb lattice with two atoms per unit cell depicted in Fig.~\ref{GrapheneExactSpectrumFig}a, which we can model with a tight-binding-to-first-neighbours Hamiltonian \cite{Wallace47, Saito98}, obtaining the energy spectrum
\beq\label{exactGrapheneEnergy}
E(\vec{k}) = \frac{\epsilon_{2p}\pm t w(\vec{k})}{1\pm s w(\vec{k})}.
\eeq
Here $\epsilon_{2p}$ is energy of the $2p$ level, $t$ the hopping amplitude between two neighbour atoms, $s$ the overlap between two neighbour wave functions, $\vec{k}$ the momentum, and $w(\vec{k})$ is given by
\beq\label{approxGrapheneEnergy}
w(\vec{k}) = \sqrt{1 + 4 \mathrm{cos}\left( \frac{\sqrt{3}k_x a}{2} \right) \mathrm{cos}\left( \frac{k_y a}{2} \right) + 4 \mathrm{cos}^2\left( \frac{k_y a}{2} \right)},
\eeq
where $a$ is the lattice constant. Expression (\ref{exactGrapheneEnergy}) is shown in Fig.~\ref{GrapheneExactSpectrumFig}c, where we see that the conduction and valence bands are not symmetric (there is not electron-hole symmetry). However, the parameters $t$ and $s$ have been estimated for graphene to be $t\approx 2.7$ eV and $s\approx 0.1$. In this case, and taking $\epsilon_{2p}=0$ (as for $s\ll1$ corresponds to an energy shift), to a good approximation the spectrum is given by
\beq\label{approxGrapheneEnergy}
E(\vec{k}) \approx \pm t w(\vec{k}).
\eeq
We then find an electron-hole symmetry for electrons in the $\pi$ band (c.f. Fig.~\ref{GrapheneSpectrumFig}), fact that was observed in 2004 for carbon-nanotube quantum dots \cite{Jarillo-Herrero04}.

As mentioned, the electronic properties of CNTs can be understood to a good degree from those of graphene if we take into account the effects produced by curvature. For example, the way the graphene sheet is rolled-up to form the nanotube determines the boundary conditions, and these whether the CNT is metallic (zero band-gap) or semiconducting (finite band-gap). Depending on how this folding is performed, a nanotube can be classified as \textit{armchair} (always metallic), \textit{zigzag} (sometimes metallic, sometimes semiconducting), or \textit{chiral} (always semiconducting).

In this work we are interested in carbon-nanotube quantum dots (CNTQDs), which can be formed by applying gate voltages on the nanotubes. Measurements of these QDs in the CB regime were first taken only six years after their discovery \cite{Bockrath97, Tans97}, and were seen to have similar electronic structure and transport properties to the lithographically-defined lateral QDs described in previous section. However, a theoretical model to correctly capture the spectrum of CNTQDs requires to include exchange effects and the orbital degeneracy \cite{Oreg00}.
In chapter~\ref{Chapter2} we will present how the electron spin of a single electron in a CNTQD can be manipulated through pulse-sequence techniques, and show how through transport measurements we can learn about the interactions affecting the coherent properties of this qubit.

\subsection{NV centers in diamond}

\begin{figure}
  \begin{center}
    \includegraphics[width=0.8\textwidth]{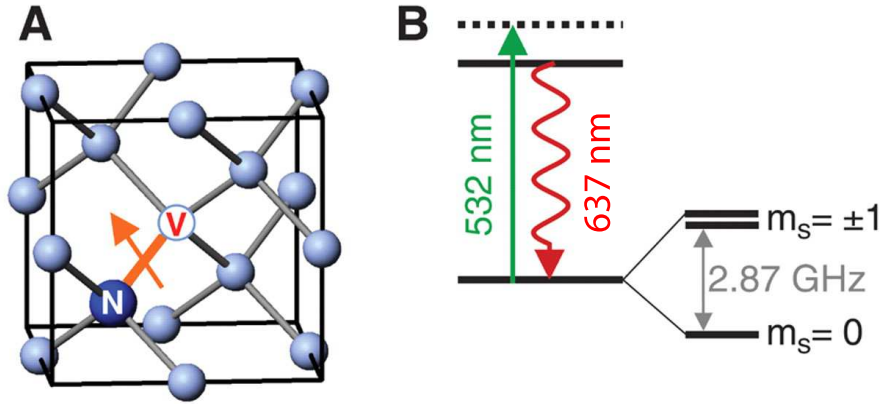}
  \end{center}
  \caption[NV center in diamond]{a) Structure of diamond and the NV-center. A missing carbon atom is adjacent to a substitutional nitrogen impurity in the lattice. b) Schematic spectrum of the NV center. The system can be excited to the phonon sideband with green light, and relaxes from the zero-phonon line to the ground state ($637$ nm line in the spectrum). The ground state is a spin triplet whose $m_S=0$ and $m_S=\pm 1$ states are split by $\sim 2.87$ GHz, and under the application of an external field permit us to isolate a two level system. Taken from \cite{Hanson08}.}
  \label{NVcenterFig}
\end{figure}

In the previous section we have discussed some of the very interesting properties of various allotropic forms of carbon. Another attractive allotrope is the $sp^3$-hybridized form: the diamond. Here we are interested in particular in the coherence properties of the electrons residing in a defect of this material, the so-called \textit{NV center} in diamond. This defect consists of a substitutional nitrogen atom adjacent to a vacancy in the carbon lattice. The structure of diamond (two interpenetrating face-centered cubic lattices) and that of the NV center can be seen in Fig.~\ref{NVcenterFig}a. The NV center behaves as a solid-state atomic system as it has strong optical transitions and an electron spin degree of freedom. Of special interest about the system is the fact that the ground state is magnetic, specifically a triplet whose $m_S=0$ and $m_S=\pm 1$ states are split by $\sim2.87$ GHz in the absence of an external field. This opens the possibility of optical manipulation by spin resonance techniques, which together with the long coherence times of the spin, to be living in a solid-state environment, make the NV center an attractive candidate as a qubit.

A basic review on spin resonance in defect centers in diamond can be found in \cite{Smith59, Loubser78}, and more specifically on NV centers in diamond in \cite{Jelezko06}. Since the realization of Rabi oscillations in the system \cite{Jelezko04} together with the possibility of reading out the state optically with a confocal microscope \cite{Gruber97}, there has been an enormous interest in the system in the context of quantum information \cite{Wrachtrup06, Greentree06, Awschalom07, Stoneham09, Wrachtrup10}. Key in this interest is the fact that even at room temperature the good coherence properties of the system are preserved. Experiments from cryogenic temperatures to room temperature have been performed, showing coherent dynamics \cite{Gaebel06, Childress06, Hanson08, Balasubramanian09, Fuchs10}, quantum registering \cite{Dutt07, Neumann10b}, single photon emission \cite{Babinec10}, nanoscale imaging magnetometry \cite{Maze08, Balasubramanian08}, entanglement \cite{Jelezko04b, Neumann08}, single-shot readout \cite{Jelezko02, Neumann10}, and dynamical decoupling techniques \cite{Lange10}.

The precise level structure of the NV center is quite complicated and still under some debate. However, the basic structure can be derived using group theory considerations \cite{Lenef96}. The NV center exists in two charge states, $\mathrm{NV}^0$ and $\mathrm{NV}^{-}$, of which we will be considering only the second species. The $6$ free electrons associated with this NV center have a spectrum with a zero-phonon line at $637$ nm (see Fig.~\ref{NVcenterFig}) and a ground state which is a spin triplet. Due to spin-spin interactions, the $m_S=0$ and $m_S=\pm 1$ states of this triplet are split by $\sim 2.87$ GHz. Now, the application of an external magnetic field allows to break the $m_S=1$ and $m_S=-1$ degeneracy, forming a $V$-system. In particular, we can identify the $m_S=0$ and $m_S=1$ (alternatively $m_S=0$ and $m_S=-1$) states as a two-level system and use it as a qubit, whose energy splitting is given by $\sim 2.87 \; {\rm GHz} \pm g_e\mu_B B^\mathrm{ext}$, where $g_e$ is the electron $g$-factor, $\mu_B$ the Bohr magneton and $B^\mathrm{ext}$ the applied external field.

In our study we will use this system to couple it magnetically to a superconducting flux qubit. As we will see, the hybrid system presents diverse advantages which arise from the combination of an electrical system with an optical system. In particular, we will show how it can be used as a magnetic memory, as a quantum bus to transmit a quantum state over long distances, and more generally, as a way to integrate quantum communication with quantum computation.

\subsection{Superconducting qubits}

The phenomenon of superconductivity is an example of quantum effect that can be observed at a macroscopic level. In this case, all the quasiparticles (Cooper pairs) forming the superconductor can be described by a single wave function. It is therefore natural to ask ourselves whether it is possible to use superconductors to observe a quantum superposition of these wave functions and therefore of macroscopic objects. By `macroscopic' we mean a collection from hundreds to millions of particles, having in mind examples such as supercurrents, superfluids or Bose-Einstein condensates. This question was studied in detail by Leggett \cite{Leggett87, Leggett02}, who proposed using superconducting rings to observe the superposition of currents flowing clockwise and anti-clockwise. Such dispositive was achieved in the late nineties \cite{Mooij99, Mooij00} and it is now known as flux qubit. In the following we review the various types of qubits based on superconducting (SC) circuits. These are used to prove quantum mechanical effects, as we will see in chapter~\ref{Chapter6}, and are good candidates for quantum processors: state-of-the-art experiments have already shown two-qubit algorithms \cite{diCarlo09} (Grover and Deutsch-Jozsa \cite{NielsenChuang}), violation of Bell inequalities \cite{Ansmann09, Palacios-Laloy10} and entanglement of up to three qubits \cite{diCarlo10, Neeley10}.

\begin{figure}
  \begin{center}
    \includegraphics[width=\textwidth]{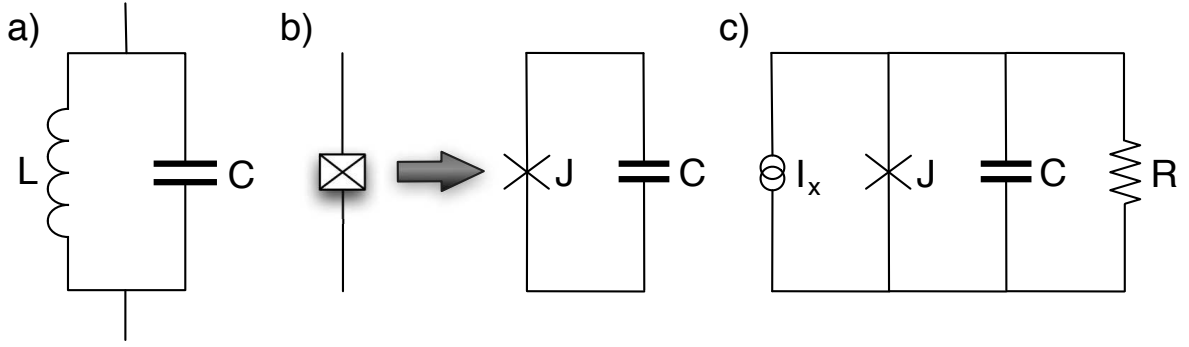}
  \end{center}
  \caption[LC resonator, Josephson junction, and RCSJ model]{a) LC circuit. At low temperatures this acts as a quantum harmonic oscillator, with resonant frequency $\omega=1/\sqrt{LC}$. b) Josephson junction. This can be modeled as an irreducible Josephson element (which follows the Josephson relations) in parallel with a capacitor. c) More generally, the so-called RCSJ model describes dissipation in a circuit with a Josephson junction.}
  \label{LCJJfig}
\end{figure}

An integrated circuit is composed of different elements such as inductors, capacitors, resistors, etc. In circuit theory \cite{Yurke84} it is useful to adopt a general definition for the variables flux and charge between two nodes:
\begin{equation} \label{fluxVchargeI}
\begin{array}{c}
\Phi(t) := \int_{-\infty}^{t} V(t') dt', \\
Q(t) := \int_{-\infty}^{t} I(t') dt',
\end{array}
\end{equation}
where $V(t)$ and $I(t)$ are the voltage difference and current between the two nodes. At temperatures smaller than the electrostatic energies characterizing the elements of the circuit, these variables become quantized. The simplest example of a quantum circuit consists of a \textit{LC resonator} (see Fig.~\ref{LCJJfig}a), which is the analog to the harmonic oscillator in quantum mechanics. This is easily seen starting from the associated Lagrangian, ${\cal L} = \frac{1}{2}C \dot{\Phi}^2 - \frac{1}{2L} \Phi^2$, where $L$ is the inductance and $C$ the capacitance. The Hamiltonian of the circuit is then ${\cal H}= \dot{\Phi}\frac{\partial {\cal L}}{\partial\dot{\Phi}} - {\cal L} = \frac{1}{2}C \dot{\Phi}^2 + \frac{1}{2L} \Phi^2$, where we notice that the flux $\Phi$ and the charge $Q=C\dot{\Phi}$ correspond to position and momentum in mechanics, and $C$ to the mass. After canonical quantization, $\Phi$ and $Q$ become operators fulfilling $[\hat{\Phi},\hat{Q}]=i\hbar$. The Hamiltonian can be then readily diagonalized through the usual unitary transformation $\hat{\Phi} = \sqrt{\frac{L\hbar\omega}{2}} (\hat{a}^{\dagger} + \hat{a})$, $\hat{Q} = i \sqrt{\frac{C\hbar\omega}{2}} (\hat{a}^{\dagger} - \hat{a})$, where $\omega\equiv 1/\sqrt{LC}$ is the resonator frequency, obtaining the single-mode harmonic oscillator Hamiltonian, ${\cal \hat{H}} = \hbar\omega (\hat{a}^{\dagger}\hat{a} + 1/2)$. We also notice that the Euler-Lagrange equation gives $C\ddot{\Phi} + \frac{1}{L}\Phi = 0$, which is the equation of motion (EOM) given by Kirchhoff's law. Therefore we see that given a circuit, we can start from the Lagrangian or conversely proceed as: Kirchhoff's laws $\to$ Lagrangian $\to$ Hamiltonian $\to$ Quantization.

The basic element of a superconducting qubit is the \textit{Josephson junction} (JJ). This is a simple way to introduce a non-linearity in the circuit. A JJ consists of two supercondutors separated by a thin insulating layer. As depicted in Fig.~\ref{LCJJfig}b, it can be modeled by a capacitance in parallel with an irreducible Josephson element, this last being characterized by the Josephson relations, $I=I_c \mathrm{sin}(2\pi\Phi/\Phi_0)$ and $\dot{\Phi}=V$, where $I$ and $V$ are the current and voltage drop through the junction, $I_c$ is the critical current of the junction and $\Phi_0:=h/2e$ the magnetic flux quantum. The energy stored in a JJ is given by $-E_J \mathrm{cos}(2\pi\Phi/\Phi_0)$ \cite{Orlando91}, where $E_J\equiv \frac{\Phi_0 I_c}{2\pi}$ is known as Josephson energy. Therefore, in this case the Lagrangian reads ${\cal L} = \frac{1}{2} C \dot{\Phi}^2 + E_J \mathrm{cos} (2\pi\Phi/\Phi_0)$, the Hamiltonian ${\cal H} = \frac{1}{2} C \dot{\Phi}^2 - E_J \mathrm{cos} (2\pi\Phi/\Phi_0)$, and the EOM describing the circuit is $C\ddot{\Phi} - I_c \mathrm{sin} (2\pi\Phi/\Phi_0)=0$. More generally, dissipation through the JJ can be taken into account, having the so-called resistively and capacitively shunted junction (RCSJ) model. This consists of an irreducible Josephson element in parallel with a capacitor and a resistor (see Fig.~\ref{LCJJfig}c), which, driven with a current $I_x$, is described by the EOM:
\beq
I_x = C\ddot{\Phi} + \frac{1}{R} \dot{\Phi} + I_c \mathrm{sin}(2\pi\Phi/\Phi_0).
\eeq
This equation is analogous to the EOM of a forced and damped pendulum in the gravitational field, that is, a particle moving with friction under the influence of the potential $U(\Phi)=-E_J \mathrm{cos(2\pi\Phi/\Phi_0)}-I_x\Phi$. Notice that here we have chosen to express the EOM in terms of the flux $\Phi$, but alternatively we could have chosen the phase $\varphi$, as the relation between these two is simply $\varphi=2\pi\Phi/\Phi_0$.
Once the Hamiltonian of a circuit is expressed in terms of the variables flux and charge (or alternatively phase and charge), the quantization is straightforward, as it is enough to consider these variables as operators fulfilling canonical commutation relations. This will become clear in the examples of the three basic types of SC qubits, which we briefly discuss below. For a more extensive description of superconducting qubits, we refer the reader to the reviews \cite{Devoret95, Makhlin01, Devoret04, Devoret07, Schoelkopf08, Clarke08, Girvin09}.

\begin{figure}
  \begin{center}
    \includegraphics[width=\textwidth]{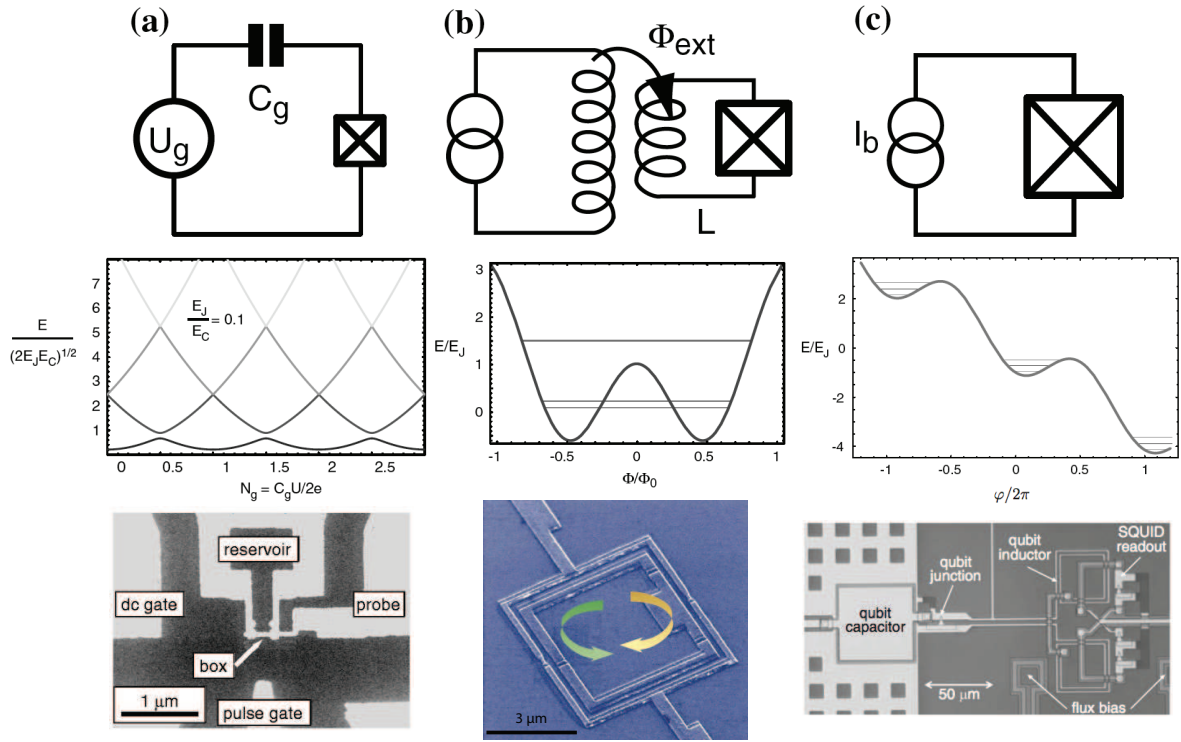}
  \end{center}
  \caption[Basic types of SC qubits]{Three basic types of SC qubits. a) Cooper pair box. The two-level system split by a few GHz determines a qubit formed by the ground and first excited charge states at the point $N_g\approx 1/2$. b) Flux qubit. The qubit eigenstates correspond to symmetric and antisymmetric superpositions of clockwise and anti-clockwise circulating currents around the loop. c) Phase qubit. The potential corresponding to the phase variable forms quantum wells for values of $\varphi$ multiples of $2\pi$. Here an anharmonic set of discrete levels is formed. Images taken from \cite{Devoret04, Nakamura99, Mooij99, Martinis09}.}
  \label{CPBFQPQfig}
\end{figure}

\subsubsection{Cooper pair box}

The Cooper pair box (CPB), see Fig.~\ref{CPBFQPQfig}a, first introduced by B\"uttiker \cite{Buttiker87}, consists of a Josephson junction in series with a capacitor ($C_g$) and a voltage source ($U_g$). The Hamiltonian describing the circuit is
\beq
{\cal \hat{H}} = E_C\left(\hat{N} - \frac{Q_{\mathrm{ext}}}{2e}\right)^2 - E_J \mathrm{cos}(\hat{\varphi}),
\eeq
where the operators $\hat{N}$ and $\hat{\varphi}$ describe the number of Cooper pairs in the superconducting island and the phase difference between superconductors, respectively, and fulfill $[\hat{\varphi},\hat{N}]=i$. The quantity $E_C\equiv \frac{(2e)^2}{2(C_g+C_J)}$ is the charging energy of the CPB, and $Q_{\mathrm{ext}}$ denotes the external charge with which the box is biased. Notice the similarity of the first term of this Hamiltonian with the charging energy of a quantum dot (c.f. equation \ref{UchargeQD}), which means that the CPB also behaves as an artificial atom. In particular, in a situation in which $E_C\gg E_J$, the JJ is in the charging regime, and a two-level system can be isolated between the ground and first excited states around $N_g\equiv \frac{C_gU_g}{2e}=\frac{1}{2}+k$, where $k$ is an integer number. The two levels form an anticrossing at these points whose splitting is given by $E_J$. This can be easily seen writing the Hamiltonian in the basis of charge states, defined by $\hat{N}\ket{N} = N\ket{N}$. In this basis the Hamiltonian takes the form $\hat{\cal H} =E_C\sum_N \left( N - N_g \right)^2 \ket{N}\bra{N} - \frac{E_J}{2} \sum_N \left( \ket{N}\bra{N+1} + \ket{N+1}\bra{N} \right)$, whose spectrum is plotted in Fig.~\ref{CPBFQPQfig}a. If $k_B T\ll E_C$, and around $N_g=1/2$, the CPB can be described by the usual two-level system Hamiltonian
\beq \label{H2LS}
\hat{\cal H}_{2LS} = \frac{\varepsilon}{2} \left( \ket{1}\bra{1} - \ket{0}\bra{0} \right) + \frac{\lambda}{2} \left( \ket{0}\bra{1} + \ket{1}\bra{0} \right),
\eeq
with $\varepsilon \equiv E_C (1-2N_g)$ and $\lambda\equiv -E_J$, and where $\ket{0}$ and $\ket{1}$ describe the ground and first excited states.
The first CPB was realized by the Saclay group \cite{Bouchiat98}, and coherent Rabi oscillations in the system were first achieved by the NTT group \cite{Nakamura99}. Since then, a number of groups have used the CPB for experiments on quantum coherence, such as strong coupling with a microwave cavity \cite{Wallraff04}.

\subsubsection{Flux qubit}

Flux qubits (FQs) were proposed by Leggett in the eighties as a test of macroscopic quantum coherence, and a decade ago groundbreaking experiments showing superpositions of current states were performed \cite{Mooij99, Mooij00, Friendman00, Mooij03}. FQs consist on a superconducting ring including one or more Josephson junctions. The Hamiltonian describing the system with one JJ can be written as \cite{Orlando99}
\beq \label{HamiltonianFQ}
{\cal \hat{H}} = \frac{\hat{Q}^2}{2C_J} + \frac{(\hat{\Phi}-\Phi_{\mathrm{ext}})^2}{2L} - E_J \mathrm{cos}(2\pi\hat{\Phi}/\Phi_0).
\eeq
Here $C_J$ is the capacitance of the junction, $L$ the self-inductance of the loop and $\Phi_{\mathrm{ext}}$ the external flux with which the FQ is biased. In contrast to the CPB, the system is in the phase regime ($E_J\gg E_C$), and thus the flux is a good quantum number\footnote{More accurately, it is the number of fluxoids what is quantized in a flux qubit.}. If $\beta:=2\pi L I_c/\Phi_0>1$ and $\Phi_{\mathrm{ext}}\approx\Phi_0/2$, the potential $U(\Phi) =  \frac{(\Phi-\Phi_{\mathrm{ext}})^2}{2L} - E_J \mathrm{cos}(2\pi\Phi/\Phi_0)$ has the form shown in Fig.~\ref{CPBFQPQfig}b. In this regime, the spectrum is highly non-linear: for typical FQ parameters, the first two levels are split by $\sim 3 \; \mathrm{GHz}$ near the degeneracy point ($\Phi_{\mathrm{ext}}=\Phi_0/2$), while the next level is $\sim 50 \; \mathrm{GHz}$ above the first excited state. Therefore, the first two levels form an effective two-level system described by the Hamiltonian (\ref{H2LS}), with $\varepsilon \equiv 2I_p (\Phi_{\mathrm{ext}} - \Phi_0/2)$ and $\lambda$ determined experimentally, and where $I_p=I_c\sqrt{6(\beta-1)}$ is the persistent current flowing through the loop. The eigenstates of this Hamiltonian correspond to a coherent superposition of clockwise and anti-clockwise -flowing currents.
The state of the qubit is typically measured by a DC-SQUID \cite{Orlando91}, and the FQ has usually three or four JJs. This does not change the essential properties described above. For more details we refer the reader to \cite{Orlando99, Makhlin01}.

\subsubsection{Phase qubit}

The phase qubit (PQ) consists of a JJ biased by a DC-current source (see Fig.~\ref{CPBFQPQfig}c). Similarly to the FQ, the system is in the phase regime ($E_J\gg E_C$). The dynamics of the system can be undestood as a particle moving in the washboard potential of a capacitively shunted junction (CSJ) model depicted in Fig.~\ref{CPBFQPQfig}c, and therefore the Hamiltonian of the PQ is
\beq
{\cal \hat{H}} = \frac{\hat{Q}^2}{2C_J} - E_J \mathrm{cos}(\hat{\varphi}) - E_b \hat{\varphi},
\eeq
where $E_b\equiv \frac{\Phi_0 I_b}{2\pi}$, and $I_b$ the current provided by the source. The wells formed by the potential at values of $\varphi$ multiples of $2\pi$ give rise to few quantized energy levels. The first two are well separated from the rest, and therefore we can again treat them as a two-level system described by the Hamiltonian (\ref{H2LS}).
The first observation of this energy quantization was performed by Martinis \textit{et al.} in 1985 \cite{Martinis85}. Recent experiments with PQs have demonstrated violation of Bell inequalities \cite{Ansmann09}, as well as multipartite entanglement \cite{Neeley10}.
In contrast to the CPB and the FQ, the system presents the advantage of built-in readout: because the tunneling rate outside the well increases exponentially as we go to an excited state,  the qubit state can be measured by sending a probe signal to induce a transition from the state $\ket{1}$ to a higher excited state. If the qubit is in $\ket{1}$, the particle representing the phase escapes from the well and we will therefore measure a finite voltage across the junction. If otherwise the qubit is in $\ket{0}$, it does not get excited by the external pulse and no signal is measured.

\subsection{Other systems}

Apart from the ones described above, other systems are commonly used in quantum transport and quantum computation, and will indirectly be related with our study. In some of these, the theory or conclusions we present here may be applied. In this subsection we list a number of them and the context in which they will be concerned. 

In quantum transport, a system of mayor importance is the \textit{quantum point contact} (QPC). This consists of a constriction between two conducting regions, being the QPC a ballistic conductor (the mean free path of the electrons is much larger than the length scales of the QPC). Although some previous studies had shown how to narrow two-dimensional electron gases (2DEGs) to one dimension \cite{Dean82, Thornton86, Berggren86}, the system was first used in two seminal papers \cite{Wees88, Wharam88} to show the conductance quantization in units of $2e^2/h$. An application of the QPC, and which will be used here, is for quantum detection \cite{Elzerman03, Averin05}. The current through the QPC is sensitive to a nearby charge due to Coulomb interaction between electrons. The presence of a charge produces a change in the current flowing through the constriction, and thereby a method of charge detection.

There are yet many other systems of interest in quantum transport and quantum information. For example, semiconducting \textit{nanowires} have shown the possibility of realizing transistors \cite{Lieber01}, diodes \cite{Lieber02}, and quantum effects such as supercurrent reversal by just adding a single electron to a quantum dot \cite{Dam06}.
Also, since its synthesis in 2004 by the recently Nobel awarded K. Novoselov and A. Geim, \textit{graphene} \cite{Novoselov04, Novoselov05b}, is a system that has attracted a huge attention, at both a fundamental and applied level, due to its very interesting electronic and mechanical properties \cite{Novoselov05, Zhang05, CastroNeto09}. \textit{Two-dimensional electron gases} (2DEGs) continue showing outstanding physical phenomena, such as the properties of the quasiparticles in the fractional quantum hall effect \cite{Radu08}. Quantum optics has also brought many other systems of interest. \textit{Cold ions} and \textit{atomic ensembles} are advanced candidates for quantum computation \cite{CiracZoller95, CiracZoller00} and quantum communication \cite{Duan01}; \textit{optical lattices} have proven interesting many-particle phenomena \cite{Bloch05, Bloch08}, and proposals such as single-photon transistors have arisen by using \textit{plasmonic waveguides} \cite{Chang07}. Superconducting qubits have also suffered a spectacular advance in the last years. Stripline resonators are used as microwave cavities and to read out the state of the qubit through the transmitted signal, and new types of qubits, which improve the ones described in the previous section, are now employed. For example, the original CPB has been substituted in the majority of experiments, first by the \textit{quantronium} \cite{Vion02}, and then by the \textit{transmon} \cite{Koch07} -- a system with $E_J\gg E_C$ insensitive to charge noise ($1/f$), and by the fluxonioum \cite{Manucharyan09} -- with $E_J/E_C \gtrsim 1$, and whose coherence times are comparable to those of the transmon.

\section{Quantum transport}

The outstanding technological progress in the industry of microelectronics that took place during the second half of the twentieth century, gave birth to new devices and nanoscopic heterostructues. These, instigated groundbreaking experiments revealing novel quantum effects, originated from the transport of electrons through nanostructures. For example, some of the effects that the field of quantum transport brought into light are \medskip\\
$\bullet$ The \textit{Aharonov-Bohm effect.} In quantum mechanics, the electromagnetic potentials have a physical significance. They may cause effects on charged particles even in spatial regions where the fields (and thus all the applied electromagnetic forces) are zero \cite{AharonovBohm59, Webb85}. \\
$\bullet$ The \textit{universal conductance fluctuations.} In systems whose length is comparable to the electron coherence length, reproducible conductance fluctuations as a function of an applied magnetic field are observed \cite{Altshuler85, Lee-Stone85, Umbach85}. These must not be confused with the Shubnikov-de Haas oscillations \cite{Schubnikow30}, that are produced when each Landau level crosses the Fermi level. \\
$\bullet$ The \textit{quantum Hall effect.} In two-dimensional electron systems at low temperatures and high magnetic fields, the resistance is quantized. The transversal conductivity shows a series of plateaus at integer \cite{Klitzing80, Laughlin81} or fractional \cite{Tsui82, Laughlin83} values of the conductance quantum. \\
$\bullet$ The phenomenon of \textit{weak localization.} In materials with a high density of defects, the resistivity is increased due to the quantum interference \cite{Anderson58}. In particular, a 2DEG becomes insulating at zero magnetic field as the temperature approaches absolute zero.

In quantum transport, there are various physical scales that, contrasted with the dimensions of the conductor, determine the nature of transport. The basic scales are the average distance that an electron travels before it scatters and therefore changes its momentum (\textit{mean free path}), and the average distance that an electron travels before it scatters inelastically and therefore randomizes its phase (\textit{coherence length}).
In terms of these length scales, transport can be characterized as \medskip\\
$\bullet$ \textit{Diffusive transport:} If the conductor dimensions are much larger than the mean free path and the coherence length of the electrons, so their initial momentum and phase get changed.\\
$\bullet$ \textit{Ballistic transport:} If the mean free path is larger than the dimensions of the conductor.\\
$\bullet$ \textit{Coherent trasport:} If the coherence length of the electrons is larger than the dimensions of the conductor.

Quite generally, the motion of electrons through a nanostructure can be described by the Schr\"odinger equation, which, assuming for generality a confining potential $U(\vec{r})$ (used e.g. to define a QD) and an applied magnetic field (characterized by the vector potential $\vec{A}$), reads
\beq
\left[ E_c + \frac{(i\hbar\vec{\nabla} + e \vec{A})^2}{2m^*} + U(\vec{r}) \right] \Psi(\vec{r}) = E \Psi(\vec{r}),
\eeq
where $E_c$ is the energy of the conduction electrons, $m^*$ their effective mass, 
$E$ the eigen-energies and $\Psi$ the eigenfunctions.
If the motion is confined to two dimensions ($\vec{r}=(x,y)$), such as it happens e.g. in 2DEGs, the spectrum reads
\beq
E=E_c+\varepsilon_m+\frac{\hbar^2}{2m^*} \left( k_x^2 + k_y^2 \right).
\eeq
Here $\varepsilon_m$ is the energy of sub-band $m$ due to the confinement in the $Z$ direction, and $k_x$, $k_y$, are the wave vectors characterizing the motion in the $XY$ plane. This is the spectrum of a 2DEG, where electrons typically occupy the first sub-band only, and thus $\varepsilon_m=\varepsilon_1$.
If a magnetic field $B$ is applied perpendicularly to the 2DEG (XY plane), one finds the popular \textit{Landau levels} \cite{Landau30}
\beq
E=E_c+\varepsilon_m + \left( n+\frac{1}{2} \right) \hbar\omega_C,
\eeq
where $n=0, 1, 2, \ldots$ and $\omega_C\equiv eB/m^*$ is the cyclotron frequency.
Further confinement, e.g. with a harmonic potential along the $Y$ direction ($U(y)=\frac{1}{2}m^*\omega_0^2y^2$), gives $E=E_c+\varepsilon_m+\frac{\hbar^2 k_x^2}{2m^*} \frac{\omega_0}{\tilde{\omega}} + \left( n +\frac{1}{2} \right)\hbar\tilde{\omega}$, being $\tilde{\omega}:=\omega_C+\omega_0$, and which reduces to $E=E_c+\varepsilon_m+\frac{\hbar^2k_x^2}{2m^*} + \left(n+\frac{1}{2} \right) \hbar\omega_0$ in the absence of an applied field.
Finally, when the confinement occurs in all three dimensions, and assuming a harmonic confinement along the $X$ and $Y$ directions, the energy levels form the so-called \textit{Fock-Darwin} spectrum \cite{Fock28, Darwin30}
\beq \label{FockDarwinSpectrum}
E= E_c+\varepsilon_m + \left( n_x + n_y + 1 \right) \hbar\Omega + \frac{1}{2} (n_x-n_y) \hbar\omega_C,
\eeq
where $\Omega^2:=\omega_0^2 + (\omega_C/2)^2$ and $n_x, n_y = 0, 1, 2, \ldots$
Eq.~(\ref{FockDarwinSpectrum}) corresponds in particular to the spectrum of a quantum dot confined in a 2DEG, which is a central object of study throughout the text.

\subsection{Current correlations: The noise is the signal}

Transport experiments exhibit current and voltage fluctuations. This `noise', is often originated from imperfections in the design of the circuit and regarded as undesirable. However, part of the noise is of more fundamental nature, and when that originated from a bad design is sufficiently suppressed, the remaining fluctuations can be used as a powerful tool to learn about electronic properties and different characteristics of the conductor. Let us start by giving a brief description of the fundamental types of noise that may be encountered in a mesoscopic conductor:\medskip\\
$\bullet$ \textbf{Thermal noise} (Johnson-Nyquist noise): At thermodynamic equilibrium and a finite temperature $T$, the distribution of the number of particles $n$ with a particular energy $\varepsilon$ (which is $f=[\mathrm{exp}\{-\frac{\varepsilon}{k_B T}\} + 1]^{-1}$ for fermions) presents a non-vanishing noise value (second cumulant) given by $\ew{\left(n-\ew{n}\right)^2} = f(1-f)$.\\
$\bullet$ \textbf{1/f noise} (flicker noise): Although still under debate, the nature of this noise is believed to be related in many cases with the random motion of charges trapped in the substrate of the material. This motion is slow compared to other time-scales in the system, what makes the fluctuations to be significant usually below 10kHz. The theory developed in the present work does not include the $1/f$ contribution.\\
$\bullet$ \textbf{Shot noise} (Schottky noise): This is due to the discrete character of the current, composed of electrons flowing along the conductor. Each of these particles has a wave character, and thus a probability $\mathsf{T}$ of being transmitted and $(1-\mathsf{T})$ of being reflected in the process of tunneling through the system. Shot noise may also originate from the random nature in which particles are released from the emitter. In any case, if $\ew{n_\mathsf{T}}=\mathsf{T}$ is the mean number of transmitted particles, we find $\ew{(n_\mathsf{T}-\ew{n_\mathsf{T}})^2}=\mathsf{T}(1-\mathsf{T})$. Thus, in the tunneling limit ($\mathsf{T}\ll1$) we have $\ew{(n_\mathsf{T}-\ew{n_\mathsf{T}})^2} \approx \mathsf{T} = \ew{n_\mathsf{T}}$, that is, the current noise is proportional to the current itself. This Poissonian behaviour was measured by Schottky in 1918 in the process of emission of electrons from a cathode in a vacuum tube \cite{Schottky18}. Notice that if electrons are emitted according to the Fermi distribution $f$, we have $\ew{n_\mathsf{T}}=\mathsf{T}f$ and thus $\ew{(n_\mathsf{T}-\ew{n_\mathsf{T}})^2} = \mathsf{T}f(1-\mathsf{T}f)$, which reduces to $\mathsf{T}(1-\mathsf{T})$ in the zero-temperature limit.\\
$\bullet$ \textbf{Quantum noise}: This arises from the quantum nature of the emission and absorption processes. First, the spectrum of radiation of the electromagnetic field follows Planck's law. Second, it incorporates vacuum fluctuations. Similarly to the zero-point motion of the quantum harmonic oscillator ($\ew{\hat{x}^2}_{\mathrm{vac}}\neq0, \ew{\hat{p}^2}_{\mathrm{vac}}\neq0$, being $\hat{x}$ and $\hat{p}$ position and momentum respectively and the averages taken in the vacuum state), voltage $V$ and current $I$ in an electric circuit present a zero-point variance as well, $\ew{\hat{V}^2}_{\mathrm{vac}}\neq0, \ew{\hat{I}^2}_{\mathrm{vac}}\neq0$ (which recalling Eq.~(\ref{fluxVchargeI}), can be understood as inherited from the fluctuations of flux $\hat{\Phi}$ and charge $\hat{Q}$, namely $\ew{\hat{\Phi}^2}_{\mathrm{vac}}\neq0, \ew{\hat{Q}^2}_{\mathrm{vac}}\neq0$).

We next aim to give a unified picture from which thermal, shot, and quantum noise arise. These fluctuations can be quantified from the current correlator at different times, so we define the noise spectrum as
\beq \label{noisedef}
S^{(2)} := \int_{-\infty}^{\infty} dt_1 dt_2 e^{-i\omega_1 t_1} e^{-i\omega_2 t_2} {\cal T}_S \ew{I(t_1)I(t_2)}_c = \int_{-\infty}^{\infty} dt e^{-i\omega t} {\cal T}_S \ew{I(t)I(0)}_c
\eeq
where $\omega\equiv\omega_2$ and ${\cal T}_S$ is the symmetrization operator, that sums over all possible time (or frequency) switchings, having in this case ${\cal T}_S \langle I(t_1)I(t_2) \rangle = \langle I(t_1)I(t_2) \rangle + \langle I(t_2)I(t_1) \rangle$. Because of the time-translational symmetry of the current-correlation function, the noise spectrum defined above is proportional to $\delta(\omega+\omega')$, which is implicit in the second equality of (\ref{noisedef}). We here have taken the symmetrized version of the noise spectrum. As we will see, various definitions can be adopted in this respect, and they are deeply connected with the detection of the emission or absorption spectrum. It is important to remark that the current is a quantum-mechanical operator. This means that two current operators evaluated at different times do not commute with each other, and the various definitions of the noise spectrum will produce different results. To recover the result measured classically, we have chosen the symmetrized form \cite{LandauStat}, which is Hermitian. Once we have pointed this out, we anticipate that in the theory presented here, the current will be treated as a classical stochastic variable. However, even if the current is treated this way, it can reveal quantum effects present in the system, as it will be shown in the forthcoming chapters.

Interestingly, the noise spectrum of a system in thermodynamic equilibrium fulfills a fluctuation-dissipation theorem (FDT). General considerations on the relation between the dissipation and the fluctuation close to linear response will be presented in the next section. At low frequencies, Johnson \cite{Johnson28} found that the noise $S^{(2)}$ (characterizing the fluctuations) in a linear electrical circuit is related through the temperature $T$ with the dc conductance $G$ (characterizing the dissipation) as\footnote{Here, $k$ denotes the Boltzman constant (to abbreviate the notation $k_B$ used in other places of the text). This, together with the electron charge $e$ and the Planck constant $\hbar$ will be taken equal to $1$ in the chapters discussing noise correlations. This criterium will be assumed when writing different expressions, as it has been done for Eq.~(\ref{noisedef}).}
\beq \label{FDTclassical}
S^{(2)} = 2kTG.
\eeq
A theoretical framework to understand this result was presented by Nyquist \cite{Nyquist28}, who anticipated that at high frequencies, the energy per degree of freedom should be taken to be $\hbar\omega/(e^{\frac{\hbar\omega}{kT}}-1)$, instead of $kT$, therefore arriving to a Planck's distribution form for the noise spectrum. However, it was not until more than twenty years later that a complete treatment, applicable to general dissipative systems, was given. Callen and Welton \cite{CallenWelton51} extended this FDT to properly include quantum fluctuations, relevant when the measured frequencies are larger than the temperature. The FDT takes then the form 
\beq \label{FDTquantum}
S^{(2)}(\omega) = \hbar\omega \mathrm{coth}(\frac{\hbar\omega}{2kT}) G(\omega) = 2 \left[ \frac{1}{2}\hbar\omega + \frac{\hbar\omega}{e^{\frac{\hbar\omega}{kT}}-1} \right] G(\omega), 
\eeq
where $G(\omega)$ is the ac conductance. This expression can be equivalently written in terms of the Bose-Einstein distribution $b(\omega) \equiv 1/[e^{\frac{\hbar\omega}{kT}}-1]$, since $\mathrm{coth}(\frac{\hbar\omega}{2kT})=2b(\omega)+1=b(\omega)-b(-\omega)$, and it becomes clear that the symmetrized noise, considered here, contains both absorption and emission. As mentioned, the relation incorporates vacuum fluctuations. These are described by the first term in the second expression, $\frac{1}{2}\hbar\omega$.

\begin{figure}
  \begin{center}
    \includegraphics[width=0.455\textwidth]{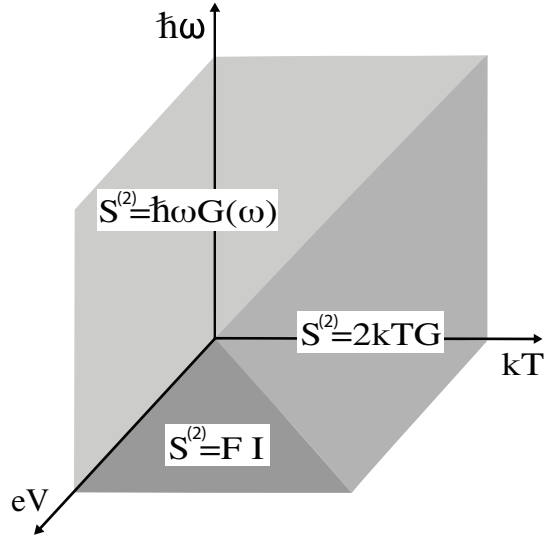}
  \end{center}  
  \caption[Regimes of the fluctuation-dissipation theorem]{Different limits of the NEFDT depending on the dominating energy scale. For $kT=0$, irrespectively of the ratio $\hbar\omega$ to $eV$, or for $eV=0$ and $\hbar\omega\gg kT$, the NEFDT reads $S^{(2)}(\omega)=\hbar\omega G(\omega)$ (light gray). When $eV=0$ and $kT\gg\hbar\omega$, we have $S^{(2)}=2kTG$ (faint gray). Finally, for $\hbar\omega=0$ and $eV\gg kT$ we find the Poisson limit $S^{(2)}=F I_{stat}$, being $F$ the Fano-factor and $I$ the current through the system (dark gray).
}
  \label{FDTlimitsFig}
\end{figure}

Out of equilibrium, a fluctuation-dissipation relation can be also found for some particular cases, such as tunnel junctions \cite{Dahm69,Rogovin-Scalapino74} or for quantum dots in the weak cotunneling regime \cite{Suk01}.
For a non-interacting two-terminal conductor driven out of equilibrium, the \textit{symmetrized} noise spectrum takes the general form \cite{Yang92, Buttiker92b, Schoelkopf97, BlanterButtiker00}
\beq \label{FDTfull}
S^{(2)}(\omega) &=& \hbar\omega \mathrm{coth}\left(\frac{\hbar\omega}{2kT}\right) \sum_n \mathsf{T_n}^2 + \left[ \frac{(\hbar\omega+eV)}{2} \mathrm{coth}\left(\frac{\hbar\omega+eV}{2kT}\right) \right. \nonumber \\ && \left. + \frac{(\hbar\omega-eV)}{2} \mathrm{coth}\left(\frac{\hbar\omega-eV}{2kT}\right) \right] \sum_n \mathsf{T_n}(1-\mathsf{T_n}),
\eeq
Here $\mathsf{T_n}$ is the transmission coefficient of the conduction channel $n$, and $kT$, $eV$ and $\hbar\omega$ refer to the temperature, voltage and frequency scales respectively. Depending on which of these scales dominates, this expression takes a different limit, talking then about thermal, voltage, or quantum noise regime.
In the tunneling limit ($\mathsf{T_n}\ll1$), Eq.~(\ref{FDTfull}) gives the non-equilibrium fluctuation-dissipation theorem (NEFDT) as reported in \cite{Dahm69, Rogovin-Scalapino74} for tunnel junctions:
\beq \label{FDTfreq} 
S^{(2)}(\omega)=\frac{1}{2}\sum_{p=\pm} I_{stat}(eV+p\hbar\omega)\mathrm{coth}\left(\frac{eV+p\hbar\omega}{2kT}\right),
\eeq
being $I_{stat}$ the stationary current through the system. The expression (\ref{FDTfreq}) is also appropriate for interacting tunnel junctions and for quantum dots in the weak cotunneling regime \cite{Suk01}, and its zero-frequency limit $S^{(2)}= I_{stat} \mathrm{coth}(\frac{eV}{2kT})$ has been derived in the context of counting statistics \cite{Lev04}. For low voltages, $eV\ll kT$, equation (\ref{FDTfull}) recovers the Callen and Welton equilibrium relation, and if also $\hbar\omega\ll kT$, it gives the Johnson-Nyquist FDT (thermal noise regime). Finally, if $eV\gg kT$ and $\hbar\omega$, (shot noise regime), we find $S^{(2)}=F I_{stat}$, where the coefficient $F\equiv \frac{2\sum_n \mathsf{T_n} (1-\mathsf{T_n})}{\sum_n \mathsf{T_n}}$ is the Fano-factor.

\begin{figure}
  \begin{center}
    \includegraphics[width=\textwidth]{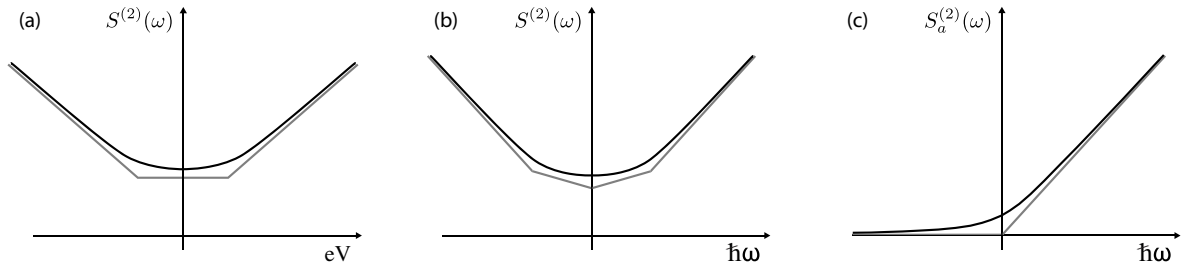}
  \end{center}  
  \caption[Noise behaviour]{Behaviour of the noise spectrum. In all the figures, the light-gray curve corresponds to the result at zero temperature, and the dark curve denotes the finite-temperature case. This last is simply the smoothed-out form of the zero-temperature solution, as the fermi functions no longer present a discontinuity at finite temperature. a)~Finite-frequency noise as a function of voltage. There is a quantum-noise floor $S^{(2)}(\omega)=\hbar\omega G$ which we clearly see at $V=0$. At $eV\sim\pm 2kT$ there is a change of behaviour to match the shot-noise limit $S^{(2)}=FI_{stat}$ at high voltage. b)~Frequency-dependent spectrum of the symmetrized noise. The slope of the curve changes abruptly at $\hbar\omega\sim\pm eV$. The behaviour of the dc conductance can be extracted from the figure, as the FDT theorem is fulfilled, and also the shot noise value $S^{(2)}=FI_{stat}$ at $\omega=0$ is clearly seen in the zero-temperature curve. c)~Non-symmetrized version of the noise spectrum. In this case, emission ($\omega>0$ side) and absorption ($\omega<0$ side) by the quantum system can be distinguished. The noise value at $\omega=0$ is $2kTG$.
}
  \label{S2behaviourFig}
\end{figure}

The different limits of the noise spectrum, depending on the dominating energy scale, are shown schematically in Fig.~\ref{FDTlimitsFig}. Also, in Fig.~\ref{S2behaviourFig} we show the behaviour of Eq.~(\ref{FDTfull}) as a function of voltage and frequency. We have mentioned that depending on the measurement scheme, the noise reflects different physics and should adopt distinct definitions in what symmetrization is concerned. As it can be seen in Fig.~\ref{S2behaviourFig} (a and b), the symmetrized noise has a change of behaviour at $eV\sim2kT$, and at $\hbar\omega\sim eV$ in the frequency-dependent spectrum. However, the non-symmetrized noise (described by the absence of ${\cal T}_S$ in the definition (\ref{noisedef})), although symmetric with voltage, is, as expected, asymmetric with frequency (Fig.~\ref{S2behaviourFig}c). This means that importantly, the asymmetric noise distinguishes emission and absorption of photons.
More details about this can be found in \cite{Lesovik97, Aguado00, Gavish00} and in the review \cite{Clerk10}.
In the following, we will adopt the symmetrized version of the noise spectrum, which will be implicit throughout the text.

As previously stated, noise measurements allow us to learn about the nature of the particles involved in transport and about the properties of the conductor \cite{BeenakkerSchonenberger}. The shot noise suppression in QPCs reveals Coulomb interactions \cite{Li90, Birk95, Reznikov95}, and the unti-bunching shown in Hanbury Brown-Twiss experiments makes evident the fermionic nature of electrons captured by the Pauli exclusion principle \cite{Schonenberger99, Yamamoto99}. A groundbreaking application of shot noise was to determine the effective charge of the quasiparticles in the fractional quantum Hall effect \cite{Kane-Fisher94, dePicciotto97, Glattli97, Reznikov99, Comforti02, Chung03, Dolev08}, and that of the Cooper pairs in superconductors \cite{Jehl00, Lefloch03}. It has also been used to investigate entanglement between electrons, which has been an active topic of research in the last decade \cite{Burkard00, Lesovik01, Martin02, Samuelsson03, Beenakker03, Samuelsson04, Beenakker04, Loss00, Lambert10}. Furthermore, there are important applications of shot noise in technology such as thermometry \cite{Spietz03}.
In the next chapters we will show how the noise spectrum reveals properties of quantum coherence and internal energy scales of a quantum system.

\section{When worlds collide}

Historically, condensed matter physics and quantum optics emerged from the study of two different worlds, the physics of materials and the physics of light. However, a number of quantum effects and techniques used in both, are in close analogy to each other. Furthermore, experiments combine these two subjects, for example, making use of lasers to manipulate electrons in solids. Another example are the hybrid quantum systems (studied in chapter~\ref{Chapter5}), which merge both worlds in the context of quantum computation. In this section we briefly discuss some basic concepts commonly used in these fields, and to which we will make reference throughout the text. 


In the next chapters we will present models of quantum systems coupled to different environments, for which the language of second quantization will be used. We recall here that an operator ${\cal \hat{O}}$ may be written in second quantization as 
\beq
{\cal \hat{O}}= \sum_{\nu\mu}\bra{\nu}{\cal \hat{O}}\ket{\mu} \hat{c}_{\nu}^{\dagger}\hat{c}_{\mu} = \sum_{\nu\mu} \left[ \int d^3\vec{r} \; \psi_{\nu}^*(\vec{r}) {\cal \hat{O}}(\vec{r})\psi_{\mu}(\vec{r}) \right] \hat{c}_{\nu}^{\dagger}\hat{c}_{\mu} = \int d^3\vec{r} \; \hat{\Psi}^{\dagger}(\vec{r}) {\cal \hat{O}}(\vec{r}) \hat{\Psi}(\vec{r}),
\eeq
where $\hat{c}_{\nu}^{\dagger}$/$\hat{c}_{\mu}$ is a creation/anihilation operator and $\left\{ \ket{\mu} \right\}$ (equivalently $\left\{\psi_{\mu}\right\}$) denotes a base of the Hilbert space. To write the last expression we have made use of the field operator $\hat{\Psi}(\vec{r}):=\sum_{\mu} \olap{\vec{r}}{\psi_{\mu}} \hat{c}_{\mu} = \sum_{\mu} \psi_{\mu}(\vec{r}) \hat{c}_{\mu}$. 
With this language, and using a basis of plane waves, we find:
\begin{description}
\item - Kinetic energy: $\frac{-\hbar^2\nabla^2}{2m} \to \sum_{\vec{k}\sigma} \frac{\hbar^2 k^2}{2m} \hat{c}_{\vec{k}\sigma}^{\dagger} \hat{c}_{\vec{k}\sigma}.$ 
\item - Potential energy: $\frac{e^2}{|\vec{r}-\vec{r}'|} \to \frac{1}{2V}\sum_{\sigma\sigma'}\sum_{\vec{k}\vec{k}'\vec{q}} \frac{4\pi e^2}{q^2} \hat{c}_{\vec{k}+\vec{q},\sigma}^{\dagger}\hat{c}_{\vec{k}'-\vec{q},\sigma'}^{\dagger}\hat{c}_{\vec{k}'\sigma'}\hat{c}_{\vec{k}\sigma}.$ 
\item - Density: $\delta(\vec{r}-\vec{r}') \to \frac{1}{V}\sum_{\vec{k}\vec{q}\sigma} e^{i\vec{q}\vec{r}} \hat{c}_{\vec{k}\sigma}^{\dagger}\hat{c}_{\vec{k}+\vec{q},\sigma}.$ 
\item - Paramagnetic current: $\frac{\hbar}{2mi}\left[ \psi_{\sigma}^*(\vec{r}) \left( \vec{\nabla} \psi_{\sigma}(\vec{r}) \right) - \mathrm{H.c.} \right] \to \frac{\hbar}{mV} \sum_{\vec{k}\vec{q}}(\vec{k}+\vec{q}/2) e^{i\vec{q}\vec{r}} \hat{c}_{\vec{k}\sigma}^{\dagger}\hat{c}_{\vec{k}+\vec{q},\sigma}.$
\end{description}
Here $V$ is the volume over which the boundary conditions are set; $\vec{k}$, $\vec{k}'$ and $\vec{q}$ label the momentum, and $\sigma$, $\sigma'$, the spin degree of freedom. We can go further and list some of the most popular models encountered in condensed matter physics, whose Hamiltonian, written in second quantization, reads:
\begin{description}
\item - Fano: ${\cal \hat{H}} = \sum_{\sigma} \varepsilon_d \hat{c}_{d\sigma}^{\dagger}\hat{c}_{d\sigma} + \sum_{\vec{k}\sigma} \varepsilon_{\vec{k}} \hat{c}_{\vec{k}\sigma}^{\dagger}\hat{c}_{\vec{k}\sigma} + \sum_{\vec{k}\sigma} ({\cal V}_{\vec{k}}\hat{c}_{d\sigma}^{\dagger}\hat{c}_{\vec{k}\sigma} + \mathrm{H.c.}).$
\item - Hubbard: ${\cal \hat{H}} = -t \sum_{\ew{i,j},\sigma} \hat{c}_{i\sigma}^{\dagger}\hat{c}_{j\sigma} + U \sum_i \hat{n}_{i\uparrow} \hat{n}_{i\downarrow}.$
\item - Anderson: ${\cal \hat{H}} = \sum_{\sigma} \varepsilon_d \hat{c}_{d\sigma}^{\dagger}\hat{c}_{d\sigma} + U \hat{n}_{d\uparrow}\hat{n}_{d\downarrow} + \sum_{\vec{k}\sigma} \varepsilon_{\vec{k}} \hat{c}_{\vec{k}\sigma}^{\dagger}\hat{c}_{\vec{k}\sigma} + \sum_{\vec{k}\sigma} ({\cal V}_{\vec{k}}\hat{c}_{d\sigma}^{\dagger}\hat{c}_{\vec{k}\sigma} + \mathrm{H.c.}).$
\item - Kondo: $ {\cal \hat{H}} = \sum_{\vec{k}\sigma} \varepsilon_{\vec{k}} \hat{c}_{\vec{k}\sigma}^{\dagger}\hat{c}_{\vec{k}\sigma} + J \hat{\vec{s}}\cdot\hat{\vec{S}}.$
\item - Ising: $-J\sum_{\ew{i,j}} \hat{S}_z^{(i)} \hat{S}_z^{(j)} - H_z\sum_i \hat{S}_z^{(i)}.$
\item - Heisenberg: $-J\sum_{\ew{i,j}} \hat{\vec{S}}^{(i)}\cdot\hat{\vec{S}}^{(j)} - \vec{H}\sum_i \hat{\vec{S}}^{(i)}.$
\end{description}
Here, $\hat{n}_{d\sigma}:=\hat{c}_{d\sigma}^{\dagger}\hat{c}_{d\sigma}$; $U$, ${\cal V}$ and $J$ denote coupling constants, and $\vec{H}$ an applied magnetic field; $\varepsilon_d$, $\varepsilon_{\vec{k}}$, are single-particle energies and $\hat{\vec{S}}$, $\hat{\vec{s}}$, the spin.
In the next chapters we will study models similar to these, more specifically, to the Fano and Heisenberg models.

Another concept recursively used in the text is that of \textit{linear response}. If a quantum system is described by a Hamiltonian of the form ${\cal \hat{H}} = {\cal \hat{H}}_0 + f(t) \hat{x}$, being $f(t)$ a time-dependent function, then the average of the variable $\hat{x}$ can be written to linear order in $f(t)$ as
\beq
\ew{\hat{x}(t)} \equiv \ew{\hat{x}(0)} + \int_{-\infty}^{\infty} \chi(t-t') f(t') dt',
\eeq
where the \textit{response function} $\chi(t-t')$ is given by $\chi(t-t')\equiv -i\ew{[\hat{x}(t),\hat{x}(t')]}\theta(t-t')$, with $\theta(t)$ the Heaviside step function. Importantly, to linear response the fluctuation-dissipation theorem is then fulfilled, and takes a similar structure to (\ref{FDTquantum}) (or (\ref{FDTclassical}) in the classical case), with $S^{(2)}(\omega)$ corresponding to the fluctuations of the variable $\hat{x}$ and $G(\omega)$ to the imaginary part of the response function. More details on linear response theory can be found for example in \cite{LandauStat, VignaleBook}.


In this thesis, we will also study systems that follow popular models in quantum optics. These typically involve the coupling between spin degrees of freedom and a bosonic bath. The most relevant in our context will be:
\begin{description}
\item - Jaynes-Cummings model: ${\cal \hat{H}} = \frac{\omega_q}{2} \hat{\sigma}_z + \omega_c \left( \hat{a}^{\dagger}\hat{a} + \frac{1}{2} \right) + g \left( \hat{a}^{\dagger}\hat{\sigma}_{-} + \mathrm{H.c.} \right).$
\item - Rabi model: ${\cal \hat{H}} = \frac{\omega_q}{2} \hat{\sigma}_z + \omega_c \left( \hat{a}^{\dagger}\hat{a} + \frac{1}{2} \right) + g \left( \hat{a}^{\dagger} + \hat{a} \right) \left( \hat{\sigma}_{+} + \hat{\sigma}_{-} \right).$
\item - Tavis-Cummings model: ${\cal \hat{H}} = \frac{\omega_q}{2} \hat{J}_z + \omega_c \left( \hat{a}^{\dagger}\hat{a} + \frac{1}{2} \right) + G \left( \hat{a}^{\dagger}\hat{J}_{-} + \mathrm{H.c.} \right).$
\item - Dicke model: ${\cal \hat{H}} = \frac{\omega_q}{2} \hat{J}_z + \omega_c \left( \hat{a}^{\dagger}\hat{a} + \frac{1}{2} \right) + G \left( \hat{a}^{\dagger} + \hat{a} \right) \left( \hat{J}_{+} + \hat{J}_{-} \right).$
\end{description}
The first two models describe the coupling between a two-level system (2LS) (c.f. appendix \ref{2LSappendix}) -- described by the Pauli operators $\hat{\sigma}_z$ and $\hat{\sigma}_{\pm}\equiv(\hat{\sigma}_x\pm i \hat{\sigma}_y)/2$, and a cavity -- described by the bosonic operators $\hat{a}^{\dagger}$, $\hat{a}$. The Jaynes-Cummings (JC) model (explained in detail in appendix \ref{JCappendix}) is a simplified version of the Rabi model, where the counter-rotating terms $\hat{a}^{\dagger}\hat{\sigma}_{+}$ and $\hat{a}\hat{\sigma}_{-}$ are neglected (\textit{rotating wave approximation} (RWA)). The second two models describe the coupling of $N$ 2LSs to a cavity mode. Here we have introduced the collective spin operators $\hat{J}_z\equiv \sum_{j=1}^N \hat{\sigma}_z^{(j)}$ and $\hat{J}_{\pm}\equiv \frac{1}{\sqrt{N}} \sum_{j=1}^N \hat{\sigma}_{\pm}^{(j)}$. The parameters $\omega_q$, $\omega_c$, $g$ and $G\equiv g \sqrt{N}$ correspond to qubit and cavity energies and to coupling constants. Again, the Tavis-Cummings model originates from the Dicke model if the RWA is performed.
We have assumed a single-mode cavity, although the generalization to a many modes is straightforward.

A cavity with infinitely many modes constitutes a basic model to account for quantum dissipation. This effect comes from the interaction of any quantum system with its environment (represented as a bath with an infinite number of degrees of freedom), which forces the former to loose its coherence properties. Generally, quantum dissipation is successfully described by the \textit{Caldeira-Leggett model} \cite{CaldeiraLeggett83}, which treats the bath as a set of harmonic oscillators. Here, we will study the particular case of a spin (quantum two-level system) coupled to a bath. This coupling can be of different nature. For example, the \textit{Wigner-Weisskopf} model considers that the interaction between 2LS (c.f. Eq.~(\ref{H2LS})) and bath takes the form \cite{ScullyBook}
\beq \label{Wigner-Weisskopf-Eq}
\hat{\cal H}_\mathrm{V}=\sum_k g_k \ket{1}\bra{0} \hat{a}_k^{\dagger} + \mathrm{H.c.},
\eeq
where $g_k$ is the coupling constant and $\hat{a}_k^{\dagger}$ excites mode $k$ in the bath. This gives the atomic spontaneous-decay rate predicted by Einstein model. A different system-bath coupling, namely a dipolar coupling, is considered by the \textit{spin-boson model}, whose Hamiltonian reads \cite{WeissBook}
\beq \label{SpinBosonHamiltonian}
\hat{\cal H}= \hat{\cal H}_{2LS} + \sum_k \omega_k \hat{a}_k^{\dagger}\hat{a}_k + (\op{1}{1}-\op{0}{0}) \sum_k \left(g_k \hat{a}_k + \mathrm{H.c.}\right),
\eeq
being $\hat{\cal H}_{2LS}$ given by Eq.~(\ref{H2LS}).
Other approaches to quantum dissipation are the Fokker-Planck and Langevin equations \cite{GardinerZollerBook}.

\begin{figure}
  \begin{center}
    \includegraphics[width=\textwidth]{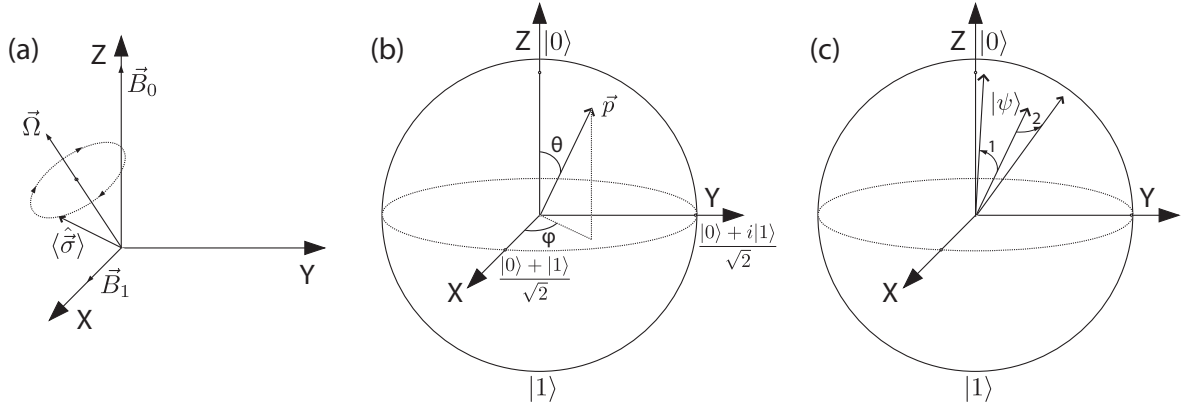}
  \end{center}  
  \caption[The Bloch sphere]{a) Motion of an electron spin in a magnetic field $\vec{\Omega}=\vec{B}_0+(\vec{B}_1 e^{i\omega t} + \mathrm{H.c.})$, which is described by equation (\ref{NMReq}). The picture is in the rotating frame, where $\vec{B}_1$ (circulating around $\vec{B}_0$ in the laboratory frame) remains static. Equivalently, this motion corresponds to the dynamics of an electric dipole coupled to the radiation field. Then $\vec{B}_0=-\vec{u}_z\delta$ and $\vec{B}_1=-2\vec{u}_x\mathrm{Re}\{g(t)\}$. b) Bloch-sphere picture. An arbitrary state of the 2LS $\ket{\psi} = \mathrm{cos}(\theta/2) \ket{0} + \mathrm{sin}(\theta/2) e^{i\varphi} \ket{1}$ (illustrated by $\vec{p}$ in the figure -- c.f. Eq.~(\ref{rhoPauli})) can be represented. At the `north' and `south' poles of the sphere we find the pure states $\ket{0}$ and $\ket{1}$ respectively. Similarly, we find the states $(\ket{0}+\ket{1})/\sqrt{2}$ and $(\ket{0}+i\ket{1})/\sqrt{2}$ in the equator. An external static field makes the wave vector to precess about the $Z$-axis. If however, a dc field is applied, the precession occurs about an axis perpendicular to $Z$. c) When relaxation and dephasing are taken into account (Eq.~(\ref{NMReqT1T2})), these cause a motion that destroys the $\ket{1}$ component (process $1$) and randomize the phase $\varphi$ (process $2$) respectively.
}
  \label{BlochNMRfig}
\end{figure}

The coupling between a 2LS and the radiation field is a fundamental problem in quantum optics. A classical field of frequency $\omega$ and momentum $\vec{k}$, represented by the electric field $\vec{E} (\vec{r},t) = {\cal E}(\vec{r},t) e^{-i(\omega t -\vec{k}\cdot\vec{r})} + \mathrm{H.c}$, couples with a two-level atom through an electric dipolar interaction $-\hat{\vec{d}}\cdot\vec{E}$, being $\hat{\vec{d}}\equiv \hat{\sigma}_x\vec{d}$ the atomic dipole moment\footnote{A magnetic dipole interacting with a similar magnetic field produces the same EOM as the electric case. The coupling constant is in this case given by the magnetic dipolar interaction, $g={\cal B} \mu$, being ${\cal B}$ the amplitude of the field and $\mu$  the magnitude of the dipole moment.}. Under the assumption that atom (with energy splitting $\omega_q$) and field are close to resonance (small detuning: $\delta:=\omega-\omega_q\ll \omega, \omega_q$) and that the applied field is weak (small coupling: $g:={\cal E}d\ll \omega$), this gives rise to the optical \textit{Bloch equations} describing the dynamics of the 2LS, that assuming the RWA read
\begin{equation}\label{BlochEqs}
\begin{array}{l}
\dot{\rho}_{00} = ig \left( \rho_{01} - \rho_{10} \right). \\
\dot{\rho}_{01} = ig \left( \rho_{00} - \rho_{11} \right) + i\omega_q\rho_{01}. \\
\dot{\rho}_{10} = -ig \left( \rho_{00} - \rho_{11} \right) - i\omega_q\rho_{10}. \\
\dot{\rho}_{11} = -ig \left( \rho_{01} - \rho_{10} \right).
\end{array}
\end{equation}
Here, $\rho_{00}$ and $\rho_{11}$ denote the populations of the ground and excited state respectively and $\rho_{01}$, $\rho_{10}$, the coherences. These equations give the popular \textit{Rabi oscillations} -- coherent oscillations of the state of the system between the two levels $\ket{0}$ and $\ket{1}$, which on resonance ($\delta=0$) achieve a perfect transfer of quantum information stored in the state $\ket{1}$ to the state $\ket{0}$ and viceversa. The probability $P_{01}$ of preparing $\ket{1}$ initially and finding the state $\ket{0}$ after a time $\Delta t$ is given by the Rabi formula,
\beq \label{RabiFormula}
P_{01}=|\rho_{01}|^2=\frac{g^2}{g^2+(\delta/2)^2}\mathrm{sin}^2\left[\Delta t \sqrt{g^2+(\delta/2)^2} \right].
\eeq
Defining $\vec{\Omega}(t) := (-2\mathrm{Re}\{g(t)\}, -2\mathrm{Im}\{g(t)\}, -\delta)_T$, the Bloch equations can be written in the interaction picture in terms of Pauli spin-two operators as
\beq \label{NMReq}
\frac{d}{dt}\ew{\hat{\vec{\sigma}}} = \vec{\Omega}(t) \times \ew{\hat{\vec{\sigma}}}.
\eeq
Notice that this is the EOM of a magnetic dipole (spin\footnote{Described by $\hat{\vec{\sigma}}\equiv(\hat{\sigma}_x,\hat{\sigma}_y,\hat{\sigma}_z)_T$, with the Pauli operators $\hat{\sigma}_x=\op{0}{1}+\op{1}{0}$, $\hat{\sigma}_y=-i(\op{0}{1}-\op{1}{0})$, and $\hat{\sigma}_z=\op{1}{1}-\op{0}{0}$.}) in the presence of an effective magnetic field $\vec{\Omega}(t)$. If the field is static, we find the so-called \textit{Larmor precession}, in which the spin is rotating with frequency $\Omega$ around the axis set by the direction of $\vec{\Omega}$. If the field depends on time, the spin precesses around the effective field, but this is itself moving in the laboratory frame. We may construct a particular situation in which we initially apply a static field, which sets the spin orientation, and we then apply a perpendicular ac-signal rotating about the dc-field (see Fig.~\ref{BlochNMRfig}a). 
The magnetic moment experiences then a precession that, if the ac-frequency equals the Larmor frequency $\Omega$, the spin is completely flipped after a time $1/\Omega$.
This is the essence of electron spin resonance commonly used to perform gate operations in quantum computation. The spin in equation (\ref{NMReq}) can be thought as to represent an arbitrary state of the 2LS, which may be written in the general form $\ket{\psi} = \mathrm{cos}(\theta/2) \ket{0} + \mathrm{sin}(\theta/2) e^{i\varphi} \ket{1}$. This state is conveniently represented in the so-called \textit{Bloch-sphere}, depicted in Fig.~\ref{BlochNMRfig}b. More specifically, the \textit{Bloch vector} $\vec{p}=\ew{\hat{\vec{\sigma}}}$ represented in the figure is defined through the relation between density and Pauli operators
\beq \label{rhoPauli}
\hat{\rho} = \frac{1}{2} \left( \mathds{1} + \vec{p}\cdot\hat{\vec{\sigma}} \right).
\eeq
The above optical Bloch equations do not account for incoherent relaxation processes. These can be included by incorporating the decay rate $\Gamma_{fi}$ in the equations, that to first order in the coupling between system and bath (causing the relaxation) is determined by the Fermi golden rule:
\beq
\Gamma_{fi}(\varepsilon) = 2\pi |\ew{f|\hat{\cal H}_{\mathrm{int}}|i}|^2 {\cal D}(\varepsilon).
\eeq
Here $i$ and $f$ denote the initial and final states, $\hat{\cal H}_{\mathrm{int}}$ is the system-bath interaction Hamiltonian, and ${\cal D}(\varepsilon)$ the bath density of states. Including similarly dephasing processes, the Bloch equations can be written in the form
\beq \label{NMReqT1T2}
\frac{d}{dt} \ew{\hat{\sigma}_z} &=& [\vec{\Omega}(t) \times \ew{\hat{\vec{\sigma}}}]_z + \frac{\ew{\hat{\sigma}_z^{\mathrm{eq}}}-\ew{\hat{\sigma}_z}}{T_1}. \label{NMReqZ}\\
\frac{d}{dt} \ew{\hat{\sigma}_{\pm}} &=& [\vec{\Omega}(t) \times \ew{\hat{\vec{\sigma}}}]_{\pm} - \frac{1}{T_2}\ew{\hat{\sigma}_{\pm}}. \label{NMReqPM}
\eeq
In equation (\ref{NMReqZ}), $\ew{\hat{\sigma}_z^{\mathrm{eq}}}\equiv \frac{\Gamma_{01}-\Gamma_{10}}{\Gamma_{01}+\Gamma_{10}}$, and $T_1$ is the \textit{relaxation} time, having $\frac{1}{T_1} = \Gamma_{01}+\Gamma_{10}$. In Eq.~(\ref{NMReqPM}), $T_2$ is the \textit{decoherence} time. This can be expressed as $\frac{1}{T_2}=\frac{1}{2T_1}+\frac{1}{T_{\varphi}}$, being $T_{\varphi}$ the \textit{(pure) dephasing} time. We therefore have the general result $T_2\leqslant2T_1$.
The processes of relaxation and dephasing can be represented in the Bloch-sphere picture as it is shown in Fig.~\ref{BlochNMRfig}c.


Throughout the thesis we will study systems that have become of great importance in the area of quantum computation. Although this field is sometimes looked skeptically, the progress realized in only two decades is enormous. To perceive this, one can take as an example the classical computer. The idea of making computations with machines is an old concept. Already in 1679 Leibniz wrote: 
``This [binary] calculus could be implemented by a machine (without wheels)... provided with holes in such a way that they can be opened and closed. They are to be open at those places that correspond to a 1 and remain closed at those that correspond to a 0. Through the open gates small cubes or marbles are to fall into tracks, through the others nothing. It [the gate array] is to be shifted from column to column as required...''. However, it was not until much later -- the twentieth century -- that computers arrived to our lives. In a parallel manner, but in the early 1980's, Feynman would defend the idea of a quantum computer \cite{Feynman82, Feynman83}: ``Now, we can, in principle make a computing device in which the numbers are represented by a row of atoms with each atom in either of the two states. That's our input. The Hamiltonian starts `Hamiltonianizing' the wave function... The ones move 
around, the zeros move around... Finally, along a particular bunch of atoms, ones and 
zeros... occur that represent the answer''. Later, a series of seminal papers in the field would follow \cite{Deutsch85, Feynman86, Deutsch89}, and different physical systems were proposed to implement a quantum computer \cite{CiracZoller95, LossDiVincenzo98, Kane98}, a growing list that has recently incorporated the proposal of hybrid quantum systems (c.f. chapter \ref{Chapter5}).

A quantum computer, regardless of the type, must satisfy five conditions. These requirements are known as DiVincenzo criteria \cite{DiVincenzo95}:\\
$\bullet$ A `scalable' physical system with well characterized qubits.\\
$\bullet$ The ability of preparing the system in a given state (initialization).\\
$\bullet$ A `universal' set of quantum logic gates (e.g. two-qubit gate CNOT) (control).\\
$\bullet$ The capability of detecting the state of the system (measurement).\\
$\bullet$ A decoherence time larger than the gate time cycle.\\
Additionally, if we want to allow for quantum communication, it is needed:\\
$\bullet$ The possibility to interconvert stationary qubits and `flying' qubits.\\
$\bullet$ The ability to faithfully transmit `flying' qubits between specified locations.\\
The importance of the race towards a quantum computer becomes more obvious when the breakdown of the Moore's law \cite{Moore65} in probably less than a decade is foreseen, and also in view of the different applications such as factorization, search algorithms, cryptography, teleportation and quantum simulation \cite{NielsenChuang, BouwmeesterBook}.

~

~

~

~

~

~

~

~

~

~

~

~

~

~

~

~

~

~

~

~

~

~


\chapter{Quantum transport in nanoscopic conductors} 
\label{Chapter2}
\lhead{Chapter 2. \emph{Quantum transport in nanoscopic conductors}} 

\begin{flushright}

\textit{``An expert is a man who has made all the mistakes\\ which can be made in a very narrow field."}

Niels Bohr

\end{flushright}

\begin{small}

In this chapter we overview the most common methods in quantum transport. In the first section we describe the three basic formalisms of interest in this thesis, namely, the scattering matrix method, the density operator approach, and the Green's functions technique. Special emphasis is made on the application of these formalisms to obtain noise correlations. In particular, we give a detailed derivation of the quantum regression theorem and the MacDonald's formula, and their possible extension to high-order correlations. Furthermore, we introduce the reader to the field of full counting statistics (FCS). Although traditionally this is based on the scattering matrix theory, we here focus on the FCS in the context of the density matrix approach, for which, as we will see, the projection or Nakajima-Zwanzig techniques are extremely useful. We study a few examples as an application of the described methods. Next, we present an experiment of electron transport and spin manipulation in carbon-nanotube quantum dots. Here, the hyperfine interaction is characterized by means of contrasting $^{13}$C and $^{12}$C nanotubes. Finally, we give a basic approach to our theory of counting statistics in quantum transport, which will be presented in more detail in the following chapters.

\end{small}

\newpage

\section{Theoretical methods in quantum transport}

Quantum transport systems are of typical size comparable to the electron coherence length and much larger than the Fermi wavelength. As it was shown in the previous chapter, examples include quantum dots attached to 2DEGs, or molecules close to metallic contacts. In this section we present the most common mathematical tools to treat quantum transport problems. Generally, these can be described in terms of a central region (quantum system) which may exchange particles with electronic reservoirs (leads). The application of a bias voltage makes the system a non-equilibrium problem. Our aim here is to develop techniques that allow us to obtain information about the central system through the study of an output signal, namely the electrical current passing through the circuit. Roughly speaking, these techniques can be divided into three families: the scattering-matrix approach, the density-operator formalism, and the Green's-functions methods. In particular, we center our attention on noise-correlation techniques. Within the three families, there is a variety of ways to tackle this problem. These will be illustrated with examples, and finally, an introduction to our theory will be presented.

\subsection{Scattering matrix approach}

\begin{figure}
  \begin{center}
    \includegraphics[width=0.7\textwidth]{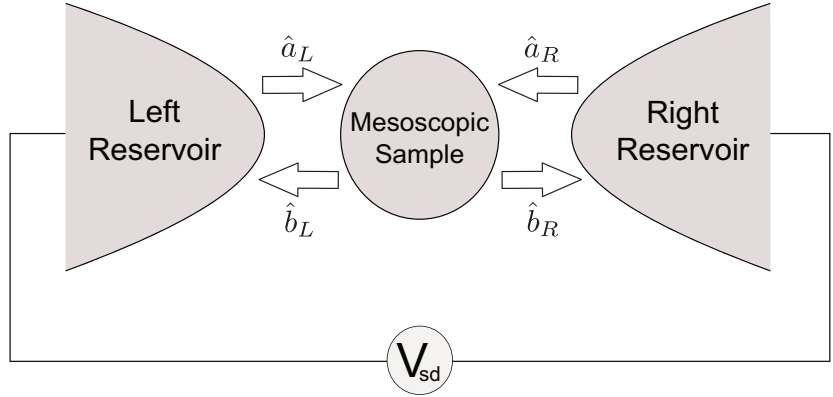}
  \end{center}  
  \caption[Scattering processes in a two-termial device]{
Two-terminal transport device. The mesoscopic sample can be treated as a scattering center, and electronic waves are transmitted or reflected from reservoir $\alpha=L,R$, with an amplitude $\hat{a}_{\alpha}$ or $\hat{b}_{\alpha}$.}
  \label{ScatteringTransportFig}
\end{figure}

A widely used technique in quantum transport is the scattering matrix formalism. This treats the mesoscopic conductor as a center of scattering (see Fig.~\ref{ScatteringTransportFig}), and assumes that transport of electrons occurs along different independent channels. Incident and reflected electrons are then represented in terms of the wave function
\begin{equation}
\begin{array}{l}
\Psi_E(\vec{r}) = 
\left\{
\begin{array}{lcl}
\sum_n a_{Ln} \psi_{LnE}^{\mathrm{(in)}}(\vec{r}) + \sum_n b_{Ln} \psi_{LnE}^{\mathrm{(r)}}(\vec{r}), &\mbox{if}& \vec{r}\in L.  \\
\Psi_{E}^{(M)}(\vec{r}), &\mbox{if}& \vec{r}\in M. \\
\sum_n a_{Rn} \psi_{RnE}^{\mathrm{(in)}}(\vec{r}) + \sum_n b_{Rn} \psi_{RnE}^{\mathrm{(r)}}(\vec{r}), &\mbox{if}& \vec{r}\in R.
\end{array}
\right.
\end{array}
\end{equation}
Here $\psi_{\alpha nE}^{\mathrm{(in)}/\mathrm{(r)}}$ is an incident/reflected electronic wave with energy $E$ in channel $n$ of lead $\alpha=L, R$, and the index $M$ denotes the central system. The amplitudes $a_{\alpha n}$, $b_{\alpha n}$ become operators after a canonical quantization, and are related through the \textit{scattering matrix} ${\cal S}$:
\begin{equation}
\begin{pmatrix}
        \hat{b}_{L1} \\
        \vdots \\
        \hat{b}_{LN_L} \\
        \hat{b}_{R1} \\
        \vdots \\
        \hat{b}_{RN_R}
\end{pmatrix} = 
        {\cal S}
\begin{pmatrix}
        \hat{a}_{L1} \\
        \vdots \\
        \hat{a}_{LN_L} \\
        \hat{a}_{R1} \\
        \vdots \\
        \hat{a}_{RN_R}
\end{pmatrix} \equiv 
\begin{pmatrix}
        r_{11} & \ldots & r_{1N_L} & t'_{11} & \ldots & t'_{1N_R} \\
        \vdots & & \vdots & \vdots & & \vdots \\
                r_{N_L1} & \ldots & r_{N_LN_L} & t'_{N_L1} & \ldots & t'_{N_LN_R} \\
        t_{11} & \ldots & t_{1N_L} & r'_{11} & \ldots & r'_{1N_R} \\
        \vdots & & \vdots & \vdots & & \vdots \\
        t_{N_L1} & \ldots & t_{N_LN_L} & r'_{N_L1} & \ldots & r'_{N_LN_R} \\
\end{pmatrix}
\begin{pmatrix}
        \hat{a}_{L1} \\
        \vdots \\
        \hat{a}_{LN_L} \\
        \hat{a}_{R1} \\
        \vdots \\
        \hat{a}_{RN_R}
\end{pmatrix},
\end{equation}
which is Hermitian: ${\cal S}{\cal S}^{\dagger} = {\cal S}^{\dagger} {\cal S} = \mathds{1}$. The transmission and reflection coefficients $r_{n'n}$, $r'_{n'n}$, $t_{n'n}$, $t'_{n'n}$, generally depend on energy. In many cases, however, it is a good approximation to neglect this energy dependence. These coefficients, or alternatively the transmission and reflection probabilities ($\mathsf{T}\equiv \mathsf{T_n}\equiv \sum_{nn'}t_{nn'}^{\dagger}t_{n'n}$ and $\mathsf{R}\equiv \mathsf{R_n} \equiv \sum_{nn'}r_{nn'}^{\dagger}r_{n'n}$ respectively), can be found e.g. by imposing continuity of the wave functions and of their first derivative at each junction. 

Using the scattering matrix, the wave function in the left and right lead can be written as
\begin{equation}
\begin{array}{l}
\Psi_{LnE}(\vec{r}) = 
\left\{
\begin{array}{lcl}
\psi_{LnE}^{\mathrm{(in)}}(\vec{r}) + \sum_{n'} r_{n'n} \psi_{Ln'E}^{\mathrm{(r)}}(\vec{r}), &\mbox{if}& \vec{r}\in L.  \\
\Psi_{E}^{(M)}(\vec{r}), &\mbox{if}& \vec{r}\in M. \\
\sum_{n'} t_{n'n} \psi_{Rn'E}^{\mathrm{(in)}}(\vec{r}), &\mbox{if}& \vec{r}\in R.
\end{array}
\right.\\ \\
\Psi_{RnE}(\vec{r}) = 
\left\{
\begin{array}{lcl}
\sum_{n'} t'_{n'n} \psi_{Ln'E}^{\mathrm{(r)}}(\vec{r}), &\mbox{if}& \vec{r}\in L.  \\
\Psi_{E}^{(M)}(\vec{r}), &\mbox{if}& \vec{r}\in M. \\
\psi_{RnE}^{\mathrm{(r)}}(\vec{r}) + \sum_{n'} r'_{n'n} \psi_{Rn'E}^{\mathrm{(in)}}(\vec{r}), &\mbox{if}& \vec{r}\in R.
\end{array}
\right.
\end{array}
\end{equation}
Here, the wave functions $\psi_{\alpha nE}$ describe a many-particle state and are typically assumed to be plane waves. The scattering matrix method is therefore particularly useful in situations where electron-electron interactions are weak. A standard system that is well described using this formalism is the quantum point contact. This system is particularly simple, since transport is ballistic, and therefore $\mathsf{T}\approx 1$. Atomic contacts behave similarly, and the number of channels in these systems, as well as the value of their transmission probabilities can be determined using the scattering matrix approach \cite{Agrait03}. This is done by means of the so-called Landauer-B\"uttiker formula, which relates the conductance through the device with the transmission probabilities. The current through a terminal $\alpha$ can be derived using the quantum-mechanical equation $\hat{I}_{\alpha}=\frac{e\hbar}{2mi} \left[ \Psi_{\alpha}^*(\vec{\nabla}\Psi_{\alpha}) - (\vec{\nabla}\Psi_{\alpha}^*)\Psi_{\alpha} \right]$, giving \cite{BlanterButtiker00}
\beq \label{currentLanButt}
\hat{I}_{\alpha}(t) = \frac{e}{h} \sum_n \int dE dE' e^{i(E-E')t/\hbar} \left[ \hat{a}_{\alpha n}^{\dagger}(E)\hat{a}_{\alpha n}(E') - \hat{b}_{\alpha n}^{\dagger}(E)\hat{b}_{\alpha n}(E') \right].
\eeq
Setting $E'=E+\hbar\omega$ and integrating over $\omega$, we get $\hat{I}_{\alpha}(t)=\frac{e}{h} \sum_n \int dE \left[ \hat{n}_{\alpha n}^{+}(E,t) - \hat{n}_{\alpha n}^{-}(E,t) \right]$, being $\hat{n}_{\alpha n}^{+}(E,t) \equiv \hat{a}_{\alpha n}^{\dagger}(E)\hat{a}_{\alpha n}(E)$ and $\hat{n}_{\alpha n}^{-}(E,t) \equiv \hat{b}_{\alpha n}^{\dagger}(E)\hat{b}_{\alpha n}(E)$ the time-dependent occupation numbers of the incoming and outgoing carriers. Therefore, Eq.~(\ref{currentLanButt}) can be interpreted as the incoming flow of particles minus the outgoing flow of particles at each channel, and summed over all channels. Defining the quantity $A_{\beta\gamma}^{mn}(\alpha;E,E'):=\delta_{mn}\delta_{\alpha\beta}\delta_{\alpha\gamma} - \sum_k s_{\alpha\beta m k}^{\dagger}(E) s_{\alpha\gamma k n}(E')$, with $s_{\alpha\beta m n}$ given by the relation $\hat{b}_{\alpha m}(E) = \sum_{\beta n} s_{\alpha\beta m n}(E) \hat{a}_{\beta n}(E)$, expression (\ref{currentLanButt}) can be rewritten in terms of incoming amplitudes and the scattering matrix:
\beq
\hat{I}_{\alpha}(t) = \frac{e}{h} \sum_{\beta\gamma m n} \int dE dE' e^{i(E-E')t/\hbar} \hat{a}_{\beta m}^{\dagger}(E) A_{\beta\gamma}^{mn}(\alpha;E,E') \hat{a}_{\gamma n}(E').
\eeq
This permits us to calculate the average current flowing from a terminal $\alpha$ at chemical potential $\mu_{\alpha}=E_F+eV/2$ to a terminal $\beta$ at chemical potential $\mu_{\beta}=E_F-eV/2$.
Using $\ew{\hat{a}_{\alpha m}^{\dagger}(E)\hat{a}_{\beta n}(E')}=\delta_{\alpha\beta}\delta_{mn}\delta(E-E')f_{\alpha}(E)$, with Fermi function $f_{\alpha}$, it yields
\beq \label{LandauerButtikerMeanCurrent}
\ew{\hat{I}} = \frac{e}{h} \sum_n \int dE \; \mathsf{T_n}(E) \left[ f(E-E_F-eV/2)-f(E-E_F+eV/2) \right].
\eeq
In the linear response regime, $V\ll E-E_F, k_BT$, this expression can be approximated by $\ew{\hat{I}} = \frac{e^2}{h} \sum_n \int dE \; \mathsf{T_n}(E) \left( -\frac{\partial f}{\partial E} \right) V$. Furthermore, at low temperatures, $k_B T \ll E-E_F$, we have $( -\partial f/\partial E )\approx \delta(E-E_F)$, arriving to the \textit{Landauer-B\"uttiker formula} for the conductance\footnote{Notice that a factor $2$ can be included in this expression to account for the spin degeneracy. We then recover the conductance quantization in multiples of $G_0\equiv 2e^2/h$ as discussed in the previous chapter.} $G=d\ew{\hat{I}}dV\vert_{V=0}$:
\beq
G=\frac{e^2}{h}\sum_n \mathsf{T_n}.
\eeq
The transmission probabilities are evaluated at the Fermi energy of the reservoirs, but at low voltages and temperatures they can be assumed to be energy-independent. We therefore see that the conduction of electrons occurs along independent transmission channels. The noise spectrum $S^{(2)}_{\alpha\beta}=\int_{-\infty}^{\infty} dt e^{-i\omega t} {\cal T}_S \ew{I_{\alpha}(t) I_{\beta}(0)}$, with ${\cal T}_S$ the symmetrization operator defined in the previous chapter, can be also evaluated using the scattering matrix formalism. For a derivation see \cite{Buttiker92, Buttiker92b, BlanterButtiker00}. In the following, the upper sign denotes the fermionic case, and the lower sign the bosonic case. We have:
\beq \label{noiseLanButt}
S^{(2)}_{\alpha\beta}(\omega) = \frac{2e^2}{h} \sum_{\gamma\delta m n} \int dE A_{\gamma\delta}^{mn}(\alpha;E,E+\hbar\omega) A_{\delta\gamma}^{nn}(\beta;E+\hbar\omega,E) \nonumber\\ \times \left\{ f_{\gamma}(E) [ 1\mp f_{\delta}(E+\hbar\omega) ] + [1\mp f_{\gamma}(E)] f_{\delta}(E+\hbar\omega) \right\},
\eeq
equation that, assuming a voltage difference $V$ between terminals $\alpha$ and $\beta$ and energy-independent transmission coefficients, gives the general expression (\ref{FDTfull}) for $S_{\alpha\alpha}^{(2)}(\omega)$ discussed in the previous chapter. At zero frequency, Eq.~(\ref{noiseLanButt}) takes the form
\beq \label{S0LanButt}
S_{\alpha\alpha}^{(2)}(0)=-S_{\alpha\beta}^{(2)}(0)=\frac{2e^2}{h} \sum_n \int dE \left\{ \mathsf{T_n}(E) [ f_{\alpha}(1\mp f_{\alpha}) + f_{\beta} (1\mp f_{\beta}) ] \right. \nonumber\\ \pm \left. \mathsf{T_n}(E) [1-\mathsf{T_n}(E)](f_{\alpha}-f_{\beta})^2 \right\},
\eeq
being here $\beta \neq \alpha$, and $f_{\alpha}=f(E-E_F-V/2)$, $f_{\beta}=f(E-E_F+V/2)$. In the zero-temperature limit, Eq. (\ref{S0LanButt}) gives $S^{(2)}_{\alpha\alpha}(0)=\frac{2e^3V}{h} \sum_n \mathsf{T_n}(1-\mathsf{T_n})$, which in the tunneling regime ($\mathsf{T_n}\ll 1$) reduces to the Schottky limit $2e\ew{\hat{I}}$. We conclude that within this non-interacting picture, the zero-frequency shot noise is always sub-Poissonian. It is therefore convenient to generally express the noise in terms of the Fano factor $F\equiv \frac{S^{(2)}}{e\ew{\hat{I}}}=\frac{2\sum_n \mathsf{T_n}(1-\mathsf{T_n})}{\sum_n \mathsf{T_n}}$, which is zero for perfectly transmitting or reflecting channels, and maximal and equal to $1$ for $\mathsf{T_n}=1/2$.

\begin{figure}
  \begin{center}
    \includegraphics[width=0.6\textwidth]{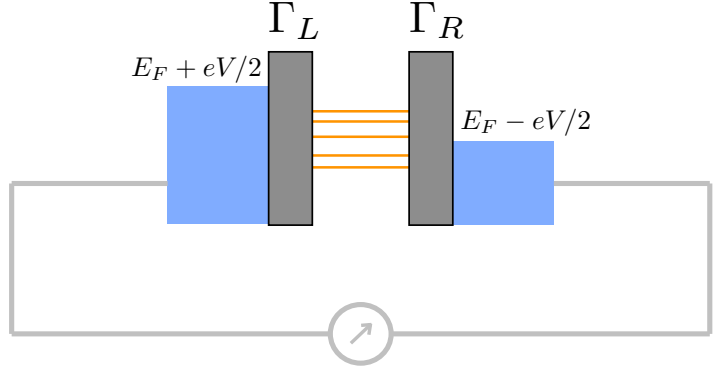}
  \end{center}  
  \caption[Quantum well]{Double-barrier structure. The system consists of a quantum well, with quantized energy levels. Each level $n$ is coupled to electronic reservoirs at different chemical potentials $E_F+eV/2$ and $E_F-eV/2$ with coupling rates $\Gamma_{Ln}$, $\Gamma_{Rn}$ respectively.
}
  \label{DoubleBarrierFig}
\end{figure}

The method is well exemplified in a resonant double-barrier structure (see Fig.~\ref{DoubleBarrierFig}). The transmission through this system can be written in the form \cite{Stone-Lee85, BlanterButtiker00}
\beq \label{TransmissionQuantumWell}
\mathsf{T}(E) \approx \sum_n \mathsf{T_n}^{\mathrm{max}} \frac{\Gamma_n^2/4}{(E-E_n^r)^2 + \Gamma_n^2/4},
\eeq
where $\mathsf{T_n}^{\mathrm{max}} \equiv 4\Gamma_{Ln}\Gamma_{Rn}/\Gamma_n^2$, being $\Gamma_n:=\Gamma_{Ln}+\Gamma_{Rn}$, and the rate $\Gamma_{Ln}$/$\Gamma_{Rn}$ accounting for the tunneling through channel $n$ from/to reservoir $L$/$R$. The energy $E_n^r$ is the $n$th. resonant energy in the well. This result, together with (\ref{S0LanButt}), will be used in the following chapters as one of the possible checks of our theory.

\subsection{The density operator} \label{SecDensityOperator}

The density operator, introduced in the previous chapter (c.f. section \ref{moreCoherence}), is also a powerful tool to study quantum-transport systems\footnote{For a more extended study of the density operator formalism, c.f. for example \cite{Cohen, BlumBook}.}. Its potential becomes apparent in quantum systems with strong interactions and coupled to a bath with many degrees of freedom. As mentioned, the density operator captures both the populations and coherences in the system, and its dynamics can be evaluated solving the von Neumann's equation\footnote{For this discussion we will use $\hbar=1$.}
\beq \label{vonNeumann}
\frac{d}{dt}\hat{\rho}(t) = -i [ \hat{{\cal H}}, \hat{\rho}(t) ].
\eeq
If the system is in thermodynamic equilibrium at temperature $T$, the density operator reads $\hat{\rho}=\mathrm{exp}\left(-\frac{\hat{{\cal H}}}{k_B T}\right)/\mathrm{Tr}\left\{ \mathrm{exp}\left(-\frac{\hat{{\cal H}}}{k_B T}\right) \right\}$ for a canonical ensemble (with well defined mean energy), or $\hat{\rho}=\mathrm{exp}\left(-\frac{\hat{{\cal H}}-\mu\hat{N}}{k_B T}\right)/\mathrm{Tr}\left\{ \mathrm{exp}\left(-\frac{\hat{{\cal H}}-\mu\hat{N}}{k_B T}\right) \right\}$ for a grand canonical ensemble (with well defined mean energy and number of particles), being $\mu$ the chemical potential and $\hat{N}$ the operator describing the number of particles.
Here, however, we will be generally interested in out-of-equilibrium situations, therefore needing to solve Eq.~(\ref{vonNeumann}) for the specific problem to determine the density operator. The typical situations of our interest can, nevertheless, be described by a Hamiltonian with no explicit time dependence and of the form
\beq \label{Hform}
{\cal \hat{H}} = {\cal \hat{H}}_\mathrm{S} + {\cal \hat{H}}_\mathrm{R} + {\cal \hat{H}}_\mathrm{V},
\eeq
where each of the terms denotes system, reservoir, and coupling between both respectively. If this system-reservoir coupling is sufficiently small, we can treat ${\cal \hat{H}}_\mathrm{V}$ perturbatively\footnote{We will be more specific about the perturbative parameter below.}. First, it is convenient to work in the interaction picture with respect to ${\cal \hat{H}}_\mathrm{S} + {\cal \hat{H}}_\mathrm{R}$, where an operator takes the form
\beq \label{InteractionPictureDef}
\hat{\tilde{{\cal O}}} (t) = e^{i ( {\cal \hat{H}}_\mathrm{S} + {\cal \hat{H}}_\mathrm{R} )t} {\cal \hat{O}}  e^{-i ( {\cal \hat{H}}_\mathrm{S} + {\cal \hat{H}}_\mathrm{R} )t}.
\eeq
Changing the density operator and the interaction Hamiltonian accordingly, the von Neumann's equation reads
\beq \label{vonNeumannInt}
\frac{d}{dt}\hat{\tilde{\rho}}(t) = -i [ \hat{\tilde{{\cal H}}}_\mathrm{V}(t) , \hat{\tilde{\rho}}(t) ],
\eeq
whose integral form $\hat{\tilde{\rho}} (t+\Delta t) = \hat{\tilde{\rho}} (t) -i \int_t^{t+\Delta t} dt' [ \hat{\tilde{{\cal H}}}_\mathrm{V} (t'), \hat{\tilde{\rho}}(t') ]$ can be iterated to second order in $\hat{\tilde{{\cal H}}}_\mathrm{V}$ to give
\beq \label{Deltarho1}
\Delta\hat{\tilde{\rho}} (t) = -i \int_t^{t+\Delta t} dt' [ \hat{\tilde{{\cal H}}}_\mathrm{V} (t'), \hat{\tilde{\rho}}(t) ] - \int_t^{t+\Delta t} dt' \int_t^{t'} dt'' [ \hat{\tilde{{\cal H}}}_\mathrm{V} (t'), [ \hat{\tilde{{\cal H}}}_\mathrm{V} (t''), \hat{\tilde{\rho}}(t'') ]],
\eeq
being $\Delta\hat{\tilde{\rho}} (t) \equiv \hat{\tilde{\rho}} (t+\Delta t) - \hat{\tilde{\rho}} (t)$. We will consider here the case in which the interaction is a bilinear product of system ($\hat{S}$) and reservoir ($\hat{R}$) operators, namely\footnote{A more general form $\hat{\cal H}_\mathrm{V} = \sum_p \hat{S}_p\hat{R}_p^{\dagger} + \mathrm{H.c.}$ can be also considered. In this case, attention must be paid to possible cross terms of the form $S_p(t'')R_p^{\dagger}(t') S_q(t'') R_q^{\dagger}(t'')$ arising in the master equation.} $\hat{\cal H}_\mathrm{V} = \hat{S}\hat{R}^{\dagger} + \mathrm{H.c.}$ Furthermore, we are interested in the system's dynamics only, for which we trace over the reservoir degrees of freedom. Finally, we consider the so-called \textit{Born approximation}, which considers the reservoir large enough so that the density operator can be written as a separable state at all times: $\hat{\tilde{\rho}}(t) = \hat{\tilde{\rho}}_\mathrm{S}(t) \otimes \hat{\tilde{\rho}}_\mathrm{R}(t)$, with $\hat{\tilde{\rho}}_\mathrm{R}$ not affected by the state of the system, thereby being in a stationary state at thermal equilibrium described by a canonical/grand canonical distribution. Doing this, we have
\beq \label{Deltarho2}
\Delta\hat{\tilde{\rho}}_\mathrm{S} (t) = &-& i \int_t^{t+\Delta t} dt' \mathrm{Tr}_\mathrm{R}[ \hat{\tilde{S}}(t')\hat{\tilde{R}}^{\dagger}(t') + \mathrm{H.c.}, \hat{\tilde{\rho}}_\mathrm{S}(t) \otimes \hat{\tilde{\rho}}_\mathrm{R} ] \nonumber\\&-& \int_t^{t+\Delta t} dt' \int_t^{t'} dt'' \mathrm{Tr}_\mathrm{R}[ \hat{\tilde{S}}(t')\hat{\tilde{R}}^{\dagger}(t') + \mathrm{H.c.}, [ \hat{\tilde{S}}(t'')\hat{\tilde{R}}^{\dagger}(t'') + \mathrm{H.c.}, \hat{\tilde{\rho}}_\mathrm{S}(t'') \otimes \hat{\tilde{\rho}}_\mathrm{R} ]]. \nonumber\\
\eeq
Notice that the first term in this equation can be written as $-i\int_t^{t+\Delta t} dt' [\hat{\tilde{{\cal H}}}'_\mathrm{S} (t'),\hat{\tilde{\rho}}_\mathrm{S}(t)]$, with $\hat{\tilde{{\cal H}}}'_\mathrm{S} (t') \equiv \hat{\tilde{S}}(t')\ew{\hat{\tilde{R}}^{\dagger}(t')} + \mathrm{H.c.}$, therefore corresponding to a shift in the system's Hamiltonian $\hat{\cal H}_\mathrm{S}$. We here will take $\ew{\hat{\tilde{R}}^{\dagger}(t)} = 0$, so the first term in Eq.~(\ref{Deltarho2}) vanishes. In the case $\ew{\hat{R}^{\dagger}(t)} \neq 0$, this term can also be eliminated redefining the system's Hamiltonian as $\hat{\cal H}_\mathrm{S} \to \hat{\cal H}_\mathrm{S} - e^{-i ( {\cal \hat{H}}_\mathrm{S} + {\cal \hat{H}}_\mathrm{R} )t} \frac{1}{\Delta t} \int_t^{t+\Delta t} dt' \left( \hat{\tilde{S}}(t')\ew{\hat{\tilde{R}}^{\dagger}(t')} + \mathrm{H.c.} \right) e^{i ( {\cal \hat{H}}_\mathrm{S} + {\cal \hat{H}}_\mathrm{R} )t}$. The second term in Eq.~(\ref{Deltarho2}) can be simplified in terms of the correlation functions
\begin{equation}\label{corrFunc}
\begin{array}{l}
\xi_{++} (t',t'') := \ew{\hat{\tilde{R}}^{\dagger}(t')\hat{\tilde{R}}^{\dagger}(t'')}, \\
\xi_{+-} (t',t'') := \ew{\hat{\tilde{R}}^{\dagger}(t')\hat{\tilde{R}}(t'')}, \\
\xi_{-+} (t',t'') := \ew{\hat{\tilde{R}}(t')\hat{\tilde{R}}^{\dagger}(t'')}, \\
\xi_{--} (t',t'') := \ew{\hat{\tilde{R}}(t')\hat{\tilde{R}}(t'')}, \\
\end{array}
\end{equation}
which obey the symmetry properties $\xi_{--}^* (t',t'') = \xi_{++}(t'',t')$, $\xi_{-+}^* (t',t'') = \xi_{-+}(t'',t')$, and $\xi_{+-}^* (t',t'') = \xi_{+-}(t'',t')$. We find
\beq \label{QMEgeneral}
\frac{\Delta\hat{\tilde{\rho}}_\mathrm{S}}{\Delta t} = - \frac{1}{\Delta t} \int_t^{t+\Delta t} dt' \int_t^{t'} dt'' \left\{ \xi_{-+}(t',t'') \left( \hat{\tilde{S}}^{\dagger}(t')\hat{\tilde{S}}(t'') \hat{\tilde{\rho}}_\mathrm{S}(t'') - \hat{\tilde{S}}(t'') \hat{\tilde{\rho}}_\mathrm{S}(t'')\hat{\tilde{S}}^{\dagger}(t') \right) \right. \nonumber\\ + \xi_{+-}(t',t'') \left( \hat{\tilde{S}}(t')\hat{\tilde{S}}^{\dagger}(t'') \hat{\tilde{\rho}}_\mathrm{S}(t'') - \hat{\tilde{S}}^{\dagger}(t'') \hat{\tilde{\rho}}_\mathrm{S}(t'')\hat{\tilde{S}}(t') \right) \nonumber\\ +
\xi_{--}(t',t'') \left( \hat{\tilde{S}}^{\dagger}(t')\hat{\tilde{S}}^{\dagger}(t'') \hat{\tilde{\rho}}_\mathrm{S}(t'') - \hat{\tilde{S}}^{\dagger}(t'') \hat{\tilde{\rho}}_\mathrm{S}(t'')\hat{\tilde{S}}^{\dagger}(t') \right) \nonumber\\ \left. + \xi_{++}(t',t'') \left( \hat{\tilde{S}}(t')\hat{\tilde{S}}(t'') \hat{\tilde{\rho}}_\mathrm{S}(t'') - \hat{\tilde{S}}(t'') \hat{\tilde{\rho}}_\mathrm{S}(t'')\hat{\tilde{S}}(t') \right) + \mathrm{H.c.} \right\}.
\eeq
This general form of the quantum master equation (QME), can be further simplified if a series of assumptions are followed.
First, let us notice that the correlation functions (\ref{corrFunc}) only depend on the time difference $\tau:=t'-t''$ or on the sum of times $T:=\frac{t'+t''}{2}$, that is $\xi_{-+}=\xi_{-+}(\tau)$, $\xi_{+-}=\xi_{+-}(\tau)$, $\xi_{--}=\xi_{--}(T)$, and $\xi_{++}=\xi_{++}(T)$, and similarly for the corresponding product of system operators. This motivates us to change to these new variables in the integrals: $\int_t^{t+\Delta t} dt' \int_t^{t'} dt'' \to \int_0^{2t+\Delta t} d\tau \int_t^{t+\Delta t} dT$. Now, let us notice that the QME (\ref{QMEgeneral}) contains three different time scales:
\begin{itemize}
\item $\tau_c$: Bath-correlation time, determined by the decay of the correlation functions as\footnote{Here $\mathrm{e}$ is the number $\mathrm{e} = 2.71828...$} $\xi_{+-}(\tau_c)=\xi_{+-}(0)/\mathrm{e}$.
\item $\Delta t$: Time needed for the system to change appreciably.
\item $T_S$: Typical system's evolution time. This time is of the order $1/T_S\sim v^2\tau_c$, where $v\equiv\sqrt{\ew{{\cal \hat{H}}_V^2}}$.
\end{itemize}
The Born approximation introduced above is based on the assumption that the initial correlations between system and bath disappear after a time $\tau_c$. Generally, the initial density operator would have the form $\hat{\tilde{\rho}}(t) = \hat{\tilde{\rho}}_\mathrm{S}(t) \otimes \hat{\tilde{\rho}}_\mathrm{R}(t) + \hat{\tilde{\rho}}_\mathrm{corr}(t)$, with the last term accounting for system-bath correlations. However, if $\Delta t \gg \tau_c$ this term can neglected, and the Born approximation is valid. This limit permits us to extend the upper limit of the integral $\int_0^{2t+\Delta t} d\tau$ to infinity, since the correlation functions $\xi_{+-}(\tau)$ and $\xi_{+-}(\tau)$ decay in a time much faster than $\Delta t$. A further approximation is to assume that the system's dynamics is local in time, which is known as \textit{Markov approximation}: If $T_S\gg \Delta t$, then it is appropriate to substitute $\hat{\tilde{\rho}}_\mathrm{S}(t'')$ by $\hat{\tilde{\rho}}_\mathrm{S}(t)$ in the integrals, since the state of the system during $\Delta t$ changes only slightly. This simplifies greatly the QME, but as we will see in the next two chapters, it has profound consequences in certain parameter regimes, such as neglecting the physics of vacuum fluctuations. This limit also allows us to replace $\frac{\Delta \hat{\tilde{\rho}}_\mathrm{S}}{\Delta t}$ by the derivative $\frac{d \hat{\tilde{\rho}}_\mathrm{S}}{d t}$ in equation (\ref{QMEgeneral}). Finally, following the assumption $\ew{\hat{\tilde{R}}^{\dagger}}=0$, we will take here the correlation functions $\xi_{++}$ and $\xi_{--}$ to vanish, since they involve the average of two $\hat{\tilde{R}}^{\dagger}$ or $\hat{\tilde{R}}$ operators. With these assumptions we find the QME
\beq \label{QMEsimplified}
\frac{d\hat{\tilde{\rho}}_\mathrm{S}(t)}{dt} = - \int_0^{\infty} d\tau \left\{ \xi_{-+}(\tau) \left( \hat{\tilde{S}}^{\dagger}(\tau/2)\hat{\tilde{S}}(-\tau/2) \hat{\tilde{\rho}}_\mathrm{S}(t) - \hat{\tilde{S}}(-\tau/2) \hat{\tilde{\rho}}_\mathrm{S}(t)\hat{\tilde{S}}^{\dagger}(\tau/2) \right) \right. \nonumber\\ \left. + 
\xi_{+-}(\tau) \left( \hat{\tilde{S}}(\tau/2)\hat{\tilde{S}}^{\dagger}(-\tau/2) \hat{\tilde{\rho}}_\mathrm{S}(t) - \hat{\tilde{S}}^{\dagger}(-\tau/2) \hat{\tilde{\rho}}_\mathrm{S}(t)\hat{\tilde{S}}(\tau/2) \right) + \mathrm{H.c.} \right\},
\eeq
where it is important to remember the premise
\beq
\tau_c \ll \Delta t \ll T_S,
\eeq
which is crucial not only for the approximations above to be well-founded, but also for the justification of the perturbation theory in $\hat{\cal H}_\mathrm{V}$, that has $v\tau_c$ as perturbative parameter.
After integration, Eq.~(\ref{QMEsimplified}) can be written as
\beq \label{SelfEnergyEquation}
\frac{d\hat{\tilde{\rho}}_\mathrm{S}(t)}{dt} = \tilde{\Sigma} \hat{\tilde{\rho}}_\mathrm{S}(t),
\eeq
where $\tilde{\Sigma}$ is the Markovian self-energy.
We note that if the integral corresponding to $\xi_{+-}$ is equal to its conjugate counterpart (and similarly for $\xi_{-+}$),\footnote{This is e.g. fulfilled given that the correlation functions are generally symmetric: $\xi_{+-}(\tau)=\xi_{+-}^*(\tau)=\xi_{+-}(-\tau)$, and similarly for $\xi_{-+}(\tau)$.} this master equation (ME) has a Lindblad form \cite{Lindblad76}, that is, the self-energy is of the type $\Sigma=-\hat{\cal O}^{\dagger}\hat{\cal O}\hat{\rho} + 2 \hat{\cal O} \hat{\rho}\hat{\cal O}^{\dagger} - \hat{\rho} \hat{\cal O}^{\dagger} \hat{\cal O}$, for each of the terms it comprises. This immediately guarantees the required positivity of the density operator at all times.

As a first example of the method, let us consider a two-level atom coupled to the radiation field. The Hamiltonian of this problem is of the form (\ref{Hform}), with $\hat{\cal H}_\mathrm{S}= \frac{\omega_q}{2} (\op{1}{1}-\op{0}{0})$, $\hat{\cal H}_\mathrm{R}= \sum_k \omega_k (\hat{a}_k^{\dagger}\hat{a}_k +\frac{1}{2})$ and, if atom and field are resonant, $\hat{\cal H}_\mathrm{V}^\mathrm{RWA}=\sum_k g_k \op{1}{0}\hat{a}_k + \mathrm{H.c.}$ Thus, taking $\hat{R}=\sum_k g_k \hat{a}_k$ and $\hat{S}=\op{0}{1}$, the Markovian dynamics corresponding to this model can be determined using Eq.~(\ref{QMEsimplified}), which, projected onto the basis $\{\ket{0},\ket{1}\}$, gives the following master equations\footnote{This set of MEs are already in the Schr\"odinger picture, since in the present case the coherent term $-i[{\cal \hat{H}}_\mathrm{S},\hat{\rho}_\mathrm{S}]$ vanishes when it is projected onto the $\{\ket{0},\ket{1}\}$ basis. From here on, we will drop the subscript $\mathrm{S}$ in the elements of the system density operator.}:
\begin{equation}\label{BlochEqsDissEB}
\begin{array}{l}
\dot{\rho}_{00} = -\gamma \ew{n} \rho_{00} + \gamma \left( 1+\ew{n} \right) \rho_{11}. \\
\dot{\rho}_{01} = -\frac{\gamma - i\Delta}{2} \left( 1+2\ew{n} \right) \rho_{01}. \\
\dot{\rho}_{10} = -\frac{\gamma + i\Delta}{2} \left( 1+2\ew{n} \right) \rho_{10}. \\
\dot{\rho}_{11} = \gamma \ew{n} \rho_{00} - \gamma \left( 1+\ew{n} \right) \rho_{11}.
\end{array}
\end{equation}
Here, $\gamma\equiv 2\pi \sum_k |g_k|^2 \delta(\omega_q - \omega_k)$ is a decay rate, $\Delta \equiv 2\sum_k |g_k|^2 P\left( \frac{1}{\omega_q-\omega_k} \right)$ (with $P$ the principal value) corresponds to an energy shift, and $\ew{n}\equiv \int_{-\infty}^{\infty} d\omega {\cal D}(\omega) b(\omega)$ (with ${\cal D}(\omega)$ the bath density of states and $b(\omega)$ the Bose distribution) is the average number of photons in the bath. For a fermionic bath, these equations have the same form, but with the factor $\left( 1+\ew{n} \right)$ in the populations substituted by $\left( 1-\ew{n} \right)$, and the factor $\left( 1+2\ew{n} \right)$ in the coherences substituted by $1$. As we will see, there is an analogy between this fermionic case and the so-called single resonant level model, which we explain below. It is important to notice that the previous model considers atom and field coupled through the dipolar operator corresponding to the diagonal basis of ${\cal \hat{H}}_\mathrm{S}$, namely $\ket{0}\bra{1}+\mathrm{H.c.}$ More interesting is the case in which atom and field are brought into resonance sufficiently fast, and the coupling is realized through the dipolar operator corresponding to the bare atom. This situation can be described with the same model, but with ${\cal \hat{H}}_\mathrm{S}$ being in this case the two-level system Hamiltonian (\ref{H2LS}). The MEs include now the coherent dynamics coming from the term $-i[{\cal \hat{H}}_\mathrm{S},\hat{\rho}_\mathrm{S}]$:
\begin{equation}\label{BlochEqsDissLB}
\begin{array}{l}
\dot{\rho}_{00} = \frac{i\lambda}{2} \left( \rho_{01} - \rho_{10} \right) -\gamma \ew{n} \rho_{00} + \gamma \left( 1+\ew{n} \right) \rho_{11}. \\
\dot{\rho}_{01} = \frac{i\lambda}{2} \left( \rho_{00} - \rho_{11} \right) + i\varepsilon\rho_{01} -\frac{\gamma - i\Delta}{2} \left( 1+2\ew{n} \right) \rho_{01}. \\
\dot{\rho}_{10} = -\frac{i\lambda}{2} \left( \rho_{00} - \rho_{11} \right) - i\varepsilon\rho_{10} -\frac{\gamma + i\Delta}{2} \left( 1+2\ew{n} \right) \rho_{10}. \\
\dot{\rho}_{11} = -\frac{i\lambda}{2} \left( \rho_{01} - \rho_{10} \right) + \gamma \ew{n} \rho_{00} - \gamma \left( 1+\ew{n} \right) \rho_{11}.
\end{array}
\end{equation}
These, correspond to the previously introduced Bloch equations (\ref{BlochEqs}), with $\lambda=2g$, $\varepsilon=\omega_q$, and generalized to include the dissipative dynamics induced by the bath. It is remarkable that the dissipative part is exactly the same in this case. In general, one would expect to find here the overlaps between dressed and bare states. However, these two are connected through an orthogonal transformation (c.f. appendix \ref{2LSappendix}), and as a consequence, the dissipative dynamics takes the same form. Below we will see that Eqs.~(\ref{BlochEqsDissLB}) are in close analogy to the MEs describing a double quantum dot connected to electronic leads.

If we project equation (\ref{QMEgeneral}) onto the eigenbasis of $\hat{{\cal H}}_\mathrm{S}$, a general form for the density-matrix dynamics can be derived. Let $\{\ket{a}\}$ denote such a basis, that is $\hat{{\cal H}}_\mathrm{S}\ket{a}=\omega_a\ket{a}$. Neglecting the terms $\xi_{--}$ and $\xi_{++}$, we can write
\beq \label{nonSecME}
\frac{\Delta\tilde{\rho}_{ab}(t)}{\Delta t} = \sum_{cd} \frac{1}{\Delta t} \int_t^{t+\Delta t} dT e^{i(\omega_{ab}-\omega_{cd})T} {\cal R}_{abcd} \tilde{\rho}_{cd}(t),
\eeq
with $\omega_{ab}\equiv \omega_a - \omega_b$, and ${\cal R}$ the so-called \textit{Bloch-Redfield} tensor, defined by
\beq \label{BlochRedfieldTensor}
{\cal R}_{abcd} = - \int_0^{\infty} d\tau \Big\{ \xi_{-+}(\tau) \Big( \sum_n e^{i\omega_{an}\tau/2} e^{-i\omega_{nc}\tau/2} S^{\dagger}_{an}S_{nc} \delta_{bd} - e^{-i\omega_{ac}\tau/2} e^{i\omega_{db}\tau/2} S_{ac}S^{\dagger}_{db} \Big) \nonumber\\ \xi_{+-}(\tau) \Big( \sum_n e^{i\omega_{an}\tau/2} e^{-i\omega_{nc}\tau/2} S_{an}S^{\dagger}_{nc} \delta_{bd} - e^{-i\omega_{ac}\tau/2} e^{i\omega_{db}\tau/2} S^{\dagger}_{ac}S_{db} \Big) + \mathrm{H.c.} \Big\}. \;
\eeq
The integral $\frac{1}{\Delta t} \int_t^{t+\Delta t} dT e^{i(\omega_{ab}-\omega_{cd})T} = e^{i(\omega_{ab}-\omega_{cd})t} f[( \omega_{ab} - \omega_{cd} ) \Delta t]$, with $f[x]:=e^{ix/2}\frac{\mathrm{sin}(x/2)}{x/2}$, is vanishingly small for $|\omega_{ab}-\omega_{cd}|\gg \frac{1}{\Delta t}$, and approximately $1$ for $|\omega_{ab}-\omega_{cd}|\ll \frac{1}{\Delta t}$. This means that we can restrict the sum in (\ref{nonSecME}) to terms satisfying $|\omega_{ab}-\omega_{cd}|=0$. This is called \textit{secular approximation}, and will be denoted with a superscript $\mathrm{(sec)}$. Going back to the Schr\"odinger picture, we then find the QME
\beq \label{BlochRedfieldEq}
\dot{\rho}_{ab}(t) = -i \omega_{ab} \rho_{ab}(t) + \sum_{c,d}{}^{\mathrm{(sec)}} {\cal R}_{abcd} \rho_{cd} (t)
\eeq
For the populations, this equation can be written in a particularly simple and intuitive form, named \textit{Pauli rate equation}:
\beq \label{PauliEq}
\dot{\rho}_{aa}(t) = \sum_{n \neq a} \Gamma_{an} \rho_{nn}(t)  -\sum_{n \neq a} \Gamma_{na} \rho_{aa}(t),
\eeq
with
\beq
\Gamma_{na} \equiv 2\pi \sum_{\alpha\nu} \bra{\alpha}\hat{\rho}_R\ket{\alpha} \Big\vert \bra{\nu,n} {\cal \hat{H}}_\mathrm{V} \ket{\alpha,a} \Big\vert^2 \delta(\omega_{\nu}+\omega_n-\omega_{\alpha}-\omega_a).
\eeq
This rate represents the probability of going from $\ket{a}$ to $\ket{n}$ in the system, and Eq.~(\ref{PauliEq}) can be therefore interpreted as gain in $\ket{a}$ minus loss from $\ket{a}$.

In the Schr\"odinger picture, Eq.~(\ref{SelfEnergyEquation}), that is, a general Markovian master equation for the system density operator, can be written in the form
\beq \label{rhoSevolution}
\frac{d\hat{\rho}_\mathrm{S}(t)}{dt} = {\cal W} \hat{\rho}_\mathrm{S}(t),
\eeq
where ${\cal W}\equiv{\cal L}_\mathrm{S}+\Sigma$, being ${\cal L}_\mathrm{S}$ the Liouvillian corresponding to ${\cal \hat{H}}_\mathrm{S}$, that is ${\cal L}_\mathrm{S}{\cal \hat{O}} \equiv -i [{\cal \hat{H}}_\mathrm{S}, {\cal \hat{O}}]$, with ${\cal \hat{O}}$ an operator. Notice that the kernel ${\cal W}$ appears after having traced out the reservoir degrees of freedom, and that Eq.~(\ref{rhoSevolution}) is local in time because of the Markovian approximation. The total system fulfills the general equation
\beq \label{LiouvilleEq}
\frac{d\hat{\rho}(t)}{dt} = {\cal L} \hat{\rho}(t), 
\eeq
with ${\cal L}$ the full system's \textit{Liouvillian super-operator}, whose action on an operator ${\cal \hat{O}}$ is defined through the von Neumann's equation: ${\cal L}{\cal \hat{O}} \equiv -i [{\cal \hat{H}}, {\cal \hat{O}}]$, so the Liouvillian components are ${\cal L}_{abcd} = -i ({\cal H}_{ac}\delta_{bd} - \delta_{ac} {\cal H}_{db})$. Taking matrix elements in Eq.~(\ref{rhoSevolution}) we find $\dot{\rho}_{ab} = \sum_{cd} {\cal W}_{abcd} {\cal \rho}_{cd}$. The kernel components can thus be related to the components of the Bloch-Redfield tensor. In Eq.~(\ref{BlochRedfieldEq}) there are three types of secular terms:
i) $a=b$, $c=d$, $a\neq c$. 
ii) $a=c$, $b=d$, $a\neq b$. 
iii) $a=b=c=d$.
This means that $\sum^{\mathrm{(sec)}} {\cal R}_{abcd} \rho_{cd} = \sum_{cd} [ \delta_{ab}\delta_{cd}(1-\delta_{ac}) + \delta_{ab}\delta_{cd}(1-\delta_{ab}) + \delta_{ab}\delta_{bc}\delta_{cd} ] {\cal R}_{abcd} \rho_{cd}$, and therefore we have
\beq
{\cal W}_{abcd} = -i \delta_{ac}\delta_{bd} \omega_{cd} + [ \delta_{ab}\delta_{cd}(1-\delta_{ac}) + \delta_{ab}\delta_{cd}(1-\delta_{ab}) + \delta_{ab}\delta_{bc}\delta_{cd} ] {\cal R}_{abcd}.
\eeq

A simple example that illustrates how the density-operator formalism is applied to quantum transport is the quantum-well model depicted in Fig.~\ref{DoubleBarrierFig}. Here we will assume for simplicity that only one level is defined in the well. This is the so-called single resonant level model and will be studied in detail below. Also, we assume here that the chemical potentials of the reservoirs are much larger than the rest of energy scales in the problem (infinite bias voltage approximation), such that transport is unidirectional. In this situation, the rate equations corresponding to this model read simply (c.f. (\ref{PauliEq}))
\begin{equation}\label{SRLrateEqs}
\begin{array}{l}
\dot{\rho}_{00} = -\Gamma_L \rho_{00} + \Gamma_R \rho_{11}. \\
\dot{\rho}_{11} = \Gamma_L \rho_{00} - \Gamma_R \rho_{11}.
\end{array}
\end{equation}
Here $\Gamma_{L/R}$ denotes the incoming/outgoing rate of electrons to the quantum dot (determined by the coupling with the reservoirs as we will see below), and it has been assumed that there is no coherence between dot and reservoirs.
The density operator formalism will be the approach mainly used in this thesis. As we will see, given the kernel ${\cal W}$ of the QME, we will be able to calculate current, noise, and high-order current-correlation functions in quantum transport problems.

\subsection{Green's functions}

Green's functions (GFs) are a widely used technique in condensed matter physics, where the dynamics dictated by the Schr\"odinger equation must be solved in many-particle systems. Their use can be extended to non-equilibrium problems (\textit{Keldysh} Green's functions) and therefore to quantum transport systems. Let us consider a problem of the type (\ref{Hform}). Defining ${\cal \hat{H}}_0 := {\cal \hat{H}}_\mathrm{S}+{\cal \hat{H}}_\mathrm{R}$, the associated Schr\"odinger equation is $\left[ i\partial_{t} - {\cal \hat{H}}_{0}(\vec{r}) - {\cal \hat{H}}_\mathrm{V}(\vec{r}) \right ] \Psi(\vec{r},t) = 0$, and the \textit{Green's function} $G$ corresponding to ${\cal \hat{H}}$ is defined through the relation
\begin{eqnarray} \label{GreenFuncDef}
\left[ i\partial_{t} - {\cal \hat{H}}_{0}(\vec{r}) - {\cal \hat{H}}_\mathrm{V}(\vec{r}) \right ] G(\vec{r},t;\vec{r}',t') = \delta(\vec{r}-\vec{r}')\delta(t-t'). 
\end{eqnarray} 
In a similar way, we define a `bare' Green's function $G_0$ as $\left[ i\partial_{t} - {\cal \hat{H}}_{0}(\vec{r}) \right ] G_{0}(\vec{r},t;\vec{r}',t') = \delta(\vec{r}-\vec{r}')\delta(t-t')$. If $\Psi^{(0)}(\vec{r},t)$ is an eigenfunction of ${\cal \hat{H}}_0$, that is, $\left[ i\partial_{t} - {\cal {H}}_{0}(\vec{r}) \right ] \Psi^{(0)}(\vec{r},t) = 0$, an eigenfunction of ${\cal \hat{H}}$ can be then expressed in terms of the integral equation
\begin{eqnarray}
\Psi(\vec{r},t)=\Psi^{(0)}(\vec{r},t) + \int d^{3}\vec{r}' \int dt' G_{0}(\vec{r},t;\vec{r}',t){\cal \hat{H}}_\mathrm{V}(\vec{r}')\Psi(\vec{r}',t'). 
\end{eqnarray}
This equation can be solved iteratively to give
\beq \label{PsiSeries}
\Psi=\Psi^{(0)} + \left( G_{0} + G_{0}{\cal \hat{H}}_\mathrm{V}G_{0} + G_{0}{\cal \hat{H}}_\mathrm{V}G_{0}{\cal \hat{H}}_\mathrm{V}G_{0} + \ldots \right){\cal \hat{H}}_\mathrm{V}\Psi^{(0)}.
\eeq
Here we have omitted the explicit spatial and time dependence, and the products should be understood with an implicit summation over the internal variables\footnote{Notice that in Fourier/Laplace space and projected on a certain basis, this equation has a matrix form, and then the product reduces to a simple matrix multiplication.}. Now, since the full Green's function $G$ fulfills $\Psi=\Psi^{(0)}+G{\cal \hat{H}}_\mathrm{V}\Psi^{(0)}$, upon comparison with (\ref{PsiSeries}), we find that $G$ can be calculated solving the \textit{Dyson equation}
\beq
G= G_0+G_0{\cal \hat{H}}_\mathrm{V}G \Longrightarrow G = \frac{1}{1-G_0{\cal \hat{H}}_\mathrm{V}} G_0. 
\eeq
It is immediate to check that $\Psi (\vec{r},t) = \int d^{3}\vec{r}' G(\vec{r},t;\vec{r}',t') \Psi(\vec{r}',t') $, so that the Green's function may be interpreted as a propagator. More specifically, it can be seen that the following six propagators also fulfill Eq.~(\ref{GreenFuncDef}):
\begin{equation}
\begin{array}{c}
G^{>}(x,x') := -i \langle \mathrm{vac} \vert \hat{\Psi}_{\sigma} (x) \hat{\Psi}_{\sigma'}^{\dagger} (x') \vert \mathrm{vac} \rangle. \\
G^{<}(x,x') := \pm i \langle \mathrm{vac} \vert \hat{\Psi}_{\sigma'}^{\dagger} (x') \hat{\Psi}_{\sigma} (x) \vert \mathrm{vac} \rangle. \\
G^{C}(x,x') := -i \theta(t-t') \langle \mathrm{vac} \vert \hat{\Psi}_{\sigma} (x) \hat{\Psi}_{\sigma'}^{\dagger} (x') \vert \mathrm{vac} \rangle \pm i \theta(t'-t) \langle \mathrm{vac} \vert \hat{\Psi}_{\sigma'}^{\dagger} (x') \hat{\Psi}_{\sigma} (x) \vert \mathrm{vac} \rangle. \\
G^{\overline{C}}(x,x') := -i \theta(t'-t) \langle \mathrm{vac} \vert \hat{\Psi}_{\sigma} (x) \hat{\Psi}_{\sigma'}^{\dagger} (x') \vert \mathrm{vac} \rangle \pm i \theta(t-t') \langle \mathrm{vac} \vert \hat{\Psi}_{\sigma'}^{\dagger} (x') \hat{\Psi}_{\sigma} (x) \vert \mathrm{vac} \rangle. \\
G^{R}(x,x') := -i \theta(t-t') \langle \mathrm{vac} \vert [ \hat{\Psi}_{\sigma} (x), \hat{\Psi}_{\sigma'}^{\dagger} (x') ]_{+,-} \vert \mathrm{vac} \rangle.  \\
G^{A}(x,x') := i \theta(t'-t) \langle \mathrm{vac} \vert [ \hat{\Psi}_{\sigma} (x), \hat{\Psi}_{\sigma'}^{\dagger} (x') ]_{+,-} \vert \mathrm{vac} \rangle.
\end{array}
\end{equation}
Here, $\ket{\mathrm{vac}}$ is the vacuum of the full system in the presence of the interaction (${\cal \hat{H}}\ket{\mathrm{vac}}=0$), $\theta(t)$ is the usual step function, $\sigma$ denotes the spin degree of freedom, the wave functions have become field operators after a canonical quantization, and the variables $(\vec{r},t)$ have been denoted with $(x)$. The symbol $+$ stands for fermions (therefore the anti-commutator) while $-$ stands for bosons (therefore the commutator). The previous GFs are named \textit{greater}, \textit{lesser}, \textit{time-ordered}, \textit{anti-time-ordered}, \textit{retarded}, and \textit{advanced}. For a free system (${\cal \hat{H}}_\mathrm{V}=0$) they can be easily calculated. To do so, we write the field operators as the eigen-decomposition $\hat{\Psi}(x) = \sum_{\lambda}\hat{c}_{\lambda}\psi_{\lambda}(\vec r) e^{-i \varepsilon_{\lambda} t}$, being $\{\varepsilon_{\lambda}\}$ a set of eigen-energies of ${\cal \hat{H}}_0$, and $\psi_{\lambda}(\vec{r})=\psi_{\vec{k}}(\vec{r})=(1/\sqrt{V})\mathrm{exp}\lbrace i\vec{k}\vec{r}\rbrace$ the corresponding eigenfunctions (with momentum $\vec{k}$ and volume $V$). The bare GFs in the momentum space and in the frequency domain then read for the fermionic case:
\begin{equation}
\begin{array}{ll}
G^{>}_{0}(\vec{k},\tau) = -i \left( 1-n_{\vec{k}} \right) e^{-i\varepsilon_{\vec{k}}\tau}; & G^{>}_{0}(\vec{k},\omega) = -2\pi i \left( 1-n_{\vec{k}} \right) \delta \left( \omega - \varepsilon_{\vec{k}} \right ); \\
G^{<}_{0}(\vec{k},\tau) = i n_{\vec{k}} e^{-i\varepsilon_{\vec{k}}\tau}; & G^{<}_{0}(\vec{k},\omega) = 2\pi i n_{\vec{k}} \delta \left( \omega - \varepsilon_{\vec{k}} \right); \\
G_{0}^{C}(\vec{k},\tau) = -i \left[ \theta(\tau)-n_{\vec{k}} \right] e^{-i\varepsilon_{\vec{k}}\tau}; & G_{0}^{C}(\vec{k},\omega) = \frac{1}{\omega - \varepsilon_{\vec{k}} +i \delta_{\vec{k}}}; \\
G_{0}^{\overline{C}}(\vec{k},\tau) = -i \left[ \theta(-\tau)-n_{\vec{k}} \right] e^{-i\varepsilon_{\vec{k}}\tau}; & G_{0}^{\overline{C}}(\vec{k},\omega) = \frac{-1}{\omega - \varepsilon_{\vec{k}} -i \delta_{\vec{k}}}; \\
G^{R}_{0}(\vec{k},\tau) = -i \theta(\tau) e^{-i\varepsilon_{\vec{k}}\tau}; & G^{R}_{0}(\vec{k},\omega) = \frac{1}{\omega - \varepsilon_{\vec{k}} +i \delta}; \\
G^{A}_{0}(\vec{k},\tau) = i \theta(-\tau) e^{-i\varepsilon_{\vec{k}}\tau}; & G^{A}_{0}(\vec{k},\omega) = \frac{1}{\omega - \varepsilon_{\vec{k}} -i \delta};
\end{array}
\end{equation}
where we have defined $\tau:=t-t'$, $n_{\vec{k}}:=\ew{c_{\vec{k}}^{\dagger}c_{\vec{k}}}$, $\delta_{\vec{k}} := \delta sign(\xi_{\vec{k}})$, $\xi_{\vec{k}} := \varepsilon_{\vec{k}}-\mu$ (with $\mu$ the system's chemical potential), and $\delta$ an infinitesimal positive number.

Notice that the zero-temperature GFs are defined in terms $\ket{\mathrm{vac}}$. This state, however, is generally unknown, since knowing it would mean to have in fact a solution to the full
problem. We therefore would like to express the GFs in terms of a well known state, namely the vacuum in the absence of interaction. To connect these two states we use the so-called $S$ matrix. This is defined in terms of the \textit{evolution operator}, which in the interaction picture reads $U(t):=e^{{\cal \hat{H}}_0 t}e^{-{\cal \hat{H}} t}$. This propagates the wave function as $\Psi(t)=U(t)\Psi(0)$, and fulfills $\frac{\partial}{\partial t}U(t)=-i{\cal \hat{H}}_\mathrm{V}(t)U(t)$. Therefore
\begin{eqnarray} \label{EvolutionOperatorEq}
U(t) &=& \mathds{1} + \sum_{n=1}^{\infty} (-i)^{n} \int_{0}^{t}dt_{1}\int_{0}^{t_{1}}dt_{2}\dots \int_{0}^{t_{n-1}}dt_{n}\hat{{\cal H}}_\mathrm{V}(t_{1})\hat{{\cal H}}_\mathrm{V}(t_{2})\dots \hat{{\cal H}}_\mathrm{V}(t_{n}) =\nonumber\\
&=& \mathds{1} + \sum_{n=1}^{\infty} \frac{(-i)^{n}}{n!} \int_{0}^{t}dt_{1}\int_{0}^{t}dt_{2}\dots \int_{0}^{t}dt_{n} {\cal T} \lbrace \hat{{\cal H}}_\mathrm{V}(t_{1})\hat{{\cal H}}_\mathrm{V}(t_{2})\dots \hat{{\cal H}}_\mathrm{V}(t_{n}) \rbrace =\nonumber\\
&=& {\cal T} \left\lbrace exp \left( -i \int_{0}^{t}dt'\hat{{\cal H}}_\mathrm{V}(t') \right) \right\rbrace,
\end{eqnarray}
where ${\cal T}$ is the time-ordering operator. The \textit{S matrix} is defined as $S(t,t') := U(t)U^{\dagger}(t')$ and is therefore given by
\beq
S(t,t')={\cal T} \left\lbrace exp \left( -i \int_{t'}^{t}dt''V(t'') \right) \right\rbrace.
\eeq
We can now write a state at time $t$ in terms of another at a previous general time $t'$ as $\Psi(t)=S(t,t')\Psi(t')$. In particular, we can think of the state $\ket{\mathrm{vac}}$ as coming from an adiabatic evolution of the free vacuum $\ket{0}$, defined by ${\cal \hat{H}}_0 \ket{0}=0$. That is, we can take the system at time $t=-\infty$ to be described by ${\cal \hat{H}}_0$, and then adiabatically switch the interaction ${\cal \hat{H}}_\mathrm{V}$, being able to write $\ket{\mathrm{vac}} = S(0,-\infty)\ket{0}$. Using the $S$ matrix, we can then express the GFs in terms of the free vacuum. For example, the causal (time-ordered) Green's function reads
\begin{eqnarray} \label{GcausalSmatrix}
G^{C}(\lambda,t;\lambda',t') = \frac{-i \langle 0 \vert {\cal T} \{\hat{\Psi}_{\lambda}(t)\hat{\Psi}_{\lambda'}^{\dagger}(t')\}S(-\infty,\infty)\vert 0 \rangle}{\langle 0 \vert S(-\infty,\infty)\vert 0 \rangle} \;\;\;\;\;\;\;\;\;\;\;\;\;\;\;\;\;\;\;\;\;\;\;\;\;\;\;\;\;\;\;\;\;\;\;\;\;\;\;\;\;\;\;\;\;\;\;\;\;\;\;\;\;\;\;\;\;\;\;\;\;\;\;\;\;\;\; \nonumber\\
= -i \sum_{n=0}^{\infty} \frac{(-i)^{n}}{n!} \int_{-\infty}^{\infty}dt_{1}\int_{-\infty}^{\infty}dt_{2}\dots \int_{-\infty}^{\infty}dt_{n}\frac{\langle 0 \vert {\cal T} \lbrace \hat{\Psi}_{\lambda}(t)\hat{\Psi}_{\lambda'}^{\dagger}(t') \hat{{\cal H}}_\mathrm{V}(t_{1})\hat{{\cal H}}_\mathrm{V}(t_{2})\dots \hat{{\cal H}}_\mathrm{V}(t_{n}) \rbrace \vert 0 \rangle}{\langle 0 \vert S(-\infty,\infty)\vert 0 \rangle} \;\;\;\;\;\;\; \nonumber\\
= -i \sum_{n=0}^{\infty} \frac{(-i)^{n}}{n!} \int_{-\infty}^{\infty}dt_{1}\int_{-\infty}^{\infty}dt_{2}\dots \int_{-\infty}^{\infty}dt_{n} \langle 0 \vert {\cal T} \lbrace \hat{\Psi}_{\lambda}(t)\hat{\Psi}_{\lambda'}^{\dagger}(t') \hat{{\cal H}}_\mathrm{V}(t_{1})\hat{{\cal H}}_\mathrm{V}(t_{2})\dots \hat{{\cal H}}_\mathrm{V}(t_{n}) \rbrace \vert 0 \rangle_{c} \;\;\;\;\;\;\; \nonumber\\
= -i \sum_{n=0}^{\infty} (-i)^{n} \int_{-\infty}^{\infty}dt_{1}\int_{-\infty}^{\infty}dt_{2}\dots \int_{-\infty}^{\infty}dt_{n} \langle 0 \vert {\cal T} \lbrace \hat{\Psi}_{\lambda}(t)\hat{\Psi}_{\lambda'}^{\dagger}(t') \hat{{\cal H}}_\mathrm{V}(t_{1})\hat{{\cal H}}_\mathrm{V}(t_{2})\dots \hat{{\cal H}}_\mathrm{V}(t_{n}) \rbrace \vert 0 \rangle_{c,d}. \;\;\;\;\;
\end{eqnarray}
In this expression the subindex $c$ refers to the connected Feynman diagrams of the perturbation series, while the subindex $d$ refers to the physically not equivalent terms/diagrams (since to order $n$ we have $n!$ equivalent diagrams). Also, the term $n=0$ in the sum gives simply the bare Green's function $G_0(\lambda,t;\lambda',t')$. Introducing the form of the perturbation ${\cal \hat{H}}_\mathrm{V}$ in (\ref{GcausalSmatrix}) and using Wick's theorem, the causal Green's function can be written in terms of a Dyson equation:
\beq \label{DysonSelfEnergy}
G= G_0+G_0\Sigma G \Longrightarrow G = \frac{1}{1-G_0\Sigma} G_0. 
\eeq
Here, $\Sigma$ is the self-energy -- an expansion including all the connected and physically not equivalent terms/diagrams allowed by the interaction.

Green's functions are useful in both equilibrium and non-equilibrium situations. They allow for example to calculate the density of states, or the current and noise spectrum through out-of-equilibrium systems. In equilibrium and at finite temperature, GFs fulfill as expected a fluctuation-dissipation relation. This can be seen projecting the GFs onto an eigenbasis $\{\varepsilon_n\}$ of ${\cal H}$ (Lehmann representation). For example, for $G^{<}$ we find
\beq
G^{<} (\lambda,t;\lambda',t') &=& i \langle \hat{\Psi}_{\lambda'}^{\dagger}(t')\hat{\Psi}_{\lambda}(t) \rangle = i \frac{1}{Z} \sum_{\mu} \langle \mu \vert e^{-\beta {\cal \hat{H}}}  \hat{\Psi}_{\lambda'}^{\dagger}(t')\hat{\Psi}_{\lambda}(t) \vert \mu \rangle \nonumber\\ &=& i \frac{1}{Z} \sum_{\mu,\nu} e^{-\beta \varepsilon_{\mu}} \langle \mu \vert e^{i{\cal \hat{H}}t'} \hat{\Psi}_{\lambda'}^{\dagger} e^{-i{\cal \hat{H}}t'} \vert \nu \rangle \langle \nu \vert e^{i{\cal \hat{H}}t} \hat{\Psi}_{\lambda} e^{-i{\cal \hat{H}}t} \vert \mu \rangle \nonumber\\ &=& i \frac{1}{Z} \sum_{\mu,\nu} e^{-\beta \varepsilon_{\mu}} \langle \mu \vert \hat{\Psi}_{\lambda'}^{\dagger} \vert \nu \rangle \langle \nu \vert \hat{\Psi}_{\lambda} \vert \mu \rangle e^{i(\varepsilon_{\nu}-\varepsilon_{\mu})(t-t')} \Longrightarrow \\ \Rightarrow G^{<}(\lambda,\lambda',\omega) &=& \frac{2\pi i}{Z} \sum_{\mu,\nu} e^{-\beta \varepsilon_{\mu}} \langle \mu \vert \hat{\Psi}_{\lambda'}^{\dagger} \vert \nu \rangle \langle \nu \vert \hat{\Psi}_{\lambda} \vert \mu \rangle \delta \left( \varepsilon_{\nu}-\varepsilon_{\mu} +\omega \right).
\eeq
Similarly, $G^{>}(\lambda,\lambda',\omega) = \frac{-2\pi i}{Z} \sum_{\mu,\nu} e^{-\beta \left( \varepsilon_{\mu}-\omega \right)} \langle \mu \vert \hat{\Psi}_{\lambda'}^{\dagger} \vert \nu \rangle \langle \nu \vert \hat{\Psi}_{\lambda} \vert \mu \rangle \delta \left( \varepsilon_{\nu}-\varepsilon_{\mu} + \omega \right)$, so that $G^{>}(\lambda,\lambda',\omega) = -e^{\beta \omega} G^{<}(\lambda,\lambda',\omega)$, a relation known as \textit{detailed balance}. In these expressions $Z:=\mathrm{Tr}\{e^{-\beta{\cal \hat{H}}}\}$ is the partition function and $\beta:=\frac{1}{k_B T}$. Proceeding in a similar manner for the \textit{spectral function} $A(\lambda,\omega):=i \left[ G^{>}(\lambda,\omega) - G^{<}(\lambda,\omega) \right ] = i \left[ G^{R}(\lambda,\omega) - G^{A}(\lambda,\omega) \right ] = -2 Im \lbrace G^{R}(\lambda,\omega) \rbrace$, we arrive to the identity
\beq \label{GFsFDT}
G^{<}(\lambda,\omega) = i f(\omega) A(\lambda,\omega),
\eeq
where $f(\omega)\equiv\left(e^{\beta\omega}+1\right)^{-1}$ is the Fermi function. This relation is the fluctuation-dissipation theorem, which may be also written as $G^{>}(\lambda,\omega) = i \left[ f(\omega) -1 \right] A(\lambda,\omega)$. Using Eq.~(\ref{GFsFDT}) we realize that the spectral function is nothing but the density of states at a given energy, since the average occupation number is given by $\ew{\hat{n}_{\lambda}}= \langle \hat{\Psi}_{\lambda}^{\dagger} \hat{\Psi}_{\lambda} \rangle = -i G^{<}(\lambda,t=0) = -i \int_{-\infty}^{\infty} \frac{d\omega}{2\pi}G^{<}(\lambda,\omega) = \int_{-\infty}^{\infty} \frac{d\omega}{2\pi}f(\omega) A(\lambda,\omega)$.

The presented GFs formalism is valid at finite temperatures and out of equilibrium provided that we use the appropriate integration contour in the integrals appearing in (\ref{GcausalSmatrix}) \cite{HaugJauho}. 
In particular, we will focus on transport problems (c.f. Fig.~\ref{ScatteringTransportFig}) described by a Hamiltonian of the form (\ref{Hform}), with  ${\cal \hat{H}}_\mathrm{S}=\sum_m E_m \ket{m}\bra{m}$, ${\cal \hat{H}}_\mathrm{R}=\sum_{\lambda\alpha}\varepsilon_{\lambda\alpha}\hat{c}_{\lambda\alpha}^{\dagger}\hat{c}_{\lambda\alpha}$ and ${\cal \hat{H}}_\mathrm{V}=\sum_{\lambda\alpha m} {\cal V}_{\lambda\alpha m}\hat{c}_{\lambda\alpha}^{\dagger}\hat{d}_m + \mathrm{H.c.}$, corresponding to system, reservoirs and coupling respectively. Here $\hat{c}_{\lambda\alpha}^{\dagger}$ creates an electron with quantum numbers $\lambda$ in reservoir $\alpha$, and $\hat{d}_m$ annihilates an electron from site $m$ of the central region. $\{\ket{m}\}$ and $\{E_m\}$ are the eigenstates and eigen-energies of the central region, $\{\varepsilon_{\lambda\alpha}\}$ are the eigen-energies of reservoir $\alpha$, and ${\cal V}_{\lambda\alpha m}$ the site-reservoir couplings. The current through the system can be expressed in terms of the lesser Green's function $G^{<}_{k\alpha,m}(t-t')\equiv i \ew{\hat{c}^{\dagger}_{k\alpha}(t')\hat{d}_m(t)}$. This can be easily seen if we generate the equation of motion for the mean current, that is $\ew{\hat{I}}=\ew{\hat{I}_L} =\ew{\dot{\hat{n}}_L} = i\ew{[{\cal \hat{H}},\hat{n}_L]}$, which, using $\hat{n}_L\equiv \sum_{\lambda} \hat{c}_{\lambda L}^{\dagger}\hat{c}_{\lambda L}$, for the previous Hamiltonian gives\footnote{Here we take spinless particles. An extra factor 2 should be added to these formulae if spin-two particles are considered.}
\beq
\ew{\hat{I}} = \ew{\hat{I}_L} = \frac{2e}{\hbar} \mathrm{Re} \Big\{ \sum_{k,m} {\cal V}_{kLm} G^{<}_{kL,m} (t,t) \Big\}.
\eeq
For non-interacting leads, $G^{<}_{k\alpha,m}(t-t')$ can be written in terms of the GFs corresponding to central region and reservoirs, $G_{mm'}$ and $g_{k\alpha}$ respectively. In the frequency space, this relation reads $G^{<}_{k\alpha,m}(\varepsilon)=\sum_{m'} {\cal V}^*_{k\alpha m'} \left[ G^R_{mm'}(\varepsilon) g^{<}_{k\alpha}(\varepsilon) + G^{<}_{mm'}(\varepsilon) g^{A}_{k\alpha}(\varepsilon) \right]$ \cite{MeirWingreen92}, and leads to the expression for the current through the system 
\cite{MeirWingreen92, Jauho94, HaugJauho}
\beq \label{MeirWingreenFormula}
\ew{\hat{I}} = \frac{ie}{2h} \int d\varepsilon \mathrm{Tr} \Big\{ \left[ \Gamma_L(\varepsilon) - \Gamma_R(\varepsilon) \right] G^{<}(\varepsilon) +\left[ f_L(\varepsilon)\Gamma_L(\varepsilon) - f_R(\varepsilon)\Gamma_R(\varepsilon) \right] \left[ G^R(\varepsilon) - G^A(\varepsilon) \right] \Big\}.~~~
\eeq
This formula was derived by Meir and Wingreen in a seminal paper, and has been widely used to study transport problems. Here, $\left[ \Gamma_{\alpha}(\varepsilon) \right]_{mn}= 2\pi {\cal D}(\varepsilon) {\cal V}^*_{\alpha m}(\varepsilon) {\cal V}_{\alpha n}(\varepsilon)$ is an energy-dependent rate corresponding to reservoir $\alpha$, being ${\cal D}(\varepsilon)$ its density of states and $f_{\alpha}(\varepsilon)$ its distribution (Fermi) function. 
Interestingly, if the rates are proportional, $\Gamma_L \propto\Gamma_R$, Eq.~(\ref{MeirWingreenFormula}) adopts the form of a Landauer-B\"uttiker formula, namely $\ew{\hat{I}} = \frac{e}{h} \int d\varepsilon \mathsf{T}(\varepsilon)\left[ f_L(\varepsilon)-f_R(\varepsilon) \right] $, with $\mathsf{T}(\varepsilon)\equiv i\mathrm{Tr}\left\{ \frac{\Gamma_L(\varepsilon)\Gamma_R(\varepsilon)}{\Gamma_L(\varepsilon)+\Gamma_R(\varepsilon)} \left[ G^R(\varepsilon) - G^A(\varepsilon) \right] \right\}$. An equation for the zero-frequency noise spectrum in terms of GFs of the central region can be similarly found \cite{KiesslichThesis}. However, a frequency-dependent version of this result has, to our knowledge, not yet been reported.

The quantum-well model of Fig.~\ref{DoubleBarrierFig}, used used above to briefly exemplify the scattering matrix and density operator formalisms, can be also treated with the non-equilibrium GFs approach. The Green's functions $G^{R}$ and $G^{A}$ in the expression for the current (\ref{MeirWingreenFormula}) can be obtained through the Dyson equation (\ref{DysonSelfEnergy}). Assuming that only one level is defined in the well, the self-energy reads $\Sigma^{R/A}(\varepsilon)=\sum_{k,\alpha=L,R} \vert {\cal V}_{k\alpha} \vert^2  g_{k\alpha}^{R/A}=\sum_{k,\alpha=L,R} \frac{\vert {\cal V}_{k\alpha} \vert^2}{\varepsilon-\varepsilon_{k\alpha}\pm i\eta}=\Delta(\varepsilon) \mp \frac{i}{2}\Gamma(\varepsilon)$, where $\Delta\equiv\Delta_L+\Delta_R$ and $\Gamma\equiv\Gamma_L+\Gamma_R$ are the real and imaginary parts of the self-energy respectively.
The lesser GF, however, needs to be obtained through the Keldysh equation $G^{<}=G^{R}\Sigma^{<}G^{A}$ \cite{HaugJauho}, being $\Sigma^{<}(\varepsilon)=\sum_{k,\alpha=L,R} \vert {\cal V}_{k\alpha} \vert^2 g_{k\alpha}^{<}(\varepsilon)=i\left[ \Gamma_L(\varepsilon)f_L(\varepsilon)+ \Gamma_R(\varepsilon)f_R(\varepsilon) \right]$. Inserting these results in Eq.~(\ref{MeirWingreenFormula}), the current through the resonant level (with energy $\varepsilon_0$) reads
\beq
\ew{\hat{I}} = \frac{e}{h} \int d\varepsilon \frac{\Gamma_L(\varepsilon)\Gamma_R(\varepsilon)}{[\varepsilon-\varepsilon_0-\Delta(\varepsilon)]^2 + [\Gamma(\varepsilon)/2]^2} \left[ f_L(\varepsilon)-f_R(\varepsilon) \right],
\eeq
which has the form of a Landauer-B\"uttiker formula, and whose transmission coefficient should be compared with Eq.~(\ref{TransmissionQuantumWell}).
In the following, we present in detail a more complicated example of a quantum transport, solved with the non-equilibrium GFs formalism.

\subsubsection{Example: Tight-Binding model}

\begin{figure}
  \begin{center}
    \includegraphics[width=0.8\textwidth]{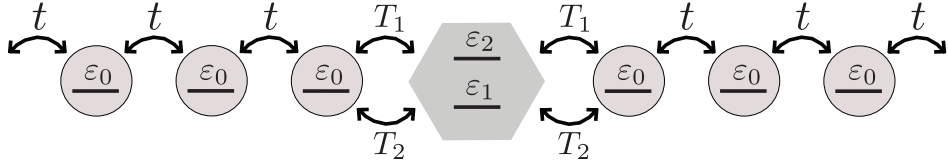}
  \end{center}  
  \caption[Tight-binding model]{A molecule with two levels (not coupled to each other) is attached to two (left and right) semi-infinite tight-binding chains. If these chains are at different chemical potentials, transport through the molecule is possible. The levels have energies $\varepsilon_1$ and $\varepsilon_2$, and whose couplings to the chains are given by $T_1$ and $T_2$ respectively. The chains are composed of atoms with energy $\varepsilon_0$ and coupled to each other with energy $t$.
}
  \label{TightBindingFig}
\end{figure}

We here illustrate how the GFs formalism is applied in quantum transport problems, let us consider the model depicted in Fig.~\ref{TightBindingFig}. It consists of a molecule with two levels (not coupled to each other) attached to contacts modeled as a tight-binding chain. The Hamiltonian of the total system, can be written in the matrix form
\beq
{\cal H} =
\begin{pmatrix}
{\cal H}_L & {\cal H}_{LM} & 0 \\
{\cal H}_{ML} & {\cal H}_M & {\cal H}_{MR} \\
0 & {\cal H}_{RM} & {\cal H}_R \\
\end{pmatrix},
\eeq
where
\begin{equation}
\begin{array}{c}
\begin{array}{lr}
{\cal H}_M=\begin{pmatrix} \varepsilon_1 & 0 \\ 0 & \varepsilon_2 \end{pmatrix}; & {\cal H}_L={\cal H}_R=\begin{pmatrix} \varepsilon_0 & t & 0 & \ldots \\ t & \varepsilon_0 & t & \ldots \\ 0 & t & \varepsilon_0 & \ldots \\ \vdots & \vdots & \vdots \end{pmatrix};
\end{array}\\ \\
\begin{array}{lr}
{\cal H}_{ML} = ({\cal H}_{LM})_T = \begin{pmatrix} 0 & \ldots & 0 & T_1 \\ 0 & \ldots & 0 & T_2 \end{pmatrix}; & {\cal H}_{MR} = ({\cal H}_{RM})_T = \begin{pmatrix} T_1 & 0 & \ldots & 0 \\ T_2 & 0 & \ldots & 0 \end{pmatrix};
\end{array}
\end{array}
\end{equation}
we assume that the system is driven with a voltage such that there is a net current of electrons flowing through the two levels of the central molecule, and we aim to calculate this current. According to the GFs formalism, it will be given by\footnote{Here we take spinless particles. An extra factor 2 should be added to these formulae if spin-two particles are considered.}
\beq \label{currentGFsEquality}
\ew{\hat{I}}=\ew{\hat{I}_L}= \frac{ie}{\hbar} \sum_{m=1,2} \left\{ T_m \ew{\hat{c}^{\dagger}_L \hat{d}_m} - \mathrm{H.c.} \right\} = \frac{e}{\hbar} \sum_{m=1,2} \left\{ T_m G_{Lm}^{<} + \mathrm{H.c.} \right\},
\eeq
with $\hat{c}^{\dagger}_L$ creating an electron in chain $L$, on the site coupled to level $m$ of the molecule, and with $\hat{d}_m$ destroying an electron from this level. To calculate $G_{Lm}^{<}$ we use \cite{HaugJauho}
\beq \label{GlesserEquality}
G^< = G^R (G_0^R)^{-1} G_0^< (G_0^A)^{-1} G^A.
\eeq
In our case, the bare GFs can be written as
\begin{equation}
\begin{array}{lr}
G_0^{R/A}=\begin{pmatrix} (g_L^{R/A})^{-1} & 0 & 0 & 0 \\ 0 & \omega -\varepsilon_1 & 0 & 0 \\ 0 & 0 & \omega- \varepsilon_2 & 0 \\ 0 & 0 & 0 & (g_R^{R/A})^{-1} \end{pmatrix}; & G_0^{<}=\begin{pmatrix} i f_L(\omega)A(\omega) & 0 & 0 & 0 \\ 0 & 0 & 0 & 0 \\ 0 & 0 & 0 & 0 \\ 0 & 0 & 0 & i f_R(\omega)A(\omega) \end{pmatrix}; \nonumber
\end{array}
\end{equation}
where we have used the fluctuation-dissipation relation (\ref{GFsFDT}). Here, $g_{L/R}^{R/A}$ is the retarded/advanced (R/A) GF corresponding to the left/right (L/R) semi-infinite tight-binding chain. Let us denote this simply as $g$. To relate it to the GF of an infinite tight-binding chain, say $g_{TB}$, we notice that if the last fulfills $(\omega - \varepsilon_0 - 2\Sigma) g_{TB}(\omega) = \mathds{1}$, then $(\omega - \varepsilon_0 - \Sigma) g(\omega) = \mathds{1}$, so that the relation between $g$ and $g_{TB}$ is
\beq
g(\omega) = \frac{2g_{TB}(\omega)}{\omega g_{TB}(\omega) + 1}.
\eeq
Now, $g_{TB}(\omega)$ can be calculated as
\beq
g_{TB}(\omega)=\sum_k \frac{\ket{\phi_k}\bra{\phi_k}}{\omega-\varepsilon_k} = \frac{|c_k|^2}{N} \sum_{nn'}\sum_k \frac{e^{ka(n-n')}\ket{\varphi_n}\bra{\varphi_{n'}}}{\omega- {\cal H}_k/s_k} = \sum_{nn'} \underbrace{ \frac{1}{N} \sum_k \frac{e^{ika(n-n')}}{\omega s_k - {\cal H}_k} }_{[g_{TB}(\omega)]_{nn'}} \ket{\varphi_n}\bra{\varphi_{n'}}. \nonumber
\eeq
Here we have expanded the orbital wavefunction in terms of atomic wavefunctions: $\ket{\phi_k}=\frac{c_k}{\sqrt{N}} \sum_n e^{ikna}\ket{\varphi_n}$, being $N$ the number of atoms, $a$ the interatomic distance, and $k$ the momentum. Also, we have used the definitions of $s_k$ and ${\cal H}_k$ that arise from the normalization condition and the Schr\"odinger equation:
\beq
1&=&\olap{\phi_k}{\phi_k} = \frac{|c_k|^2}{N} \sum_{nn'} e^{ik(n-n')a} \olap{\varphi_{n'}}{\varphi_n} \simeq |c_k|^2 \underbrace{\left( 1+2s_1 \cos(ka) \right)}_{:=s_k},\\
0&=& \bra{\varphi_{n'}} (\hat{{\cal H}}_{\alpha}-\varepsilon_k) \ket{\phi_k} = \frac{c_k}{\sqrt{N}} \sum_n \left( \bra{\varphi_{n'}} \hat{{\cal H}}_{\alpha} \ket{\varphi_n} - \varepsilon_k \olap{\varphi_{n'}}{\varphi_n} \right) \nonumber\\ &\simeq& \frac{c_k}{\sqrt{N}} \Big[ \underbrace{ \varepsilon_0 + t (e^{ika} + e^{-ika}) }_{:={\cal H}_k} - \varepsilon_k \underbrace{ \left( 1 + s_1 (e^{ika} + e^{-ika}) \right) }_{s_k} \Big],
\eeq
with $s_1\equiv \olap{\varphi_{n+1}}{\varphi_n}$, $\alpha=L, R$, and where in the approximation $\simeq$ we have taken the sum to first neighbours only. Therefore we have the tight-binding GF
\beq
[g_{TB}(\omega)]_{nn'} = \frac{1}{N} \sum_k  \frac{e^{ika(n-n')}}{\omega s_k - {\cal H}_k} \approx \frac{a}{2\pi} \int_{-\pi/a}^{\pi/a} \frac{e^{ika(n-n')} dk}{\omega[1 +s_1 (e^{ika}+e^{-ika})] - \varepsilon_0 - t(e^{ika}+e^{-ika})}. \nonumber\\
\eeq
Performing the integral, we find
\beq
\left[g_{TB}(\omega)\right]_{nn}&=& \frac{1}{\sqrt{(\omega-\varepsilon_0)^2 - [2(\omega s_1-t)]^2}}. \\ 
\left[g_{TB}(\omega)\right]_{n,n+1}&=&\frac{1}{2\pi i} \frac{1}{\omega s_1 -t} \ln \left\{ \frac{2(\omega s_1 - t)+(\omega - \varepsilon_0)}{2(\omega s_1 -t) - (\omega - \varepsilon_0)} \right\} - \frac{(\omega-\varepsilon_0)/[2(\omega s_1 -t)]}{\sqrt{(\omega-\varepsilon_0)^2 - [2(\omega s_1 -t)]^2}}. \nonumber\\
\eeq
The GFs $G^R$ and $G^A$ are obtained by adding the proper infinitesimal convergence factor to
\beq
G=
\begin{pmatrix}
g^{-1} & \omega s_1 - T_1 & \omega s_1 - T_2 & 0 \\
\omega s_1 - T_1 & \omega-\varepsilon_1 & 0 & \omega s_1 - T_1 \\
\omega s_1 - T_2 & 0 & \omega-\varepsilon_2 & \omega s_1 - T_2 \\
0 & \omega s_1 - T_1 & \omega s_1 - T_2 & g^{-1}
\end{pmatrix}.
\eeq
Using the derived GFs together with Eq.~(\ref{GlesserEquality}) and Eq.~(\ref{currentGFsEquality}), and assuming that a potential difference $eV$ is applied between both tight-binding chains, we finally find a Landauer-B\"uttiker formula (\ref{LandauerButtikerMeanCurrent}) for the current, with
\beq
\mathsf{T_n}(E)= \frac{(\omega-\varepsilon_{\bar{n}})(\omega s_n - T_n) \left[ (\omega-\varepsilon_n)(\omega s_n - T_{\bar{n}})^2 + (\omega-\varepsilon_{\bar{n}})(\omega s_n - T_n)^2 \right] T_n A^2(E)}{2 \left\{ \left[ (\omega-\varepsilon_n)(\omega-\varepsilon_{\bar{n}}) \right]^2 + 4 |g(E)|^2 \left[ (\omega-\varepsilon_n) (\omega s_n - T_{\bar{n}})^2 + (\omega-\varepsilon_{\bar{n}})(\omega s_n - T_n)^2 \right]^2 \right\}}, \nonumber\\
\eeq
where $n=1,2$ and $\bar{n}=\left\{\begin{array}{ccc} 2 & \mbox{if} & n=1 \\ 1 & \mbox{if} & n=2 \end{array} \right.$, and which reduces to $\mathsf{T_n}(E)=\frac{T_n^2 A^2(E)}{(\omega-\varepsilon_n)^2 + 16|g(E)|^2 T_n^4}$ in the limit $\varepsilon_{\bar{n}}=\varepsilon_n$, $T_{\bar{n}}=T_n$, $s_n=0$.

\section{Experiments in quantum transport}

In the last years, experiments on quantum transport have experienced an enormous advance with the development of novel nanoscopic conductors. In particular, new types of quantum dots are now being fabricated, for example using nanowires, nanotubes, and graphene. This opens the possibility of exploring new physics, such as strong spin-orbit and hyperfine effects. In this section we focus our interest on carbon-nanotube quantum dots, a very rich system where important effects such as Klein tunneling have been recently demonstrated \cite{Steele09b}, and currently drawing a lot of attention in the context of nano-mechanics \cite{Sazonova04, Steele09, Lassagne09} and optoelectronics \cite{Avouris06, Dresselhaus07, Kuemmeth10}.
Here we present an experiment on the hyperfine interaction in carbon-nanotube double quantum dots (CNDQDs). The results shown in this section characterize the nuclear spin environment to which quantum dots are typically coupled, and the advantages of using carbon-nanotube quantum dots -- a system free of nuclear spin -- in the context of quantum computation.

\subsection{Electron transport and spin manipulation in CNDQDs\footnote{The results presented in this subsection have been published in \cite{Churchill09}.}} \label{CNDQDsSection}

For coherent electron spins, the hyperfine coupling to nuclei in the
host material can either be a dominant source of unwanted
spin decoherence \cite{Khaetskii02, Petta05, Koppens06} or, if controlled effectively, a resource
enabling storage and retrieval of quantum information \cite{Kane98, Taylor03, Dutt07, Hanson08}. Here we investigate the effect of a controllable nuclear environment on
the evolution of confined electron spins. The system we consider is a gate-defined double quantum dots with integrated charge sensors made from single-walled carbon nanotubes
with a variable concentration of $^{13}$C (nuclear spin $I=1/2$)
among the majority zero-nuclear-spin $^{12}$C atoms. We observe
strong isotope effects in spin-blockaded transport, and from
the magnetic field dependence estimate the hyperfine coupling
in $^{13}$C nanotubes to be of the order of $100$ $\mu$eV, two orders of
magnitude larger than anticipated \cite{Yazyev08, Pennington96}. $^{13}$C-enhanced nanotubes
are an interesting system for spin-based quantum information
processing and memory: the $^{13}$C nuclei differ from those
in the substrate, are naturally confined to one dimension,
lack quadrupolar coupling and have a readily controllable
concentration from less than one to $10^5$ per electron.

\subsubsection{The hyperfine problem}

Nuclear magnetism is typically weaker than electronic magnetism, since the nuclear mass, $m_n$, is three orders of magnitude larger than the electron mass, $m_e$, and thus the respective magneton, $\mu_I\equiv \frac{e\hbar}{2m_n} \gg \mu_B\equiv \frac{e\hbar}{2m_e}$. Nuclear magnetism is nonetheless a very relevant source of decoherence in the spin-qubit proposals \cite{LossDiVincenzo98, Kane98}. Here, a large number of nuclei ($10^5-10^6$) in the host material interact with the electronic spin and induce relaxation and dephasing. Mathematically, both fine and hyperfine (HF) interactions are obtained from the \textit{Dirac Hamiltonian}\footnote{In this section we obviate the hat in the notation for operators.}
\beq
{\cal H}_\mathrm{Dirac} = \vec{\alpha}\cdot \left(\vec{p} - e\vec{A}(\vec{R})\right) + eU(\vec{R}) + \beta m_ec^2,
\eeq
where $\vec{p}$ is the electron momentum, $\vec{A}$ the vector potential associated with the nuclear magnetic field, $c$ the speed of light, $U(\vec{R})$ the nuclear potential\footnote{Notice that this potential can contain different terms from the multipolar expansion. For the two cases of interest in this thesis, namely $^{13}$C with nuclear spin $1/2$ and $^{14}N$ with nuclear spin $1$, $U(\vec{R})$ contains only the dipolar term and the dipolar and quadropolar terms respectively.}, $\vec{R}$ a vector from the nucleus to the electron, and 
\begin{equation}
\begin{array}{lr}
\vec{\alpha}\equiv\begin{pmatrix} 0 & \vec{\sigma} \\ \vec{\sigma} & 0 \end{pmatrix}; & \beta\equiv\begin{pmatrix} \mathds{1} & 0 \\ 0 & -\mathds{1} \end{pmatrix}; \nonumber
\end{array}
\end{equation}
being $\vec{\sigma}\equiv(\sigma_x,\sigma_y,\sigma_z)_T$ a vector of Pauli operators denoting the electron spin. In the non-relativistic limit, $|E-eU|\ll mc^2$, with $E$ the electron energy, this Hamiltonian reduces to the \textit{Pauli Hamiltonian}
\beq
{\cal H}_\mathrm{Pauli} = \frac{1}{2m_e}\left(\vec{p} - e\vec{A}(\vec{R})\right)^2 + eU(\vec{R}) -\mu_B \left(\vec{\nabla} \times \vec{A}(\vec{R})\right) \cdot \vec{\sigma}.
\eeq
Notice that the first two terms correspond to kinetic and potential energy, while the third term captures the magnetic field - spin interaction. Expanding $\left(\vec{p} - e\vec{A}(\vec{R})\right)^2$ and neglecting the quadratic term in the vector potential (which is legitimate due to the very small energy correction that it produces), and decoupling the two spinors through a \textit{Foldy-Wouthuysen transformation}, we have ${\cal H}_\mathrm{Pauli} = {\cal H}_0 + {\cal H}_{F} + {\cal H}_{HF}$, with ${\cal H}_0\equiv\frac{\vec{p}^2}{2m_e} + eU(\vec{R})$, and the hyperfine Hamiltonian
\beq \label{HF1stEq}
{\cal H}_{HF} = - \frac{e}{2m_e} \left( \vec{p}\cdot \vec{A}(\vec{R}) + \vec{A}(\vec{R})\cdot \vec{p} \right) -\mu_B \left(\vec{\nabla} \times \vec{A}(\vec{R})\right) \cdot \vec{\sigma}.
\eeq
The fine-structure Hamiltonian ${\cal H}_{F}$ contains effects of the electric field $\vec{E}=-\vec{\nabla}U(\vec{R})$ created by the nucleus, such as the familiar \textit{spin-orbit coupling} $-\frac{e\hbar}{4m_e^2c^2}(\vec{E}\times\vec{p})\cdot \vec{\sigma}$. Although in principle these can be very relevant, spin-orbit effects are negligible in highly confined structures \cite{Halperin01}, and we thus here center our attention on the HF interaction. The first term in Eq.~(\ref{HF1stEq}), simplified properly, gives a `nuclear spin-orbit': $-\frac{\mu_0}{4\pi} \frac{2\mu_B}{\hbar}\frac{\vec{L}\cdot\vec{m}_I}{R^3}$, where $\mu_0$ is the vacuum magnetic permeability constant, $\vec{m}_I$ the nuclear magnetic dipole moment, and $\vec{L}\equiv \vec{R}\times\vec{p}$ the electron angular momentum. The second term in Eq.~(\ref{HF1stEq}) gives a different contribution depending on whether $\vec{R}$ lies outside our inside the nucleus. In the first case, the term simplifies to the dipole-dipole interaction, while in the second case, it gives the so-called Fermi contact term. Although of finite size, the nuclear radius can be taken to be zero, as this is typically much smaller than the atomic size ($\sim$ Bohr radius $a_0$). Altogether, the \textit{hyperfine interaction} can be written as
\beq \label{HFhamiltonian}
{\cal H}_{HF} = -\frac{\mu_0}{4\pi} \Big[ \underbrace{ \frac{2\mu_B}{\hbar} \frac{\vec{L}\cdot\vec{m}_I}{R^3} }_{\mbox{nuclear spin-orbit}} + \underbrace{ \frac{1}{R^3}\left[ 3(\vec{m}_S\cdot\vec{e}_R)(\vec{m}_I\cdot\vec{e}_R) - \vec{m}_S\cdot\vec{m}_I \right] }_{\mbox{dipole-dipole}} + \underbrace{ \frac{8\pi}{3} \vec{m}_S\cdot\vec{m}_I \delta(\vec{R}) }_{\mbox{contact term}} \Big]. \nonumber\\
\eeq
In this expression, $\vec{e}_R$ is a unit vector in the $\vec{R}$ direction and $\vec{m}_S$ is the electron magnetic dipole moment. In terms of the electron spin $\vec{S}$, this reads $\vec{m}_S=g_e\mu_B\vec{S}=\hbar\gamma_e\vec{S}$, and similarly, the nuclear magnetic moment is related with the nuclear spin $\vec{I}$ as $\vec{m}_I=g_n\mu_n\vec{I}=\hbar\gamma_n\vec{I}$, where $g_{e/n}$ and $\gamma_{e/n}$ are the electron/nuclear g-factor and gyromagnetic ratio respectively. 

The contact-term in the Hamiltonian (\ref{HFhamiltonian}) gives a non-zero contribution only for s-wave electrons, and in this case it is indeed the most relevant contribution to the HF interaction. Most of the present studies on nuclear HF interaction focus on this term, which, including the effect of an externally-applied uniform magnetic field along the $z$ direction, $B_z^\mathrm{ext}$, and considering an ensemble of nuclear spins, takes the form
\beq
{\cal H}_{HF}^\mathrm{(Fermi)} = \underbrace{ g_e\mu_B B_z^\mathrm{ext} S_z + g_n B_z^\mathrm{ext} \sum_j \mu_n^{(j)} I_z^{(j)} }_{\mbox{external field}} + \underbrace{ \sum_j a_j I_z^{(j)} S_z }_{\mbox{Overhauser field}} + \underbrace{ \sum_j 2a_j (I_{+}^{(j)}S_{-} + \mathrm{H.c.}) }_{\mbox{flip-flop term}}, \nonumber\\
\eeq
where we have defined $S_{\pm}:=\frac{1}{2}(S_{+}\pm S_{-})$, $I_{\pm}:=\frac{1}{2}(I_{+}\pm I_{-})$, and used the rotating wave approximation. The second term produces the so-called \textit{Overhauser shift} in the electronic spectrum, or alternatively it can be viewed as a shift in the nuclear frequency -- the \textit{Knight shift}. The third term produces `flip-flop' processes between electrons and nuclei. This term dominates the electron dynamics if it is of larger magnitude than the effective field $B_z^\mathrm{eff}=B_z^\mathrm{ext}+\frac{1}{g_e\mu_B}\sum_j A_j I_z^{(j)}$, and it can be used to store the quantum information in the nuclei \cite{Taylor03}. The quantity $A = \frac{a_j}{v_0|\psi(\vec{r}_j)|^2}$ (with $v_0$ the atomic volume and $\psi(\vec{r}_j)$ the atomic wavefunction evaluated at the nuclear position) is the hyperfine-interaction constant, which in GaAs takes the value $A\simeq 90\; \mu$eV \cite{Paget77}. In the work presented in this section, we have measured the HF constant in $^{13}$C CNDQDs, finding $A\simeq 100 \; \mu$eV. This is a remarkable large value. The spin-orbit coupling, expected to be three orders of magnitude larger than the HF coupling, as noticed above, has been reported to be $\sim 400 \; \mu$eV in CNTs \cite{Kuemmeth08}; moreover, the measured value of the HF constant is in discrepancy with the existing theory \cite{Pennington96, Saito98, Yazyev08, Fischer09}.

In this work we also present the first existing measurements of spin blockade \cite{Ono02} in CNTs, together with those of Buitelaar \textit{et al.} \cite{Buitelaar08}. This effect is used as a means of spin-to-charge conversion, and in this way being able to read out the electron spin state. In a later work \cite{Churchill09b}, spin blockade has been used to measure relatively good relaxation and dephasing times in CNDQDs. In subections \ref{IntroCNDQDs} -- \ref{ConclusionsCNDQDs} the experiment is described. For additional reading about the hyperfine interaction, we refer the reader to \cite{AbragamBook, SlichterBook, CohenQMbook}. A review of the HF interaction in the context of quantum dots can be found e.g. in \cite{Fischer09b}.

\subsubsection{Introduction to the experiment} \label{IntroCNDQDs}

Techniques to prepare, manipulate and measure few-electron
spin states in quantum dots have advanced considerably in
recent years, with the leading progress in III-V semiconductor
systems \cite{Petta05, Koppens06, Ono02, Hanson07}. 
All stable isotopes of III-V semiconductors, such
as GaAs, have non-zero nuclear spin, and the hyperfine coupling
of electron spins to host nuclei is a dominant source of spin
decoherence in these materials \cite{Khaetskii02, Petta05, Merkulov02, Coish08}. 
To eliminate this source of decoherence, group-IV 
semiconductors--various forms of carbon, 
silicon and silicon-germanium--which have predominantly
zero nuclear spin, are being vigorously pursued as the basis of
coherent spin electronic devices. Double quantum dots have
recently been demonstrated in carbon nanotubes \cite{Biercuk05, Sapmaz06, Graber06}, 
including the investigation of spin effects \cite{Buitelaar08, Jorgensen08}.

\begin{figure}
\center
\includegraphics[width= 0.9\textwidth]{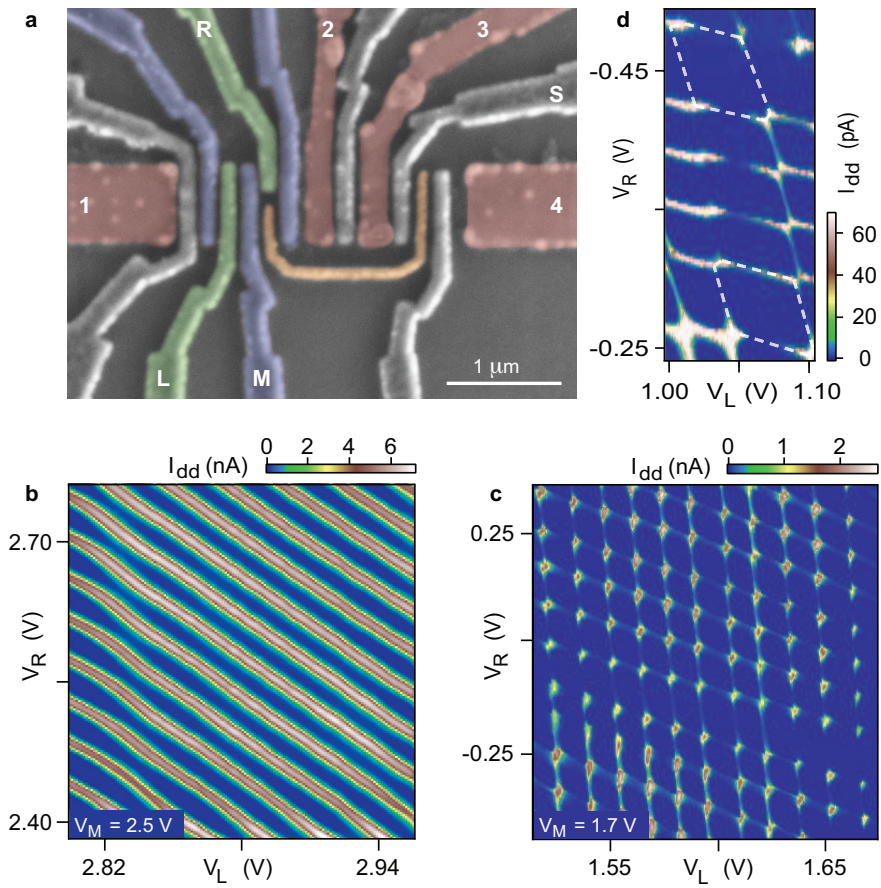}
\caption[Nanotube double dot with integrated charge sensor]{\footnotesize{Nanotube double dot with integrated charge sensor. a) SEM micrograph (with false color) of a device similar to the measured $^{12}$C~and $^{13}$C~devices.  The carbon nanotube (not visible) runs horizontally under the four Pd contacts (red).  Top-gates (blue) create voltage-tunable tunnel barriers allowing the formation of a single or double quantum dot between contacts 1 and 2.  Plunger gates L and R (green) control the occupancy of the double dot.  A separate single dot contacted by Pd contacts 3 and 4 is controlled with gate plunger gate S (gray) and is capacitively coupled to the double dot via a coupling wire (orange). b) Current through the double dot, $I_{\rm dd}$, (color scale) with the top-gates configured to form a large single dot. c) When carriers beneath the middle gate, M, are depleted, $I_{\rm dd}$ shows typical double-dot transport behavior, demarcating the honeycomb charge stability pattern. d) Within certain gate voltage ranges, honeycomb cells with larger addition energy and fourfold periodicity (outlined with dashed lines) indicate the filling of spin and orbital states in shells. 
Source-drain bias is $-1.0$ mV for (b), (c), and (d).}}
\label{CNDQDfig1}
\end{figure}

The devices reported are based on single-walled carbon
nanotubes grown by chemical vapour deposition using methane
feedstock containing either 99\% $^{13}$C (denoted $^{13}$C devices) or 99\%
$^{12}$C (denoted $^{12}$C devices) \cite{Liu01}. The device
design (Fig.~\ref{CNDQDfig1}a) uses two pairs of Pd contacts on the same nanotube;
depletion by top-gates (blue, green and grey in Fig.~\ref{CNDQDfig1}a) forms a
double dot between one pair of contacts and a single dot between
the other. Devices are highly tunable, as demonstrated in Fig.~\ref{CNDQDfig1},
which shows that tuning the voltage on gate M (Fig.~\ref{CNDQDfig1}a) adjusts the
tunnel rate between dots, enabling a crossover from large single-dot
behaviour (Fig.~\ref{CNDQDfig1}b) to double-dot behaviour (Fig.~\ref{CNDQDfig1}c). Left and
right tunnel barriers can be similarly tuned using the other gates
shown in blue in Fig.~\ref{CNDQDfig1}a.

\subsubsection{Spin blockade}

\begin{figure}
\center
\includegraphics[width= 0.69\textwidth]{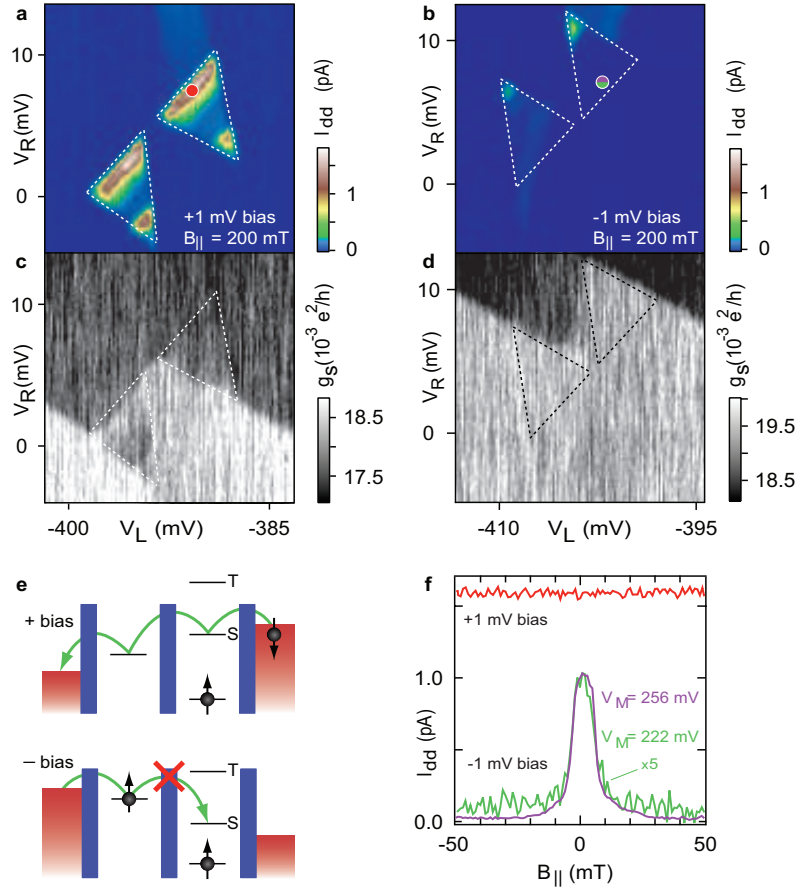}
\caption[Spin blockade in a $^{13}$C~nanotube double dot]{\footnotesize{Spin blockade in a $^{13}$C~nanotube double dot. a) Current $I_{\rm dd}$ (color scale) at $+1.0$ mV source-drain bias, the non-spin-blockaded bias direction. Transport is dominated by resonant tunneling through the ground state at the base of the finite bias triangles and through an excited state at a detuning of 0.7 meV. b) $I_{\rm dd}$ (color scale) at $-1.0$ mV source-drain bias, the spin-blockaded bias direction. $I_{\rm dd}$ is suppressed except near the tips of the transport triangles, where an excited state of the right dot becomes accessible. Suppressed transport for one bias direction is the signature of spin blockade. c) Charge sensing signal, $\rm{g_s}$, (conductance of the sensing dot between contacts 3 and 4 in Fig.~\ref{CNDQDfig1}a), acquired simultaneously with (a) detects the time-averaged occupation of the right dot.  d) Charge sensing signal $\rm{g_s}$~for $-1.0$ mV bias (blockade direction).  The transfer of charge from the left dot to the right is delayed until the excited state is reached at high detuning.  In (a)--(d) dashed lines indicate allowed regions for current flow in the absence of blockade.  e) Schematic of spin-blockaded transport.  Any spin may occupy the left dot, but only a spin singlet is allowed in the right dot, suppressing negative bias current once an electron enters the left dot and forms a triplet state.
f) Current $I_{\rm dd}$ at zero detuning as a function of magnetic field  for positive bias (non-blockade, red trace) and negative bias (blockade, purple trace).}}
\label{CNDQDfig2}
\end{figure}

A notable feature of nanotube quantum dots that is not shared by GaAs dots is that the energy required to add each subsequent electron, the addition energy, often shows shell-filling structure even in the many-electron regime \cite{Jorgensen08}.
An example of a shell-filling pattern, with larger addition energy every fourth electron in the right dot, is seen in Fig.~\ref{CNDQDfig1}d. We find, however, that evident shell filling is not necessary to observe spin blockade at finite bias. Figures ~\ref{CNDQDfig2}a and ~\ref{CNDQDfig2}b show current through the double dot, $I_{\rm dd}$, as a function of gate voltages $V_{\rm R}$~and $V_{\rm L}$~for a weakly coupled, many-electron $^{13}$C~double dot at $+1$ and $-1$ mV source-drain bias, respectively, in a range of dot occupancy that does not show shell structure in the addition spectrum of either dot.  
With a magnetic field $B_{||}=200$~mT applied along the tube axis, current flow is observed throughout the finite-bias triangles at positive bias, but is suppressed at negative bias for detuning below 0.8 meV, which presumably indicates where an excited state of the right dot enters the transport window.

Current rectification of this type is a hallmark of spin blockade \cite{Ono02} (Fig.~\ref{CNDQDfig2}e). At positive bias, current flows freely as electrons of appropriate spin are drawn from the right lead to form the singlet ground state; at negative bias, current is blocked whenever a triplet state is formed between separated electrons, as the excess electron on the left can neither reenter the left lead nor occupy the lowest orbital state on the right without flipping its spin. Spin blockade was identified in all four devices measured, two each of $^{12}$C~and $^{13}$C.
Spin blockade was occasionally found to follow a regular even-odd filling pattern, as seen in few-electron GaAs dots \cite{Johnson05b}, though no pattern was seen adjacent to the area in Fig.~\ref{CNDQDfig2}. 

Electrostatic sensing of the double-dot charge state is provided by a gate-defined quantum dot formed on a separately contacted portion of the same nanotube. The sensing dot is capacitively coupled to the double dot by a $\sim 1~\mu$m coupling wire \cite{Hu07} (orange gate in Fig.~\ref{CNDQDfig1}a) but electrically isolated by a depletion gate between the Pd contacts.
Charge sensor conductance ${\rm g_s}$ as a function of $V_{\rm R}$~and $V_{\rm L}$, acquired simultaneously with transport data in Fig.~\ref{CNDQDfig2}a,b, is shown in Fig.~\ref{CNDQDfig2}c,d. 
The location of the coupling wire makes ${\rm g_s}$ especially sensitive to occupancy of the right dot. 
Inside the positive-bias triangles (Fig.~\ref{CNDQDfig2}c), ${\rm g_s}$ is intermediate in value between their bordering regions, indicating that the excess electron is rapidly shuttling between the dots as current flows through the double dot.
In contrast, inside the negative-bias triangles (Fig.~\ref{CNDQDfig2}d), ${\rm g_s}$ shows no excess electron on the right dot as a result of spin blockade. These sensor values are consistent with models of finite-bias charge sensing in the spin-blockade regime \cite{Johnson05b}.

The magnetic field dependence of spin blockade provides important information about electron spin relaxation mechanisms \cite{Koppens05,Jouravlev06}.
A first look at field dependence (Fig.~\ref{CNDQDfig2}f) for a $^{13}$C~device shows that for negative bias (purple), spin-blockade leakage current is strongly peaked at ${B}_{||}=0$, while for positive bias, the unblockaded current does not depend on field. 
As discussed below, this field dependence can be understood in terms of hyperfine-mediated spin relaxation.

\subsubsection{$^{12}$C vs $^{13}$C}

\begin{figure}
\center
\includegraphics[width= 0.7\textwidth]{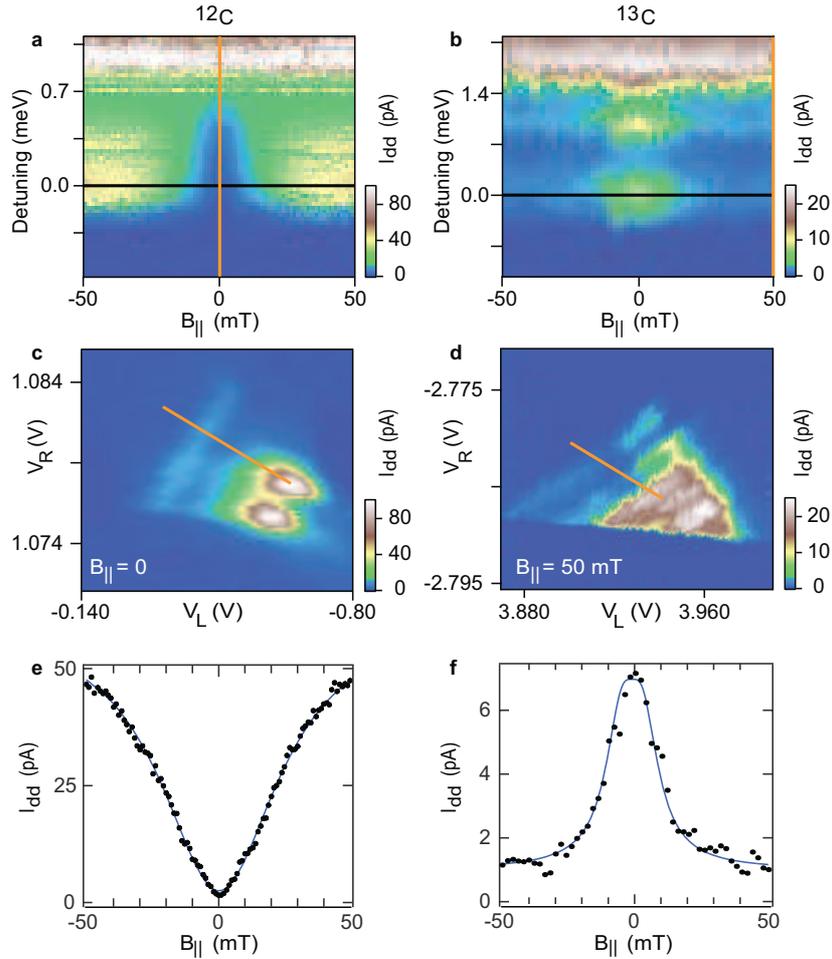}
\caption[Magnetic field dependence of leakage current for $^{12}$C~and $^{13}$C~devices]{\footnotesize{Contrasting magnetic field dependence of leakage current for $^{12}$C~and $^{13}$C~devices.  Leakage current through spin blockade (color scale) as a function of detuning and magnetic field, $B_{||}$, for (a) $^{12}$C~and (b) $^{13}$C~devices.  The vertical axes in (a) and (b) are interdot detuning as indicated by the orange lines in (c) and (d), respectively.  In (a) $B_{||}$~was swept and detuning stepped, while in (b) detuning was swept and $B_{||}$~stepped.  Bias is $-1.5$ mV in (c) and $-4$ mV in (d).  (e) and (f) show cuts along $B_{||}$~at the detunings indicated by the red lines in (a) and (b), respectively.  The fit in (e) is a Lorentzian with a width of 30 mT, and the fit in (f) is to the theory of Jouravlev and Nazarov \cite{Jouravlev06}, providing a measure of $B_{\rm nuc}=6.1$ mT.}}
\label{CNDQDfig3}
\end{figure}

The striking difference in field dependence of spin-blockade leakage current between $^{12}$C~and $^{13}$C~devices is illustrated in Fig.~\ref{CNDQDfig3}a,b. These data show that for negative (spin-blockaded) bias, leakage current is a minimum at $B_{||}=0$ for the $^{12}$C~device  and a maximum at $B_{||}=0$ for the $^{13}$C~device. In fourteen instances of spin blockade measured in four devices (two $^{13}$C~and two $^{12}$C), we find that leakage current minima can occur at $B_{||}=0$ in both $^{12}$C~and $^{13}$C~devices, particularly for stronger interdot tunneling. For weaker interdot tunneling, however, only the $^{13}$C~devices show maxima of spin-blockade leakage at $B_{||}=0$. In all cases, the positive bias (non-spin-blockade) current shows no appreciable field dependence.

Figure ~\ref{CNDQDfig3}e shows spin-blockade leakage current as a function of $B_{||}$~at fixed detuning (the detuning value is shown as a black line in Fig.~\ref{CNDQDfig3}a), along with a best-fit lorentzian, for the $^{12}$C~device. The lorentzian form was not motivated by theory, but appears to fit rather well.
The width of the dip around $B_{||}=0$ increases with interdot tunneling (configuration Fig.~\ref{CNDQDfig3}e has $t\sim50~\mu$eV, based on charge-state transition width \cite{Hu07}). 
We note that a comparable zero-field dip in spin-blockade leakage current was recently reported in a double dot formed in an InAs nanowire \cite{Pfund07}, a material system with strong spin-orbit coupling. 
In the present system, the zero-field dip may also be attributable to spin-orbit coupling \cite{Kuemmeth08}, resulting in phonon-mediated relaxation that vanishes at $B_{||}=0$.

Hyperfine coupling appears to the confined electrons as an effective local Zeeman field (the Overhauser field) that fluctuates in time independently in the two dots, driven by thermal excitation of nuclear spins. 
The difference in local Overhauser fields in the two dots will induce rapid mixing of all two-electron spin states whenever the applied field is less than the typical difference in fluctuating Overhauser fields. (At higher fields, only the $m=0$ triplet can rapidly mix with the singlet). How hyperfine-mediated spin mixing translates to a field dependence of spin-blockade leakage current was investigated experimentally in GaAs devices \cite{Koppens05}, with theory developed by Jouravlev and Nazarov \cite{Jouravlev06}. 

Field dependence of spin-blockade leakage current for the $^{13}$C~device is shown in Fig.~\ref{CNDQDfig3}f, along with a theoretical fit (Eq.~(11) of Ref.~\cite{Jouravlev06}, with a constant background current added), from which we extract a root mean square amplitude of fluctuations of the local Overhauser fields, $B_{\rm nuc}=6.1$~mT.  
Assuming gaussian distributed Overhauser fields and uniform coupling, $B_{\rm nuc}$ is related to the hyperfine coupling constant $A$ by $g\mu_BB_{\rm nuc}=A/\sqrt{N},$ where $g$ is the electron g-factor and $N$ is the number of $^{13}$C~nuclei in each dot \cite{Jouravlev06}.
Taking $N\sim3$--$10 \times 10^4$ and $g=2$, yields $A\sim1$--$2\times 10^{-4}$~eV, a value that is two orders of magnitude larger than predicted for carbon nanotubes \cite{Yazyev08} or measured in fullerenes \cite{Pennington96}.

\subsubsection{Hyperfine constant}

\begin{figure}
\center
\includegraphics[width= 0.55\textwidth]{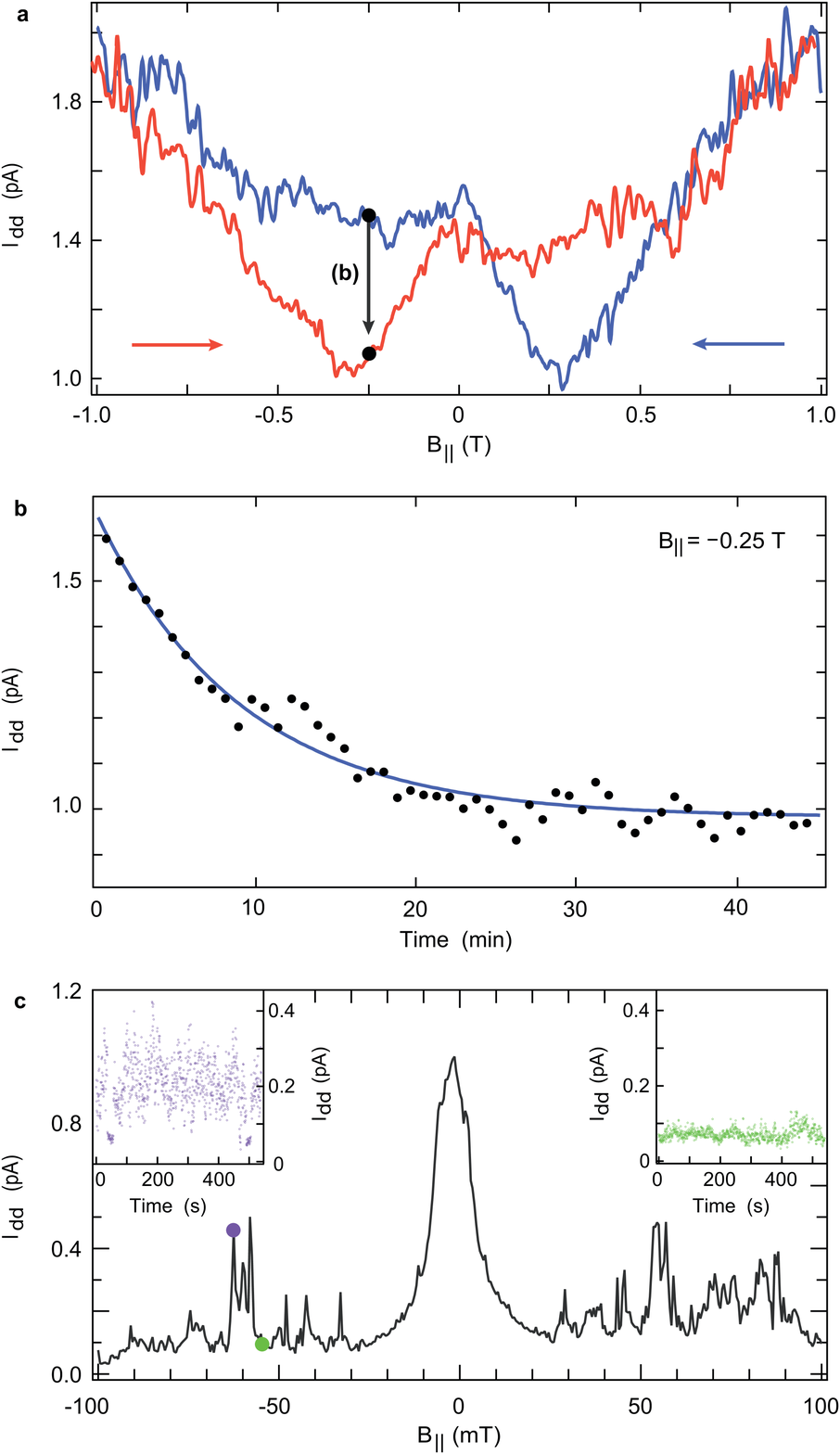}
\caption[Hysteresis and fluctuations in leakage current]{\footnotesize{Hysteresis and fluctuations in leakage current.  a) The spin-blockade leakage current for a $^{13}$C~device measured for decreasing (increasing) magnetic field (sweep rate 0.4 mT/s), shown in blue (red), after waiting at $+1$ T ($-1$ T) for 10~minutes. Hysteresis is seen on a field scale $>0.5$~T for both sweep directions.  b) Decay of leakage current over time measured by stopping a downward sweep at $-0.25$ T.  The fit is to an exponential decay with a time constant of 9 min.  c) Dependence of leakage current on $B_{||}$~near zero detuning in a second $^{13}$C~device.  The leakage current fluctuates over time at some values of $B_{||}$, while remaining steady at others (insets).}}
\label{CNDQDfig4}
\end{figure}

Signatures of dynamic nuclear polarization provide further evidence of a strong hyperfine interaction in $^{13}$C~double dots.
Hysteresis in the spin-blockade leakage current near zero detuning is observed when the magnetic field is swept over a tesla-scale range, as shown in Fig.~\ref{CNDQDfig4}a.  
The data in Fig.~\ref{CNDQDfig4}a,b are from the same $^{13}$C~device as in Fig.~\ref{CNDQDfig3}, but with the barriers tuned such that cotunneling processes provide a significant contribution to the leakage current.

We interpret the hysteresis in Fig.~\ref{CNDQDfig4}a as resulting from a net nuclear polarization induced by the electron spin flips required to circumvent spin blockade \cite{Baugh07}.
This nuclear polarization generates an Overhauser field felt by the electron spins that opposes $B_{||}$~once $B_{||}$~passes through zero.
The value of the coercive field, $B_c\sim0.6$ T, the external field at which the two curves rejoin, places a lower bound for the hyperfine
coefficient, $A\geq g\mu_BB_c\sim0.7 \times 10^{-4}$~eV (equality corresponding to full polarization), independent of the value inferred from the width of the leakage current peak around zero field (Fig.~\ref{CNDQDfig3}c).
If we instead use the value of $A$ inferred from the current peak width (Fig.~\ref{CNDQDfig3}c), the size of $B_c$ implies a  $\sim50\%$ polarization for the data in Fig.~\ref{CNDQDfig4}a.
Hysteresis is not observed for non-spin-blockaded transport in the $^{13}$C~devices and is not observed in the $^{12}$C~devices, suggesting that this effect cannot be attributed to sources such as the Fe catalyst particles or interaction with nuclei in the substrate or gate oxide. 
 
Figure ~\ref{CNDQDfig4}b shows that the induced nuclear polarization persists for $\sim10$~minutes, two orders of magnitude longer than similar processes in GaAs double dots \cite{Reilly10}. The long relaxation time indicates that nuclear spin diffusion is extremely slow, due both to the one-dimensional geometry of the nanotube and the material mismatch between the nanotube and its surroundings. 
Field and occupancy dependence of relaxation were not measured.

Large fluctuations in $I_{\rm dd}$ are seen at some values of magnetic field, but not at others (Fig.~\ref{CNDQDfig4}c), similar to behavior observed in GaAs devices \cite{Koppens05}. This presumably reflects an instability in nuclear polarization that can arise when polarization or depolarization rates themselves are polarization dependent \cite{Rudner07, Baugh07}.
 
\subsubsection{Conclusions of the experiment} \label{ConclusionsCNDQDs} 
 
An important conclusion of this work is that the hyperfine coupling constant, $A\sim 1$--$2 \times 10^{-4}$ eV, in the $^{13}$C~devices (for both electron and holes) is much larger than anticipated \cite{Yazyev08,Pennington96}.
It is possible that the substrate or gate oxide may enhance the degree of $s$-orbital content of conduction electrons,  thus strengthening the contact hyperfine coupling. We also note that a recent theoretical study of electron-nuclear spin interactions in $^{13}$C~nanotubes found that the one-dimensional character of charge carriers greatly enhances the effective electron-nuclear interaction \cite{Braunecker09}. Finally, we note that a large value of $A$ motivates the fabrication of isotopically enriched $^{12}$C~nanotubes to reduce decoherence and the use of $^{13}$C~tubes as a potential basis of electrically addressable quantum memory. 

\section{Beyond DC transport. Noise-correlation techniques}

As we have mentioned before, the investigation of shot noise is important as it contains information about the particle statistics and the intrinsic properties of a conductor. Photons do not interact and obey Bose-Einstein statistics, while electrons obey Fermi-Dirac statistics and interact strongly. These interactions may contribute to create correlations contrary to the sub-Poissonian behaviour expected from the Pauli principle. The common association anti-bunching $\Leftrightarrow$ sub-Poissonian, bunching $\Leftrightarrow$ super-Poissonian is not correct in general, and a detailed study to make this kind of statements is required \cite{KiesslichThesis}. We have also stressed the importance of the noise spectrum to investigate the decoherence properties of atomic systems coupled to different environments \cite{Schoelkopf-NazarovBook, AguadoBrandes04, Clerk10}. In this section, we provide the mathematical tools to derive the correlation functions in which we are interested. The techniques presented below are based on the density operator approach. The most straightforward procedure to obtain the second-order current-correlation function is the \textit{master equation method}, as given by Hershfield \textit{et al.} \cite{Hershfield93}. In this approach, the (Markovian) noise correlation function is given by
\beq
S_{\alpha\beta}^{(2)}(\tau)=\ew{I_{\alpha}(t)I_{\beta}(t+\tau)}_c &=& \theta(\tau) \mathrm{Tr_S}\left\{ {\cal J}_{\alpha}\Omega_0(\tau){\cal J}_{\beta} \rho_\mathrm{S}^{\mathrm{stat}} \right\} \nonumber\\ &+& \theta(-\tau) \mathrm{Tr_S}\left\{ {\cal J}_{\beta}\Omega_0(-\tau){\cal J}_{\alpha} \rho_\mathrm{S}^{\mathrm{stat}} \right\} \nonumber\\ &+& \delta_{\alpha\beta}\delta(\tau) \mathrm{Tr_S} \left\{ {\cal J}_{\alpha} \rho_\mathrm{S}^{\mathrm{stat}} \right\}.
\eeq
Here, $\Omega_0(\tau)\equiv e^{{\cal W}\tau}$, with ${\cal W}$ the kernel introduced in Eq.~(\ref{rhoSevolution}). ${\cal J}_{\alpha}$ is a `jump' super-operator, given by the terms in ${\cal W}$ that concern a change in the number of particles in lead $\alpha$, and $\mathrm{Tr_S}$ denotes the trace over the system degrees of freedom. The mean current through lead $\alpha$ reads simply $\ew{I_{\alpha}}=\mathrm{Tr_S}\{ {\cal J}_{\alpha} \rho_\mathrm{S}^{\mathrm{stat}} \}$.
Notice that although the current is a quantum-mechanical operator, it can be treated as a classical stochastic variable, and still it does contain quantum effects present in the system under study. Equation (\ref{rhoSevolution}) is indeed equivalent to the result given by our formalism, and can also be derived from MacDonald's formula, as shown below. We nevertheless present it in the general context of full counting statistics, where the noise spectrum, and higher-order correlation functions, are derived from a cumulant generating function.

\subsection{The quantum regression theorem}

The most popular method to calculate correlation functions in quantum optics is the \textit{quantum regression theorem}. Below we present its derivation to arbitrary order of the correlation function we want to calculate. Let us start by considering the following correlation function
\beq \label{corr2QRT}
\langle {\cal \hat{O}}_1(t){\cal \hat{O}}_2(t') \rangle &=& \mathrm{Tr} \left\lbrace \hat{\rho} (0) {\cal \hat{O}}_1(t){\cal \hat{O}}_2(t') \right\rbrace  \nonumber \\ 
&=& \mathrm{Tr} \left\lbrace e^{i{\cal \hat{H}}t} \hat{\rho} (t) e^{-i{\cal \hat{H}}t} e^{i{\cal \hat{H}}t} {\cal \hat{O}}_1(0) e^{-i{\cal \hat{H}}t} e^{i{\cal \hat{H}}t'} {\cal \hat{O}}_2(0) e^{-i{\cal \hat{H}}t'} \right\rbrace \nonumber \\
&=& \mathrm{Tr_S} \left\lbrace {\cal \hat{O}}_2(0) \mathrm{Tr_R} \left\lbrace e^{-i{\cal \hat{H}}\tau} \hat{\rho} (t) {\cal \hat{O}}_1(0) e^{i{\cal \hat{H}}\tau} \right\rbrace \right\rbrace,
\eeq
where ${\cal \hat{H}}$ is the Hamiltonian governing the dynamics of the system and $\hat{\rho}$ the corresponding density operator. In the last line we have defined $\tau:=t'-t$. We further define $\hat{\rho}^{\cal O}(0):=\hat{\rho}(t){\cal \hat{O}}(0)=\hat{\rho}_\mathrm{R}\hat{\rho}_\mathrm{S}(t){\cal \hat{O}}(0)\equiv\hat{\rho}_\mathrm{R}\hat{\rho}_\mathrm{S}^{\cal O}(0)$, where we have used the Born approximation to factorize the total density operator in terms of system and reservoir density operators. The density operator $\hat{\rho}^{\cal O}(\tau)\equiv e^{-i{\cal \hat{H}}\tau}\hat{\rho}(t){\cal \hat{O}}(0)e^{i{\cal \hat{H}}\tau}$ satisfies the von Neumann's equation (\ref{vonNeumann}), and similarly, $\hat{\rho}_\mathrm{S}^{\cal O}(0)\equiv \hat{\rho}_\mathrm{S}(t){\cal \hat{O}}(0)$ satisfies (\ref{rhoSevolution}). We thus have the formal solution $\hat{\rho}^{\cal O} (\tau) = e^{{\cal W}\tau} \hat{\rho}^{\cal O}(0) = e^{{\cal W}\tau}( \hat{\rho}(t){\cal \hat{O}}(0))$. Using this result, and assuming that ${\cal \hat{O}}$ is a system operator, we can write Eq.~(\ref{corr2QRT}) as
\beq 
\ew{{\cal \hat{O}}_1(t){\cal \hat{O}}_2(t+\tau)}= \mathrm{Tr_S} \left\lbrace {\cal \hat{O}}_2(0) e^{\mathcal{W}\tau} \hat{\rho}_\mathrm{S}(t){\cal \hat{O}}_1(0) \right\rbrace. \label{QRT2results} \\
\ew{{\cal \hat{O}}_1(t+\tau){\cal \hat{O}}_2(t)}= \mathrm{Tr_S}\left\lbrace {\cal \hat{O}}_1(0) e^{\mathcal{W}\tau} {\cal \hat{O}}_2(0) \hat{\rho}_\mathrm{S}(t) \right\rbrace.
\eeq
In a similar way, defining $\hat{\rho}^{{\cal O}_1{\cal O}_2}(\tau):=e^{-i{\cal \hat{H}}\tau}{\cal \hat{O}}_1(0)\hat{\rho}(t){\cal \hat{O}}_2(0)e^{i{\cal \hat{H}}\tau}$ and $\hat{\rho}_\mathrm{S}^{{\cal O}_1{\cal O}_2}(\tau):=\mathrm{Tr_R}\{\hat{\rho}^{{\cal O}_1{\cal O}_2}\}$, we find
\beq
\ew{{\cal \hat{O}}_1(t){\cal \hat{O}}_2(t+\tau){\cal \hat{O}}_3(t)}= \mathrm{Tr_S} \left\lbrace {\cal \hat{O}}_2(0) e^{\mathcal{W}\tau} {\cal \hat{O}}_3(0) \hat{\rho}_\mathrm{S}(t) {\cal \hat{O}}_1(0) \right\rbrace.
\eeq
Now, there exists a set of operators $\{\hat{A}_{\mu}\}$ such that for every operator ${\cal \hat{O}}$, we have\footnote{The operators $\hat{A}_{\mu}$, given by $\hat{A}_{\mu}=\hat{A}_{nm}= \op{n}{m}$, verify: \newline $\mathrm{Tr_S}\lbrace A_{nm}(\mathcal{W}{\cal \hat{O}})\rbrace = \sum_{n'm'}M_{nm,n'm'}\mathrm{Tr_S}\lbrace \hat{A}_{n'm'}{\cal \hat{O}}\rbrace$, with $M_{nm,n'm'}\equiv\bra{m} ( \mathcal{W}\op{m'}{n'} )\ket{n}$.}
\begin{eqnarray} \label{AmuEq}
&&\mathrm{Tr_S} \left\lbrace \hat{A}_{\mu}(\mathcal{W}{\cal \hat{O}})\right\rbrace = \sum_{\lambda} M_{\mu\lambda} \mathrm{Tr_S} \left\lbrace \hat{A}_{\lambda}{\cal \hat{O}} \right\rbrace \Rightarrow \nonumber \\ &\Rightarrow& \ew{\dot{\hat{A}}_{\mu}} = \mathrm{Tr_S} \left\lbrace \hat{A}_{\mu}\dot{\hat{\rho}}_\mathrm{S} \right\rbrace = \mathrm{Tr_S} \left\lbrace \hat{A}_{\mu}\left(\mathcal{W}\hat{\rho}_\mathrm{S}\right) \right\rbrace = \sum_{\lambda} M_{\mu\lambda} \mathrm{Tr_S} \left\lbrace \hat{A}_{\lambda}\hat{\rho}_\mathrm{S} \right\rbrace = \sum_{\lambda} M_{\mu\lambda} \ew{\hat{A}_{\lambda}} \Rightarrow \nonumber \\ &\Rightarrow& \frac{d}{dt}\ew{\hat{\vec{A}}}=M\ew{\hat{\vec{A}}}.
\end{eqnarray}
Using this result, together with (\ref{QRT2results}), we find
\beq
\frac{d}{d\tau}\ew{{\cal \hat{O}}_1(t)\hat{A}_{\mu}(t+\tau)} &=& \frac{d}{dt} \mathrm{Tr_S} \left\lbrace \hat{A}_{\mu}(0)\left(e^{\mathcal{W}\tau}\hat{\rho}_\mathrm{S}(t){\cal \hat{O}}_1(0)\right)\right\rbrace = \mathrm{Tr_S} \left\lbrace \hat{A}_{\mu}(0)\left(\mathcal{W}e^{\mathcal{W}\tau}\hat{\rho}_\mathrm{S}(t){\cal \hat{O}}_1(0)\right)\right\rbrace \nonumber\\ &=& \sum_{\lambda} M_{\mu\lambda} \mathrm{Tr_S} \left\lbrace \hat{A}_{\lambda}(0)e^{\mathcal{W}\tau}\hat{\rho}_\mathrm{S}(t){\cal \hat{O}}_1(0)\right\rbrace = \sum_{\lambda} M_{\mu\lambda} \ew{{\cal \hat{O}}_1(t)\hat{A}_{\lambda}(t+\tau)}.
\eeq
We therefore arrive to
\begin{equation}
\begin{array}{l} \label{QRTresults2order}
\frac{d}{d\tau}\ew{{\cal \hat{O}}_1(t)\hat{\vec{A}}(t+\tau)} = M \ew{{\cal \hat{O}}_1(t)\hat{\vec{A}}(t+\tau)} \Longrightarrow \ew{{\cal \hat{O}}_1(t)\hat{\vec{A}}(t+\tau)} = e^{M\tau} \ew{{\cal \hat{O}}_1(t)\hat{\vec{A}}(t)}. \\
\frac{d}{d\tau}\ew{\hat{\vec{A}}(t+\tau){\cal \hat{O}}_2(t)} = M \ew{\hat{\vec{A}}(t+\tau){\cal \hat{O}}_2(t)} \Longrightarrow \ew{\hat{\vec{A}}(t+\tau){\cal \hat{O}}_2(t)} = e^{M\tau} \ew{\hat{\vec{A}}(t){\cal \hat{O}}_2(t)}. \\
\frac{d}{d\tau}\ew{{\cal \hat{O}}_1(t)\hat{\vec{A}}(t+\tau){\cal \hat{O}}_2(t)} = M \ew{{\cal \hat{O}}_1(t)\hat{\vec{A}}(t+\tau){\cal \hat{O}}_2(t)} \Longrightarrow \\ \;\;\;\;\;\;\;\;\;\;\;\;\;\;\;\;\;\;\;\;\;\;\;\;\;\;\;\;\;\;\;\;\;\;\;\;\;\;\;\;\;\;\;\;\;\;\; \Rightarrow \ew{{\cal \hat{O}}_1(t)\hat{\vec{A}}(t+\tau){\cal \hat{O}}_2(t)} = e^{M\tau} \ew{{\cal \hat{O}}_1(t)\hat{\vec{A}}(t){\cal \hat{O}}_2(t)}.
\end{array}
\end{equation}
In these expressions, it is important to remember that the matrix $M$ acts only on $\hat{\vec{A}}$.

Let us now turn to the most general problem of calculating a correlation function of an arbitrary number of system operators. To illustrate the procedure we give here the derivation of the quantum regression theorem for the third-order correlator
\beq
&&\ew{{\cal \hat{O}}_1(t){\cal \hat{O}}_2(t'){\cal \hat{O}}_3(t'')} = \mathrm{Tr} \left\lbrace \hat{\rho} (0) {\cal \hat{O}}_1(t){\cal \hat{O}}_2(t'){\cal \hat{O}}_3(t'') \right\rbrace =  \nonumber \\ &=& \mathrm{Tr} \left\lbrace e^{i{\cal \hat{H}}t} \hat{\rho} (t) e^{-i{\cal \hat{H}}t} e^{i{\cal \hat{H}}t} {\cal \hat{O}}_1(0) e^{-i{\cal \hat{H}}t} e^{i{\cal \hat{H}}t'} {\cal \hat{O}}_2(0) e^{-i{\cal \hat{H}}t'} e^{i{\cal \hat{H}}t''} {\cal \hat{O}}_3(0) e^{-i{\cal \hat{H}}t''} \right\rbrace = \nonumber \\
&=& \mathrm{Tr_S} \left\lbrace {\cal \hat{O}}_3(0) \mathrm{Tr_R} \left\lbrace e^{-i{\cal \hat{H}}\tau'} \hat{\rho} (t) {\cal \hat{O}}_1(0) e^{i{\cal \hat{H}}\tau} {\cal \hat{O}}_2(0) e^{i{\cal \hat{H}}(\tau'-\tau)} \right\rbrace \right\rbrace,
\eeq
where we have defined $\tau:=t'-t$ and $\tau':=t''-t$. Furthermore, we define
\beq
\hat{\rho}^{{\cal O}_1{\cal O}_2(\tau)}(\tau') := e^{-i{\cal \hat{H}}\tau'} \underbrace{\Big[ \hat{\rho}(t){\cal \hat{O}}_1 (0) \overbrace{e^{i{\cal \hat{H}}\tau}{\cal \hat{O}}_2(0)e^{-i{\cal \hat{H}}\tau}}^{{\cal \hat{O}}_2(\tau)} \Big]}_{\hat{\rho}^{{\cal O}_1{\cal O}_2(\tau)}(\tau'=0)} e^{i{\cal \hat{H}}\tau'},
\eeq
which fulfills the von Neumann's equation (\ref{vonNeumann}). Similarly, $\hat{\rho}_\mathrm{S}^{{\cal O}{\cal O}_2(\tau)}(\tau'=0)\equiv \hat{\rho}_\mathrm{S}(t){\cal \hat{O}}(0){\cal \hat{O}}(\tau)$ satisfies (\ref{rhoSevolution}), with formal solution $\hat{\rho}_\mathrm{S}^{{\cal O}_1{\cal O}_2(\tau)}(\tau') = e^{\mathcal{W}\tau'}( \hat{\rho}_\mathrm{S}^{{\cal O}_1{\cal O}_2(\tau)}(\tau'=0))$. We thus have the result for the three-time correlator
\beq
\ew{{\cal \hat{O}}_1(t){\cal \hat{O}}_2(t'){\cal \hat{O}}_3(t'')} = \mathrm{Tr_S} \left\lbrace {\cal \hat{O}}_3(0) e^{\mathcal{W}\tau'}\hat{\rho}_\mathrm{S}(t) {\cal \hat{O}}_1(0){\cal \hat{O}}_2(\tau) \right\rbrace.
\eeq
Now, if $\{A_{\mu}\}$ and $\{B_{\mu}\}$ are two sets of operators satisfying (\ref{AmuEq}), we find
\beq
&&\frac{d}{d\tau'}\ew{{\cal \hat{O}}_1(t)\hat{B}_{\nu}(t+\tau)\hat{A}_{\mu}(t+\tau')} \nonumber \\ &=& \frac{d}{d\tau'} \mathrm{Tr_S} \left\lbrace \hat{A}_{\mu}(0) e^{\mathcal{W}\tau'}\hat{\rho}_\mathrm{S}(t) {\cal \hat{O}}_1(0)\hat{B}_{\nu}(\tau) \right\rbrace = \mathrm{Tr_S} \left\lbrace \hat{A}_{\mu}(0) \mathcal{W}e^{\mathcal{W}\tau'}\hat{\rho}_\mathrm{S}(t) {\cal \hat{O}}_1(0)\hat{B}_{\nu}(\tau) \right\rbrace \nonumber \\ &=& \sum_{\lambda} M_{\mu\lambda}^{(A)} \mathrm{Tr_S} \left\lbrace \hat{A}_{\lambda}(0)e^{\mathcal{W}\tau'}\hat{\rho}_\mathrm{S}(t){\cal \hat{O}}_1(0)\hat{B}_{\nu}(\tau) \right\rbrace = \sum_{\lambda} M_{\mu\lambda}^{(A)} \ew{{\cal \hat{O}}_1(t)\hat{B}_{\nu}(t+\tau)\hat{A}_{\lambda}(t+\tau')} \Longrightarrow \nonumber \\ &\Rightarrow& \frac{d}{d\tau'}\ew{{\cal \hat{O}}_1(t)\hat{\vec{B}}(t+\tau)\otimes\hat{\vec{A}}(t+\tau')} = M^{(A)} \ew{{\cal \hat{O}}_1(t)\hat{\vec{B}}(t+\tau)\otimes\hat{\vec{A}}(t+\tau')} \Longrightarrow \nonumber\\ &\Rightarrow& \ew{{\cal \hat{O}}_1(t)\hat{\vec{B}}(t+\tau)\otimes\hat{\vec{A}}(t+\tau')} = e^{M^{(A)} \tau'} \ew{{\cal \hat{O}}_1(t)\hat{\vec{B}}(t+\tau)\otimes\hat{\vec{A}}(t)}.
\eeq
Now, applying Eq.~(\ref{QRTresults2order}) we arrive to our final result:
\beq
\ew{{\cal \hat{O}}_1(t)\hat{\vec{B}}(t+\tau)\otimes\hat{\vec{A}}(t+\tau')} = e^{M^{(A)} \tau'}e^{M^{(B)} \tau} \ew{{\cal \hat{O}}_1(t)\hat{\vec{B}}(t)\otimes\hat{\vec{A}}(t)}.
\eeq
This equations can be easily generalized to give
\beq
\ew{\mathcal{\hat{O}}(t)\hat{\vec{A}}_1(t+\tau_1)\otimes \ldots \otimes\hat{\vec{A}}_k(t+\tau_k)} = e^{M^{(k)} \tau_k} \ldots e^{M^{(1)} \tau_1} \ew{\mathcal{\hat{O}}(t)\hat{\vec{A}}_1(t)\otimes \ldots \otimes\hat{\vec{A}}_k(t)},
\eeq
where it is important to remember that the $M^{(i)}$ operator, defined through $\frac{d}{dt}\ew{\hat{\vec{A}}_i} = M^{(i)} \ew{\hat{\vec{A}}_i}$, acts only on the corresponding $\hat{\vec{A}}$ operator, that is
\beq
\ew{\mathcal{\hat{O}}(t)\hat{\vec{A}}_1(t+\tau_1)\otimes \ldots \otimes\hat{\vec{A}}_k(t+\tau_k)} = \ew{\mathcal{\hat{O}}(t)e^{M^{(1)} \tau_1}\hat{\vec{A}}_1(t)\otimes \ldots \otimes e^{M^{(k)} \tau_k}\hat{\vec{A}}_k(t)}.
\eeq

\subsection{MacDonald's formula}

Following the studies by D. K. C. MacDonald on spontaneous fluctuations \cite{MacDonald49}, a simplified formula for the frequency-dependent noise spectrum in terms of single-time averages can be derived \cite{MacDonaldBook}. This is called MacDonald's formula. Here we give its derivation and comment on a possible extension to higher orders. Let us consider the number of charge carriers $n$ flowing along a circuit element. After time $\tau$, this gives the net charge $e n(\tau)$, and creates a (fluctuating) current $I$ given by $e n(\tau)=\int_t^{t+\tau} I(t') dt'$, which gives the average $\ew{n(\tau)}=\tau\ew{I}/e$. Therefore, the displaced current $\delta I(t) := I(t) - \ew{I}$ between times $\tau_1$ and $\tau_2$ can be expressed as
\beq
\int_{\tau_1}^{\tau_2} \delta I(t') dt' = en(\tau_2) - en(\tau_1) - (\tau_2 - \tau_1)\ew{I} := e N(\tau_1,\tau_2)
\eeq
Let us consider the correlator
\beq
&&C^{(2)}(\tau_1,\tau_2,\tau_1',\tau_2'):=e^2 {\cal T}_S \ew{N(\tau_1,\tau_2)N(\tau_1',\tau_2')} = \int_{\tau_1}^{\tau_2} dt_1 \int_{\tau_1'}^{\tau_2'} dt_2 {\cal T}_S \ew{\delta I(t_1)\delta I(t_2)} \nonumber\\
&&=\int_{\tau_1}^{\tau_2} dt_1 \int_{\tau_1'}^{\tau_2'} dt_2 {\cal T}_S \int_{-\infty}^{\infty} d\omega_1 d\omega_2 \left(\frac{1}{2\pi}\right)^2 e^{i\omega_1 t_1}e^{i\omega_2 t_2} S^{(2)}(\omega_1,\omega_2) 
\\
&&=\left(\frac{1}{2\pi}\right)^2 {\cal T}_S \int_{-\infty}^{\infty} d\omega_1 d\omega_2 \left(\frac{1}{i\omega_1}\right)\left(\frac{1}{\omega_2}\right) \left( e^{i\omega_1\tau_2} - e^{i\omega_1\tau_1} \right)  \left( e^{i\omega_2\tau_2'} - e^{i\omega_2\tau_1'} \right) S^{(2)}(\omega_1,\omega_2),\nonumber
\eeq
where ${\cal T}_S$ is the symmetrization operator introduced in the previous chapter, and where we have used the definition (\ref{noisedef}) of the noise spectrum. Taking $\tau_1=\tau_1'=0$, $\tau_2=\tau_2'=\tau$, and using that $S^{(2)}(\omega_1,\omega_2) = 2\pi \delta(\omega_1+\omega_2) S^{(2)}(\omega)$, with $\omega\equiv\omega_2$, we find
\beq
C^{(2)}(0,\tau,0,\tau)= \frac{1}{2\pi} {\cal T}_S \int_{-\infty}^{\infty} \frac{2}{\omega^2} \left( 1-\cos(\omega\tau) \right) S^{(2)} (\omega) d\omega.
\eeq
Taking the derivative of this equation with respect to $\tau$ and inverting the result we get
\beq
S^{(2)}(\tau) = \frac{1}{2}\frac{\partial C^{(2)}(0,\tau,0,\tau)}{\partial \tau} = \frac{e^2}{2} {\cal T}_S \frac{\partial}{\partial\tau} \ew{N(0,\tau)N(0,\tau)},
\eeq
which, assuming $n(0)=0$, in the frequency space gives the \textit{MacDonald's formula}:
\beq \label{MacDonaldEq}
S^{(2)}(\omega) &=& 2e^2\omega \int_0^{\infty} \sin(\omega\tau) \frac{\partial}{\partial\tau} \left[ \ew{n^2(\tau)} - \tau^2\ew{I}^2/e^2 \right] d\tau \nonumber\\ &=& 2e^2\omega \int_0^{\infty} \sin(\omega\tau) \frac{\partial}{\partial\tau} \ew{n^2(\tau)}_c d\tau,
\eeq
where the subscript $c$ means cumulant. The importance of this equation is based on the fact that the frequency-dependent power spectrum can be calculated from the knowledge of a single-time average, namely $\ew{n^2(\tau)}_c$. Generalizing this property to higher orders is, to the best of our knowledge, not possible in general. For example, the third-order current correlator would involve the function $C^{(3)}(\tau_1,\tau_2,\tau_1',\tau_2',\tau_1'',\tau_2''):=e^3{\cal T}_S \ew{N(\tau_1,\tau_2)N(\tau_1',\tau_2')N(\tau_1'',\tau_2'')}$. The limit $\tau_1=\tau_1'=\tau_1''=0$, $\tau_2=\tau_2'=\tau_2''=\tau$, needed to express the third-order cumulant in terms of single-time averages, only eliminates one frequency integral in the expression for $C^{(3)}$, and it is thereby not possible to isolate $S^{(3)}(\omega,\omega')$, as it was done for $S^{(2)}(\omega)$. There are, however, possible ways to express $S^{(3)}(\omega,\omega')$ in terms of correlation functions of the form $\ew{n^2(\tau_2)n(\tau_1)}$. The impossibility of deriving a MacDonald's formula for the frequency-dependent skewness, arises from the need to consider at least two different times, $\tau$ and $\tau'$, in $C^{(3)}$, and which leads to equations of the type
\beq
S^{(3)}(\omega,\omega') = -\frac{\omega+\omega'}{4} \int_{-\infty}^{\infty} d\tau d\tau' \sin(\omega\tau+\omega\tau') \frac{\partial^2 C^{(3)}(\tau,\tau',\tau,\tau',0,\tau+\tau')}{\partial\tau \partial\tau'}.
\eeq

\subsection{Projection techniques}

A very useful method complementing the density operator approach are the projection or Nakajima-Zwanzig techniques \cite{Nakajima58, Zwanzig60, Mori65, ZwanzigBook, BreuerPetruccioneBook}. As mentioned in subsection \ref{SecDensityOperator}, one is usually interested in the dynamics of a part of the whole system under study. In system-bath problems, the bath degrees of freedom are typically traced out and we solve the dynamics of the system density operator only. The projection techniques introduce the projector $P$, whose action on the density operator is defined as\footnote{Alternatively, $P$ can be defined as $P\hat{\rho}=\mathrm{Tr_B}\{\hat{\rho}\}=\hat{\rho}_\mathrm{S}$.} $P\hat{\rho}=\mathrm{Tr_B}\{\hat{\rho}\}\otimes\hat{\rho}_\mathrm{B}=\hat{\rho}_\mathrm{S}\otimes\hat{\rho}_\mathrm{B}$, so that the trace operation can be written in a simple way. The QME (\ref{LiouvilleEq}) can be projected onto the relevant and irrelevant parts of the Liouville space using $P$ and $Q:=\mathds{1}-P$, giving rise to
\beq
\frac{d}{dt} P \hat{\rho}(t) = P {\cal L} \hat{\rho}(t). \label{projectedrho1}\\
\frac{d}{dt} Q \hat{\rho}(t) = Q {\cal L} \hat{\rho}(t). \label{projectedrho2}
\eeq
Introducing $\mathds{1}=P+Q$ between ${\cal L}$ and $\hat{\rho}$ in both equations, and substituting the formal solution of (\ref{projectedrho2}), namely $Q\hat{\rho}(t)=e^{Q{\cal L}t}Q\hat{\rho}(0)+\int_0^t e^{Q{\cal L}(t-t')}Q{\cal L}P\hat{\rho}(t')dt'$, into (\ref{projectedrho1}), we obtain the \textit{Nakajima-Zwanzig} equation \cite{Nakajima58, Zwanzig60}:
\beq \label{NakajimaZwanzigEq}
\frac{d}{dt} P\hat{\rho}(t)=P{\cal L}P\hat{\rho}(t) + \int_0^t P{\cal L} e^{Q{\cal L}(t-t')}Q{\cal L}P\hat{\rho}(t')dt' + P{\cal L}e^{Q{\cal L}t}Q\hat{\rho}(0).
\eeq
Defining $\hat{\rho}_{\cal O}:={\cal O}\hat{\rho}$ and ${\cal L}_{{\cal O}{\cal O}'}:={\cal O}{\cal L}{\cal O}'$, with ${\cal O}=P, Q$, this equation can be equivalently written as\footnote{The projectors $P$ and $Q$ fulfill $P^2=P$, $Q^2=Q$, $PQ=QP=0$.} $\frac{d}{dt} \hat{\rho}_P(t)={\cal L}_{PP}\hat{\rho}_P(t) + \int_0^t {\cal L}_{PQ} e^{{\cal L}_{QQ}(t-t')}{\cal L}_{QP}\hat{\rho}_P(t')dt' + {\cal L}_{PQ}e^{{\cal L}_{QQ}t}\hat{\rho}_Q(0)$. The first term in (\ref{NakajimaZwanzigEq}) can be identified with the first term of equation (\ref{Deltarho2}), which, following the arguments given in that section, can be taken to zero in the interaction picture assuming that the average of an odd number of interaction terms vanishes. Thus, in the Schr\"odinger picture it just reduces to ${\cal L}_\mathrm{S}\hat{\rho}_\mathrm{S}$. The second term in (\ref{NakajimaZwanzigEq}) matches the second term of Eq.~(\ref{Deltarho2}), after having neglected $\hat{\rho}_\mathrm{corr}$ introduced in that section. This \textit{inhomogeneous term} is in correspondence with the last term of the Nakajima-Zwanzig equation. It describes the system-bath correlations built up to time $t=0$. As we will see, although it can be neglected within the Markovian approximation, it is of vital importance for the proper description of the non-Markovian dynamics of an open quantum system, as well as for reproducing correctly the physics of quantum noise.

The projection techniques introduced in this subsection are of particular interest for the calculation of correlation functions \cite{Becker88, Becker89, FuldeBook}. They have been used to evaluate non-Markovian corrections \cite{Breuer99, Breuer01, BreuerPetruccioneBook}, and in the field of shot noise, Flindt and collaborators have made a great effort to formulate a theory of counting statistics using this formalism \cite{Flindt04, Novotny04, Jauho05, Flindt05c, Flindt05b, Flindt08, Flindt09, Flindt10}. One of the main aims of this thesis is to give a simplistic formulation of frequency-dependent current correlations making use of the projection techniques. Equations for the cumulant generating function and finite-frequency current cumulants will be derived in both the Markovian and non-Markovian situations.

\subsection{Full counting statistics} \label{FCSsubsectionCH2}

The theory of \textit{full counting statistics} (FCS) of electron transport centers its interest in the probability distribution $P(n,t)$ of having $n$ transfers of charge through the system after time $t$. To be more specific, let us consider the example of a tunnel barrier connected to electrodes at different chemical potentials. The current flow through the system, determined by the number of charges that tunnel through the junction, will follow a probability distribution. At the junction, each particle is transmitted with probability $\mathsf{T}$, say, and reflected with probability $1-\mathsf{T}$. Therefore, the charge flow (and so the current) follows a binomial distribution in this particular example. For a most general conductor, this distribution can be arbitrary complicated. However, the knowledge of $P(n,t)$ is of great importance. As we have learned, a great deal of information about the particle statistics and about the conductor can be gained from the study of current correlations. It looks thus apparent that having the full distribution will provide us with a broader knowledge of the transport properties than that acquired from the noise spectrum (\ref{noisedef}) -- second cumulant of the current distribution. These cumulants are connected with $P(n,t)$ through the \textit{cumulant generating function} (CGF) ${\cal F}$, defined as the Fourier transform of the probability distribution:
\beq \label{MGF-CGF}
e^{{\cal F}(\chi,t)} := {\cal G}(\chi,t) := \sum_n P(n,t) e^{in\chi}.
\eeq
Here, the variable $\chi$ is the so-called counting field, and ${\cal G}$ the \textit{moment generating function} (MGF). Derivatives of ${\cal F}$ with respect to $\chi$ generate the cumulants of the number distribution,
\beq \label{ncumulantCGF}
\ew{n^N(t)}_c = \frac{\partial^N{\cal F}(\chi,t)}{\partial(i\chi)^N}\Big|_{\chi=0},
\eeq
and these are connected with the current cumulants taking the time derivative, $\ew{I(t)}_c=\frac{d}{dt}\ew{n(t)}_c$. Similarly, the MGF generates the moments, $\ew{n^N(t)} = \frac{\partial^N{\cal G}(\chi,t)}{\partial(i\chi)^N}\Big|_{\chi=0}$. It is important to notice that the first three cumulants coincide with their central-moments counterparts, $\ew{(n-\ew{n})^N}$, with $N$ the cumulant order. However, for $N>3$, these two are different. Also, we notice that the number cumulants grow linearly with time, which can be inferred from $\ew{n(t)n(t+\tau)}_c=\int_0^t \int_0^{t+\tau}\ew{I(0)I(t_1)}_c dt_0 dt_1 \sim t S^{(2)}(\omega=0)$, where we have used the translational-invariance property of the current cumulants and the definition (\ref{noisedef}), and this property holds similarly for higher orders. The more commonly used probability distributions and respective CGFs are captured in table \ref{tabDistributions}.

\begin{table}
\setcounter{table}{0}
\begin{center}
\begin{tabular}{| c || c | c |}
	\hline
\textit{Distribution}  & $P(n)$ & ${\cal F}(\chi)$ \\ \hline \hline
Delta   & $\delta_{n\ew{n}}$ & $i\chi\ew{n}$ \\ \hline
Gaussian   & $e^{ -\frac{(n-\ew{n})^2}{2\sigma} }$ & $i\chi\ew{n}-\sigma\chi^2/2$ \\ \hline
Binomial   & $\binom{m}{n}p^n(1-p)^{m-n}$ & $m\ln\left\{ 1+p(e^{i\chi}-1) \right\}$ \\ \hline
Poisson & $\frac{\ew{n}^n}{n!}e^{-\ew{n}}$  & $\ew{n} (e^{i\chi}-1)$ \\ \hline
\end{tabular}
\caption[Typical probability distributions and respective cumulant generating functions]{Typical probability distributions and respective cumulant generating functions. Notice that the binomial distribution reduces to the Poisson distribution for $p\ll 1$. This implies that a tunnel junction or quantum point contact has Poissonian statistics in this tunnel limit.}
\label{tabDistributions}
\end{center}
\end{table}

From equations (\ref{MacDonaldEq}) and (\ref{ncumulantCGF}) it is obvious that the frequency-dependent noise can be expressed in terms of the CGF/MGF. Using a convergence factor in the integrals ($\omega\to\omega\pm i\eta$), one can easily see that \cite{Flindt08}
\beq \label{MacDonaldFlindtEq}
S^{(2)}(\omega) = -\frac{\omega^2}{2}  \frac{\partial^2}{\partial (i\chi)^2} \left[ {\cal G}(\chi, z=i\omega) + (\omega\to -\omega) \right] \Big\vert_{\chi\to 0},
\eeq
where ${\cal G}(\chi,z)\equiv \int_0^{\infty} e^{-zt} {\cal G}(\chi,t)$.
Levitov and coworkers derived a general equation for the CGF, describing problems of the type shown in figure \ref{ScatteringTransportFig}. This is known as \textit{Levitov's formula}, and reads \cite{Levitov92, Levitov93, Levitov96}
\beq
{\cal F}(\chi) = \det \left[ \mathds{1} + \bar{f}(E) \left( \tilde{{\cal S}}^{\dagger}\tilde{{\cal S}} - \mathds{1} \right) \right],
\eeq
where $\tilde{{\cal S}}$ is the $\chi$-dependent scattering matrix, defined as $\tilde{{\cal S}}_{jk}\equiv e^{i(\chi_j-\chi_k)} {\cal S}_{jk}$, and $[\bar{f}(E)]_{jk} \equiv f_j(E)\delta_{jk}$, with $f_j(E)$ and $\chi_j$ Fermi distribution and counting field, respectively, corresponding to lead $j$. This result was later generalized to include interacting problems by Belzig and Nazarov \cite{BelzigNazarov01}, who found
\beq
{\cal F}(\chi,t) = \frac{-t}{2\pi} \sum_n \int dE \; \mathrm{Tr}\left\lbrace \ln\left[ 4 + \mathsf{T_n} \left( \lbrace \check{G}_1(\chi,E),\check{G}_2(\chi,E) \rbrace - 2 \right) \right] \right\rbrace.
\eeq
In this equation, $\check{G}_i$ is the $i$th.-contact GF in the Keldysh-Nambu notation, and $\mathsf{T_n}$ the transmission eigenvalues caracterizing the system. A theory of FCS in the framework of the density matrix approach was established in \cite{BagretsNazarov03}. Here, the authors arrive to the equation for the CGF
\beq \label{BagretsNazarovEq}
e^{{\cal F}(\chi)} = {\cal T}_S \ew{e^{\int_{-\infty}^{\infty}{\cal W}[\chi(t')] dt'}}, 
\eeq
where the operator ${\cal T}_S$ ensures that we obtain the symmetrized cumulant of the distribution, and the kernel ${\cal W}(\chi)$ has a term $e^{i\chi}$ in the entries concerning a change in the number of particles that we are `counting' (a detailed explanation will be given in section \ref{FCSwPaper1Sec} and in the next chapter). Below we will show how this equation is derived to account for multi-time current cumulants. A technique will be developed to obtain frequency-dependent correlation functions up to arbitrary order in both the Markovian and non-Markovian cases, and this will be applied to study the counting statistics in quantum-dot systems. Related experiments have been performed in the last years, where the counting of electrons tunneling through a quantum dot is possible using a nearby quantum point contact \cite{Fujisawa04, Fujisawa06, Gustavsson06prl, Gustavsson06prb, Gustavsson07, Sukhorukov07, Fricke07, Flindt09}. In these experiments, high-order cumulants and the full probability distribution has been measured. As mentioned in the previous chapter, these correlations cannot only be measured by `real time' detection, but also with transport measurements, probing the current signal with a spectrum analyzer (see e.g. \cite{Reulet03, Bomze05}). The difference between these two types of detection is contained in our theory and will be explained in chapter \ref{Chapter3}.

\subsection{Examples}

In this section we present two paradigmatic models of quantum transport, and of central interest in the thesis. Some of the techniques presented above will be exemplified in this subsection. Although the noise theory presented in the following chapters is based on the density operator approach, it is useful to see how the different techniques can be applied to obtain current and noise spectrum in these models.

\subsubsection{Single resonant level model and other configurations} \label{SRLandothers}

\begin{figure}
  \begin{center}
    \includegraphics[width=0.6\textwidth]{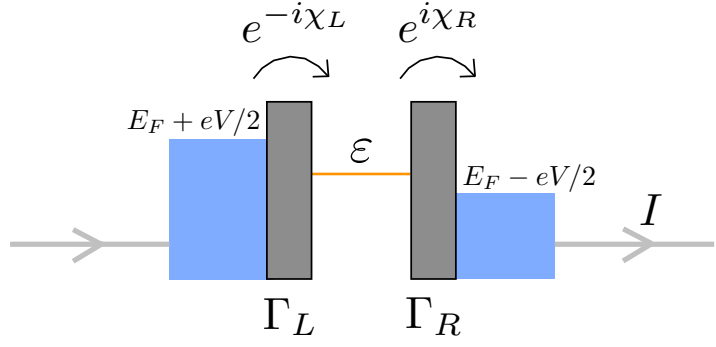}
  \end{center}  
  \caption[Single Resonant Level model]{Single Resonant Level model. A discrete level with energy $\varepsilon$ is coupled with rates $\Gamma_L$ and $\Gamma_R$ to a continuum of states in two different electronic reservoirs at chemical potentials $E_F\pm eV/2$. In the figure, counting fields $e^{-i\chi_L}$ and $e^{i\chi_R}$ corresponding to the infinite-bias limit have been included to illustrate the counting statistics of electrons tunneling through the left and right barriers.
}
  \label{SRLFig}
\end{figure}

The so-called single resonant level (SRL) model represents a basic paradigm of quantum transport. It consists of a single level -- formed for example in a quantum dot (QD) -- coupled to electronic reservoirs at different chemical potentials (see Fig.~\ref{SRLFig}). Transport of electrons through the system is thus possible when an electron of the leads is resonant with the discrete level in the QD. The model is described by the Hamiltonian
\beq \label{SRLhamiltonian} 
\hat{\mathcal{H}} =
\varepsilon {\ket 1}{\bra 1} + \sum_{k,\alpha\in L,R} \left(
\varepsilon_{k\alpha} + \mu_{\alpha} \right) \hat{c}_{k\alpha}^{\dagger}
\hat{c}_{k\alpha} + \sum_{k,\alpha\in L,R} {\cal V}_{k\alpha}
\hat{c}_{k\alpha}^{\dagger} {\ket 0}{\bra 1} \; + \mathrm{H.c.}.
\eeq 
Here, ${\cal V}_{k\alpha}$ are system-reservoir coupling constants, where $k$ is the momentum and $\alpha$ the lead index. $\mu_{\alpha}$ denotes the chemical potential and $\varepsilon_{k\alpha}$ the energies of the electrons in the leads. ${\ket 0}$ and ${\ket 1}$ are the only two possible QD states (referring to empty and occupied level) due to Coulomb blockade, with respective energies $0$ and $\varepsilon$. Finally, the operators $\hat{c}_{k\alpha}^{\dagger}$/$\hat{c}_{k\alpha}$ create/destroy an electron in the leads.
In the infinite-bias limit (voltage much larger than the other
energy scales, excepting the bandwidth of the Fermi contacts) the
Hamiltonian (\ref{SRLhamiltonian}) leads to the kernel (c.f. Eqs.~(\ref{SRLrateEqs}))
\beq\label{kernelinfty} \mathcal{W}(\chi) =
\begin{pmatrix}
    - \Gamma_L &\Gamma_R  e^{i\chi_R} \\
    \Gamma_L  e^{-i\chi_L} & -\Gamma_R
\end{pmatrix},
\eeq 
expressed in the basis $\lbrace {\ket 0}, {\ket 1} \rbrace$, and where $\Gamma_{\alpha} \approx \Gamma_{\alpha}(E) := \frac{2\pi}{\hbar} \sum_k |{\cal V}_{k\alpha}|^2 \delta(E-\varepsilon_{k\alpha})$ are rates corresponding to the system-reservoir coupling. In the kernel (\ref{kernelinfty}), the counting fields $\chi_L$ and $\chi_R$ -- responsible for the counting at the left and right reservoirs respectively -- have been introduced according to the scheme presented in subsection \ref{FCSsubsectionCH2} (more details on how to construct this kernel will be given in chapter~\ref{Chapter3}).
The current and noise through the system are exactly solvable using the GFs approach \cite{Averin93, EngelThesis} and the scattering matrix formalism \cite{BlanterButtiker00}. Below we will reproduce these results using our theory. We can also use the QRT to derive these results. It is important to remark that this theorem refers to correlation functions of system operators only. In the SRL case, the current operator can be expressed in terms of QD operators and therefore the QRT can be applied to obtain the noise spectrum. Let us, for example, calculate the charge noise
\beq \label{chargenoiseDef}
S^{(2)}_{Q}(\omega) \equiv 2 \int_0^{\infty} \cos(\omega\tau) \ew{\hat{Q}(t+\tau)\hat{Q}(t)}d\tau,
\eeq
for this model. If $\hat{n}_0:=\op{0}{0}$ and $\hat{n}_1:=\op{1}{1}$, in the infinite-bias limit we have
\begin{equation} \label{EOMsrlQRT}
\begin{pmatrix}
        \dot{\hat{n}}_{0}(t) \\
        \dot{\hat{n}}_{1}(t)
\end{pmatrix}
=
\underbrace{\begin{pmatrix}
        -\Gamma_L & \Gamma_R \\
        \Gamma_L & -\Gamma_R
        \end{pmatrix}}_{M}
\underbrace{\begin{pmatrix}
        \hat{n}_{0}(t) \\
        \hat{n}_{1}(t)
        \end{pmatrix}}_{\hat{\vec{A}}},
\end{equation}
whose solution is
\beq
\hat{\vec{A}} (t) = e^{M(t-t_0)} \hat{\vec{A}}(t_0) = \frac{1}{\Gamma} 
\begin{pmatrix}
        \Gamma_R + \Gamma_L e^{-\Gamma (t-t_0)} & \Gamma_R \left( 1 - e^{-\Gamma (t-t_0)} \right) \\
        \Gamma_L \left( 1 - e^{-\Gamma (t-t_0)} \right) & \Gamma_L + \Gamma_R e^{-\Gamma (t-t_0)} \\
        \end{pmatrix}
\hat{\vec{A}} (t_0),
\eeq
where $\Gamma\equiv \Gamma_L+\Gamma_R$. In terms of the QRT, we can calculate correlators as
\begin{equation}
\langle \hat{n}_i(t) \hat{\vec{A}} (t+\tau) \rangle = e^{M\tau} \langle \hat{n}_i(t) \hat{\vec{A}} (t) \rangle,
\end{equation}
with $i=0,1$, and the matrix $M$ acting on $\hat{\vec{A}}$. Using $\hat{n}_0^2=\hat{n}_0$, $\hat{n}_0\hat{n}_1=\hat{n}_1\hat{n}_0=0$, $\hat{n}_1^2=\hat{n}_1$, and $\ew{\hat{n}_0}=\Gamma_R/\Gamma$, $\ew{\hat{n}_1}=\Gamma_L/\Gamma$, we find
\begin{equation}
\langle \hat{n}_i(t) \hat{\vec{A}} (t+\tau) \rangle =
\frac{1}{\Gamma^2}
\begin{pmatrix}
        \Gamma_R \left( \Gamma_R + \Gamma_L e^{-\Gamma \tau} \right) \delta_{i0} + \Gamma_L\Gamma_R \left( 1 - e^{-\Gamma \tau} \right) \delta_{i1} \\
        \Gamma_L\Gamma_R \left( 1 - e^{-\Gamma \tau} \right) \delta_{i0} + \Gamma_L \left( \Gamma_L + \Gamma_R e^{-\Gamma \tau} \right) \delta_{i1}
\end{pmatrix}.
\end{equation}
In particular, $\ew{\hat{Q}(t)\hat{Q}(t+\tau)}=e^2\langle n_1(t) n_1(t+\tau) \rangle = \frac{\Gamma_L}{\Gamma^2} \left( \Gamma_L + \Gamma_R e^{-\Gamma\tau} \right)$, so the charge noise (\ref{chargenoiseDef}) reads
\beq
S^{(2)}_Q(\omega) = \frac{2e\langle I\rangle}{\Gamma^2 + \omega^2},
\eeq
being $\ew{I}\equiv \frac{e\Gamma_L\Gamma_R}{\Gamma}$.

Higher-order correlators can be also calculated with the version of the QRT presented above. For example, to obtain the third order correlation function $\ew{\hat{n}_1(t)\hat{n}_1(t+\tau_1)\hat{n}_1(t+\tau_2)}$, we use $\ew{\hat{n}_1(t)\hat{\vec{A}}(t+\tau_1)\otimes\hat{\vec{A}}(t+\tau_2)} = \ew{\hat{n}_1(t)e^{M \tau_1}\hat{\vec{A}}(t)\otimes e^{M \tau_2}\hat{\vec{A}}(t)}$. Taking into account that $\hat{n}_i(t)\hat{n}_j(t)\hat{n}_k(t)=\delta_{ij}\delta_{ik}\hat{n}_i(t)$, we obtain
\beq
\ew{\hat{n}_1(t)\hat{n}_1(t+\tau_1)\hat{n}_1(t+\tau_2)} = \frac{\Gamma_L^3}{\Gamma^3} + \frac{\Gamma_L^2\Gamma_R}{\Gamma^3}\left( e^{-\Gamma\tau_1} + e^{-\Gamma\tau_2} \right) + \frac{\Gamma_L\Gamma_R^2}{\Gamma^3} e^{-\Gamma(\tau_1+\tau_2)},
\eeq
which has the appropriate limits $\ew{\hat{n}_1(t)\hat{n}_1(t+\tau)\hat{n}_1(t)} =\ew{\hat{n}_1(t)\hat{n}_1(t+\tau)} = \frac{\Gamma_L}{\Gamma^2} \left(\Gamma_L + \Gamma_R e^{-\Gamma \tau}\right)$ and $\langle \hat{n}_1^3(t)\rangle=\langle \hat{n}_1(t)\rangle=\Gamma_L/\Gamma$.

The current noise spectrum corresponding to a capacitive system has two contributions. One is the charge noise calculated above, and the other one corresponds to the particle noise generated at each barrier \cite{BlanterButtiker00} (more details about this will be given in chapter \ref{Chapter3}). The total noise accounting for these two contributions can be expressed as \cite{AguadoBrandes04}
\beq\label{totalnoiseEq}
S^{(2)}_{tot}(\omega) = \alpha S^{(2)}_L(\omega) + \beta S^{(2)}_R(\omega) - \alpha\beta\omega^2 S^{(2)}_Q(\omega),
\eeq
where $S^{(2)}_L$ and $S^{(2)}_R$ refer to the particle noise through the left and right barrier respectively, and $\alpha$ and $\beta$ are coefficients accounting for the relative capacitances of the barriers. If we use $\hat{{\cal I}}_L=\Gamma_L \hat{n}_0 - \alpha \dot{\hat{n}}_0$, $
\hat{{\cal I}}_R=\Gamma_R \hat{n}_1 + \beta \dot{\hat{n}}_1$, and $S^{(2)}_{tot}(\tau)=\frac{1}{2}[\langle \delta\hat{{\cal I}}_L(t) \delta\hat{{\cal I}}_L(t') \rangle + \langle \delta\hat{{\cal I}}_R(t) \delta\hat{{\cal I}}_R(t') \rangle]$, with $\delta\hat{\cal I}_{\alpha}(t):=\hat{\cal I}_{\alpha}(t)-\hat{\cal I}_{\alpha}(0)$, the QRT gives
\beq \label{totalFanonoiseSRL}
S^{(2)}_{tot}(\omega) = e \langle I \rangle \left[ \frac{\Gamma_L^2 + \Gamma_R^2 + (1-2\alpha\beta)\omega^2}{\Gamma^2 + \omega^2} \right],
\eeq
which is the correct result for the total noise spectrum of the SRL model \cite{BlanterButtiker00}.

\begin{figure}
  \begin{center}
    \includegraphics[width=\textwidth]{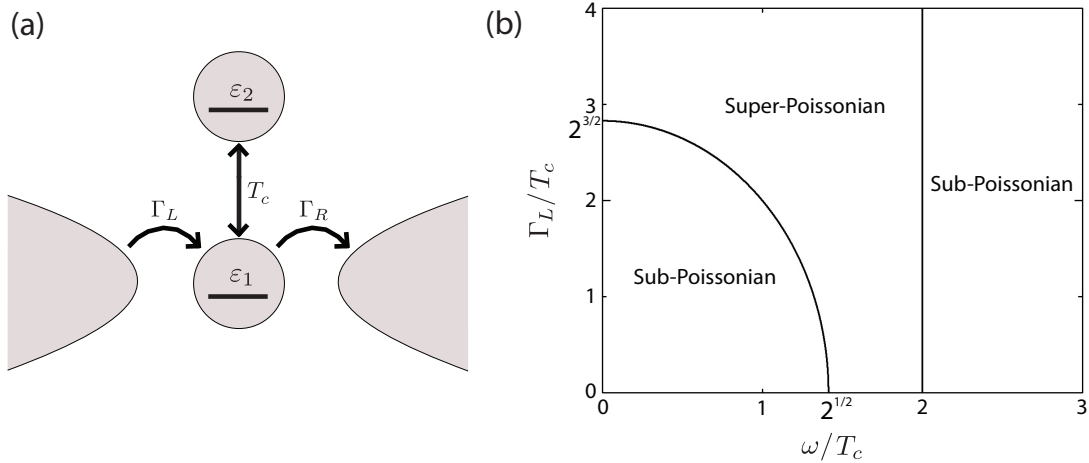}
  \end{center}  
  \caption[On-site dot model]{a) A single-level quantum dot, with energy $\varepsilon_1$ is coherently coupled ($T_c$) to an on-site dot with energy $\varepsilon_2$. Only the first dot is connected to leads, with chemical potentials such that the current flows in the direction indicated by the arrows. b) `Phase diagram' corresponding to the on-site dot model shown in (a). A super-Poissonian behaviour is obtained for certain values of the rates and frequency. Here we have taken $\Gamma_R/T_c=1$.
}
  \label{BingDongFig}
\end{figure}

\begin{figure}
  \begin{center}
    \includegraphics[width=0.5\textwidth]{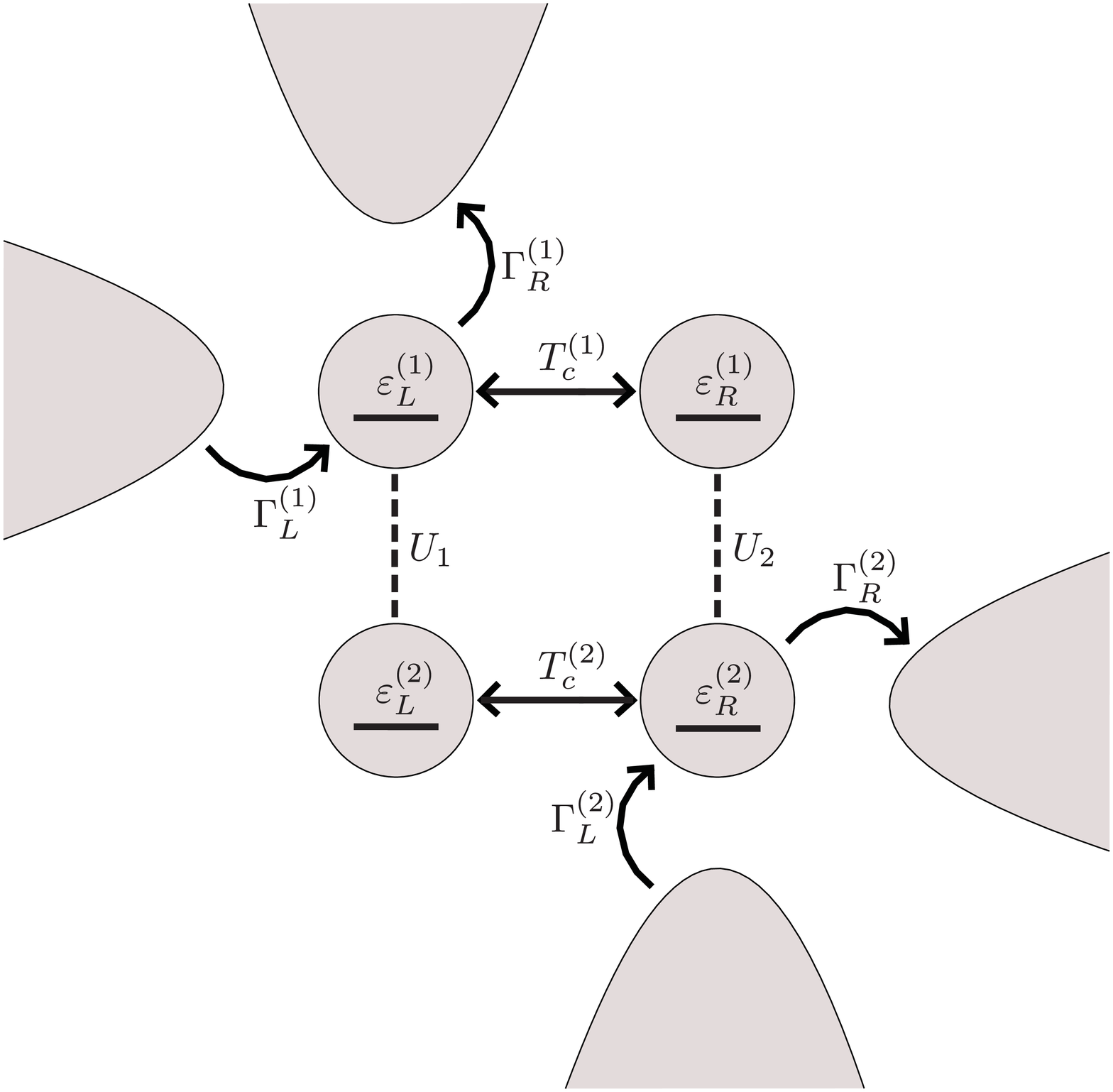}
  \end{center}  
  \caption[Model to prove entanglement via noise correlations]{Model to prove entanglement through noise cross correlations. The total system consists of four quantum dots. The upper dots are coherently coupled via $T_c^{(1)}$ and only unidirectional transport is possible through one of the dots. These dots are coupled via a Coulomb interaction with a similar system below.
}
  \label{CrossCorrFig}
\end{figure}

Once we have understood the SRL model, we can study more complicated situations. For example, let us consider the system depicted in Fig.~\ref{BingDongFig}a. The projection techniques explained in the previous section and the expressions derived in Ref.~\cite{Flindt05c} can be used to easily calculate current and noise through the system. Here we just cite the results, as the technique will be presented in detail in chapter \ref{Chapter3}. It is important to remark that this study was already presented in \cite{Djuric05} using a different method. This model is interesting because it is one of the simplest examples showing how the noise reveals information about the system that is absent in the current. In terms of the rates shown in the figure, the current through the system is $\ew{I}=\frac{e\Gamma_L\Gamma_R}{2\Gamma_L+\Gamma_R}$, and the zero-frequency Fano-factor $F^{(2)}(0)=\frac{S^{(2)}(0)}{e\ew{I}}$ reads
\beq
F^{(2)}(0)=1-\frac{4\Gamma_L\Gamma_R}{(2\Gamma_L+\Gamma_R)^2} + \frac{\Gamma_L^2\Gamma_R^2}{2T_c^2(2\Gamma_L+\Gamma_R)^2}.
\eeq
Notice, that while the current does not contain the energy scale $T_c$, this is present in the zero-frequency noise spectrum. At finite frequencies one obtains the `phase diagram' shown in Fig.~\ref{BingDongFig}b. Interestingly, a super-Poissonian behaviour is obtained for large values of the incoming rate and certain frequencies. This simple model is also interesting since it is a part of the much richer model shown in Fig.~\ref{CrossCorrFig}. This last is an interesting candidate to prove entanglement properties through the violation of noise Bell inequalities \cite{Yamamoto00, Loss00, Martin02, Burkard03, Samuelsson03, Faoro04, Lambert07, Emary09, Lambert10}. 

\subsubsection{Double quantum dot}

\begin{figure}[t]
  \begin{center}
    \includegraphics[width=0.6\textwidth]{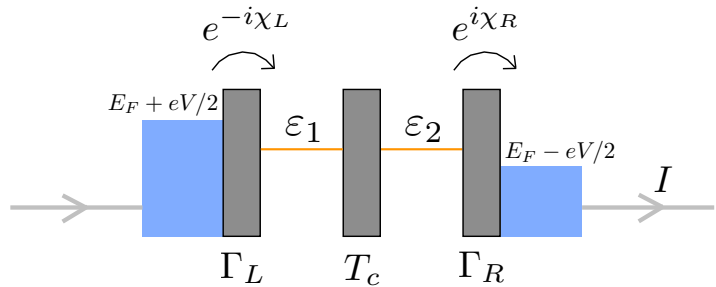}
  \end{center}  
  \caption[Double quantum dot]{Double quantum dot. Two dots with energies $\varepsilon_1$ and $\varepsilon_2$ are coupled with rates $\Gamma_L$ and $\Gamma_R$ to a continuum of states in two different electronic reservoirs at chemical potentials $E_F+ eV/2$ and $E_F- eV/2$ respectively. They are also coupled coherently with each other via $T_c$. In the figure, counting fields $e^{-i\chi_L}$ and $e^{i\chi_R}$, corresponding to the infinite-bias limit, have been included to illustrate the counting statistics of electrons tunneling through the left and right barriers.
}
  \label{DQDFig}
\end{figure}

The double quantum dot (DQD) is the paradigm of two-level system (c.f. appendix \ref{2LSappendix}) in quantum transport. It is shown schematically in Fig.~\ref{DQDFig}. Transport through the system is possible due to an externally applied voltage. Additionally, charge can tunnel coherently between both left and right dots. The Hamiltonian describing the system is of the form (\ref{Hform}), with
\beq
\hat{\cal H}_\mathrm{S} &=& \varepsilon(\ket{L}\bra{L} - \ket{R}\bra{R}) + T_c (\ket{L}\bra{R} + \ket{R}\bra{L}). \label{HSdqd} \\
\hat{\cal H}_\mathrm{R} &=& \sum_{k,\alpha\in L,R} \varepsilon_{k\alpha} \hat{c}_{k\alpha}^{\dagger}\hat{c}_{k\alpha}. \label{HRdqd} \\
\hat{\cal H}_\mathrm{V} &=& \sum_{k,\alpha\in L,R} V_{k\alpha} \hat{c}_{k\alpha}^{\dagger} \ket{0}\bra{\alpha} + \mathrm{H.c.} \label{HVdqd}
\eeq
Here $\varepsilon\equiv (\varepsilon_L-\varepsilon_R)/2$; $\varepsilon_{k\alpha}$ and $V_{k\alpha}$ are lead eigen-energies and tunnel couplings respectively. The three possible states of the system are $\ket{0}\equiv \ket{N_L,N_R}$, $\ket{L}\equiv \ket{N_L+1,N_R}$ and $\ket{R}\equiv \ket{N_L,N_R+1}$ (with $N_L$ and $N_R$ being the number of electrons in the left/right dot), or the system can be in a quantum superposition of $\ket{L}$ and $\ket{R}$.
An expression for the current through the system in the infinite-bias-voltage limit was first derived using the density operator approach by Stoof and Nazarov \cite{Stoof96}. This reads
\beq \label{StoofNazarovCurrent}
\ew{I} = \frac{T_c^2\Gamma_R}{T_c^2(2+\Gamma_R/\Gamma_L) + \Gamma_R^2/4 + 4\varepsilon^2},
\eeq
equation that can be also obtained using for example a perturbative approach to the many-body Schr\"odinger equation \cite{Gurvitz96}. Interestingly, the infinite-bias limit is exact. This means that the result (\ref{StoofNazarovCurrent}), although obtained to second order in perturbation theory with respect to $\hat{\cal H}_\mathrm{V}$, corresponds to the result that one obtains to all orders. The reason for this is that the problem fulfills the so-called singular coupling limit \cite{Spohn80}. 
Also, the Markovian approximation is exact in this limit, since in this case the large chemical potentials give a bath correlation function $\xi(\tau)\propto \delta(\tau)$ in the Bloch-Redfield tensor (\ref{BlochRedfieldTensor}).
Later we will study the transport noise through this system. We will see that the noise spectrum develops a resonance at the frequency of the qubit, namely $\Delta:=2\sqrt{\varepsilon^2+T_c^2}$, and whose width is proportional to the qubit dephasing time.

\begin{figure}[t]
  \begin{center}
    \includegraphics[width=\textwidth]{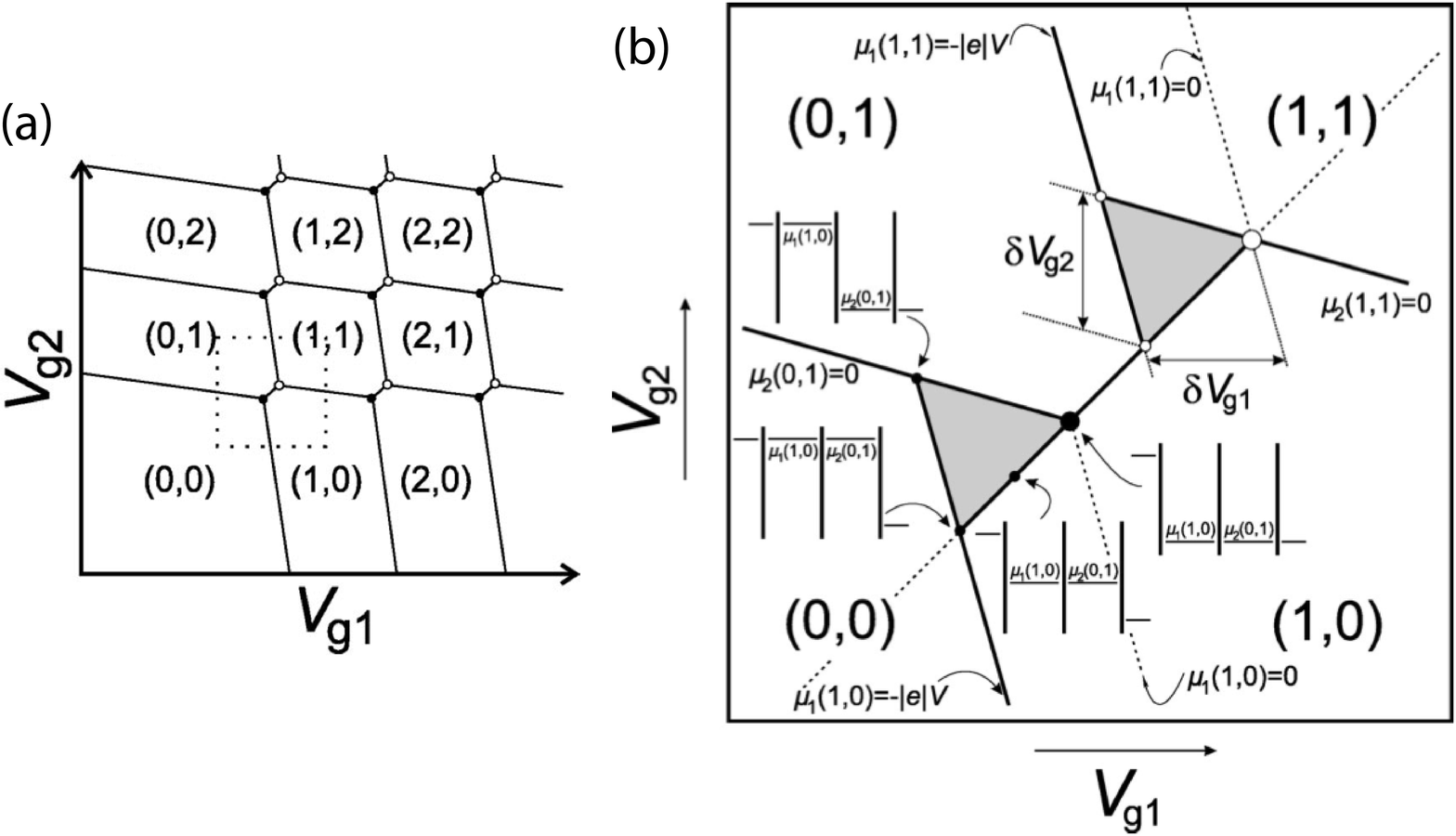}
  \end{center}  
  \caption[Double quantum dot charge-stability diagram]{Double quantum dot charge-stability diagram at intermediate interdot tunnel couplings. a) The charge in the DQD is denoted by $(N_L,N_R)$, with $N_L$ and $N_R$ the number of electrons in the left and right dot respectively. The black (white) dots denote the triple points in which an electron (hole) is transferred sequentially through the DQD. b) Zoom in corresponding to the dotted region in (a). Here we assume a finite bias voltage is applied. This produces the shaded triangles in the figure. The different resonant points as a function of the chemical potentials $\mu_1$, $\mu_2$ are also indicated. Taken from \cite{vanderWiel03}.
}
  \label{VanderWielFig}
\end{figure}

The charge-stability diagram of the DQD is shown in Fig.~\ref{VanderWielFig}a. Varying the gate voltages $V_{g1}$ and $V_{g2}$ applied to left and right dot respectively, a stable state with a well defined number of electrons in the dots can be isolated as a consequence of the Coulomb blockade. An extensive study of this physics can be found in \cite{vanderWiel03}. At finite voltages (see Fig.~\ref{VanderWielFig}b), transport through the DQD is possible, and a series of resonance peaks occur in the current within the shaded triangles in the figure.

As we will see, the current-noise spectrum of this system can be derived with the methods explained in the previous section. For example, the solution in the infinite-bias limit has been calculated by Aguado and Brandes \cite{AguadoBrandes04} using the MacDonald's formula, by Flindt and collaborators \cite{FlindtThesis} using the projection techniques, and by Kiesslich \textit{et al.}\cite{KiesslichThesis} using various approaches, such as the master equation method. In section~\ref{FCSwPaper1Sec} we derive the infinite-bias noise spectrum of the DQD by means of the full counting statistics scheme, and in chapters \ref{Chapter3} and \ref{Chapter4} we will extend the study to the general case in which voltage, frequency and temperature are competing energy scales.

\section{Frequency-dependent full counting statistics\footnote{The results presented in this section have been published in \cite{Emary07}.}} \label{FCSwPaper1Sec}

In this section we present a formalism to calculate finite-frequency current
correlations in interacting nanoscopic conductors based on the full counting statistics method. 
We use the n-resolved density matrix, and obtain a multi-time cumulant
generating function that provides the fluctuation statistics 
from the spectral decomposition of the kernel of the quantum master equation. 
We apply the method to the frequency-dependent third cumulant
of the current through a single resonant level and through a
double quantum dot. Our results, which show that deviations from
Poissonian behaviour strongly depend on frequency, demonstrate the
importance of finite-frequency higher-order cumulants in
fully characterizing transport.

\subsection{General context}

Following the considerable success of shot noise in the
understanding of transport through mesoscopic systems \cite{BlanterButtiker00},
attention is now turning towards the higher-order statistics of
electron current.  The so-called Full Counting Statistics (FCS) of
electron transport yields all moments (or cumulants) of the
probability distribution $P(n,t)$ of the number of transferred
electrons during time $t$. Despite their difficulty, measurements of
the third moment of voltage fluctuations have been made
\cite{Reulet03,Bomze05}, and recent developments in single electron
detection \cite{Fujisawa06, Gustavsson06prl, Gustavsson06prb, Gustavsson07, Sukhorukov07} 
promise to open new horizons on the experimental side.

The theory of FCS is now well established in the zero-frequency limit
\cite{Levitov93, Levitov96, Bagrets03prl, BagretsNazarov03}.  
However, this is by no means the full picture,
since the higher-order current correlators at finite frequencies
contain much more information than their zero-frequency
counterparts. Already at second order (shot noise), one can extract
valuable information about transport time scales and correlations.
When the conductor has various intrinsic time scales like, for
example, the charge relaxation time and the dwelling time of a chaotic
cavity \cite{Nagaev04}, one needs to go beyond second-order in order
to fully characterize electronic transport.  Apart from this
example, and some other notable exceptions
\cite{Pilgram04, Galaktionov03, Salo06}, the behaviour of finite-frequency
correlators beyond shot noise is still largely unexplored.

\subsection{Introduction to the theory}

Here we develop a theory of frequency-dependent current
correlators of arbitrary order in the context of the $n$-resolved
density matrix (DM) approach, --- a Quantum Optics technique
\cite{Cook81} that has recently found application in mesoscopic
transport \cite{Gurvitz98}. Within this approach, the DM of the system,
$\rho_S(t)$, is unravelled into components $\rho_S^{(n)}(t)$ in which
$n=n(t)=0,1,\ldots$ electrons have been transferred to the
collector. Considering a generic mesoscopic system with
Hamiltonian\footnote{In this section we obviate the hat in the notation for operators.}
 $\mathcal{H}=\mathcal{H}_S + \mathcal{H}_L +
\mathcal{H}_V$, where $\mathcal{H}_S$ and $\mathcal{H}_L$ refer to
the system and leads respectively, and provided that the Born-Markov
approximation with respect to the tunnelling term $\mathcal{H}_V$ is
fulfilled, the time-evolution of this $n$-resolved system DM can be written
quite generally as
\beq
  \dot{\bm{\rho}}_S^{(n)}(t) = {\cal W}_0 \bm{\rho}_S^{(n)}(t)
  + {\cal W}_J \bm{\rho}_S^{(n-1)}(t)
  \label{eomFCSpaper1}
  ,
\eeq where the vector $\bm{\rho}_S^{(n)}(t)$ contains the nonzero
elements of the DM, written in a suitable many-body basis.
The kernel ${\cal W}_0$ describes the `continuous' evolution of
the system, whereas ${\cal W}_J$ describes the quantum jumps of the
transfer of an electron to the collector.  We make the infinite-bias-voltage approximation so that the electron transfer is unidirectional. By
construction, this method is very powerful for studying interacting
mesoscopic systems that are weakly coupled to the reservoirs, such
as coupled quantum dots (QDs) in the Coulomb Blockade (CB) regime
\cite{Gurvitz98, Bagrets03prl, BagretsNazarov03, AguadoBrandes04} or Cooper-pair boxes \cite{Choi01}.
Within this framework, our theory of frequency-dependent FCS is of
complete generality and therefore of wide applicability.

In this picture, electrons are transferred to the leads via quantum
jumps and there exists no quantum coherence between states within
the system and those in the leads.  Thus, although the system itself
may be quantum, the measured current may be considered as a purely
classical stochastic variable and is therefore amenable to classical
counting\footnote{This is in contrast with quantum
detection, where the non-commutativity of current at different times must be explicitly
considered \cite{Levitov93}.}. This observation allows us to
derive various generalizations of results in classical stochastic
methods, and to obtain a multi-time cumulant generating function in terms of local
propagators.
We illustrate our method by calculating the frequency-dependent third
cumulant (skewness) for two paradigms of mesoscopic transport:
the single resonant level (SRL) and the double quantum dot (DQD).

\subsection{Theory of frequency-dependent counting statistics}

Equation (\ref{eomFCSpaper1}) can be solved by Fourier transformation.
Defining $ \bm{\rho}_S(\chi,t) = \sum_n \bm{\rho}_S^{(n)}(t) e^{i n
\chi} $, we obtain $
  \dot{\bm{\rho}}_S(\chi,t) = {\cal W}(\chi) \bm{\rho}_S(\chi,t)
  \label{eom2FCSpaper1}
$, with ${\cal W}(\chi)\equiv{\cal W}_0+ e^{i\chi}{\cal W}_J$.
Let $N_v$ be the dimension of ${\cal W}(\chi)$, and
$\lambda_i(\chi);~i=1,\ldots,N_v$, its eigenvalues.  In the
$\chi\rightarrow 0$ limit, one of these eigenvalues,
$\lambda_1(\chi)$ say, tends to zero and the corresponding
eigenvector gives the stationary DM for the system. This single
eigenvalue is sufficient to determine the zero-frequency FCS
\cite{BagretsNazarov03}.  In contrast, here we need all $N_v$ eigenvalues.
Using the spectral decomposition, $ {\cal W}(\chi) = V(\chi) \Lambda(\chi)
V^{-1}(\chi)$, with $\Lambda(\chi)$ the diagonal matrix of
eigenvalues and $V(\chi)$ the corresponding matrix of eigenvectors,
the DM of the system at an arbitrary time $t$ is given by \beq
  \bm{\rho}_S(\chi,t) = \Omega(\chi,t-t_0) \bm{\rho}_S(\chi,t_0),
\eeq where $
  \Omega(\chi;t)
  \equiv
  e^{{\cal W}(\chi)t}
  =
  V(\chi) e^{\Lambda(\chi)t)} V^{-1}(\chi)
$ is the propagator in $\chi$-space, and $\bm{\rho}_S(\chi,t_0)$ is the
(normalized) state of the system at $t_0$, at which time we assume
no electrons have passed, so that $\bm{\rho}_S^{(n)}(t_0) =
\delta_{n,0} \bm{\rho}_S(t_0)$, and thus $\bm{\rho}_S(\chi,t_0) \equiv
\bm{\rho}_S(t_0)$.
The propagator in $n$-space, $
  G(n,t)\equiv \int \frac{d\chi}{2\pi} e^{-in\chi} \Omega(\chi,t)
$, such that $ \bm{\rho}_S^{(n)}(t) = G(n,t-t_0)  \bm{\rho}_S(t_0) $,
for $ t>t_0$,
fulfills the property:
$  G(n-n_0,t-t_0)=
  \sum_{n'}G(n-n',t-t')
G (n'-n_0,t'-t_0)$,
for $t>t'>t_0$, ($n'\equiv n(t')$, $n_0\equiv n(t_0)$). This is
an operator version of the Chapman-Kolmogorov equation
\cite{vanKampenBook}.

The joint probability of obtaining $n_1$ electrons after $t_1$ and
$n_2$ electrons after $t_2$, namely $P^>(n_1,t_1;n_2,t_2)$ (the
superscript `$>$' implies $t_2 > t_1$), can be written in terms of
these propagators by evolving the local probabilities\footnote{In our notation here with $\rho_S$ as a vector, the trace operation corresponds to a sum over all the diagonal elements of the original density matrix.}
$P(n,t)=\mathrm{Tr_S}\lbrace \bm{\rho}_S^{(n)}(t) \rbrace$, and taking
into account $P(n_2,t_2)=\sum_{n_1} P^>(n_1,t_1;n_2,t_2)$,
such that
\beq
  P^>(n_1,t_1;n_2,t_2) &=& \mathrm{Tr_S}
  \left\{
    G(n_2-n_1,t_2-t_1)
    G (n_1,t_1-t_0)\bm{\rho}_S(t_0)
  \right\}
  \label{P>2}
  .
\eeq
The total joint probability reads $P(n_1,t_1;n_2,t_2) =
\mathcal{T}P^>(n_1,t_1;n_2,t_2) = P^>(n_1,t_1;n_2,t_2)
\theta(t_2-t_1) + P^<(n_1,t_1;n_2,t_2) \theta(t_1-t_2)$ where $\mathcal{T}$ is the time-ordering operator and $\theta(t)$ the
unit-step function. It should be noted that, in contrast to the
local probability $P(n,t)$, the joint probability
$P(n_1,t_1;n_2,t_2)$ contains information about the correlations at \textit{different times}.

Equation (\ref{P>2}) may be alternatively derived using the Bayes
formula for the conditional density operator \cite{Korotkov01}: \beq
  P^>(n_1,t_1;n_2,t_2)
  &=& P(n_1,t_1)P^>(n_2,t_2\vert n_1,t_1)
  = \mathrm{Tr_S}\Big\lbrace \bm{\rho}_S^{(n_1)}(t_1)
  \Big\rbrace \mathrm{Tr_S}\Big\lbrace \bm{\rho}_S^{(n_2\vert n_1)}(t_2)\Big\rbrace
  \nonumber \\ &=&
  \mathrm{Tr_S}\Big\lbrace \bm{\rho}_S^{(n_1)}(t_1) \Big\rbrace \mathrm{Tr_S}\Bigg\lbrace
  G(n_2-n_1,t_2-t_1)\frac{\bm{\rho}_S^{(n_1)}(t_1)}{\mathrm{Tr_S}\Big\lbrace
  \bm{\rho}_S^{(n_1)}(t_1)\Big\rbrace}\Bigg\rbrace.\nonumber
\eeq The normalization in the denominator accounts for the collapse
$n=n_1$ at $t=t_1$ using von Neumann's projection postulate
\cite{Korotkov01}. The result (\ref{P>2}) is recovered when
$\bm{\rho}_S^{(n_1)}(t_1)$ is written as a time evolution from $t_0$.

The two-time cumulant generating function (CGF) associated with
these joint probabilities is \beq
e^{\mathcal{F}(\chi_1,\chi_2;t_1,t_2)}
  &\equiv& \sum_{n_1,n_2}
  P(n_1,t_1;n_2,t_2)
  e^{i n_1 \chi_1 + i n_2 \chi_2},\nonumber
\eeq which, using Eq.~(\ref{P>2}), and $e^{\mathcal{F}}=e^{{\cal
T}\mathcal{F}^>}={\cal T} e^{\mathcal{F}^>}$, gives
\beq
  e^{{\cal F}(\chi_2,\chi_1;t_2,t_1)}
  &=& {\cal T}~
  \mathrm{Tr_S}
  \left\{
    \Omega(\chi_2,t_2-t_1)
 \Omega(\chi_1+\chi_2,t_1-t_0)\bm{\rho}_S(t_0)
  \right\}
  .
\eeq
The above procedure can be easily generalized to obtain the N-time
CGF. We find
\beq\label{genF}
  e^{\mathcal{F}(\bm{\chi};\bm{t})}
  &=&
  \mathcal{T}~\mathrm{Tr_S}
  \left\{
  \prod_{k=1}^N
    \Omega
    (\sigma_k; \tau_{N-k})
    \bm{\rho}_S(t_0)
  \right\},
\eeq
where\footnote{One can also use the general expressions for the generating functions 
$\mathcal{F}$ (cumulant) and $\mathcal{G}$ (moment), defined as
$e^{\mathcal{F}[\chi]}:=\mathcal{G}[\chi] \equiv \Big\langle e^{i\int
n(t) \chi(t) dt} \Big\rangle_\mathcal{T}$,
where the average $\langle...\rangle_\mathcal{T}$ denotes a
time-ordered path integral with the joint probability as the weight
function. Established the biyection between $t$ and $n$, one
recovers the CGF in terms of products of evolution operators by
dividing the time interval into subintervals $\tau_k:=t_{k+1}-t_k$,
$k=0,\dots,N-1$, with associated time-independent counting fields
$\chi_k$ and variables $n_k$.} $\sigma_k \equiv \sum_{i=N+1-k}^N \chi_i$,
$\bm{\chi}\equiv(\chi_1,\dots,\chi_N)_T$, $\bm{t}\equiv(t_1,\dots,t_n)_T$ and
$\tau_k \equiv t_{k+1}-t_k$. The multi-time CGF in Eq.~(\ref{genF}) is
a product of local-time propagators, and expresses the Markovian character of the
problem. Importantly, it allows one to obtain all the
frequency-dependent cumulants from the spectral decomposition of
${\cal W}(\chi)$.
The $N$-time current cumulant (setting the electron charge $e= 1$) reads then\footnote{In this section the frequency-dependent cumulants, denoted by $S^{(N)}$, refer to the particle-current contribution only.}
\beq
  &&S^{(N)}(t_1,\dots,t_N)
  \equiv \langle I(t_1)\dots I(t_N) \rangle_c = \nonumber \\
  &&=
  \partial_{t_1}\dots
  \partial_{t_N} \langle n(t_1)\dots n(t_N)\rangle_c = \nonumber \\ &&= (-i)^N
  \partial_{t_1}\dots \partial_{t_N} \partial_{\chi_1}\dots
  \partial_{\chi_N}
  \mathcal{F}(\bm{\chi};\bm{t})\Big|_{\bm{\chi}=\bm{0}}.
  \label{corr}
\eeq
Here the subscript $c$ denotes cumulant. The Fourier transform of $S^{(N)}$ with respect to the time
intervals $\tau_k$, gives the $N$th-order correlation function, 
dependent of $N-1$ frequencies. In particular, the
frequency-dependent skewness is a function of two frequencies, which,
as a consequence of time-symmetrization and the Markovian
approximation, has the symmetries $S^{(3)}(\omega,\omega') =
S^{(3)}(\omega',\omega) =
S^{(3)}(\omega,\omega-\omega')=S^{(3)}(\omega'-\omega,\omega')=
S^{(3)}(-\omega,-\omega')$, and is therefore real.  The
$N$th-order Fano-factor is defined as $F^{(N)} \equiv S^{(N)}/\ew{I}$.

If the jump matrix ${\cal W}_J$ contains a single
element, $({\cal W}_J)_{ij} =
\Gamma_R\delta_{i\alpha}\delta_{j\beta}$, which is the situation for a
wide class of models, including our two examples below, all the
correlation functions can be expressed solely in terms of the
eigenvalues $\lambda_k$ of ${\cal W}(0)$, and the $N_v$ coefficients $
  c_k \equiv (V^{-1}{\cal W}_J V)_{kk} = \Gamma_R V_{\beta k }V^{-1}_{k \alpha}
$. The second-order Fano factor then has the simple, general form
\beq
  F^{(2)}(\omega)
 & =& 1 - 2\sum_{k=2}^{N_v} \frac{c_k
  \lambda_k}{\omega^2+\lambda_k^2},
 \label{F2FCSpaper1}
\eeq
which can also be derived via different approaches \cite{Hershfield93, Flindt05c}.
The skewness has the form $
  F^{(3)}(\omega,\omega') =
  -2
  +\sum_{i=1}^3 F^{(2)}(\nu_i)+
  \widetilde{F}^{(3)}(\omega,\omega')
$, with  $\nu_1 = \omega$, $\nu_2 = \omega'$, and $\nu_3 =
\omega-\omega'$. $\widetilde{F}^{(3)}(\omega,\omega')$ is an
irreducible contribution. The high-frequency limit of the skewness is
$F^{(3)}(\omega,\infty) = F^{(2)}(\omega)$.

\subsection{Application to a single resonant level model}

\begin{figure}
\center
\includegraphics[width= 0.6\textwidth]{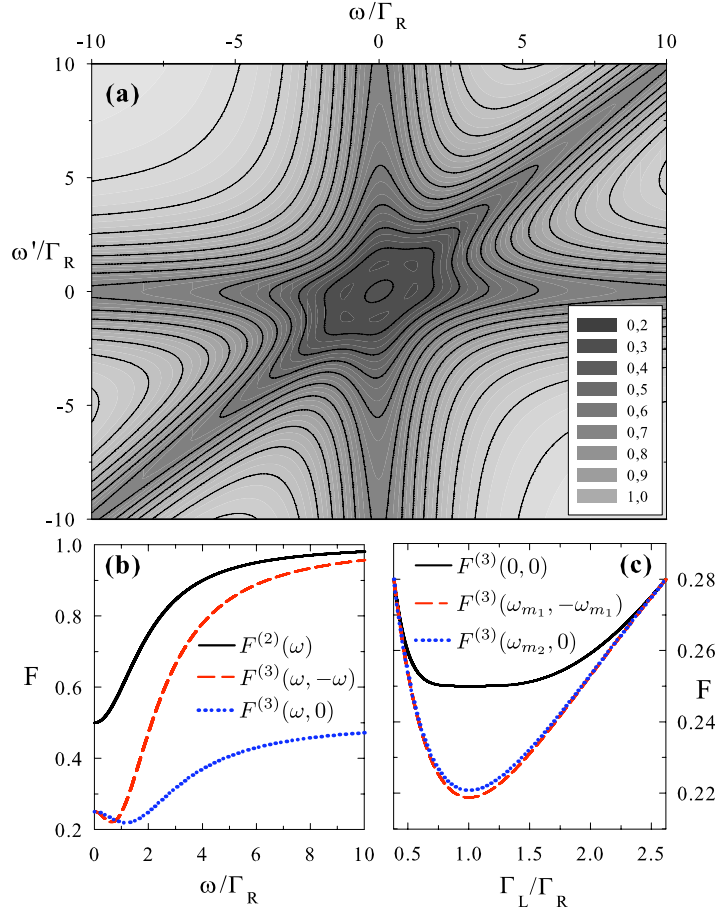}
\caption[Particle noise and skewness in the single resonant level model]{The third-order frequency-dependent Fano factor $F^{(3)}(\omega,\omega')$ for the single resonant level.
     a) Contour plot of the skewness as function of its frequency arguments for $\Gamma_R=\Gamma_L$.
     b) Sections $F^{(3)}(\omega,0)$ and $F^{(3)}(\omega,-\omega)$
    show that the skewness is suppressed throughout frequency space both with respect to the Poissonian value of unity and to
    $F^{(2)}(\omega)$. In contrast with the shotnoise, the skewness has a minimum at a finite frequency $\omega_m$, which exists in the coupling range
    $\left( 3-\sqrt{5} \right)/2 \leq \Gamma_L/\Gamma_R \leq \left( 3+\sqrt{5} \right)/2$.
    Along direction $\omega=-\omega'$, this
minimum occurs, for $\Gamma_L=\Gamma_R$, at $\omega_{m_1}/\Gamma_R=\sqrt{\sqrt{10}-2}/\sqrt{3}$
(dashed curve) whereas for $\omega'=0$ the position of the minimum
shifts slightly  to $\omega_{m_2}/\Gamma_R=2/\sqrt{3}$ (dotted curve).
    c) The maximum suppression of skewness occurs at $\Gamma_L=\Gamma_R$.}
\label{FCSpaper1Fig1}
\end{figure}

As a first example we consider a SRL, described by $\bm{\rho}=\left(
\rho_{00},\rho_{11} \right)_T$ and
\beq \label{kernelSRLrightFCS}
  {\cal W}(\chi) =
  \begin{pmatrix}
    - \Gamma_L & e^{i\chi}\Gamma_R \\
    \Gamma_L & -\Gamma_R
  \end{pmatrix},
  \eeq
in the basis of `empty' and `populated' states, $\left\lbrace\ket{0}, \ket{1}\right\rbrace$.
Using Eq.~(\ref{genF}), we obtain the known results for the current
and noise, and arrive at our result for the skewness:
\beq
  F^{(3)}(\omega,\omega')
  &=&
   1
   -2\Gamma_L \Gamma_R
   \frac{
     \prod_{i=1}^2
     \left(\gamma_i + \omega^2 - \omega \omega' + \omega'^2\right)
   }
   {
     \prod_{j=1}^3
     \left(\Gamma^2+\nu_j^2\right)
   },
   \label{F3wwFCSpaper1}
\eeq
with $\gamma_1 =\Gamma_L^2 + \Gamma_R^2$, $\gamma_2
=3\Gamma^2$, and $\Gamma=\Gamma_L+\Gamma_R$.
Expression (\ref{F3wwFCSpaper1}) yields the zero-frequency limit $F^{(3)}(0,0)$ in accordance with Ref.~\cite{Jong96}.

The result for the skewness is plotted in Fig.~\ref{FCSpaper1Fig1}a, from which the
six-fold symmetry of $F^{(3)}$ is readily apparent.
The third-order Fano factor gives, in accordance with the noise,
a sub-Poissonian behaviour for all frequencies.
This can be easily understood as a CB suppression of the long tail in the
probability distribution of instantaneous current: due to the infinite bias,
the distribution is bounded on the left by zero, but, in principle, it is not bounded on
the right (large, positive skewness). CB suppresses large current
fluctuations which explains a sub-Poissonian skewness.

Along the symmetry lines in frequency space corresponding to
$\omega'=0$, $\omega'=\omega$ and $\omega=0$, the skewness is highly
suppressed. In contrast with the noise, the minimum in the skewness
occurs at finite frequency (Fig.~\ref{FCSpaper1Fig1}b) with the strongest
suppression occuring at $\Gamma_L=\Gamma_R$ (Fig.~\ref{FCSpaper1Fig1}c).

\subsection{Application to a double quantum dot}

As a second example we consider a DQD in the strong CB regime
\cite{Stoof96, Brandes05}. In the basis of `left' and `right' states
$\ket{L}$ and $\ket{R}$, which denote states with one excess
electron with respect to the many body `empty' state $\ket{0}$, the
Hamiltonian is of the form (\ref{Hform}), with ${\cal H}_\mathrm{S}$, 
${\cal H}_\mathrm{R}$ and ${\cal H}_\mathrm{V}$ given by (\ref{HSdqd}), 
(\ref{HRdqd}) and (\ref{HVdqd}), respectively.
The two levels $\ket{L}$, $\ket{R}$, are coupled to their respective leads with
rates $\Gamma_L$ and $\Gamma_R$.
The DM vector is now $\bm{\rho}=\left(\rho_{00},
\rho_{LL},\rho_{RR}, \mathrm{Re}(\rho_{LR}),
\mathrm{Im}(\rho_{LR})\right)_T$, and the kernel in this basis
reads: \beq
  {\cal W}(\chi) =
\begin{pmatrix}
    -\Gamma_L & 0 & e^{i\chi} \Gamma_R & 0 &0\\
    \Gamma_L & 0 & 0 & 0 & - 2 T_c \\
    0 & 0 & -\Gamma_R & 0 & 2 T_c \\
    0 & 0 & 0 & -\frac{1}{2}\Gamma_R & 2 \varepsilon\\
    0 & T_c & - T_c &  -2 \varepsilon &  - \frac{1}{2}\Gamma_R
\end{pmatrix}.
\eeq

\begin{figure}
\center
\includegraphics[width= 0.8\textwidth]{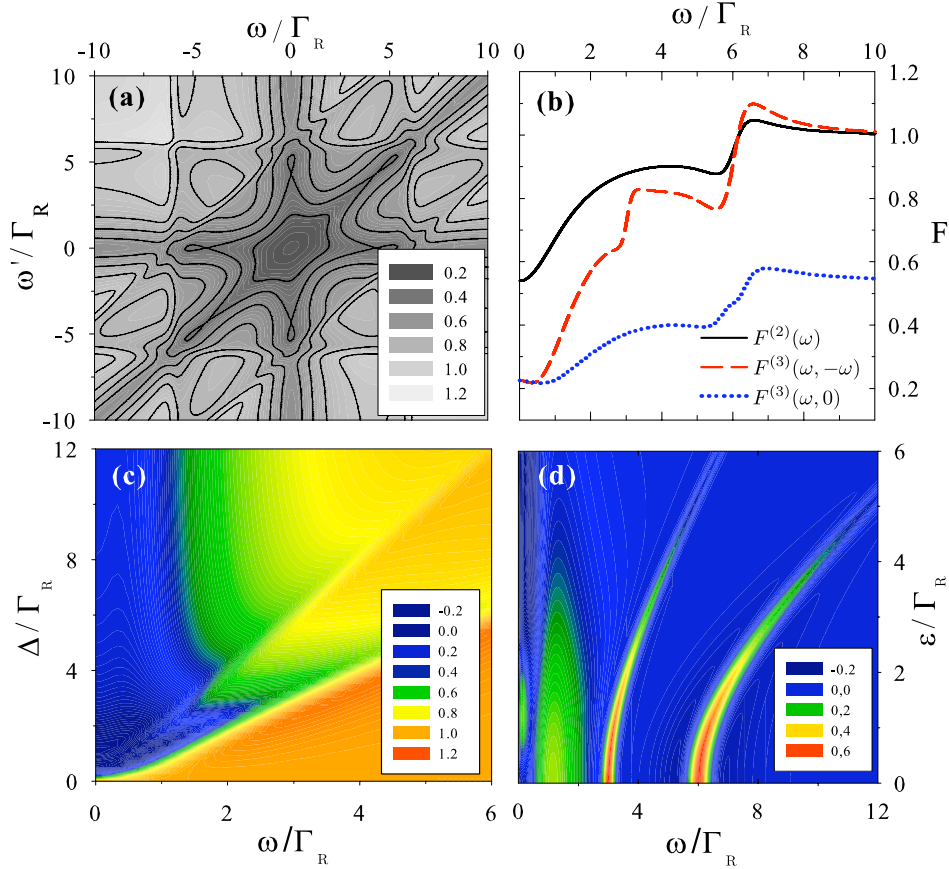}
\caption[Particle noise and skewness in a double quantum dot]{Third-order frequency-dependent Fano factor $F^{(3)}(\omega,\omega')$, for the double quantum dot in Coulomb blockade.
    a) Contour plot in the strong coupling regime, $T_c=3\Gamma_R$,  with $\Gamma_L=\Gamma_R$ and $\varepsilon=0$.
    b) Sections $F^{(3)}(\omega,0)$ and $F^{(3)}(\omega,-\omega)$, and
    second-order Fano factor $F^{(2)}(\omega)$ show a series of abrupt increases with increasing $\omega$.
    Both the noise and skewness exhibit both sub- and super- Poissonian behaviour.
   c) Varying the internal coupling $T_c$, the skewness shows rapid increases along the lines $\omega=\Delta$ and $\omega=\Delta/2$.
   For $\omega >\Delta$ the system is Poissonian (slightly super-Poissonian for $\omega \gtrsim \Delta$), while for $\omega<\Delta$ the transport is always
   sub-Poissonian. In particular, the skewness is strongly suppressed at low frequencies.
   d) The derivative $dF^{(3)}(\omega,-\omega)/d\omega$ as a function of the frequency and the detuning $\varepsilon$, for $T_c=3\Gamma_L=3\Gamma_R$. Resonances around $\omega=\Delta$, $\Delta/2$ and $\Gamma_R$ are observed.}
\label{FCSpaper1Fig2}
\end{figure}

Comparison of the quantum-mechanical level-splitting $\Delta
\equiv 2\sqrt{T_c^2 + \varepsilon^2}$ with the incoherent rates
$\Gamma_{L,R}$ divides the dynamical behaviour of the system into
two distinct regimes.
For $\Delta \ll \Gamma_{L,R}$, all eigenvalues of ${\cal W}(0)$ are real\footnote{For example, with $\Gamma_L=\Gamma_R$ and $\varepsilon=0$, all eigenvalues are real if $(T_c/\Gamma_R)^2 \le (3\sqrt{5}-11)/32$.}, and correspondingly, the noise and skewness are
slowly-varying functions of their frequency arguments.
In this regime, the dephasing induced by the leads suppresses the
interdot coherence, and the transport is largely incoherent. In the
opposite regime, $\Delta \gg \Gamma_{L,R}$, two of the eigenvalues
form a complex conjugate pair, $ \lambda_\pm \approx \pm i \Delta -
\Gamma_R/2 +O(\Gamma/\Delta)^2$,
which signals the persistence of coherent oscillations in the
current.  The finite-frequency correlators then show resonant
features at
$\Delta$ since these eigenvalues enter into the denominators, as in
Eq.~(\ref{F2FCSpaper1}), giving rise
to poles such as $\omega \mp \Delta - i\Gamma_R/2$. 
The consequent structure of the skewness is similar to the SRL for
weak coupling ($\Delta \ll \Gamma_{L,R}$), but much richer in the
strong coupling regime ($\Delta \gg \Gamma_{L,R}$). The results for
this case are illustrated in Fig.~\ref{FCSpaper1Fig2}. Now the skewness
exhibits a series of rapid increases. Moving from the origin
outwards in the $\omega$-$\omega'$ plane, we first observe the
minimum at finite frequency and then inflexion points at
$\omega\sim\Gamma_R$, $|\omega|=\Delta$, $|\omega'|=\Delta$ and
$|\omega-\omega'|=\Delta$. Fig.~\ref{FCSpaper1Fig2}b shows different
sections along the $\omega$-$\omega'$ plane. The resonant behaviour
is more pronounced in $F^{(3)}(\omega,-\omega)$ in the form of Fano
shapes. Starting from high frequencies, the onset of antibunching
occurs at $\omega=\Delta$. At higher frequencies (shorter times) the
system has no information about correlations and is thus Poissonian.
The overall behaviour is clearly seen in Fig.~\ref{FCSpaper1Fig2}c, where
we plot $F^{(3)}(\omega,-\omega)$ as a function of both $T_c$ and
$\omega$. The line $\omega=\Delta$ delimits two regions: at
high frequencies the skewness is Poissonian. At resonance, and after
a small super-Poissonian region at $\omega \gtrsim \Delta$, the
system becomes sub-Poissonian (for certain internal couplings it
can even become negative). 
In the limit $\omega\rightarrow 0$, our results qualitatively agree
with the ones in Ref.~\cite{Kiesslich06} for a noninteracting DQD: as a
function of $T_c$, the skewness presents two minima and a maximum
(where the noise is minimum, not shown). In our case, however, the
position of this maximum occurs at around $\Delta=\Gamma_R/2$, half
the value of the noninteracting case. Finally, we plot
$dF^{(3)}(\omega,-\omega)/d\omega$ as a function of both $\varepsilon$
and $\omega$ (Fig.~\ref{FCSpaper1Fig2}d) where the resonances at
$\omega=\Delta$, $\omega=\Delta/2$ and $\omega\sim\Gamma_R$ are
clearly resolved. In contrast to Fig.~\ref{FCSpaper1Fig2}d, the
derivative $dF^{(3)}(\omega,0)/d\omega$ (not shown), exhibits a
minimum at $\omega=\Delta$ for small $\varepsilon$, which transforms into a
maximum for $\varepsilon \gg T_c$. As expected, transport tends to be
Poissonian, signaling lost of coherence in the DQD, as $\varepsilon$
increases.

\subsection{FCS of the total current}

\begin{figure}[b]
\center
\includegraphics[width=\textwidth]{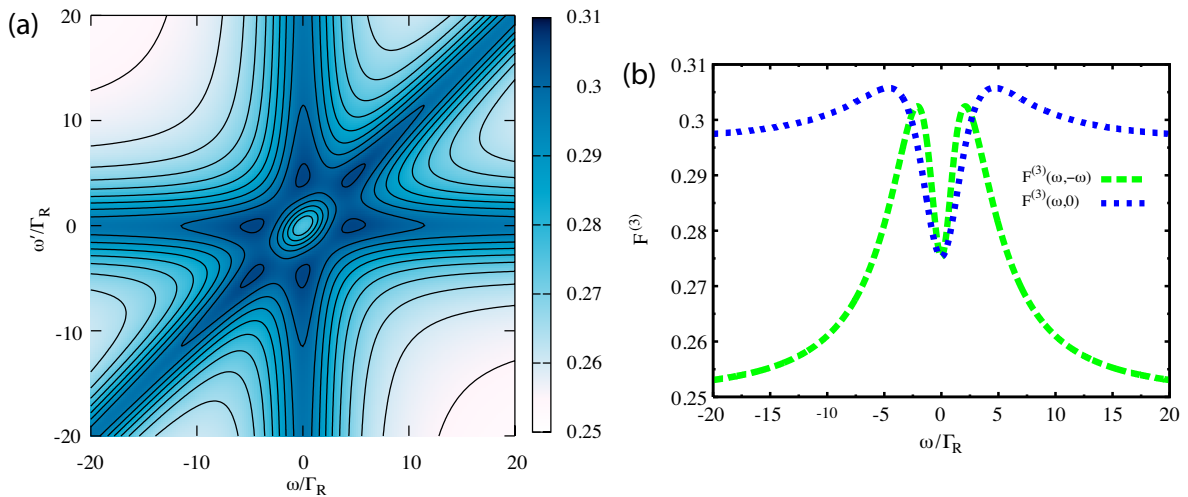}
\caption[Total skewness in the single resonant level model]{Total third-order Fano factor, $S^{(3)}/\ew{I}$, in the single resonant level model at infinite bias voltage. a) Contour map corresponding to $\Gamma_L/\Gamma_R=2.5$ (for $\Gamma_L=\Gamma_R$ the skewness is flat, as we will see in detail in chapter \ref{Chapter3}). The symmetries of the skewness become apparent. b) Cuts along the directions set by $\omega'=0$ and $\omega'=-\omega$. The skewness shows a local minimum at zero frequency and a local maximum at finite frequencies.}
\label{skewnessTotmapSRL}
\end{figure}

As we have mentioned before, at finite frequencies one needs to consider both the contribution from particle currents (charge variation in the leads) and that from displacement currents in the calculation of current correlations. 
According to Maxwell's equations, the electric field or polarization along the sample creates a displacement current that also fluctuates in time. This gives equation (\ref{totalnoiseEq}) for the noise spectrum; although, as it will be explained in chapter \ref{Chapter3}, our theory allows the calculation of `total' (particle plus displacement) cumulants by considering different counting fields in the kernel of the QME. In this section we present results for the total skewness (c.f. appendix \ref{SuppToFCSwPaper1Sec} or chapter \ref{Chapter3} for different formulations) corresponding to the SRL model considered above.
The total third-order Fano factor corresponding to this system is shown in Fig.~\ref{skewnessTotmapSRL}a. Similarly to the previous examples, the behavior is sub-Poissonian, reaching the Poissonian value at high frequencies. Cuts along the representative directions $\omega'=-\omega$ and $\omega'=0$ are shown in Fig.~\ref{skewnessTotmapSRL}b. Here we clearly see that the total skewness develops a local minimum at $\omega=\omega'=0$, and that a local maximum appears at a finite frequency. 

\begin{figure}[t]
\center
\includegraphics[width=0.85\textwidth]{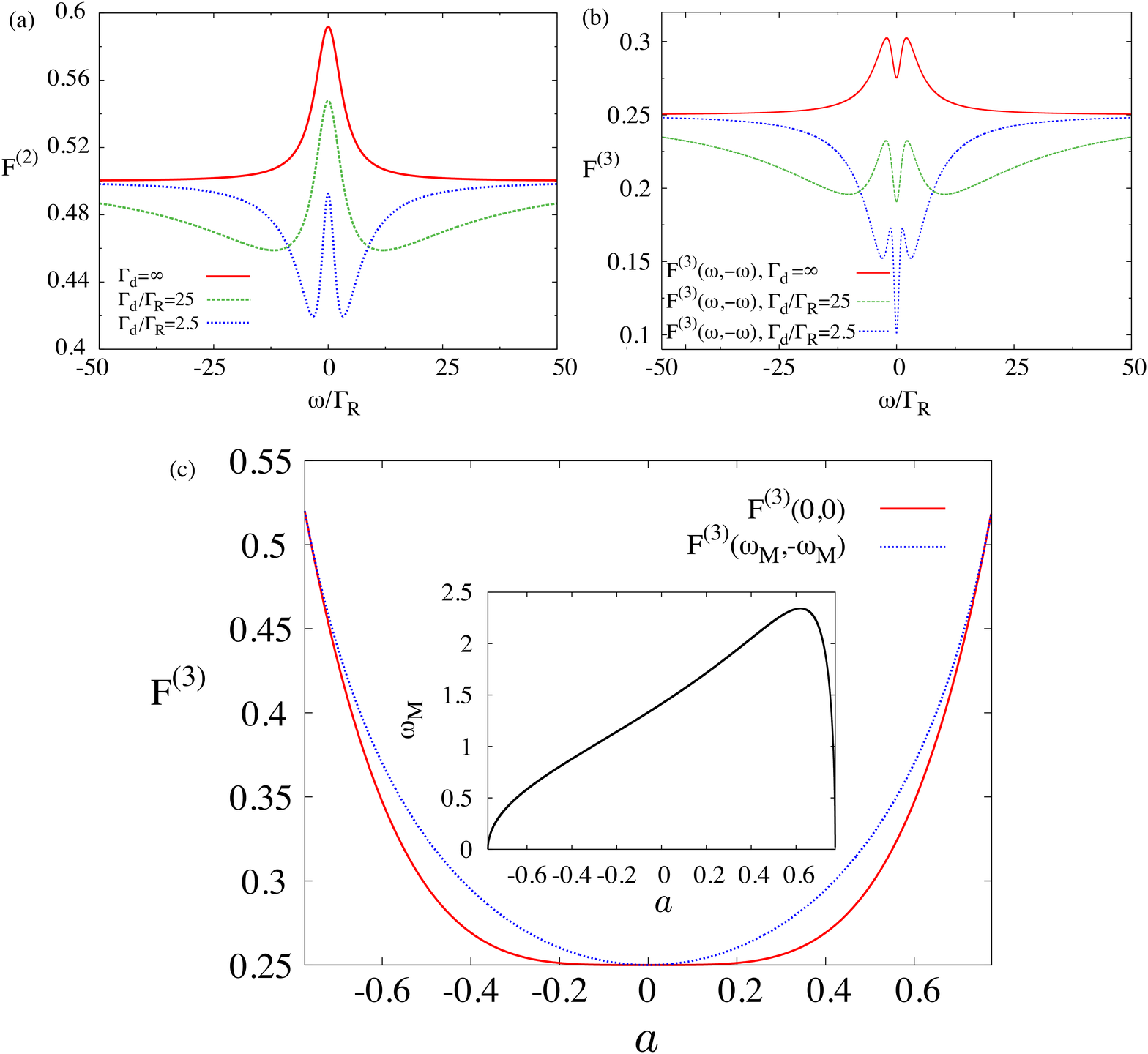}
\caption[Detector model]{Fano factor for the detector model of Ref.~\cite{NaamanAumentado06}. a) Second-order Fano factor as a function of frequency for $\Gamma_L/\Gamma_R=2.5$ and three different detector bandwidths $\Gamma_d$. An ideal detector has $\Gamma_d=\infty$. When this bandwidth becomes comparable with the internal energy scales of the system, the cumulant changes drastically. b) Third-order Fano factor as a function of frequency for $\Gamma_L/\Gamma_R=2.5$. In the symmetric case, $\Gamma_L=\Gamma_R$, the second-order Fano factor is flat with frequency and equal to $1/2$, while the third-order Fano factor is flat with frequency and equal to $1/4$. c) Third-order Fano factor as a function of the asymmetry $a:=\frac{\Gamma_L-\Gamma_R}{\Gamma_L+\Gamma_R}$ evaluated at $\omega=\omega'=0$ (red full line) and at $\omega=-\omega'=\omega_M$ (blue dotted line), where $\omega_M$ is the position of the local maximum in $F^{(3)}(\omega,-\omega)$. This maximum exists for $4-\sqrt{15}<\Gamma_L/\Gamma_R<4+\sqrt{15}$ and is equal to $\sqrt{2}\Gamma_R$ for $\Gamma_L=\Gamma_R$. In the inset we show the position of this maximum as a function of $a$.}
\label{noiseGdFig}
\end{figure}

The theory developed here can incorporate the effects of finite-bandwidth detection that inevitably appear in the experiment. To do this, we consider the detector model introduced by Naaman and Aumentado \cite{NaamanAumentado06}, and that has been successfully applied in counting statistics experiments at zero frequency \cite{Gustavsson07}. The model assumes that the detector only captures events (charging or discharging of the central island) that occur within a certain bandwidth, $\Gamma_d$, set by the detector. As explained in Refs.~\cite{NaamanAumentado06, Gustavsson07}, we need to consider a larger Hilbert space, accounting for the possible states: $\{ \ket{0,0}, \ket{1,0}, \ket{0,1}, \ket{1,1} \}$, where $\ket{n,m}$ denotes a situation with $n$ electrons in the system but $m$ electrons registered by the detector. Projected onto this basis, the $\chi$-dependent QME kernel of the total system, in the infinite-bias limit, takes the form
\beq \label{chikerneldetector}
{\cal W}(\chi) =
\begin{pmatrix}
-\Gamma_L & \Gamma_R & \Gamma_d e^{i\chi_R} & 0 \\
\Gamma_L & -(\Gamma_R + \Gamma_d) & 0 & 0 \\
0 & 0 & -(\Gamma_R+\Gamma_d) & \Gamma_R \\
0 & \Gamma_d e^{-i\chi_L} & \Gamma_L & -\Gamma_R \\
\end{pmatrix},
\eeq
where $\Gamma_L$ and $\Gamma_R$ denote the incoming and outgoing rates of the SRL, and the counting fields $\chi_L$, $\chi_R$, refer to the tunneling events through the left and right barrier respectively. In the decomposition ${\cal W}={\cal W}_0+{\cal W}_J$, we can identify ${\cal W}_J$ as the terms in (\ref{chikerneldetector}) containing a counting field, and the remaining part being ${\cal W}_0$. The total noise and skewness can be then obtained similarly to the previous, simpler case. In Fig.~\ref{noiseGdFig}a we show the total second-order Fano factor corresponding to this model for three different detector bandwidths. We notice that the effect of the detector is to underestimate the noise value, which similarly occurs for the third-order Fano factor (Fig.~\ref{noiseGdFig}b). This effect, which is more pronounced for frequencies $|\omega|<\Gamma_d$, is due to the detector missing the events within this frequency range. In the limit $\omega\to\infty$ all curves coincide. Figure \ref{noiseGdFig}c shows $F^{(3)}\equiv S^{(3)}/\ew{I}$ as a function of the asymmetry $a:=\frac{\Gamma_L-\Gamma_R}{\Gamma_L+\Gamma_R}$, evaluated at two different frequencies. The skewness is symmetric with $a$ as a consequence of the symmetry of the system in the infinite-bias limit.
Finally, we notice the analogy of this model with that of two coupled parallel dots in contact with source and drain leads \cite{KiesslichThesis}.

\subsection{Conclusions of the theory}

Despite the simplicity of the models we have presented, our results
demonstrate the importance of finite-frequency studies. Deviations
from Poissonian behaviour of high-order cumulants are found to be
frequency-dependent, such that a comprehensive analysis in the
frequency domain is needed to fully characterize
correlations and statistics in quantum transport. In the next chapters,
we will extend the theory to study high-order frequency-dependent cumulants
of the total current at finite bias voltage (chapter \ref{Chapter3}),
and non-Markovian corrections (chapter \ref{Chapter4}). In appendix \ref{SuppToFCSwPaper1Sec}
we give details of the calculations leading to the results presented in this section.

~

~

~

~

~

~

~

~

~

~

~

~

~

~

~

~

~

~

~

~

~

~

~

~

~

~

~

~

~

~

~

~

~

~

~

~

~

~

~

~

~

~

~

~


\chapter{Markovian counting statistics of electron transport\footnote{The results presented in this chapter have been published in \cite{Marcos10Mark}.}} 
\label{Chapter3}
\lhead{Chapter 3. \emph{Markovian counting statistics of electron transport}} 

\begin{flushright}

\textit{``It is very certain that, when it is not in our power to determine\\ what is true, we ought to act according to what is most probable.''}

R. Descartes

\end{flushright}

\begin{small}

We present a theory of frequency-dependent counting statistics of electron transport through nanostructures within the framework of Markovian quantum master equations. Our method allows the calculation of finite-frequency current cumulants of arbitrary order, as we explicitly show for the second- and third-order cumulants.  Our formulae generalize previous zero-frequency expressions in the literature and  can be viewed as an extension of MacDonald's formula beyond shot noise. When combined with an appropriate treatment of tunneling, using, e.~g. Liouvillian perturbation theory in Laplace space, our method can deal with arbitrary bias voltages and frequencies, as we illustrate with the paradigmatic example of transport through a single resonant level model. We discuss various interesting limits, including the recovery of the fluctuation-dissipation theorem near linear response, as well as some drawbacks inherent of the Markovian description arising from the neglect of quantum fluctuations.

\end{small}

\newpage

\section{Introduction}

Transport of electrons through nanoscopic conductors is a very powerful tool to learn about interactions and to characterize quantum systems \cite{Kouw97}. Examples include the quantum Hall effect \cite{Klitzing80}, weak localization \cite{Anderson58}, or universal conductance fluctuations \cite{Altshuler85, Lee-Stone85}. Transport processes are governed by tunneling events, which are stochastic in nature.  It is therefore natural to expect that the statistics of these tunneling events will be strongly influenced by interactions and quantum effects. Interestingly, these statistics, which can be analyzed by studying current fluctuations, contain a great deal of new information beyond that provided by dc transport \cite{BlanterButtiker00, NazarovNoiseBook}. In particular, the second-order current-correlation function (noise), can be used to determine the effective charge \cite{Kane-Fisher94,dePicciotto97, Glattli97} and the statistics of the quasiparticles \cite{Schonenberger99, Yamamoto99}, and to reveal information on the transmission properties of the conductor \cite{BlanterButtiker00}. Moreover, current correlations can be used to learn about entanglement \cite{Klich-Levitov09}, quantum coherence \cite{Kiesslich07}, and the deep connection that exists with fluctuation theorems \cite{EspositoRMP, Saito08, Forster08}. 
Further information can be gained from noise at high frequencies which is valuable for extracting internal energy scales in systems such as quantum dots \cite{AguadoBrandes04}, spin-valves \cite{Braun06}, Cooper Pair Boxes \cite{Choi01}, diffusive wires \cite{Nagaev98} or chaotic cavities \cite{Nagaev04}. 
While most of the work on noise is theoretical (in particular at high frequencies), the field of noise and counting statistics is producing a great deal of experimental breakthroughs, including measurements of high order cumulants \cite{Reulet03, Bomze05, Rimberg03, Gustavsson06prl, Fujisawa06, Fricke07, Timofeev07, Gershon08, Flindt09, Gabelli09}. Owing to this experimental progress, noise measurements at high frequencies, which until recently were scarce, are now possible  \cite{Clarke81, Movshovich90, Schoelkopf97, Deblock03, Billangeon06, Zakka-Bajjani07, Gabelli09b, Billangeon09, Xue09}.

A  proper treatment of fluctuations in non-equilibrium transport is needed to address the problems listed above.  While the list of theories available is too large to be given here, it is safe to say that they can be divided roughly into three families: the scattering approach \cite{BlanterButtiker00}, the Keldysh Green's functions method \cite{HaugJauho}, and the various quantum master equation (QME) treatments \cite{Cohen} (for a recent overview of transport in this last context see e.g. Ref~\cite{Timm08}). 
A theory of counting statistics of electron transport was first formulated by Levitov and Lesovik for noninteracting electrons using the scattering formalism \cite{Levitov93, Levitov96}, and later
works enabled the treatment interacting problems \cite{BelzigNazarov01}. QME
approaches followed \cite{BagretsNazarov03}, and prooved particularly useful for studying
systems in the Coulomb Blockade regime. Recent avances within this last scheme also involve studies of the counting statistics including non-Markovian dynamics \cite{Braggio06, Flindt08}.  

In the previous chapter we presented a method for calculating high-order current correlations at finite frequencies in the context of Markovian QMEs. Here we significantly extend the theory. First, we provide a detailed derivation of our multi-time generating function. Next, we present a new approach to derive finite-frequency cumulants from this expression.  
We provide analytical formulae for the second and third-order current cumulants (noise and skewness respectively). These results generalize previous zero-frequency expressions in the literature and recover the finite frequency shot-noise expressions \cite{Flindt05c, Lambert07} obtained using the MacDonald's formula. Our method can thus be viewed as a generalization of this formula, as it allows us to obtain high-order current-correlation functions. To illustrate the formalism we study the case of a single resonant level (SRL) model, and compare it with the exact solution and with the non-equilibrium version of the fluctuation-dissipation theorem, derived in various approaches, such as for tunnel junctions, or for the weak cotunneling regime in quantum dots \cite{Lev04,Dahm69,Rogovin-Scalapino74,Suk01}.

The chapter is organized as follows. In section \ref{TheorySec}, we present our formalism of finite-frequency cumulants in the context of Markovian QMEs. Subsections \ref{QMEsec} and \ref{FCSsec} are devoted to establish the general framework of full counting statistics. In subsection \ref{FCSwsec} we derive a multitime cumulant generating function. Subsection \ref{S2S3sec} shows how to obtain finite-frequency cumulants of the current distribution. Here, exact equations for the frequency-dependent noise and skewness are given. We end this part with special emphasis on how to calculate the counting statistics of the `total' and `accumulated' currents (subsection \ref{FCStotaccum}). We explicitly show how both current correlations and charge/voltage correlations can be calculated.
In section \ref{Resultssec}, we study the example of a SRL model, providing spectra for the frequency-dependent noise and skewness, and a detailed comparison with the fluctuation-dissipation theorem, the finite-frequency version of the non-equilibrium fluctuation-dissipation theorem, and the exact solution of the SRL model. First we focus on the zero-frequency case (section \ref{FCSresults}), where the general behaviour of noise and skewness is presented as a function of different system parameters. In section \ref{FCSwresults} we extend this study to the finite-frequency case. Interestingly, we show that, even though the theory does not contain quantum fluctuations, the Markovian limit is basically exact in transport configurations, with the dot level within the bias voltage window, as long as $\hbar\omega \gg eV$ or $\hbar\omega\ll eV$. In intermediate situations, where $\hbar\omega \sim eV$, or with the level outside the bias windows, the Markovian limit fails at finite frequencies due to its lack of quantum fluctuations. We also demonstrate that the noise spectra for particle currents and the ones for total currents significantly deviate from each other, even for large asymmetric coupling to the leads, namely  
$\Gamma_R/\Gamma_L \neq 1$.
In section \ref{ConclusionsSec} we summarize our results. Most of the technical details and intermediate steps of the derivations in section \ref{TheorySec} are discussed in detail in appendix \ref{appDiagrams}, where we also present a diagrammatic technique to arrive to the expressions for the cumulants shown in sec. \ref{S2S3sec}. In appendix \ref{appKernel} we describe how to calculate the kernel of the QME to lowest (sequential) order using perturbation theory in the Liouville space.

\section{Theory} \label{TheorySec}

\subsection{Quantum master equation} \label{QMEsec}

We are interested in phenomena which fulfill the general evolution given by Eq.~(\ref{LiouvilleEq}). Specifically, our theory will be useful to processes amenable to the counting of a classical stochastic variable $n$, which can be, for example, the number of particles that have undergone a particular process in the system.
We will focus in particular in transport systems, consisting on a central
region, with a known set of many-body eigenstates $\{\ket{a}\}$ and respective eigenenergies $\{E_a\}$, attached to non-interacting electronic leads at different chemical potentials.
This set-up can be described by a Hamiltonian that takes the form\footnote{In this chapter we obviate the hat in the notation for operators.}
$\mathcal{H}=\mathcal{H}_S + \mathcal{H}_R + \mathcal{H}_V$, where
$\mathcal{H}_S$ and $\mathcal{H}_R$ refer to system and leads
respectively, and $\mathcal{H}_V$ is the coupling between them. The different terms can be written as
\begin{eqnarray}
\mathcal{H}_S &=& \sum_a E_a \ket{a}\bra{a}. \label{Hs}\\
\mathcal{H}_R &=& \sum_{\eta,\alpha} \left( \varepsilon_{\eta\alpha}
+ \mu_{\alpha}
\right) c_{\eta\alpha}^{\dagger} c_{\eta\alpha}. \label{Hr}\\
\mathcal{H}_V &=& \sum_{\eta,\alpha,m} {\cal V}_{\eta\alpha m}
c_{\eta\alpha}^{\dagger} d_m + \mathrm{H.c.} \label{Hv}
\end{eqnarray}
Here $c_{\eta\alpha}^{\dagger}$ creates an electron
with quantum numbers $\eta$ in lead $\alpha$, and $d_m$ annihilates
an electron from site $m$ in the central region.  $\varepsilon_{\eta\alpha}$ are the eigenenergies of the electrons in the lead $\alpha$, ${\cal V}_{\eta\alpha m}$ is the coupling energy between a state in contact $\alpha$ and the level $m$ in the system. $\mu_{\alpha}$ is the chemical potential of lead $\alpha$, and that allows the system to be driven out of equilibrium. 
Given this Hamiltonian, the full evolution can be obtained using the von Neumann
equation (\ref{vonNeumann}, which defines the Liouvillian ${\cal L}$.
We are actually interested in the dynamics concerning the central system. We therefore trace out the reservoir degrees of freedom. Under the Markovian approximation we arrive then to equation (\ref{rhoSevolution}) for the system density operator.

The charge flow through the conductor is governed by the stochastic hopping of electrons in and out of the central region. These processes are susceptible to classical counting, and thus the reduced density operator can be unravelled into components $\rho_\mathrm{S}(n_{\alpha},t)$, corresponding to having $n_{\alpha}=...,-1,0,1,...$ extra electrons in lead $\alpha$ \cite{Cook81, Plenio98}. The kernel can also be split as ${\cal W}={\cal W}_0+\sum_{\pm}{\cal W}_{\pm}^{\alpha}$, where ${\cal W}_{\pm}^{\alpha}$ refers to the physical process in which one electron is created (+) or annihilated (-) at lead $\alpha$, and ${\cal W}_0$ corresponds to the part in which no tunneling processes take place. It can be shown that $\rho_\mathrm{S}(n_{\alpha},t)$ fulfills the equation (see e.g. \cite{Flindt05b}):
\beq \label{nQME}
\dot{\rho}_\mathrm{S}(n_{\alpha},t) = {\cal W}_0 \rho_\mathrm{S}(n_{\alpha},t) + {\cal W}_{+}^{\alpha} \rho_\mathrm{S}(n_{\alpha} - 1,t) + {\cal W}_{-}^{\alpha} \rho_\mathrm{S}(n_{\alpha} + 1,t);
\eeq
valid provided that only single particle tunneling processes occur. Although Eq. (\ref{nQME}) focuses on the single counting at a particular lead $\alpha$, it can be generalized to account for tunneling processes of $k$ particles at the different system-reservoir junctions:
\beq \label{nQMEgeneralized}
\dot{\rho}_\mathrm{S}(n_1,\ldots,n_M,t) = && {\cal W}_0 \rho_\mathrm{S}(n_1,\ldots,n_M,t) \nonumber\\ &&+ \sum_{\alpha,k,\pm} {\cal W}_{\pm}^{\alpha,k}\rho_\mathrm{S}(n_1,\ldots,n_{\alpha}\mp k,\ldots,n_M,t),
\eeq
where ${\cal W}_0$ is the part in which the number of particles is not changed in the central region, $M$ the number of leads, and $k$ labels the process in which $k$ particles `jump' at a time.

Unfortunately, solving equation (\ref{nQMEgeneralized}) in the $n$-space requires truncation to a certain $n$ and diagonalization of a tridiagonal matrix. It is therefore more convenient to solve it taking the Fourier transform. Multiplying (\ref{nQMEgeneralized}) by $e^{in_1\chi_1}\ldots e^{in_M\chi_M}$ and summing over $n_1,\ldots,n_M$ we obtain
\beq \label{EOMchit} \dot{\rho}_\mathrm{S}(\chi,t) =
\mathcal{W}(\chi)\rho_\mathrm{S}(\chi,t). \eeq
Here the counting field $\chi$ refers implicitly to all counting fields, and we have defined $\rho_\mathrm{S}(\chi,t)\equiv\sum_{n_1,\ldots,n_M} e^{in_1\chi_1}\ldots e^{in_M\chi_M} \rho_\mathrm{S}(n_1,\ldots,n_M,t)$ and ${\cal W}(\chi)\equiv{\cal W}_0+\sum_{\alpha,\pm}{\cal W}_{\pm}^{\alpha}e^{\pm i\chi_{\alpha}}$.
For a time-independent kernel, the solution to Eq.~(\ref{EOMchit}) is 
\beq \label{rhochit}
{\rho}_\mathrm{S}(\chi,t)=\Omega(\chi,t-t_0){\rho}_\mathrm{S}(\chi,t_0),
\eeq
with the time-evolution operator $\Omega(\chi,t-t_0):=e^{{\cal W}(\chi)(t-t_0)}$.

\subsection{Full Counting Statistics} \label{FCSsec}

Importantly, the knowledge of the system's density operator resolved in $n$ allows us to obtain the full counting statistics (FCS) of the system, that is, the probability distribution $P(n_1,\ldots,n_M,t)$ of the number of electrons
transmitted through the system-lead junctions. This is accomplished by noting that \beq\label{Pn} P(n_1,\ldots,n_M,t) = \mathrm{Tr_S} \{ \rho_S(n_1,\ldots,n_M,t) \},\eeq 
where $\mathrm{Tr_S}$ denotes the trace over the system degrees of freedom.
As explained in the previous chapter, transforming the probability distribution to the $\chi$-space, we have the cumulant/moment generating function (CGF/MGF) (\ref{MGF-CGF})\footnote{In Eq.~(\ref{MGF-CGF}), $n$ would refer implicitly to $n_1,\ldots,n_M$ for the present case.}. Making use of Eq.~(\ref{rhochit}) we find
\beq \label{MGF}
{\cal G}(\chi,t)=\mathrm{Tr_S} \left\lbrace \Omega(\chi,t-t_0){\rho}_\mathrm{S}(t_0) \right\rbrace,
\eeq
which is similar to equation (\ref{BagretsNazarovEq}) introduced in chapter \ref{Chapter2}.
The $N$-th. derivative of the MGF with respect to $\chi$ gives the $N$-th moment of the probability distribution of the number of particles that have tunneled in or out a particular lead $\alpha$:
\beq
\langle n_{\alpha}^N (t) \rangle=\frac{\partial^N {\cal
G}(\chi,t)}{\partial(i\chi_{\alpha})^N} \Big|_{\chi\rightarrow
0}.
\eeq
In equation (\ref{MGF}), averages with respect to the stationary state are established by taking 
${\rho}_\mathrm{S}(t_0)={\rho}_\mathrm{S}^{stat}$ (defined by
${\mathcal W}{\rho}_\mathrm{S}^{stat}=0$). This means that counting will start at
a time $t_0$ in which the system has reached its steady state, and therefore the fluctuations we study are around this state\footnote{In the following, all the averages
will be taken with respect to this steady state. An average in 
the Liouville space will be therefore written as $\langle A \rangle = \bm{t}_T\cdot A \cdot \bm{\rho}_\mathrm{S}^{stat} \equiv \langle\!\langle \tilde{0} | A | 0 \rangle\!\rangle$, where $| 0 \rangle\!\rangle \equiv \bm{\rho}_\mathrm{S}^{stat}$ is the normalized stationary system density matrix (written as a vector), and $\bm{t}_T\equiv \langle\!\langle \tilde{0} |$ is the transposed trace vector that sums over the population degrees of freedom.}.

The moments of the current distribution can be calculated as\footnote{Throughout the chapter we will use $e$ (electron charge) $=$ $k$ (Boltzman's constant) $=$ $\hbar$ (Planck's constant/$2\pi$) = 1.}
\beq
\langle I_{\alpha}^N (t) \rangle = \frac{d}{dt} \langle n_{\alpha}^N (t) \rangle.
\eeq
This relation is important as it relates the stochastic variable $n$ with the current of particles flowing through the system. Even though the current studied here is a classical variable, it contains quantum effects present in the system. In the formalism, these are inherited from the Liouvillian operator in equation (\ref{LiouvilleEq}).
Generally, we are interested in the cumulants, rather than in the moments, of the current distribution.  These can be obtained from the derivatives of the CGF. Therefore we have
\beq
\langle
I_{\alpha}^N\rangle_c=\frac{d}{dt}\frac{\partial^N {\cal
F}(\chi,t)}{\partial(i\chi_{\alpha})^N}\Big|_{\chi\rightarrow
0,t\rightarrow\infty},
\eeq
where $\langle \ldots \rangle_c$ denotes cumulant average \cite{Kubo62} and the limit $t\to\infty$ ensures the stationary state average. Also, notice that the probability distribution itself can be obtained by inverse Fourier transform of the MGF.

From the $\chi$-independent kernel of the reduced QME (\ref{rhoSevolution}), the $\chi$-dependence leading to equation (\ref{EOMchit}) can be actually introduced in a simpler way than resolving the density operator in $n$ and taking the Fourier transform. As we describe in appendix \ref{appKernel}, it is enough to include counting fields in the appropriate tunneling terms of the kernel, and this procedure is fully equivalent to solving a generalized Von Neumann equation: \beq \label{EOMchi} \dot{\rho}(\chi,t) = -\frac{i}{\hbar}
(\mathcal{H}^+(t)\rho(\chi,t)-\rho(\chi,t)\mathcal{H}^-(t)), \eeq in
which the time evolution in the forward (+) and backward (-) Keldysh
part of the real time axis is governed by \emph{different}
Hamiltonians \cite{Lev04}, specifically $\mathcal{H}^{\pm}_T =
\sum_{\eta,\alpha,m} {\cal V}_{\eta\alpha m} e^{\pm
i\chi/2}c_{\eta\alpha}^{\dagger} d_m + \mathrm{H.c.}$. Tracing out the 
reservoir degrees of freedom in equation (\ref{EOMchi}), one can get 
equation (\ref{EOMchit}) and proceed to obtain the FCS
of the system.

\subsection{Finite-frequency full counting statistics} \label{FCSwsec}

Our aim here is to study correlations at finite
frequencies, for which the scheme presented above needs to be
generalized. To this end, we consider the \emph{joint probability
distribution}, $P(n_1,t_1;\ldots;n_N,t_N)$, defined as the
probability that $n_1$ electrons have undergone a particular process
 after a time $t_1$, $n_2$ electrons after a time $t_2$, etc. Here we focus
for simplicity on a particular lead, and denote $n$ as the number of particles
transferred to (from) it, with associated counting field $\chi$ ($-\chi$). It is straightforward to include processes at different leads.

The connection between this joint probability and the density 
operator (analogue to Eq.~(\ref{Pn})) is not straightforward. To connect them
we first need to specify a prescription for the symmetrization of the cumulants and the probability distribution. 
As pointed out in chapter \ref{Chapter1}, this prescription actually depends on the detection scheme \cite{Lesovik97, Aguado00, Gavish00}. Here we assume a `classical' detection, so the detector is incapable of distinguishing emission from absorption. This means that the results we will present correspond to the fully-symmetrized version of the power spectrum
\beq \label{currentcorrfreq}
S^{(N)}(\omega_1,\ldots,\omega_N):= \int_{-\infty}^\infty dt_1\ldots dt_N  e^{-i\omega_1 t_1}\ldots e^{-i\omega_Nt_N} {\cal T}_S \langle I(t_1) \ldots
I(t_N) \rangle_c,
\eeq
where ${\cal T}_S$ is the symmetrization operator, that sums over all possible time (or frequency) switchings, that is, we have for example ${\cal T}_S \langle I(t_1)I(t_2) \rangle = \langle I(t_1)I(t_2) \rangle + \langle I(t_2)I(t_1) \rangle$.

The spectrum (\ref{currentcorrfreq}) can be derived from a $N$-time (symmetrized) CGF ${\cal F}^{(N)}$, defined by
\beq e^{\mathcal{F}^{(N)}[\bm{\chi},\bm{t}]} = {\cal G}^{(N)} [\bm{\chi},\bm{t}] :=
\sum_{n_1,\ldots,n_N} e^{in_1\chi_1+\ldots+in_N\chi_N}
P^{(N)}[\bm{n},\bm{t}], \nonumber \eeq where
$\bm{\chi}:=(\chi_1,\ldots,\chi_N)_T$, $\bm{t}:=(t_1,\ldots,t_N)_T$, $\bm{n}:=(n_1,\ldots,n_N)_T$, $P^{(N)}$ refers to the symmetrized joint probability, ${\cal G}^{(N)}$ to the multitime MGF, and the subscript $T$ to the transpose of a column vector. That is, we have
\beq
S^{(N)}(\omega_1,\ldots,\omega_N)= && \int_{-\infty}^\infty dt_1\ldots dt_N e^{-i\omega_1 t_1}\ldots e^{-i\omega_Nt_N} \nonumber\\ && \times
\partial_{t_1} \ldots \partial_{t_N}
\partial_{i\chi_1} \ldots \partial_{i\chi_N} \mathcal{F}^{(N)}[\bm{\chi},\bm{t}] \;
\big\vert_{{\bm \chi}={\bm 0}} .\label{currentcorrtime} \eeq
Using the property of the Fourier transform of a derivative we get
\beq
S^{(N)}(\omega_1,\ldots,\omega_N)= && (i\omega_1) \ldots (i\omega_N) \int_{-\infty}^\infty dt_1\ldots dt_N e^{-i\omega_1 t_1}\ldots e^{-i\omega_Nt_N} \nonumber\\ && \times
 \partial_{i\chi_1} \ldots \partial_{i\chi_N} \mathcal{F}^{(N)}[\bm{\chi},\bm{t}] \;\big\vert_{{\bm \chi}={\bm 0}}.
\eeq
Both the probability $P^{(N)}[\bm{n},\bm{t}]$ and the CGF $\mathcal{F}^{(N)}[\bm{\chi},\bm{t}]$ can be calculated from the density operator and the kernel ${\cal W}$ if we use the Markovian approximation. Within this limit we have the evolution given by (\ref{rhochit}), and also the factorization property
\beq \label{Pgreater}
P^>(n_1,t_1;\ldots;n_N,t_N) = P(n_1,t_1) P(n_2,t_2 | n_1,t_1) \ldots P(n_N,t_N | n_{N-1},t_{N-1}),
\eeq
where the symbol $>$ constraints the times to $t_k>t_{k-1}$. Notice that as we are considering the totally symmetric correlation function, we need to take $P^{(N)}[\bm{n},\bm{t}] = {\cal T} P^>(n_1,t_1;\ldots;n_N,t_N)$, where ${\cal T}$ is the time-ordering operator. $P(n,t | n',t')$ is the conditional probability of counting $n$ electrons after time $t$, provided that we counted $n'$ electrons after time $t'$, and can be computed as
\beq \label{Pcond}
P(n,t | n',t') = \frac{\mathrm{Tr_S}\lbrace \Omega(n-n',t-t')\rho_\mathrm{S}(n',t')\rbrace}{\mathrm{Tr_S}\lbrace \rho_\mathrm{S}(n',t') \rbrace},
\eeq
where the normalization in the denominator accounts for the collapse of the state due to the measurement, as given by the von Neumann's projection postulate \cite{Korotkov01}. $\Omega(n,t)$ is the propagator in the $n$-space, that is,
\beq\label{evolrho}{\rho}_\mathrm{S}(n,t)=\sum_{n'}\Omega(n-n',t-t'){\rho}_\mathrm{S}(n',t'),\eeq and can be extracted from equation (\ref{nQMEgeneralized}) or by inverse Fourier transform of the propagator in the $\chi$-space:
\beq
\Omega(n,t) = \int \frac{d\chi}{2\pi} e^{-in\chi} \Omega(\chi,t).
\eeq
An expression for the joint probability distribution in terms of propagators can be then derived using (\ref{Pgreater}) together with (\ref{Pcond}) and (\ref{evolrho}). Alternatively, it can be obtained using the Chapman-Kolmogorov property for Markovian evolutions \cite{vanKampenBook}, from which we have
\beq
P(n_N;t_N) && = \mathrm{Tr_S} \sum_{n_1,\ldots,n_{N-1}} \Omega(n_N-n_{N-1};t_N-t_{N-1}) \nonumber\\ &&\times \Omega(n_{N-1}-n_{N-2};t_{N-1}-t_{N-2}) \ldots \Omega(n_1;t_1) \rho_\mathrm{S}(t_0).
\eeq
As we also have $P(n_N;t_N)=\sum_{n_1,\ldots,n_{N-1}} P^>(n_1,t_1;\ldots;n_N,t_N)$, reminding that ${\rho}_\mathrm{S}(t_0)={\rho}_\mathrm{S}^{stat}$, we find
\beq \label{jointPn}
P^{(N)}[\bm{n},\bm{t}] = \mathcal{T}\Big\langle
  \prod_{k=1}^N
\Omega(\nu_{N-k},\tau_{N-k})
  \Big\rangle,
\eeq
where ${\nu}_{k}:=n_{k+1}-n_k$, ${\tau}_{k}:=t_{k+1}-t_k$ and $\langle \bullet \rangle := \mathrm{Tr_S}\{\bullet {\rho}_\mathrm{S}^{stat} \}$. Transforming expression (\ref{jointPn}) to the $\chi$-space, we find the CGF
\beq\label{CGF}
 {\mathcal{F}^{(N)}[\bm{\chi},\bm{t}]}
 =
 \ln \mathcal{T}\Big\langle
  \prod_{k=1}^N
\Omega(\tilde{\chi}_k,\tau_{N-k})
  \Big\rangle,
  \eeq
being $\tilde{\chi}_k:=\sum_{i=N+1-k}^N \chi_i$. The structure in Eq. (\ref{CGF}) is
encountered in many branches of physics such as statistical physics
and field theory, where \emph{connected} correlation functions are
obtained from derivatives of the logarithm of the corresponding
generating functional (the partition function, the $S$-matrix, etc). Note
in particular the analogy of Eq.~(\ref{CGF}) with the partition function given in Ref.~\cite{Kubo62}.

\subsection{Finite-frequency cumulants} \label{S2S3sec}

Equation (\ref{CGF}) allows us to obtain frequency-dependent current cumulants to arbitrary order. This is precisely Eq.~(\ref{genF}) presented in chapter \ref{Chapter2}, which was to study the second and third cumulant for various models. Explicit derivatives of (\ref{CGF}) and the eigen-decomposition of the kernel were used then to that end.
In this subsection we show that only the stationary solution of the problem (solution to an algebraic equation) is needed to compute the finite-frequency current cumulants. We give analytical expressions (valid within the Markovian approximation) for the noise (second cumulant) and skewness (third cumulant) of the distribution of charge flowing through a conductor.

Let us decompose the Fourier transform in equation (\ref{currentcorrfreq}) into a set of Laplace transforms (defined as $f(z):=\int_0^{\infty} dt e^{-zt} f(t)$), and the cumulant averages in terms of moments (c.f. for example Eq.~(2.8) in Ref~\cite{Kubo62}). Doing this we find\footnote{Notice that Eqs. (\ref{LaplaceS1})-(\ref{LaplaceS3}) can also be derived if the derivatives of the CGF are decomposed in terms of derivatives of MGFs. For example, for $N=3$ we have ${\mathcal{F}^{(3)}_{123}}={\mathcal{G}^{(3)}_{123}}-{\mathcal{G}^{(3)}_{1}}{\mathcal{G}^{(3)}_{23}}-{\mathcal{G}^{(3)}_{2}}{\mathcal{G}^{(3)}_{13}}-{\mathcal{G}^{(3)}_{3}}{\mathcal{G}^{(3)}_{12}}+2{\mathcal{G}^{(3)}_{1}}{\mathcal{G}^{(3)}_{2}}{\mathcal{G}^{(3)}_{3}}$, with $f_{i}:=\partial_{\chi_i} f\vert_{\chi_i=0}$, $f_{ij}:=\partial_{\chi_i}\partial_{\chi_j} f\vert_{\chi_i,\chi_j=0}$, $f_{ijk}:=\partial_{\chi_i}\partial_{\chi_j}\partial_{\chi_k} f\vert_{\chi_i,\chi_j,\chi_k=0}$.}
\beq \label{LaplaceS1}
S^{(1)>}(z_1) = S_m^{(1)>}(z_1).
\eeq
\beq \label{LaplaceS2}
S^{(2)>}(z_1,z_2) = S_m^{(2)>}(z_1,z_2) - \left( \frac{-1}{z_1} \right) \left( \frac{-1}{z_2} \right) \langle I \rangle^2.
\eeq
\beq \label{LaplaceS3}
S^{(3)>}(z_1,z_2,z_3)
&=& S_m^{(3)>}(z_1,z_2,z_3) \nonumber\\ &&- \left( \frac{-1}{z_1} \right) \langle I \rangle S_m^{(2)>} (z_2,z_3) \nonumber\\ &&- \left( \frac{-1}{z_2} \right) \langle I \rangle S_m^{(2)>} (z_1,z_3) \nonumber\\ &&- \left( \frac{-1}{z_3} \right) \langle I \rangle S_m^{(2)>} (z_1,z_2) \nonumber\\ &&+ 2 \left( \frac{-1}{z_1} \right)  \left( \frac{-1}{z_2} \right)  \left( \frac{-1}{z_3} \right) \langle I \rangle^3.
\eeq
The notation ``$>$'' denotes the unsymmetrized correlation function corresponding to the time ordering $t_N>\ldots>t_2>t_1$. Symmetrization in the frequency space implies adding the part corresponding to negative $z$ and summing over all the possible switchings of frequencies.
The subscript $m$ means moment. These can be obtained as
\beq
S_m^{(N)>}(z_1,\ldots,z_N) = z_1 \ldots z_N
\partial_{i\chi_1} \ldots \partial_{i\chi_N} \mathcal{G}^{(N)>}[\bm{\chi},\bm{z}] \;
\big\vert_{\bm{\chi}=\bm{0}}, \label{currentcorr}
\eeq 
with $\bm{z}\equiv(z_1,\ldots,z_N)_T$ and\footnote{In the frequency domain, the prescription $>$ can be taken similarly, that is $z_N>\ldots>z_2>z_1$, and finally symmetrize the result.} \beq \label{GN}\mathcal{G}^{(N)>} [\bm{\chi},\bm{z}] = \Big\langle \prod_{k=1}^N
\Omega(\tilde{\chi}_k,\tilde{z}_{k}) \Big\rangle^>,\eeq with $\Omega(\chi,z)\equiv [z-{\cal W}(\chi)]^{-1}$.

One advantage of having moment averages is that we can use a diagrammatic technique (see appendix \ref{appDiagrams}) to easily obtain the desired correlation functions.
Symmetrizing $S^{(N)>}[\bm{z}]$ and evaluating it at $\bm{z}=i\bm{\omega}$ (being $\bm{\omega}\equiv (\omega_1,\ldots,\omega_N)_T$), we find that $S^{(N)}[\bm{\omega}]$ is proportional to $\delta(\omega_1+\ldots+\omega_N)$, as required by time-translational invariance. Defining the jump super-operators ${\cal
J}_{\chi}:=[{\cal W}(\chi)-{\cal W}(\chi=0)]$ and their derivatives $\mathcal{J}_0^{(n)}:=\partial_{\chi}^n\mathcal{J}_{\chi}|_{\chi=0}$, we arrive to the following expressions for the current, noise and skewness of the current distribution (c.f. appendix \ref{appDiagrams} for details):
\beq
\label{currentformula} i I_{stat} = \langle \mathcal{J}_0^{(1)}
\rangle, \eeq
\beq
\label{noiseformula} i^2 S^{(2)} (\omega) = \langle {\cal
J}_{0}^{(2)} \rangle + \langle {\cal J}_0^{(1)} {\Omega}_0(i\omega)
{\cal J}_0^{(1)} \rangle + \langle {\cal J}_0^{(1)} {
\Omega}_0(-i\omega) {\cal J}_0^{(1)} \rangle, \eeq
\beq
\label{skewnessformula} i^3 S^{(3)} (\omega,\omega') &=&
\langle {\cal J}_{0}^{(3)} \rangle + \langle{\cal
J}_0^{(2)}{\Omega}_0(i\omega){\cal J}_0^{(1)} \rangle +
\langle{\cal J}_0^{(2)}{\Omega}_0(i\omega'-i\omega){\cal J}_0^{(1)}
\rangle \nonumber\\ &&+ \langle{\cal J}_0^{(2)}{\Omega}_0(-i\omega'){\cal J}_0^{(1)} \rangle +
 \langle{\cal J}_0^{(1)}{\Omega}_0(-i\omega){\cal J}_0^{(2)}
\rangle \nonumber\\ &&+ \langle{\cal J}_0^{(1)}{\Omega}_0(i\omega'){\cal J}_0^{(2)} \rangle + \langle{\cal J}_0^{(1)}{\Omega}_0(i\omega-i\omega'){\cal J}_0^{(2)} \rangle \nonumber \\ &&+
\langle {\cal J}_0^{(1)}{\Omega}_0(-i\omega){\cal J}_0^{(1)}{\Omega}_0(-i\omega'){\cal J}_0^{(1)}\rangle \nonumber \\ &&+ \langle {\cal
J}_0^{(1)}{\Omega}_0(i\omega'){\cal J}_0^{(1)}{\Omega}_0(i\omega){\cal J}_0^{(1)}\rangle \nonumber \\ &&+ \langle
{\cal J}_0^{(1)}{\Omega}_0(-i\omega){\cal J}_0^{(1)}{\Omega}_0(i\omega'-i\omega){\cal J}_0^{(1)}\rangle \nonumber \\ &&+
\langle {\cal J}_0^{(1)}{\Omega}_0(i\omega'){\cal J}_0^{(1)}{\Omega}_0(i\omega'-i\omega){\cal J}_0^{(1)}\rangle \nonumber \\ &&+
\langle {\cal J}_0^{(1)}{\Omega}_0(i\omega-i\omega'){\cal
J}_0^{(1)}{\Omega}_0(-i\omega'){\cal J}_0^{(1)}\rangle \nonumber\\ &&+
\langle {\cal J}_0^{(1)}{\Omega}_0(i\omega-i\omega'){\cal
J}_0^{(1)}{\Omega}_0(i\omega){\cal J}_0^{(1)}\rangle, \eeq being
$\Omega_0(z):=[z-{\cal W}(\chi=0)]^{-1}$.
These equations generalize the zero-frequency results found in Ref. \cite{Flindt05b} -- c.f. their Eqs. Eqs. (7) and (8) -- to finite frequencies. The zero-frequency limit of (\ref{currentformula})-(\ref{skewnessformula}) is presented in appendix \ref{appDiagrams}.
Results for higher-order cumulants can be similarly obtained.

The relation between cumulants and moments can be formally expressed more generally at the level of the generating function. To do this one should follow the derivation by Kubo \cite{Kubo62}, making use of the property $\langle \exp(\sum_i n_i\chi_i)\rangle=\exp\{\langle \exp(\sum_i n_i\chi_i)-1\rangle_c\}$ in our context, arriving to a similar result to (7.25) in Ref.~\cite{Kubo62}.
This allows for the calculation of frequency-dependent cumulants of the current distribution up to any order, reproducing in particular the results presented above. If a diagrammatic expansion in the Liouvillian space is used \cite{Schoeller09, Leijnse08, Emary09}, cumulant averages become particularly useful, since one can then keep only \textit{connected} diagrams as those contributing to the average, in a similar way that this is done in quantum field theory.

\subsection{FCS of total and accumulated currents}
\label{FCStotaccum}

As discussed in the previous chapter, at finite frequencies it is essential to include the so-called displacement currents to have a theory consistent with current conservation \cite{BlanterButtiker00}. This point is of vital importance to reproduce correctly the noise spectra measured experimentally. Although our discussion has focused, by construction,  on particle
currents so far, we show here how to include the effect of displacement
currents in our formalism. 
Let us illustrate this point by considering a quantum well with two terminals ($L$ and $R$) in contact with Fermi leads at different chemical potentials. There will be then a net current flowing through both terminals, but also, charge can `accumulate' in the well for some time. Therefore, charge conservation may be expressed as
\beq
\dot{Q}(t) = I_L - I_R \equiv I_{accum},
\eeq
with $Q$ the charge in the well and $I_L$, $I_R$ referring to the currents through the left and right contacts respectively. $\dot{Q}$ represents the displacement current, $I_{dis}$, which can be partitioned as $I_{dis}=(\alpha+\beta)I_{dis}=I^R_{dis}+I^L_{dis}$, where $\alpha$ and $\beta$ describe how the displacement current is
divided between left and right reservoirs (obviously
$\alpha+\beta=1$). This partitioning allows us to write the 
current conservation as $I_L-I^L_{dis}-(I_R+I^R_{dis})=0$.
Equivalently, the total current (particle plus displacement) reads $I_{tot}=I_L-I^L_{dis}=I_R+I^R_{dis}=\alpha I_L+\beta
I_R$, which is the so-called Ramo-Shockley theorem\footnote{In this chapter we mean `total' cumulant when a subscript is ommited.}. In the simplest wide-band limit, the partition coefficients can be
written in terms of tunnel rates only:
$\alpha:=\Gamma_R/(\Gamma_L+\Gamma_R)$ and
$\beta:=\Gamma_L/(\Gamma_L+\Gamma_R)$ \cite{Wang99}\footnote{In a
Coulomb Blockade model, the partition coefficients read
$\alpha=C_R/(C_L+C_R)$ and $\beta=C_L/(C_L+C_R)$, where $C_L$, $C_R$
are the capacitances of each barrier and we have neglected
capacitive effects from the gate. See for instance
\cite{Bruder94}.}, and this will be the partitioning we will use throughout the thesis. 

Experimentally, one can measure correlations of the current through
the device by transport measurements \cite{Reulet03, Bomze05}, or
indirectly by studying the current through a charge sensor, such as
a quantum point contact \cite{Rimberg03, Gustavsson06prl, Fujisawa06}, that reveals
whether the well is `charged' or `uncharged'. The second method
gives the statistics of the transport current only for very large
bias voltages (unidirectional counting) but, in general the
time-dependent transport current and the charge statistics are
different. Morevover, when the device itself is used as a detector, the
difference between transport fluctuations and charge fluctuations
leads to profound physical consequences. Unlike the inelastic
backaction induced by current fluctuations of the detector \cite{Aguado00},
the one induced by charge fluctuations is the fundamental
Heisenberg backaction associated with the measurement \cite{Young10}.
Both transport and charge
fluctuations can be accounted for in our formalism by considering respective counting fields
\beq
\chi_{tot} &:=& \chi_L + \chi_R, \label{chitot} \\
\chi_{accum} &:=& \beta \chi_L - \alpha \chi_R, \label{chiaccum} \eeq which lead to
the jump super-operators \beq
\mathcal{J}_{\chi,tot}^{(n)}\equiv \alpha^n \mathcal{J}_{\chi,L}^{(n)} + \beta^n \mathcal{J}_{\chi,R}^{(n)}, \label{Jtot} \\
\mathcal{J}_{\chi,accum}^{(n)} \equiv \mathcal{J}_{\chi,L}^{(n)} +
(-1)^n \mathcal{J}_{\chi,R}^{(n)}, \label{Jaccum} \eeq where $\mathcal{J}_{\chi,L}$
and $\mathcal{J}_{\chi,R}$ refer to the two independent tunneling
processes occurring at each barrier. 

The `total' cumulant through a two terminal device can be then calculated performing derivatives of the CGF with respect to $\chi_{tot}$ defined in (\ref{chitot}). This leads to expressions (\ref{currentformula})-(\ref{skewnessformula}) with ${\cal J}_{0}^{(n)}$ substituted by ${\cal J}_{0,tot}^{(n)}$ everywhere. Also, the spectrum of charge fluctuations
\beq
S^{(N)}_Q(\omega_1,\ldots,\omega_N):= \int_{-\infty}^\infty dt_1\ldots dt_N e^{-i\omega_1 t_1}\ldots e^{-i\omega_Nt_N} {\cal T}_S \langle Q(t_1) \ldots
Q(t_N) \rangle_c,\eeq
follows from 
\beq
S^{(N)}_Q[\bm{\omega}] = \frac{\delta(i\omega_1+\ldots+i\omega_N)}{(i\omega_1)\ldots(i\omega_N)} S^{(N)}_{accum}[\bm{\omega}],
\eeq
with $S^{(N)}_{accum}[\bm{\omega}]$ obtained using Eqs.~(\ref{currentformula})-(\ref{skewnessformula}) upon the change ${\cal J}_0 \to {\cal J}_{0,acumm}$. For example, $S^{(2)}_Q(\omega) = (1/\omega^2) S^{(2)}_{accum}(\omega)$. Notice that for a capacitive conductor, due to the relation between charge and voltage, this charge noise is proportional to the voltage noise.
Finally the `left' and `right' cumulants can be computed with ${\cal J}_0 \to {\cal J}_{0,L}$ and ${\cal J}_0 \to {\cal J}_{0,R}$ respectively in (\ref{currentformula})-(\ref{skewnessformula}).

\section{Results} \label{Resultssec}

To illustrate our method, we analyze the transport
statistics of the prototypical example of spinless electrons passing
through a SRL model. As explained in the previous chapter, the system consists of a two-terminal conductor with a discrete energy level in the central region, and is described
by the Hamiltonian (\ref{SRLhamiltonian}).
In the infinite-bias limit, this Hamiltonian leads to the kernel (\ref{kernelinfty}).
Using (\ref{currentformula}), (\ref{noiseformula}) and
(\ref{skewnessformula}), the simplicity of the model allows us to
derive analytic results in this limit; for example the current gives
$I_{stat}=\Gamma_L\Gamma_R/\Gamma$, where
$\Gamma:=\Gamma_L+\Gamma_R$, and the `total' noise
Fano-factor, $F^{(2)}:=S^{(2)}/I_{stat}$, found using (\ref{noiseformula}) 
is precisely the correct and known result (\ref{totalFanonoiseSRL}).

At finite bias voltages, the kernel in Eq. (\ref{kernelinfty}) is no longer valid. Among the various choices to calculate $\mathcal{W}(\chi)$ in this case, we use Schoeller's approach \cite{Schoeller09,Leijnse08,Emary09} (c.f. appendix \ref{appKernel}),
which allows us to calculate the kernel to the desired order (sequential tunneling in our case) without
further uncontrolled approximations (such as the secular approximation). It is important to mention that the frequency-dependent shot noise of the SRL model can be solved exactly \cite{Averin93}, and therefore one does not need to use the above approximations. However, to the best of our knowledge, a finite-frequency study for this model beyond the second-order current correlator is yet lacking. Here we use the exact solution to the noise spectrum as a benchmark of the Markovian approximation in order to identify the regions of validity of our theory. This benchmark is important because most of the papers in the literature discussing noise in the context of QMEs make use of the Markovian approximation.

Another important check for the theory is to reproduce the fluctuation-dissipation theorem (FDT) in the appropriate regimes. Near linear response, that is, for applied voltages $V$ much smaller than the temperature $T$, the low-frequency noise spectrum should follow the Jonhson-Nyquist relation (\ref{FDTclassical}). Still in the linear response regime, if the measured frequencies are larger than the temperature, the proper form is the Callen and Welton formula (\ref{FDTquantum}), that accounts for quantum fluctuations.
As discussed in chapter \ref{Chapter1}, out of equilibrium, Eq.~(\ref{FDTfull}) holds in general, and for some particular cases, such as tunnel junctions or quantum dots in the weak cotunneling regime,  it takes the form (\ref{FDTfreq}), which has the zero-frequency limit $S^{(2)}= I_{stat} \mathrm{coth}(\frac{eV}{2kT})$.

In this section we show the solution given by our theory for the SRL model, as well as how it recovers the FDT and the NEFDT in the appropriate limits. By contrasting these results with the exact solution, we will be able to show that the Markovian approximation does not contain quantum fluctuations, thereby needing a non-Markovian approach to capture the physics of quantum noise (c.f. chapter \ref{Chapter4}).
It is therefore interesting to see to what extend the four results coincide, and in what regimes the Markovian approach is valid and captures the correct physics. We will see that when $kT\gg eV,\; \hbar\omega$, the theory captures well both the exact solution and the FDT and NEFDT. Also, in transport configurations, with the level within the bias window, the Markovian approximation agrees well with the exact solution, reproducing in particular previous studies with $eV\gg kT$ \cite{Gurvitz96, AguadoBrandes04, Braun06, Choi01}. However, in a situation with the level outside the bias window, the Markovian approach presented here does not capture quantum noise physics, effect that we observe at high frequencies ($\hbar\omega\gtrsim kT, \; eV$). Although in this situation transport due to cotunneling processes becomes more relevant, the difference between our results and the exact solution is due to the Markovian assumption as it will be demonstrated in chapter \ref{Chapter4}.
We also study the finite-frequency skewness of the current distribution as given by equation (\ref{skewnessformula}). This shows to be insensitive to thermal fluctuations near equilibrium, therefore revealing the `shot' contribution in a situation in which thermal fluctuations dominate in the noise spectrum ($kT \gg eV, \; \hbar\omega$). 

\begin{figure}
 \begin{center}
\includegraphics[width= 0.85\textwidth]{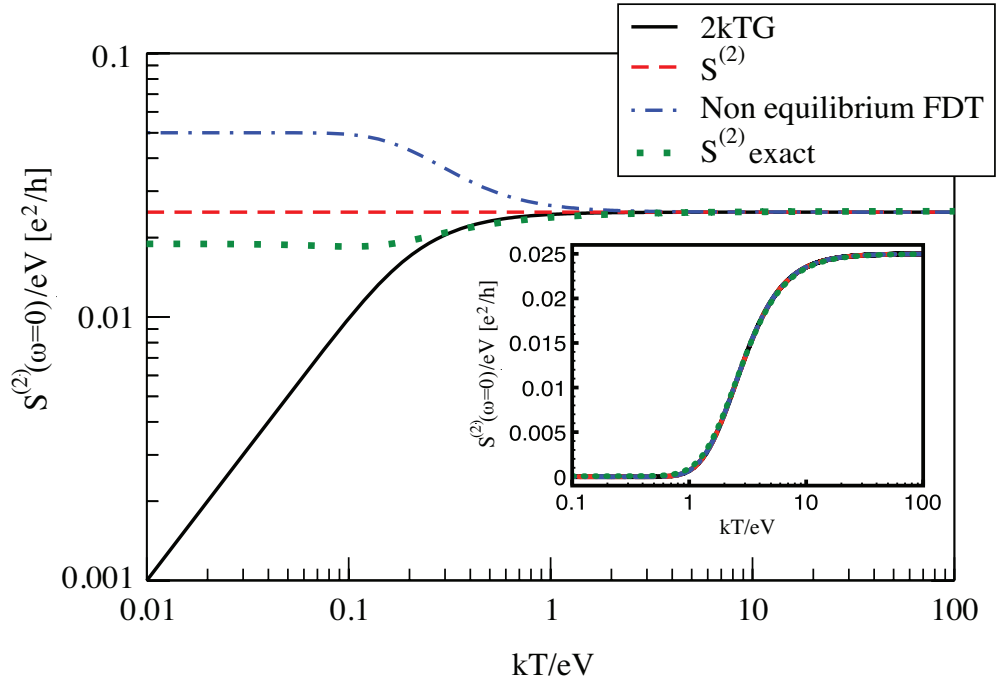} 
 \caption[Zero-frequency noise vs temperature in the single resonant level model]{Zero-frequency noise $S^{(2)}(\omega=0)$ at finite voltage for the single resonant level model as a function of temperature. For comparison we also show the FDT, the NEFDT, and the exact solution. 
  In the main figure the level is within the bias window $\varepsilon=0$. The inset  shows a regime with the level outside the bias window $\varepsilon/eV=5$.
  Rest of parameters: $\Gamma_L/eV=\Gamma_R/eV=0.1$.  In the main figure all the quantities coincide when $kT\geq eV$, as expected. 
  In the inset all fluctuations are thermal and therefore, all quantities coincide in the whole range of temperatures. The typical physical units are $T\sim 10-100\;\mathrm{mK}$, $V\sim 10-100\;\mathrm{\mu V}$, $\Gamma\sim 10-100\;\mathrm{MHz}$.
   \label{Fig1FCSpaper2}
}
 \end{center}
\end{figure}

\subsection{Zero-frequency counting statistics} \label{FCSresults}

\begin{figure}
 \begin{center}
  \includegraphics[width= 0.75\textwidth]{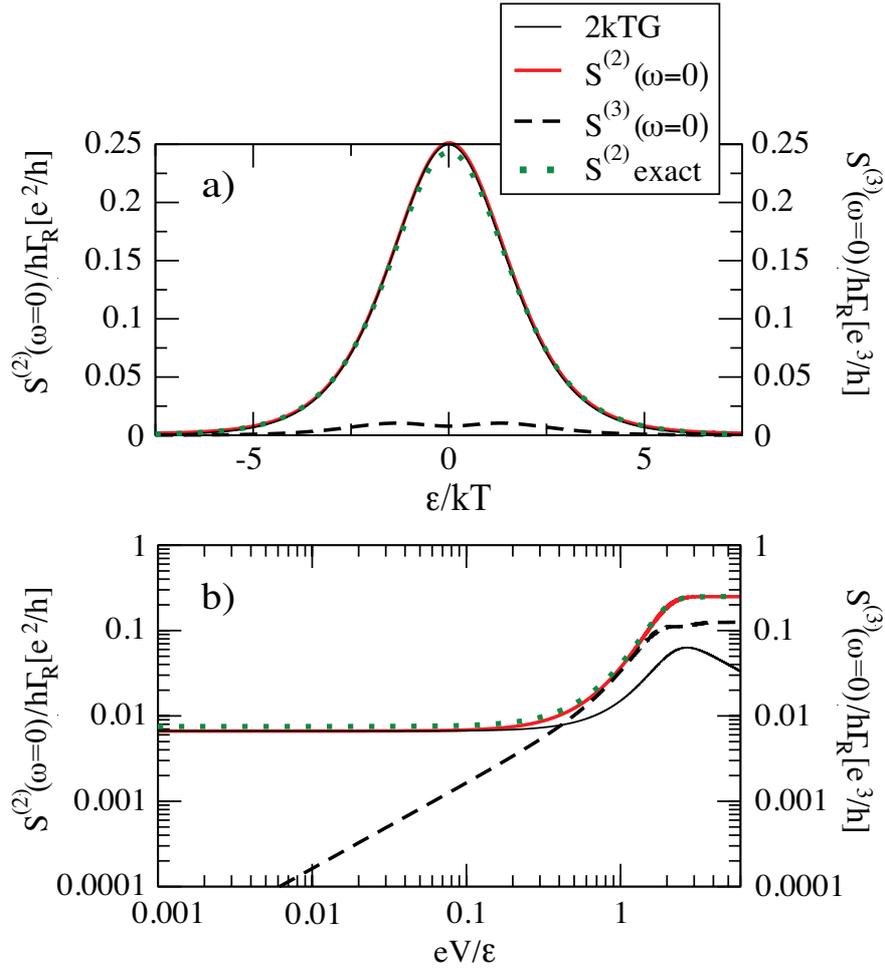} 
 \caption[Zero-frequency noise and skewness vs dot energy and bias voltage in the single resonant level model]{a) Noise and skewness near linear response, for $h\Gamma_L/kT=h\Gamma_R/kT=eV/kT=0.05$, as a function of $\varepsilon$. b) Noise and skewness as a function of voltage. We also show the equilibrium fluctuation-dissipation expression $2kTG$ and the exact solution for comparison.
   \label{Fig2FCSpaper2}
}
 \end{center}
\end{figure}

Let us start by showing the zero-frequency noise spectra corresponding to the SRL model. Although this limit has already been studied in detail, a full comparison between our theory and the exact solution will help us to understand the finite-frequency case. Particularly important is the linear response regime, at which studies of this model are scarce. As mentioned before, in this regime the noise should exhibit thermal fluctuations in order to fulfill a fluctuation-dissipation
relation, while the skewness, on the other hand, should go to zero as the voltage $V$ goes to zero \cite{Lev04}.
Fig.~\ref{Fig1FCSpaper2} shows how our calculation captures correctly the
FDT, $S^{(2)}=2kTG$, in the proximity of linear response, $kT
\gtrsim eV$. For comparison, we also plot the zero-frequency limit of the NEFDT in Eq. (\ref{FDTfreq}), namely
$S^{(2)}= I_{stat} \mathrm{coth}(\frac{eV}{2kT})$. 
In the opposite regime, $kT\lesssim eV$, the Markovian approximation is larger than the exact solution, discrepancy that can be understood as originated from the lack of cotunneling contributions in our calculation \cite{Emary09}. As expected, below $kT/eV\sim1$ the FDT is not fulfilled. We can also see that the NEFDT, exact for tunnel junctions, for a two-terminal device performs quite badly when $kT\lesssim eV$, but correctly in the opposite limit. This failure of the NEFDT at low temperatures disappears when the level is outside the bias window. This is a low-current regime, and thus a tunneling limit. The inset of Fig.~\ref{Fig1FCSpaper2} shows this situation, where all fluctuations are thermal and the four curves coincide in the whole range of temperatures. At finite frequencies we will expect a quantum noise step in the spectrum at frequencies $\hbar\omega\sim\varepsilon$, effect that will be studied in the next subsection.

\begin{figure}
 \begin{center}
  \includegraphics[width= 0.7\textwidth]{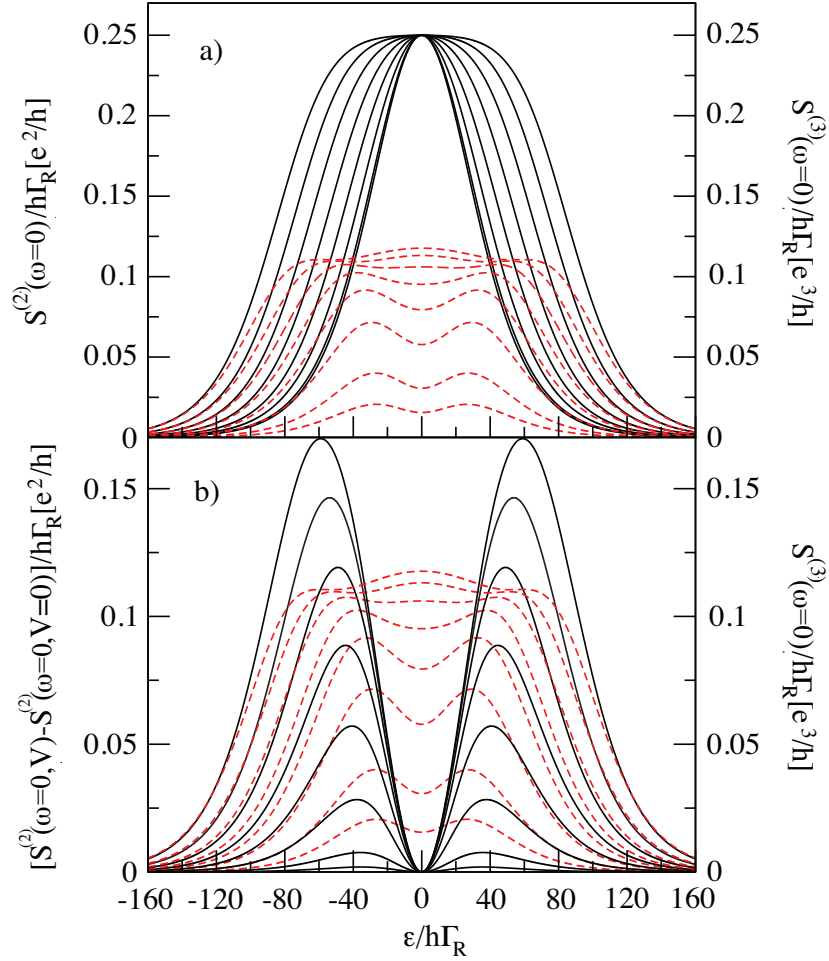} 
 \caption[Zero-frequency excess noise and skewness vs dot energy in the single resonant level model]{a) Increasing noise (black full lines going up) and skewness (red dashed lines going up) as a function of $\varepsilon$ for increasing voltages $eV/h\Gamma_R=1,10,20,40,60,80,100,120$, (parameters $kT/h\Gamma_R=20, \Gamma_L/\Gamma_R=1$). b) Excess noise  $S^{(2)}(V)-S^{(2)}(V=0)$ and skewness versus $\varepsilon$ for increasing voltages.
   \label{Fig3FCSpaper2}
}
 \end{center}
\end{figure}

The behaviour of the zero-frequency noise spectrum close to equilibrium with respect to $\varepsilon$ is shown in Fig.~\ref{Fig2FCSpaper2}a. Here we see how the FDT is fulfilled by our theory and the good agreement of this with the exact solution. We also plot the zero-frequency skewness, that although of small magnitude in the same scale, is nonzero in a situation where the noise spectrum is completely dominated by thermal fluctuations. This insensitivity of the skewness to temperature allows us to extract intrinsic correlation effects in near-equilibrium conditions.
In Fig.~\ref{Fig2FCSpaper2}b we show the same quantities as a function of voltage. Interestingly, the Markovian result coincides with the exact solution in the whole range of voltages. The FDT, however, starts to disagree with these for voltages $eV/\varepsilon\gtrsim 0.2$. As anticipated, the skewness vanishes as the voltage goes to zero.
In Fig.~\ref{Fig3FCSpaper2}a  we plot noise and skewness as a function of $\varepsilon$ for increasing voltages. As the bias increases, the skewness (dashed lines) shows peaks evolving into plateaus at values of $\varepsilon$ corresponding to the chemical potentials of the reservoirs. This effect, which is due to non-equilibrium fluctuations, is completely masked in the noise (solid lines) even at the highest voltages due to thermal fluctuations. This is clearly seen in Fig.~\ref{Fig3FCSpaper2}b, where we show the same comparison after substracting thermal fluctuations to the noise value (excess noise defined as $S^{(2)}(V)-S^{(2)}(V=0)$). Here it is clear that at low detuning, $\varepsilon\lesssim eV/2$ (position of the peaks in the figure), and at $kT\gtrsim eV$, the skewness can reveal the `shot' contribution, while this is masked by thermal fluctuations in the noise spectrum.

\subsection{Finite-frequency counting statistics} \label{FCSwresults}

 \begin{figure}
 \begin{center}
  \includegraphics[width= 0.85\textwidth]{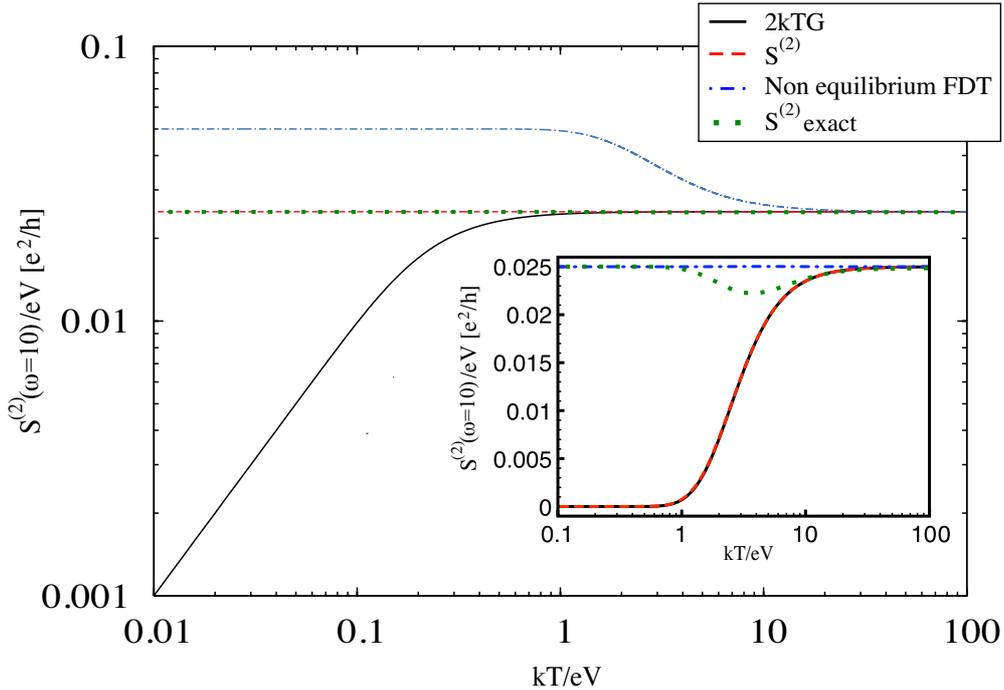} 
 \caption[Finite-frequency noise vs temperature in the single resonant level model]{Noise, FDT and NEFDT as a function of temperature for $\omega/eV=10$.
  In the main figure the level is within the bias window $\varepsilon=0$. The inset  shows a regime with the level outside the bias window $\varepsilon/eV=5$.
  Rest of parameters: $h\Gamma_L/eV=h\Gamma_R/eV=0.1$.  In the main figure all quantities coincide when $kT\geq eV+\hbar\omega$.
  In the inset all fluctuations are thermal and, therefore, the shot noise and the FDT coincide in the whole range of temperatures. When $kT\leq eV+\hbar\omega$, the NEFDT is above due to quantum fluctuations. The exact solution contains also corrections due to cotunneling processes, which are dominating in this regime.
   \label{Fig4FCSpaper2}
}
 \end{center}
\end{figure}

To study the case of finite frequencies, we use our formulae (\ref{currentformula})-(\ref{skewnessformula}) applied to the SRL model. In a situation with the level within the bias window, we find a similar behaviour to Fig.~\ref{Fig1FCSpaper2}. However, now the NEFDT -- equation (\ref{FDTfreq}) -- is fulfilled for temperatures $kT\gtrsim eV+\hbar\omega$. This is shown in Fig.~\ref{Fig4FCSpaper2}. Remarkably, at finite frequencies the Markovian approximation is basically exact in this direct transport regime.
In the high-bias regime, $eV\gg \hbar\omega, \; kT$, we also find that the Markovian approximation agrees perfectly with the exact result (not shown), in accordance with previous studies \cite{Gurvitz96, AguadoBrandes04, Braun06, Choi01}. When the level lies outside the voltage window, the situation changes drastically (see inset of Fig.~\ref{Fig4FCSpaper2}). Here the Markovian approximation is no longer appropriate when $kT\lesssim eV+\hbar\omega$. Both the exact result and the NEFDT contain quantum fluctuations, while the Markovian calculation only captures the thermal contribution (and therefore fulfills the FDT). The exact solution presents a small structure at temperatures of the order of $\varepsilon$. This cannot be resolved with the NEFDT. As expected, when $kT\gtrsim eV+\hbar\omega$, all curves coincide.

 \begin{figure}
 \begin{center}
  \includegraphics[width= 0.95\textwidth]{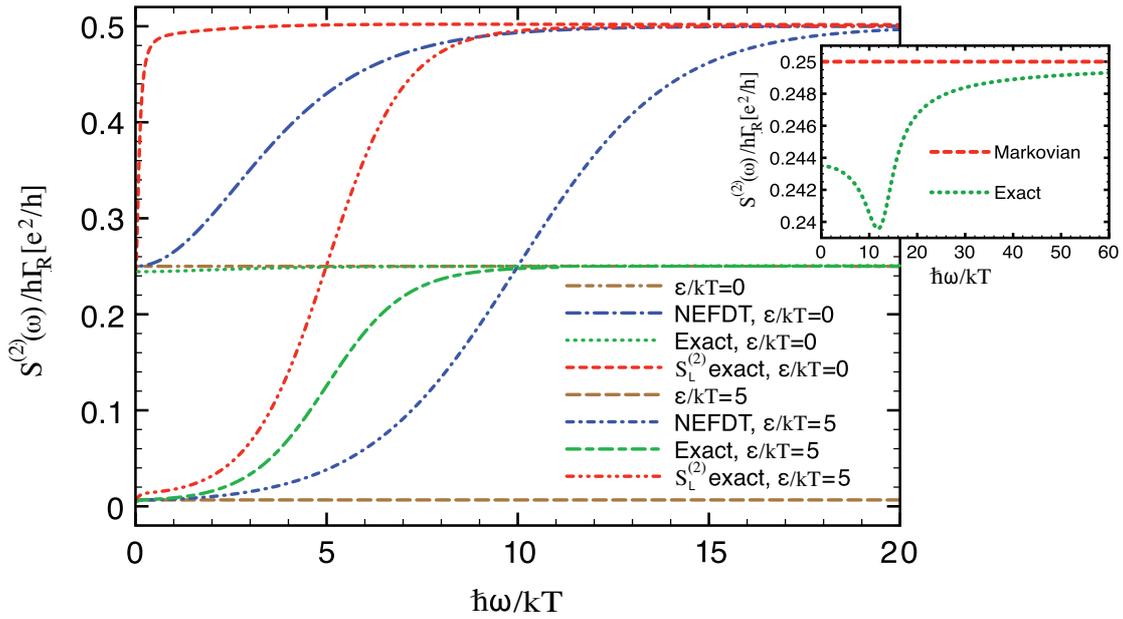} 
 \caption[Finite-frequency noise vs frequency near linear response in the single resonant level model]{$S^{(2)}$ near linear response ($eV/kT=0.0005$) as a function of frequency. For comparison we also show the NEFDT (Eq.~(\ref{FDTfreq})) and the exact solution. $S^{(2)}(\omega)$ is flat for the whole range of frequencies, and coincides with the equilibrium FDT as expected. The NEFDT however disagrees with these two, showing also quantum fluctuations which are absent in the Markovian noise spectrum. The quantum noise steps shown by the NEFDT are however at $\hbar\omega=2\varepsilon$, in contrast to the exact solution, which shows steps at $\hbar\omega=\varepsilon$. This is due to the fact that the NEFDT works well for tunnel junctions, but does not capture partition noise. This becomes clear also from the saturation value at large frequencies, as described in the text. Rest of parameters: $h\Gamma_L/kT=h\Gamma_R/kT=0.05$. The inset compares the exact solution with the Markovian approximation for a different regime, namely $eV/kT = 25$. We see that while the Markovian limit is flat for all frequencies, the exact solution presents a dip at $\hbar\omega=\pm|\varepsilon \pm eV/2|$. Rest of parameters: $\varepsilon=0, h\Gamma_L/kT=h\Gamma_R/kT=0.25$.
\label{Fig5FCSpaper2}
}
 \end{center}
\end{figure}

 \begin{figure}
 \begin{center}
  \includegraphics[width= 0.9\textwidth]{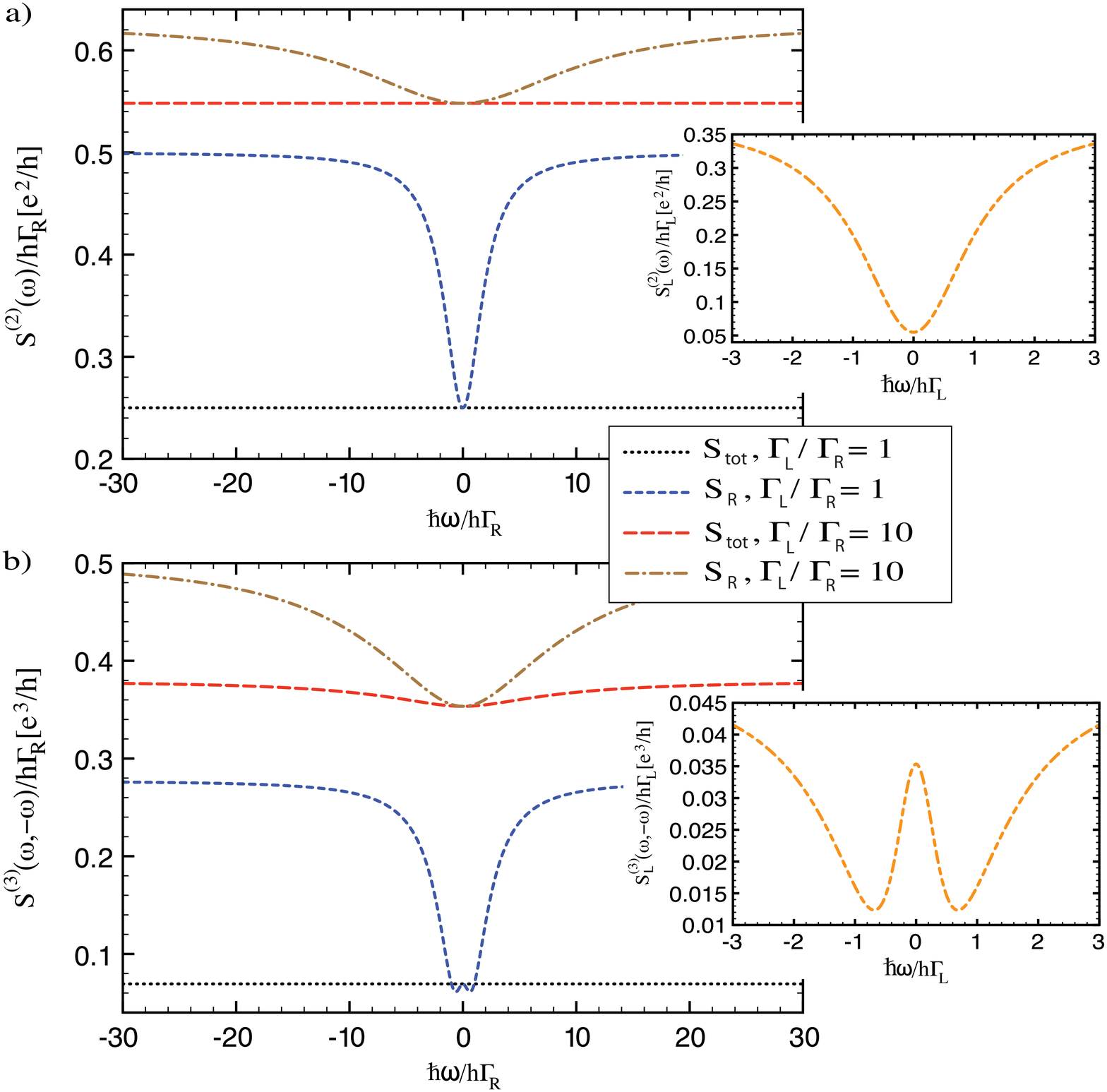} 
 \caption[Finite-frequency noise and skewness vs frequency in the single resonant level model]{a) $S^{(2)}(\omega)$ and b) $S^{(3)}(\omega,-\omega)$ for $eV/h\Gamma_R=50$, $kT/h\Gamma_R=20$, and $\varepsilon=0$. The spectra for particle currents and for total currents significantly deviate from each other, even for large asymmetry. The insets correspond to noise and skewness through the left barrier for $\Gamma_L/\Gamma_R=10$.
   \label{Fig6FCSpaper2}
}
 \end{center}
\end{figure}

The general trends explained so far become more evident in Fig.~\ref{Fig5FCSpaper2}, where we plot noise and the fluctuation-dissipation theorem near linear response as a function of frequency. Here the Markovian noise is always flat and equals $S^{(2)}=2kTG$, whereas the NEFDT and the exact solution lie above and show quantum noise steps. 
Let us start by considering the case $\varepsilon=0$. In the whole range of frequencies, the Markovian approximation is basically exact in this situation of direct transport. The NEFDT shows the correct zero-frequency limit, since in this case the fluctuations are purely thermal. At high frequencies, however, the NEFDT converges to the Poisson value of a single barrier with tunneling rate $\bar{\Gamma}/2$, being $\bar{\Gamma}:=(\Gamma_L+\Gamma_R)/2$, (c.f. $S_L^{(2)}$ in the plot). This is in agreement with the validity of equation (\ref{FDTfreq}) for a tunnel junction, and in contrast with the exact solution, which contains partitioning, and therefore its $\omega\to\infty$ limit is $\bar{\Gamma}/4$. In the case in which the energy lies outside the bias window ($\varepsilon/kT=5$ in the figure), transport is possible because of the finite temperature as well as due to quantum fluctuations. The Markovian noise only contains the former and is flat with frequency, while the exact result contains both and shows a quantum noise step centered at $\hbar\omega/kT=\varepsilon/kT=5$. Although in this regime cotunneling contributions are important, the difference lies in the Markovian approximation, as it will be seen in the next chapter. The NEFDT shows in this situation a quantum noise step centered at $\hbar\omega=2\varepsilon$. This discrepancy with respect to the exact solution can be understood in terms of the tunneling approximation leading to Eq. (\ref{FDTfreq}), which presents a step located at an effective chemical potential $2\varepsilon$. Again, the high frequency limit coincides with that of $S_L^{(2)}$.
The intermediate regime where $\hbar\omega \sim eV$ is studied in the inset. Here, we set $eV/kT=25$ and observe a flat behaviour for the Markovian solution, whereas the exact solution presents a dip at $\hbar\omega = \pm | \varepsilon \pm eV/2|$ (coinciding with the position of the chemical potentials with respect to the energy level). This clearly illustrates how the Markovian approximation captures well the physics in the linear response regime and a direct-transport configuration, but when the frequency is comparable to the applied bias, it fails to capture quantum noise.

We now proceed to discuss the finite-frequency noise and skewness spectra of the total- and particle- current distribution. 
In the previous discussion, the Markovian noise was always flat as a function of frequency, a fact that is well known for symmetric systems ($\Gamma_L=\Gamma_R$) -- see for instance \cite{BlanterButtiker00}. In the case $\Gamma_L\neq\Gamma_R$, a proper partitioning of displacement currents (see discussion in subsection \ref{FCStotaccum}) becomes essential as we will show next, and the way this is made affects significantly the spectrum. 
Fig.~\ref{Fig6FCSpaper2}a shows $S^{(2)}(\omega)$ in a transport configuration, $eV/h\Gamma_R=50$ and $\varepsilon=0$. As in the results shown previously, the total noise spectrum is flat for a symmetric configuration. Interestingly, this flat behaviour persists even when the system is made asymmetric ($\Gamma_L\neq\Gamma_R$). This is due to the current-partitioning model assumed here: $I_{tot} = \alpha I_L + \beta I_R$ with $\alpha = \Gamma_R/\Gamma$ and $\beta = \Gamma_L/\Gamma$. The noise spectrum corresponding to particle currents displays information about the rates; in contrast to the total noise, it shows a dip with half-width $2\Gamma$. In Fig.~\ref{Fig6FCSpaper2}b we show the skewness along the representative direction $\omega'=-\omega$. Interestingly, the skewness corresponding to the total current starts to develop a dip that shows the asymmetry of the system. The particle-current skewness presents a similar behaviour to the noise counterpart. However, for $(3-\sqrt{5})/2\leqslant \Gamma_L/\Gamma_R\leqslant (3+\sqrt{5})/2$, it develops a minimum whose position depends on the value of the rates. In the asymmetric case, $\Gamma_L\neq\Gamma_R$, the particle-current noise and skewness can even present different lineshapes. This can be seen contrasting the insets of Figs. \ref{Fig6FCSpaper2}a and \ref{Fig6FCSpaper2}b. In the linear response regime the curves for the noise look similar (not shown). The skewness, on the contrary, goes to zero in magnitude and shows a structure that depends on temperature, and that changes from a dip to a peak as $\varepsilon$ is increased from zero to a finite value. In summary, we see that the spectra for total and particle currents differ significantly from each other even for large asymmetry. This means that the assumption of calculating noise spectra using particle currents only, used commonly in the literature, is flawed. Here we have assumed the current partitioning given by $\alpha=\Gamma_R/\Gamma$, $\beta=\Gamma_L/\Gamma$. If the more simplistic partitioning $\alpha=\beta=1/2$ is assumed, the results for the total cumulants in the asymmetric case change drastically (not shown). In particular, the noise is then no longer flat but has a dip structure, and the skewness shows a peak around zero frequency.

\section{Conclusions} \label{ConclusionsSec}
In this chapter, we have developed a theory of frequency-dependent counting statistics of electron transport through nanostructures, within the framework of Markovian quantum master equations. We have illustrated our method with calculations of noise and skewness in a single resonant level model at finite bias voltages and frequencies. By comparing with both the exact solution and the finite-frequency version of the non-equilibrium fluctuation-dissipation theorem, Eq.~(\ref{FDTfreq}), we have identified the regimes of validity of our Markovian theory at finite frequencies. In particular, we have shown that the Markovian limit is basically exact in transport configurations (level within the bias voltage window), as long as $\hbar\omega \gg eV$ or $\hbar\omega\ll eV$. In intermediate situations, where $\hbar\omega \sim eV$, or with the level outside the bias window, the Markovian limit fails  at finite frequencies due to the lack of quantum fluctuations. 

We have also discussed how the noise spectra for particle currents and for total currents significantly deviate from each other, even for large asymmetries $\Gamma_R/\Gamma_L \neq 1$. This demonstrates that calculating spectra using particle currents only leads to incorrect results in general.
Our method allows for the calculation of finite-frequency current cumulants of arbitrary order, as we have explicitly shown for the second and third order cumulants, Eqs. (\ref{noiseformula}) and (\ref{skewnessformula}). These formulae generalize previous zero-frequency expressions and can be viewed as an extension of MacDonald's formula beyond shot noise. Recently, this has been extended to study the time-averaged shot noise spectrum in the presence of periodic ac fields \cite{Wu-Timm10}. Interesting extensions of our study along these lines would allow us to study frequency-dependent high-order cumulants in nanostructures driven by time-dependent fields, or, even more challenging, in systems showing nontrivial non-linear dynamics such as self-sustained oscillations without external time-dependent driving \cite{Kurth10}.


\chapter{Non-Markovian noise correlations of interacting electrons\footnote{The results presented in this chapter have been published in \cite{Marcos10NM}.}} 
\label{Chapter4}
\lhead{Chapter 4. \emph{Non-Markovian noise correlations of interacting electrons}} 

\begin{flushright}



\textit{``Everything should be made as simple as possible, but not simpler.''}

A. Einstein

\end{flushright}

\begin{small}

We present a theory of finite-frequency noise in non-equilibrium conductors.
It is shown that Non-Markovian correlations are essential to describe the physics of quantum noise. In particular, we show the importance of a correct treatment of the initial system-bath correlations, and how these can be calculated using the formalism of quantum master equations. Our method is particularly important in interacting systems, and when the measured frequencies are larger than the temperature and the applied voltage. In this regime, quantum-noise steps are expected in the power spectrum due to vacuum fluctuations. This is illustrated in the current noise spectrum of a double quantum dot --charge qubit-- attached to electronic reservoirs. Furthermore, the method allows for the calculation of the single-time counting statistics in quantum dots, measured in recent experiments.

\end{small}

\newpage

\section{Introduction}

Vacuum fluctuations are one of the most intriguing consequences of the quantum theory. In electronic systems, they manifest as electron-hole creation/annihilation processes in a time given by the Heisenberg uncertainty relation, $t\sim1/\omega$, being $\omega$ the measuring frequency. 
In order for these processes to be seen, other types of fluctuations must be overcome. For example, a system in thermodynamic equilibrium must be at a temperature $T$ much smaller than this frequency, and in a system driven out of equilibrium, such as a mesoscopic conductor subject to an applied voltage $V$, the quantum-noise regime (QNR) reads $\hbar\omega\gg k_B T, eV$.
Zero-point fluctuations in quantum-transport systems were first measured by Schoelkopf and collaborators \cite{Schoelkopf97} through the current-noise spectrum 
(\ref{noisedef}), which reveals valuable information beyond that contained in the dc current \cite{BlanterButtiker00, Galaktionov03, Nagaev04, Pilgram04, Salo06, AguadoBrandes04}. Among the various methods presented in chapter \ref{Chapter2} to calculate $S^{(2)}(\omega)$, quantum master equations (QMEs) are particularly attractive because of their simplicity and generality for treating dissipative dynamics of interacting systems \cite{Choi01, Gurvitz03, Ruskov03, AguadoBrandes04, Emary07}. Typically, the Markovian approximation (MA) in the system-reservoir coupling is employed. However, as it was shown in the previous chapter, this fails in describing the noise spectrum in the QNR. Although there have been a few attempts to go beyond the MA in the context of QMEs \cite{Engel-Loss04}, a complete noise theory is yet lacking.

In this chapter we present such a theory. Our method allows the calculation of the current and voltage noise spectrum of a system described by a generic non-Markovian QME, and can be applied to the increasing number of experiments exploring the QNR \cite{Deblock03, Zakka-Bajjani07, Gabelli08, Xue09}. The theory naturally contains the physics of vacuum fluctuations, for which a proper inclusion of initial system-bath correlations is essential. Furthermore, the method enables to determine the charge-noise spectrum 
\beq \label{chargenoisespectrum}
S^{(2)}_Q(\omega)= \int_{-\infty }^{\infty }d\tau
e^{-i\omega\tau}\langle\{Q(\tau),Q(0)\}\rangle_c,
\eeq 
as it is shown for a single resonant level (SRL) model. This noise dictates the back-action when the conductor is used as a detector of another quantum system \cite{Clerk10}. The technique is used to study the full noise spectrum of a double quantum dot charge qubit in the hitherto unexplored QNR. As we will see, in this regime transport fluctuations are mediated by the zero-point dynamics, showing a series of steps at frequencies corresponding to resonant processes in the system.

\section{Theory}

\begin{figure}[t]
\center
\includegraphics[width=0.8\textwidth]{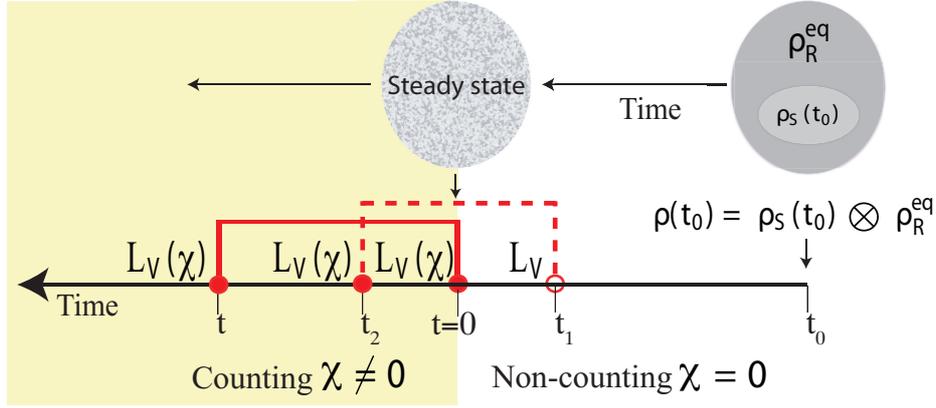}
\caption[Non-Markovian time evolution]{\label{Fig1paperNM} 
Schematics of counting: The density operator evolves from the initial separable state at time $t_0$ (represented by two distinct ellipses) until it reaches a steady state at time $t=0$, where it is no longer in a product state (single ellipse). At time $t=0$ counting begins. The shading highlights the time interval where counting is effective. Full circles denote tunnel vertices with counting factors $\chi\neq 0$, empty circles denote standard tunneling vertices ($\chi=0$). Contractions between tunneling events in counting and non-counting intervals (dashed over-line) give rise to $\Gamma(\chi,z)$, while contractions within the counting interval (solid over-line) give rise to the self-energy $\Sigma(\chi,z)$.}
\end{figure}

Here we consider phenomena that can be described by the general QME\footnote{In this chapter we obviate the hat in the notation for operators.} $\dot{\rho}(t)={\cal L}\rho(t)$, where ${\cal L}$ is the Liouvillian, that governs the evolution of the density operator (DO), $\rho$, describing the dynamics of the total system. More specifically, we focus on the case in which a central system exchanges particles with a bath, and this exchange is amenable to the counting of particles. We will take here the case of transport through a central quantum coherent system, attached to fermionic contacts. The Hamiltonian of the system is of the form (\ref{Hform}), where each of the terms can be written as in Eqs.~(\ref{Hs})-(\ref{Hv}). The left/right reservoirs will be taken to be at equilibrium, with chemical potentials $\mu_{L/R}=E_F\pm\frac{eV}{2}$. Under this Hamiltonian, the DO evolves according to equation (\ref{LiouvilleEq}), with ${\cal L}\bullet \equiv -i
\left[{\cal H}_\mathrm{S}+{\cal H}_\mathrm{R}+{\cal H}_\mathrm{V},\bullet\right]\equiv({\cal
L}_\mathrm{S} + {\cal L}_\mathrm{R} + {\cal L}_\mathrm{V})\bullet$. 
We are interested in the central-system dynamics, for which we consider the reduced system DO $\rho_\mathrm{S}(t)\equiv\mathrm{Tr}_\mathrm{R}\{\rho(t)\}$. 
If we choose $t_0$ to be the time at which system and reservoirs are in a separable state,
$\rho(t_0) =
\rho_\mathrm{S}(t_0) \otimes \rho_\mathrm{R}^\mathrm{eq}$, with
$\rho_\mathrm{S}(t_0)$ arbitrary and $
\rho_\mathrm{R}^\mathrm{eq}$ the equilibrium bath state, the evolution of $\rho_\mathrm{S}(t)$ in the Laplace space is given by
\beq \label{rhoz}
  {\rho}_\mathrm{S}(z)
  =\mathrm{Tr}_\mathrm{R}
  \left\{
    \left[z-{\cal L}\right]^{-1}
    \rho_\mathrm{S}(t_0) \otimes \rho_\mathrm{R}^\mathrm{eq}
  \right\}
  =\Omega_0(z){\rho}_\mathrm{S}(t_0).
\eeq 
Here, we find the propagator $\Omega_0(z) \equiv \left[z-{\cal W}(z)\right]^{-1}$, with kernel ${\cal W}(z) = {\cal L}_\mathrm{S} +
\Sigma(z)$, being $\Sigma(z)$ the non-Markovian (NM) self-energy, and whose form can be derived using the expansion
\beq
\frac{1}{z-{\cal L}}= \frac{1}{z-{\cal L}_\mathrm{S} -{\cal L}_\mathrm{R}}\sum_{k=0}^{\infty} \left( {\cal L}_\mathrm{V}\frac{1}{z-{\cal L}_\mathrm{S} -
{\cal L}_\mathrm{R}}\right)^k.
\eeq
This gives 
\beq
\Sigma(z)=\mathrm{Tr}_\mathrm{R}
  \left\{ {\cal L}_\mathrm{V}\frac{1}{z-{\cal L}_\mathrm{S} -
{\cal L}_\mathrm{R}}{\cal L}_\mathrm{V}\rho_\mathrm{R}^\mathrm{eq}\right\} +
  \ldots
\eeq 
Technical details on how to evaluate this expression \cite{Schoeller09, Emary09b} are not relevant for the main discussions and can be found in appendix \ref{appKernel}.

\subsection{Cumulant generating function}

Our goal here is, given Eq.~(\ref{rhoz}), to derive a formula for the cumulant generating function (CGF) in terms of known quantities such as the self-energy. This will allow us to calculate NM current correlations up to arbitrary order at zero frequency. Furthermore, we aim to give an expression for the NM finite-frequency noise correlation function. If the transfer of electrons between system and reservoirs is amenable to counting, the full counting statistics of the number of transferred electrons $n$ can be studied with the DO formalism. To do this, we unravel $\rho_\mathrm{S}(t)$ in terms of this continuous projective measurement: $\rho_\mathrm{S}(t)=\sum_n \rho_\mathrm{S}^{(n)}(t)$, similarly to how it was done in the previous chapters. The probability distribution of having $n$ transfers after time $t$ is given by $P(n,t)=\mathrm{Tr}_\mathrm{S}\{\rho_\mathrm{S}^{(n)}(t)\}$, and the corresponding CGF is ${\cal F}(\chi;t)\equiv\ln\sum_{n=-\infty}^{\infty}P(n,t) e^{in\chi}$. Let us try to relate this CGF (or alternatively the moment generating function ${\cal G}\equiv e^{\cal F}$) with a general NM evolution. As explained in chapter \ref{Chapter3}, counting in lead $\alpha$ can be effected by adding $\chi_{\alpha}$ to the tunneling Liouvillian ${\cal L}_\mathrm{V}$ through the replacement $V_{k \alpha m}\to V_{k \alpha m} e^{i p \chi_{\alpha}/2}$,
where $p=+/-$ is the Keldysh index corresponding to the forward/backward time branch. Derivatives with respect to different counting fields, e.g. $\chi_L$, $\chi_\mathrm{R}$, allow us to obtain also cross correlations of currents flowing through different contacts. In the following, the lead-dependence of the counting field will be considered implicit. In the $\chi$-space, the density operator
$ \rho_\mathrm{S}(\chi,z) \equiv\sum_{n=-\infty}^{\infty} \int_0^{\infty} dt \rho_\mathrm{S}^{(n)}(t)e^{i\chi n-zt}$ follows the evolution $\rho_\mathrm{S}(\chi,z)
  =\Omega(\chi,z)\rho_\mathrm{S}(0)
$, with $\Omega(\chi,z)\equiv [z-{\cal W}(\chi,z)]^{-1}$, and ${\cal W}(\chi,z)= {\cal L}_\mathrm{S} +
\Sigma(\chi,z)$. To lowest order we have
\beq \label{1pointSE}
\Sigma(\chi,z)=\mathrm{Tr}_\mathrm{R}
  \left\{ {\cal L}_\mathrm{V}(\chi)\frac{1}{z-{\cal L}_\mathrm{S} -
{\cal L}_\mathrm{R}}{\cal L}_\mathrm{V}(\chi)\rho_\mathrm{R}^\mathrm{eq}\right\}.
\eeq
For later use, we also introduce the two-point self-energy 
\beq \label{2pointSE}
\Pi(\chi_2,\chi_1,z)=\mathrm{Tr}_\mathrm{R}
  \left\{ {\cal L}_\mathrm{V}(\chi_2)\frac{1}{z-{\cal L}_\mathrm{S} -
{\cal L}_\mathrm{R}}{\cal L}_\mathrm{V}(\chi_1)\rho_\mathrm{R}^\mathrm{eq}\right\}.
\eeq
Obviously, we have $\Pi(\chi,\chi,z)=\Sigma(\chi,z)$, and $\Sigma(\chi=0,z)= \Sigma(z)$. Explicit expressions for Eqs. (\ref{1pointSE}) and (\ref{2pointSE}) are given in appendix \ref{appKernel}.

In the widely used Born-Markov approximation, the state at which counting begins (say $t=0$) can be taken to be $\rho_\mathrm{S}(0)\otimes \rho_\mathrm{R}^\mathrm{eq}$. However, to consider NM corrections, the state at time $t=0$ can no longer be considered as a separable state, as it contains initial system-bath correlations.
To account for these, we explicitly divide the time evolution into two intervals (see Fig.~\ref{Fig1paperNM}). The evolution from $t_0$ (time at which system and reservoirs are separable) to $t=0$ is given by $\frac{1}{z_0-{\cal L}}$, while the evolution from $t=0$ to $t$ is given by $\frac{1}{z-{\cal L}(\chi)}$. Doing this we obtain the moment generating function (MGF):
\beq
  \label{inhomo1} {\cal G}(\chi;z)
  =
  z_0\mathrm{Tr} \left\{
    \frac{1}{z-{\cal L}(\chi)}\frac{1}{z_0-{\cal L}}
    \rho_\mathrm{S}(t_0) \otimes \rho_\mathrm{R}^\mathrm{eq}
  \right\}.
\eeq
Here $z$ is the conjugate frequency to $t$, and $z_0$ to $-t_0$. We will take $t_0\to -\infty$, which implies $z_0\to 0^-$ (henceforth implicit). The trace in (\ref{inhomo1}) refers to the full trace (system plus bath degrees of freedom).
Using geometric expansions of $\frac{1}{z-{\cal L}(\chi)}$ and $\frac{1}{z_0-{\cal L}}$, and performing the trace over the reservoirs, we get 
\beq\label{inhomo2}
  {\cal G}(\chi;z)
  =
   \Big\langle
\frac{1}{z-{\cal L}_\mathrm{S}-\Sigma(\chi,z)}\left(\mathds{1}+ \Gamma(\chi,z)\right)\Big\rangle.
\eeq
In this equation, $\ew{\ldots} \equiv \mathrm{Tr}_{\mathrm{S}}\left\{\ldots \rho_\mathrm{S}^\mathrm{stat}\right\}$, where we have taken $\rho_{\mathrm{S}}(0)=\rho_{\mathrm{S}}^{\mathrm{stat}}$, as we are interested in fluctuations around the stationary state. This can be obtained either as $\rho_\mathrm{S}^\mathrm{stat}=\lim_{z\to 0} z \rho_\mathrm{S}(z)$ in equation (\ref{rhoz}), or solving ${\cal W}(0)\rho_\mathrm{S}^\mathrm{stat}=0$. The inhomogeneous term $\Gamma(\chi,z)$ in Eq.~(\ref{inhomo2}) is given by
\beq
  \Gamma(\chi;z) =\frac{1}{z}\{\Pi(\chi,0,z_0)-\Pi(\chi,0,z)\}
  +\ldots
  \label{Gamma}
  \eeq
With Eqs. (\ref{inhomo2}) and (\ref{Gamma}) we have reached our first goal of obtaining a NM MGF. As we shall show below, the inclusion of $\Gamma(\chi,z)$ in the MGF is crucial to account for NM physics and quantum noise. Importantly, $\Gamma(\chi,z)$ cannot, in general, be cast in the form of a self-energy, since only \emph{one} of the two vertices (i.e. tunneling Liouvillians) contains a counting field $\chi$. 
Notice that Eq.~(\ref{Gamma}) extends the particular form of the inhomogeneity $\Gamma(\chi; z)= \frac{1}{z}\{\Sigma(0,0)-\Sigma(0,z)\}$, which appears in \cite{Flindt08}. This is only valid 
for a system with NM dynamics but with Markovian coupling with the bath in which counting is performed, and as a result, quantum fluctuations due to the Fermi contacts are not captured in this case. Additional details on how to obtain Eqs. (\ref{inhomo2}) and (\ref{Gamma}) with Liouvillian perturbation theory will be presented in Clive Emary's Habilitation thesis.

\subsection{Noise spectrum}

From the MGF (\ref{inhomo2}), together with (\ref{Gamma}), we can derive a general equation for the noise spectrum. To this end we use Eq.~(\ref{MacDonaldFlindtEq}), and obtain
\beq
 i^2  S^{(2)}(\omega) = \left[
   \ew{
     {\cal J}^{II}(i\omega,i\omega_0)
   }
   +
   \ew{
     {\cal J}^{I}(i\omega,i\omega) \Omega_0(i\omega){\cal J}^{I}(i\omega,i\omega_0)
   } \right]
   +(\omega \leftrightarrow -\omega)
  \label{Sz}
  ,
\eeq 
with $\omega_0\to 0$ and
\beq
{\cal J}^{II}(z,z_0) &\equiv&
    \frac{\partial^2}{\partial \chi_2 \partial \chi_1}
    \Pi(\chi_2,\chi_1,z)
  \Big|_{\chi_2,\chi_1\to 0}
  +
  \ldots
  \label{J2nonMarkovian}
\\
  {\cal J}^{I}(z,z') &\equiv&  
  \frac{\partial}{\partial \chi}
  \left[
    \Pi(0,\chi,z)+\Pi(\chi,0,z')
  \right]\Big|_{\chi\to 0}
  +
  \ldots
    \label{J1nonMarkovian}
\eeq
Eq. (\ref{Sz}) is the desired result for the NM noise spectrum. 
It is exact and agrees with previous approaches in the literature in the appropriate limits \cite{Engel-Loss04, Braun06}. In particular, the Markovian result (\ref{noiseformula}) is recovered by neglecting the frequency dependence of the jump super-operators: $ {\cal J}^{II}(z,z_0)\rightarrow {\cal J}^{II}(0,0)\equiv {\cal J}_0^{(2)}$, $ {\cal J}^{I}(z,z')\rightarrow {\cal J}^{I}(0,0)\equiv{\cal J}_0^{(1)}$. 
The correct NM zero-frequency limit \cite{Flindt08} is also recovered. It is interesting to notice that Eq.~(\ref{inhomo2}) not only allows us to obtain the NM noise spectrum, but also single-time NM correlations to arbitrary order, $\ew{I^N(t)}_c$, $\ew{n^N(t)}_c$, $\ew{Q^N(t)}_c$, by simply taking derivatives with respect to the counting field.

Notice that the above derivation has focused on particle currents flowing through the barriers separating central system and leads. As explained in detail in subsection \ref{FCStotaccum}, at finite frequencies this particle current is not conserved due to charge accumulations in the system, and the total current (particle plus displacement) needs to be considered to obtain the noise spectrum.
However, our results are general, and current conservation can be considered by the inclusion of the proper counting fields in $\Sigma(\chi,z)$ and $\Pi(\chi_2,\chi_1,z)$. Thus, particle, total, and charge noise (equivalently voltage noise for a capacitive system), can be calculated from Eq.~(\ref{Sz}). To this end, it is enough to consider respectively $\chi_L/\chi_R$, or $\chi_{\mathrm{tot}}$, $\chi_{\mathrm{accum}}$, given by equations (\ref{chitot}), (\ref{chiaccum}), which in the NM case produce jump super-operators similar to (\ref{Jtot}), (\ref{Jaccum}). 

\section{Results}

\subsection{Single resonant level model}

We now use the formalism presented in the previous section to calculate the NM noise spectrum of a single resonant level model (equivalently of a single electron transistor with $E_C\gg k_BT$, being $E_C$ the charging energy, and with only two relevant charge states). This noise spectrum is exactly solvable and has been calculated with a variety of techniques \cite{Averin93, BlanterButtiker00, EngelThesis, Johansson02}. In the following we show the good agreement between our theory and the exact solution. In the QNR, these two, in contrast to the Markovian result, show quantum-noise steps due to vacuum fluctuations, as we will see. The Markovian and non-Markovian results we present correspond to first order in perturbation theory (sequential tunneling) and in the following $S^{(2)}(\omega)$ refers to the `total' noise.

The SRL model, introduced in subsection \ref{SRLandothers}, is an example of interest for us not only because of the possibility of solving the noise spectrum exactly, but also because of the great deal of interesting physics that it contains, despite its simplicity. It captures the physics of a quantum dot in which only one single level participates in transport (strong Coulomb Blockade regime), and it can be shown that there is an exact mapping between the SRL model and the spin-boson model (namely a quantum two-level system coupled with strength $\nu$ to an ohmic dissipative bosonic bath) at  $\nu=1/2$. This mapping is actually a special case of the more general relation between the spin-boson model and the anisotropic Kondo model, for which $\nu=1/2$ is the exactly solvable point, the so-called Toulouse limit of the Kondo problem \cite{Cedraschi01}.

\begin{figure}[t]
\center
\includegraphics[width=\textwidth]{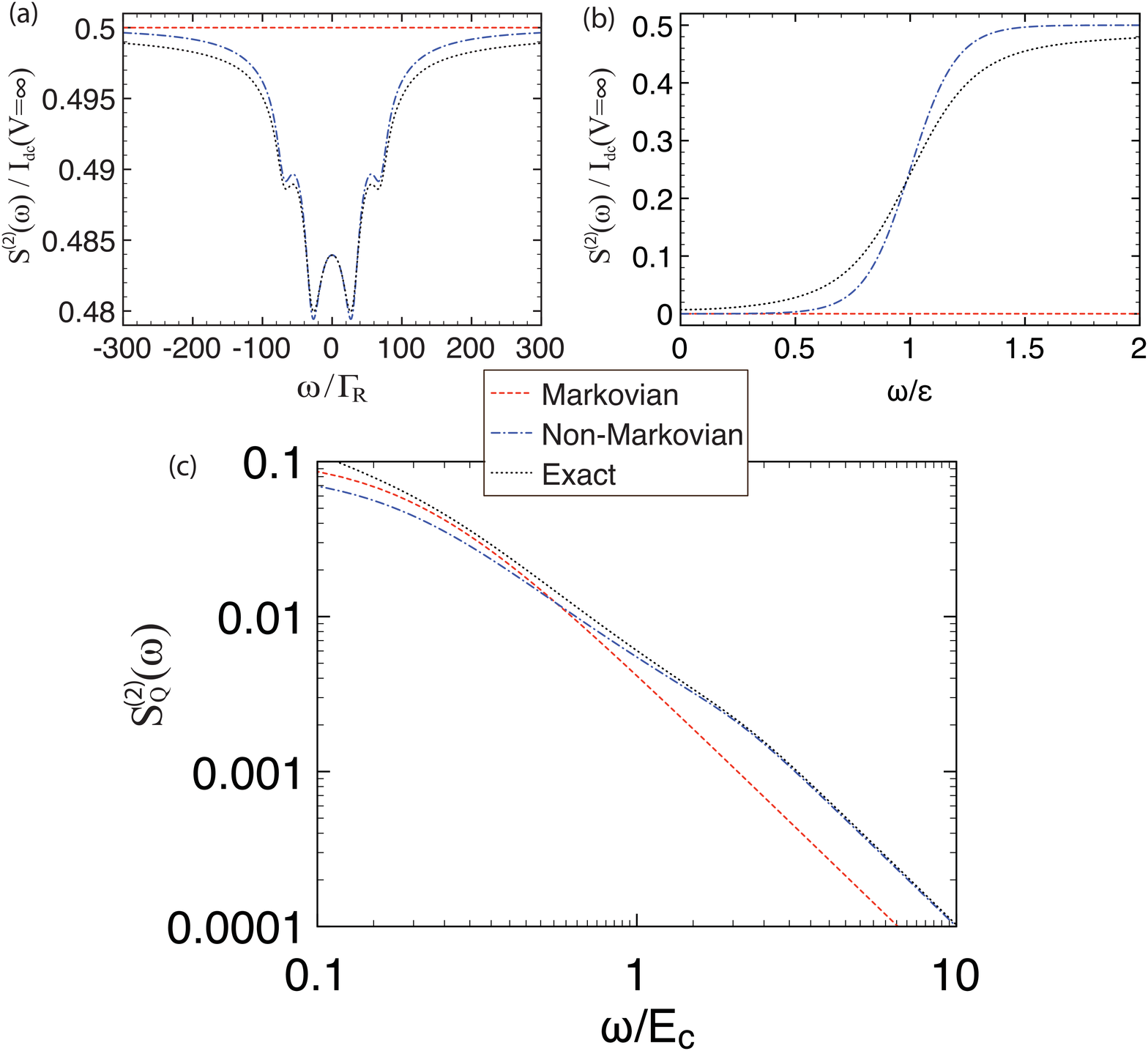}
\caption[Quantum noise spectra of a single resonant level model]{Quantum noise spectra of the SRL model ($\Gamma_L=\Gamma_R=1$ in all figures). a) $S^{(2)}(\omega)$ as a function of frequency $\omega$ in the shot noise regime ($\varepsilon=20$, $V=100$, $T=4$). In this limit, the noise develops dips at $\omega=\pm|\varepsilon\pm\frac{eV}{2}|$ . b) $S^{(2)}(\omega)$ as a function of frequency $\omega$ in the quantum noise regime ($\varepsilon=10$, $V\rightarrow 0$, $T=1$). In this limit, $S^{(2)}(\omega)$  develops a quantum noise step at $\omega=\varepsilon$. c) Charge noise $S^{(2)}_Q(\omega)$ as a function of frequency $\omega$ of a single electron transistor acting as a detector ($E_C=V=10$).  When $\omega>E_C$, $S^{(2)}_Q(\omega)$ contains extra quantum noise contributing to backaction.}
\label{Fig1paperNMsup}
\end{figure}

Fig.~\ref{Fig1paperNMsup}a shows\footnote{Throughout the chapter we will use $e$ (electron charge) $=$ $k$ (Boltzman's constant) $=$ $\hbar$ (Planck's constant/$2\pi$) = 1.} the shot noise spectrum $S^{(2)}(\omega)$ of the total current through the system obtained with the non-Markovian formalism discussed in the previous section (blue dashed-dotted curve). We also plot the exact result (black dotted curve), as given by \cite{Averin93}, and the one obtained after a Markovian approximation (red dashed curve). The agreement between the exact solution and the NM calculation is extremely good. Both develop dips at frequencies 
$\omega=\pm|\varepsilon\pm\frac{eV}{2}|$, and show a strong frequency dependence. As expected, and due to the mapping aforementioned, the shot noise spectrum in Fig.~\ref{Fig1paperNMsup}a agrees well with the one of a non-equilibrium Kondo model in the Toulouse limit (c.f. with Fig.~11 in Ref.~\cite{Schiller98}). In stark contrast, the Markovian solution is markedly different: it is frequency-independent and equals $S^{(2)}(\omega\rightarrow\infty)=\frac{\Gamma_L\Gamma_R}{2(\Gamma_L+\Gamma_R)}=\frac{\langle I\rangle}{2}$. Even at $\omega=0$, the MA deviates from the NM and exact solutions, which here fall practically on top of each other. In Fig.~\ref{Fig1paperNMsup}b, we explore the linear-response regime when the level is outside the bias voltage window. In this situation shot-noise is negligible, and quantum fluctuations are dominant in the spectrum for $\hbar\omega\gg k_BT$. The quantum noise step expected at $\omega=\varepsilon$ is fully captured by our NM approach, while here it becomes clear that the MA does not capture quantum noise physics. 

The richness of the SRL model can be further explored by noting that it also describes the physics of a single electron transistor (SET) with charging energy $E_C\gg k_BT$, and voltage such that only two charge states $|N\rangle$ and $|N+1\rangle$ are relevant. One can describe a SET in this regime with Eq. (\ref{SRLhamiltonian}) by just making the substitutions $\varepsilon\rightarrow E_C$, $|0\rangle\rightarrow|N\rangle$ and $|1\rangle\rightarrow|N+1\rangle$, see e.g. Ref.~\cite{Schoelkopf-NazarovBook}. Let us derive the charge-noise spectrum (\ref{chargenoisespectrum}) of the SET. This problem has already been studied by Johansson {\it et al.} using a different formalism \cite{Johansson02}. As discussed in the previous section, $S^{(2)}_Q(\omega)$ can be found by considering the jump operators arising from the counting field (\ref{chiaccum}). Alternatively, one can 
use Eq.~(\ref{totalnoiseEq}) together with (\ref{totalnoiseLR}) to obtain
\beq
S^{(2)}_Q(\omega)=\frac{1}{\omega^2}\left[ S^{(2)}_L(\omega)+S^{(2)}_R(\omega)-S^{(2)}_{LR}(\omega)-S^{(2)}_{RL}(\omega)\right].
\eeq
The cross correlations $S^{(2)}_{LR/RL}(\omega):=\int_{-\infty}^{\infty} d\tau e^{-i\omega\tau} \ew{\left\{ I_{L/R}(\tau)I_{R/L}(0) \right\} }_c$, can be easily calculated taking the derivative of the CGF with respect to counting fields $\chi_L$ and $\chi_R$, while the particle-noise contributions $S^{(2)}_{L/R}$ involve a double derivative with respect to $\chi_L/\chi_R$ of the CGF.
Fig.~\ref{Fig1paperNMsup}c shows the noise associated with the charge fluctuations in the central island of an SET, $S^{(2)}_Q(\omega)$. Interestingly, if the SET is used as a detector of another quantum system, this noise governs the measurement backaction \cite{Schoelkopf-NazarovBook, Clerk10, Young10}. When $\hbar\omega\geq E_C$, the charge-noise spectrum contains extra quantum noise contributing to backaction, in full agreement with the calculations in \cite{Johansson02, Schoelkopf-NazarovBook} (c.f. Fig.~5 in these references).

\begin{figure}[t]
\center
\includegraphics[width=\textwidth]{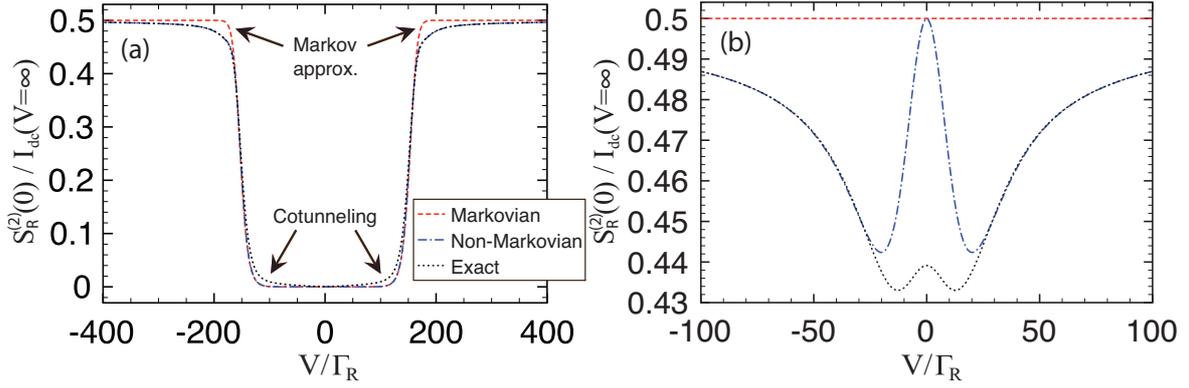}
\caption[Non-Markovian zero-frequency limit of the single resonant level model]{Zero-frequency limit of the non-Markovian theory. a) Particle-current noise for $\Gamma_L=\Gamma_R$, $\varepsilon/\Gamma_R=80$, $T/\Gamma_R=4$ The noise suddenly increases when the level enters the voltage bias window. b) $S^{(2)}_R(0)$ as a function of bias voltage $V$ for $\Gamma_L=\Gamma_R$, $\varepsilon=0$, $T/\Gamma_R=4$. While the Markovian approximation is flat at all voltages, the NM and exact solution show a structure that strongly differs in both for low voltages due to cotunneling processes.}
\label{Fig2paperNMsup}
\end{figure}

In Fig.~\ref{Fig2paperNMsup} we investigate this zero-frequency limit given by our NM theory. Fig.~\ref{Fig2paperNMsup}a shows the particle noise $S_R^{(2)}(\omega=0)$ as a function of voltage for a configuration such that $\varepsilon/\Gamma_R=80\gg T/\Gamma_R=4$. We observe a resonant step in the noise spectrum at precisely $V=\pm 2\varepsilon$. Above this step, there is a discrepancy of the Markovian solution with the NM and exact results, while right below the step, Markovian and non-Markovian limits differ from the exact solution. This last discrepancy is due to cotunneling contributions, only captured by the exact result. The difference is better observed in Fig.~\ref{Fig2paperNMsup}b, where we set $\varepsilon=0$ and vary the bias voltage again. Remarkably, the Markovian solution is flat for all voltages, while both NM and exact solutions show certain structure capturing system-bath memory effects. Only for low voltages these two disagree, when cotunneling contributions become important. At zero voltage, the Markovian and NM curves coincide as expected (since the only contribution to noise should originate from equilibrium fluctuations). For large enough voltages, the exact and NM results fall on top of each other, and we remark that, as noticed previously, the limit $V\to\infty$ is exact in both Markovian and non-Markovian approaches, and thus all three curves converge to the same value in this limit.

\subsection{Double quantum dot}

\begin{figure}[t]
\center
\includegraphics[width=0.6\textwidth]{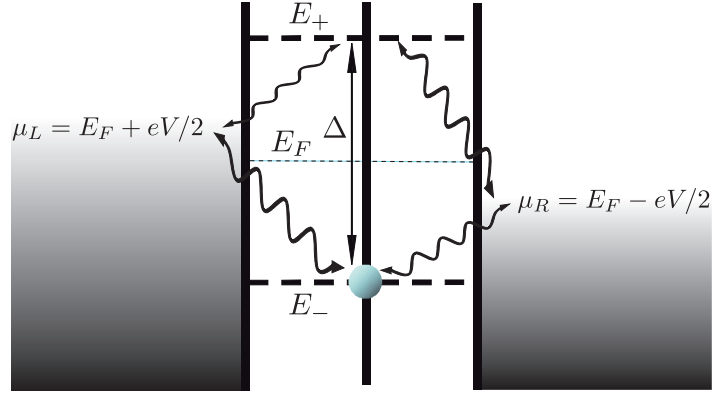}
\caption[Quantum noise processes in a double quantum dot]{Quantum noise processes in a double quantum dot. In the QNR, quantum fluctuations 
can discharge the system through the left/right reservoir if $\hbar\omega\geq |\mu_{L/R}-\Delta/2|$. These correspond to the steps in Fig.~\ref{Fig3paperNM}a. When $\omega=\Delta$, quantum interference between the eigenstates gives a noise suppression.}
\label{Fig2paperNM}
\end{figure}

\begin{figure}
\center
\includegraphics[width=0.93\textwidth] {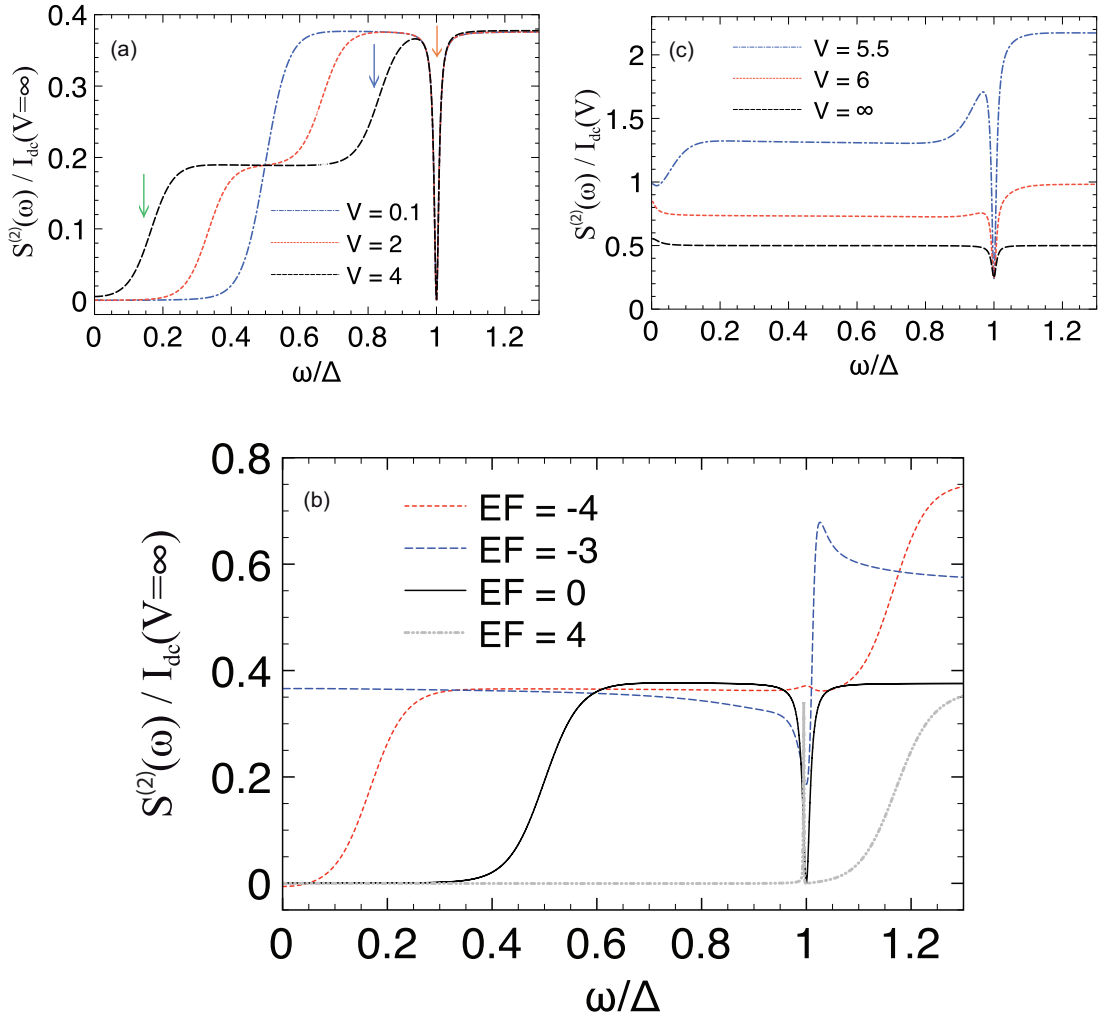}
\caption[Quantum noise spectra of a double quantum dot]{Finite-frequency noise of a double quantum dot (results normalized to the dc current  in the large-voltage limit, $I_{\mathrm{dc}}(V=\infty)$, given by expression (\ref{StoofNazarovCurrent}). a) Near linear response, $S^{(2)}(\omega)$ shows quantum noise steps at $\omega=|\Delta/2\pm V/2|$ and a dip centered at $\omega=\Delta$ (indicated with arrows for the case $V=4$ in the figure). b) The feature at $\omega=\Delta$ originates from quantum interference between the bonding and anti-bonding qubit states. Here we set $V=0.1$ and vary the reservoir Fermi energy $E_F$, observing a displacement of the quantum noise step, as well as a modification of the resonance form at the qubit frequency $\Delta$ (see text). c) Shot noise limit. Quantum noise steps are only visible for $V\lesssim \Delta, k_BT$, otherwise the contribution from shot noise or thermal noise are dominant. Parameters: $\varepsilon=0$, $\Delta=2T_c=6$, 
$\Gamma_L=\Gamma_R=T/2=0.1$.
}
\label{Fig3paperNM}
\end{figure}

To further illustrate the non-Markovian theory developed in this chapter, we now consider the example of a double quantum dot (DQD). To the best of our knowledge, a complete study of this model in the different regimes of $V$, $T$ and $\omega$, and in the NM limit is yet lacking. The following results are also applicable to a Cooper pair box qubit. Again, the Markovian and NM solutions shown here correspond to first order in perturbation theory (sequential tunneling) and $S^{(2)}(\omega)$ refers to the `total' noise. In the Coulomb blockade regime, the possible DQD states are $|0\rangle\equiv|N_L,N_R\rangle$, $|L\rangle=|N_L+1,N_R\rangle$ and $|R\rangle=|N_L,N_R+1\rangle$, with $N_L$/$N_R$ being the number of electrons in the left/right dot. The qubit, with Hamiltonian ${\cal H}_\mathrm{S}=\varepsilon\left( \ket{L}\bra{L}-\ket{R}\bra{R} \right)+T_c \left( \ket{L}\bra{R}+\ket{R}\bra{L} \right)$, has eigenvalues $E_{\pm}=\pm{\Delta\over 2}$, being $\Delta\equiv2\sqrt{\varepsilon^2+T_c^2}$.
Near linear response ($eV\ll k_B T, \hbar\omega$), the only noise contribution originates from equilibrium fluctuations -- either thermal noise for $k_BT\gg\hbar\omega$, or quantum noise for $\hbar\omega\gg k_BT$. In Fig.~\ref{Fig2paperNM} we sketch the physical processes due to quantum fluctuations, which give rise to the noise spectrum in Fig.~\ref{Fig3paperNM}a. 
For $eV\lesssim\Delta$, the conductance is zero and therefore $S^{(2)}(0)=0$, as dictated by the fluctuation-dissipation theorem. 
Quantum fluctuations, on the other hand, give rise to a finite noise for $\omega> 0$ (steps at $\hbar\omega=|\mu_{L/R}\pm\frac{\Delta}{2}|$ in Fig.~\ref{Fig3paperNM}a). 
Importantly, this physics is not captured with the MA, neither by other models for the inhomogeneity, such as $\Gamma(\chi; z)= \frac{1}{z}\{\Sigma(\chi,0)-\Sigma(\chi,z)\}$.
The spectrum also contains a strong dip centered at $\omega=\Delta$. This dip, which is voltage-independent and reaches $S^{(2)}(\omega=\Delta)=0$, can be understood as resulting from coherent destructive interference between the qubit eigenstates. This is demonstrated in Fig.~\ref{Fig3paperNM}b, where we investigate how this feature at $\omega=\Delta$ changes as we move the Fermi energy, $E_F$, of the reservoirs. For $V=0.1$ and $E_F=0$ (black solid curve), $E_{+/-}$ is above/below the chemical potentials and we find a dip shape, as discussed. When $E_F$ is aligned with the lowest level, namely $E_F=E_-=-\frac{\Delta}{2}$, the resonance changes to a Fano shape, as one expects from interference between a discrete level (the one above the chemical potentials at $E_+=\frac{\Delta}{2}$) and one strongly coupled to a continuum (the one at $E_F=E_-=-\frac{\Delta}{2}$). When both levels are above $E_F$, the interference at $\omega=\Delta$ is suppressed (red dotted curve). However, if both levels lie below $E_F$ (light grey curve), quantum interference still occurs, giving in this case a narrow resonant peak in the noise spectrum, since now we have a qubit weakly coupled to the leads -- therefore with a low dephasing rate. A very important remark of this figure, is that the situation corresponding to $E_F=-4$ gives a different result from that corresponding to $E_F=4$. In the former, the peak at $\omega=\Delta$ has been suppressed, while in the last, the resonance occurs. This we understand in terms of coherent oscillations only taking place when the levels lie below the chemical potentials. Most importantly, the light-grey curve only presents one quantum noise step, corresponding to the anti-bonding state. As the charge oscillates fast between both eigenstates, this can decay to the reservoirs via quantum noise processes only from the lowest level. However, in the situation with both eigenstates above the chemical potentials, charge can decay to the reservoirs from both levels through quantum noise processes. If $eV\gtrsim\Delta$, transport is possible and shot noise is finite, therefore $S^{(2)}(0)\neq 0$. This limit is discussed in Fig.~\ref{Fig3paperNM}c. Interestingly, quantum noise is progressively overcome by shot noise as $V$ increases. As a result, for large voltages, the quantum noise steps disappear and the noise is of smaller magnitude.
In this case an incomplete destructive interference is found at $\omega=\Delta$: $S^{(2)}(\omega=\Delta)/I_{\mathrm{dc}}(V)$ is greater than zero and does not depend on $V$. 
The width, on the other hand, increases with the voltage, which can be understood as a decrease of the dephasing time (inverse of the width) due to the coupling with the reservoirs \cite{AguadoBrandes04}. The MA is recovered as $V\rightarrow\infty$, with features at $\omega=0$ and $\omega=\Delta$ on top of a background of sub-Possonian partition noise, Fano-factor $S^{(2)}(\omega)/I_\mathrm{dc}(V)=1/2$.

\begin{figure}
\center
\includegraphics[width=\textwidth] {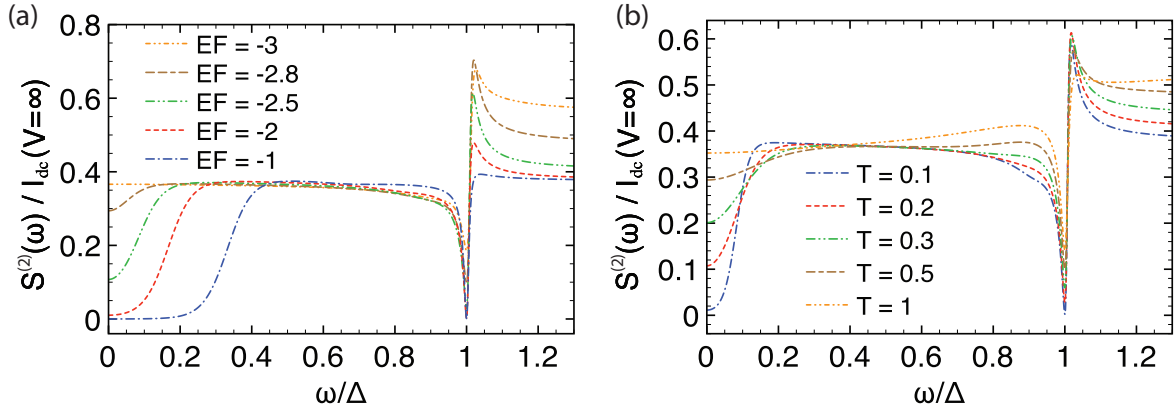}
\caption[Gate voltage and temperature effects on the quantum noise of a double quantum dot]{a) Effect of a gate voltage. As the relative distance between the dot levels and the lead chemical potentials is varied (here illustrated decreasing the Fermi energy $E_F$), a quantum noise step, absent when the bonding state is aligned with both chemical potentials, appears at the corresponding frequency difference. The Fano shape, however, gives an anti-resonance at the qubit frequency $\Delta$. Here $T=0.2$. b) Effect of the temperature. As $T$ is increased, the quantum noise step is lost, since thermal noise overcomes quantum noise, giving a finite $S^{(2)}$ value at zero frequency. The Fano shape is however preserved for high temperatures. Here $E_F=-2.5$. In both figures $V=0.1$, $\varepsilon=0$, $\Delta=2T_c=6$, $\Gamma_L=\Gamma_R=0.1$.}
\label{Fig4paperNMad}
\end{figure}

The transition from a Fano shape to an anti-resonance in the noise spectrum encountered in Fig.~\ref{Fig3paperNM}b is further investigated in figure \ref{Fig4paperNMad}a. Here we show how the quantum noise step progressively appears as the bonding state comes below the chemical potentials. At the same time, the Fano resonance gives rise to the destructive-interference feature at the qubit frequency. The effect of temperature is shown in Fig.~\ref{Fig4paperNMad}b. Still in the linear response regime, where the `shot' contribution is negligible, we see how quantum noise is overcome by thermal noise, giving a finite $S^{(2)}$ value at zero frequency for increasing temperature, as dictated by the fluctuation-dissipation theorem. The Fano shape, consequence of having the lowest level strongly coupled to the reservoirs, but also coupled to the anti-bonding state, persists at high temperatures.

\section{Conclusions}

In this chapter we have presented a general non-Markovian theory of frequency-dependent noise based on QME. The importance of NM correlations to correctly capture the physics of vacuum fluctuations has been shown through the study of a single resonant level model and a double quantum dot in the quantum noise regime. Our equations for the CGF and noise spectrum open the possibility to investigate this physics in a variety of systems where NM corrections are of vital importance, such as electromechanical resonators close to the zero-point motion \cite{Connell10}, or strongly correlated cold atoms in optical lattices \cite{Braungardt08}.

~

~

~

~

~

~

~

~

~

~

~

~

~

~

~

~

~

~

~

~

~

~


\chapter{Hybrid quantum processors: Coupling superconducting qubits and atomic systems\footnote{The results presented in this chapter have been published in \cite{Marcos10diamond}.}} 
\label{Chapter5}
\lhead{Chapter 5. \emph{Hybrid quantum processors}} 

\begin{flushright}

\textit{``These dark days in Stockholm emphazise the importance of light in our lives.''}

Roy Glauber

\end{flushright}

\begin{small}

We propose a method to achieve coherent coupling between Nitrogen-vacancy (NV) centers  in diamond and superconducting (SC) flux qubits. The resulting coupling can be used  to create a coherent interaction between the spin states of distant NV centers mediated by the flux qubit. Furthermore, the magnetic coupling can be used to achieve a coherent transfer of quantum information between the flux qubit and an ensemble of NV centers. This enables a long-term memory for a SC quantum processor and possibly an interface between SC qubits and light.

\end{small}

\newpage

\section{Introduction}

Among the many different approaches to quantum information processing, each
has its own distinct advantages. For instance, atomic systems~\cite{CiracZoller95} present
excellent isolation from their environment, and can be interfaced with 
optical photons for quantum communication. In contrast, condensed 
matter systems~\cite{Kane98, LossDiVincenzo98} offer strong interactions, 
and may benefit from the stability, robustness,
and scalability associated with modern solid-state engineering.
In the last years, much effort is being devoted to coupling
atomic and solid-state qubits to form hybrid systems, combining 
`the best of two worlds'.
One attractive approach to hybrid systems involves the integration of atomic ensembles 
with superconducting (SC) stripline resonators.  Strong coupling between SC qubits
and such resonators has already been achieved~\cite{Wallraff04}, and
approaches to extend the coupling to atomic systems
have been proposed~\cite{Tian04, Sorensen04, Rabl06}. To achieve an
appreciable coupling, these proposals often use an electric
interaction, but magnetic interactions are more desirable, since
long coherence times are mainly achieved in systems where spin
states are used to store the information. Magnetic interactions are, however, inherently weaker but recently it has been proposed theoretically \cite{Rabl06, Molmer08, Imamoglu09, Schmiedmayer09, Molmer09} and shown experimentally \cite{Esteve10, Schuster10} that strong coupling to ensembles of spin systems can be achieved. 

In this chapter, we propose a novel hybrid system, consisting of a SC flux qubit magnetically coupled to Nitrogen-vacancy centers (NVs) in diamond. The
latter system shares many of the desirable properties of atoms, such as
extremely long coherence times and narrow-band optical transitions \cite{Jelezko06}, but at the same time, 
the integration with solid state systems
can be relatively easy, as it eliminates the need for complicated trapping procedures. Additionally, much higher densities can be achieved with very limited decoherence \cite{Taylor08, Imamoglu09, Molmer09}. 
As we show below, the magnetic coupling between a 
SC flux qubit and a single NV center can be about three orders of magnitude stronger than that associated with stripline resonators, thereby making the system an attractive building block for quantum information processing.

\section{Description of the system}

\begin{figure}[t]
 \begin{center}
\includegraphics[width=0.9\textwidth]{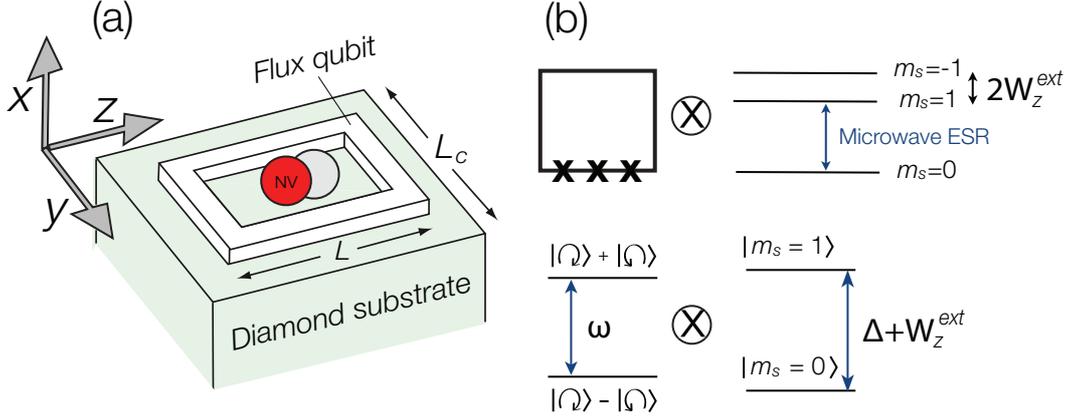}
 \caption[Schematics of the coupling between NV centers and a flux qubit]{
 (a) Schematic setup. An isotopically pure ${}^{12}\rm{C}$ diamond crystal, doped with Nitrogen-vacancy color centers, is located in the proximity of a flux qubit of size $L \times L $ and cross section $h\times h$. The two persistent-current quantum states of the flux qubit have different associated magnetic fields, which give rise to a state-dependent interaction with the electron spin in the NV center(s). (b) The combined system of a flux qubit and a NV center. The eigenstates of the flux qubit are superpositions of left- and right-circulating currents. An external magnetic field splits the $m_S=\pm1$ states of the NV, resulting in a two-level system with the $m_S=0 \leftrightarrow m_S=1$ transition close to resonance with the flux qubit.
   \label{figure1paperDiamond}
}
 \end{center}
\end{figure}

Flux qubits (FQs) form superpositions of persistent currents of
hundreds of nano-Amperes, flowing clockwise and anti-clockwise through
micrometer-sized superconducting loops~\cite{Mooij99, Mooij00}. The
magnetic field associated with this current, of the order of a $\mu$T, enables a magnetic dipole coupling to the electron spins associated with crystalline impurities such as the NV
center in diamond.  Of particular interest is the coincidence of
energy splittings: the two states of the 
FQ are typically separated by a few GHz, while NV centers have a $S=1$ ground
state, with zero-field splitting $\Delta = 2 \pi \times 2.87$ GHz between
the $m_S=0$ and $m_S=\pm 1$ states.
By the application of a mT magnetic field, one of the spin transitions of the NV center can be tuned into resonance
with the FQ (see Fig.~\ref{figure1paperDiamond}b). 
This, together with the large magnetic moment of the FQ and the 
relatively long coherence times of both systems, opens the possibility 
of achieving coherent transfer between them.

Let us consider a {\em single} NV center in a diamond crystal, located
near a square FQ of size $L$ and thickness $h$ (see
Fig.~\ref{figure1paperDiamond}a). A static external field $\vec{B}^{\rm ext}$ is
applied, whose component perpendicular to the 
FQ provides half a flux quantum, and brings the qubit
near the degeneracy of the clockwise and counter-clockwise current states.  The zero-field spin splitting of the NV center $\Delta$
sets a preferred axis of quantization to be along the axis between the
Nitrogen and the vacancy; thus, the field parallel to this axis sets
the small additional Zeeman splitting between $m_S=\pm 1$ states, and
allows to isolate a two-level subsystem comprised by $m_S=0,1$.

For convenience, we denote as the $z$ axis the crystalline axis of the
NV center. The Pauli operators for the FQ system, not
tied to a particular spatial axis, will be denoted by
$\hat{\tau}_{1},\hat{\tau}_{2},\hat{\tau}_{3}$, with $\hat{\tau}_3$ describing the population difference between the two persistent-current states. 
The interaction of the total magnetic field $\vec{B}$
(external and from the FQ) with the NV center can be written 
as $\vec{S}\cdot \vec{W}$, where $ \vec{W} \equiv g_{e} \mu_B \vec{B}$, 
$g_{e}$ is the electron g-factor and $\mu_B$ is the Bohr magneton; the two
persistent-current quantum states of the FQ give rise to different
anti-aligned magnetic fields: $\hat{\tau}_3 \vec{W}^{\rm FQ}$.
The Hamiltonian for the system is then
\begin{equation}
  \hat{H} = \varepsilon \hat{\tau}_3/2 + \lambda \hat{\tau}_1+ \Delta S_z^2 + W^{\rm ext}_z S_z +
  \hat{\tau}_3 \vec{W}^{\rm FQ} \cdot \vec{S}.
  \label{Hbare}
\end{equation}
Here the magnitude of $\vec{W}^{\rm FQ}$ corresponds to a Larmor frequency shift due
to the circulating or counter-circulating currents in the 
FQ, $\lambda$ is the coupling between these two current states, and
$\varepsilon$ is the bias in the two-well limit of the FQ, which
depends on the external field perpendicular to the loop. If the flux
qubit's plane is not perpendicular to the $z$ axis of the NV center, both
systems can be tuned on resonance by changing independently the $z$
and, e.g., $x$ components of $\vec{B}^{\rm ext}$.

\section{Individual direct coupling}

Due to the coupling $\lambda$ in equation (\ref{Hbare}), the eigenstates of the FQ Hamiltonian are not  left- and right- circulating current states. This means that there is a magnetic transition between the dressed states of the FQ which couples to the electronic spin of the NV.  To describe this, we rotate the FQ via a
unitary transformation by an angle $\cos \theta \equiv
\varepsilon / 2\omega$, giving two FQ dressed states with a transition frequency $\omega \equiv \sqrt{\varepsilon^2/4 + \lambda^2}$.  When $\Delta + W^{\rm ext}_z - \omega =
\delta$ is small, we can transform to a rotating frame and make the
rotating-wave approximation (RWA) to describe the near-resonance interaction between the NV and the FQ.  Neglecting the state
$m_S = -1$, due to the external field moving it far out of resonance
($| \delta | \ll |W^{\rm ext}_z|$), the effective Hamiltonian describing the dynamics is
\begin{equation}
  \hat{H}_{\rm RWA}  =   \frac{\delta}{2} \hat{\sigma}_z + \frac{\cos \theta}{2}  W^{\rm
    FQ}_z \hat{\tau}_3 \hat{\sigma}_z
    + \frac{\sin \theta}{\sqrt{2}} W^{\rm FQ}_\perp { \hat{\tau}_- \hat{\sigma}_+ + {\rm H.c.}},
\end{equation}
where $\bm{\hat{\sigma}}$ are Pauli operators describing   the NV $m_S=0,1$ electron-spin states, and $\vec\tau$ is in a rotated basis so that, e.g., $\hat{\tau}_3$ describes the difference  in the populations of the FQ dressed states. 

\begin{figure}[t]
 \begin{center}
 \includegraphics[width=0.7\textwidth]{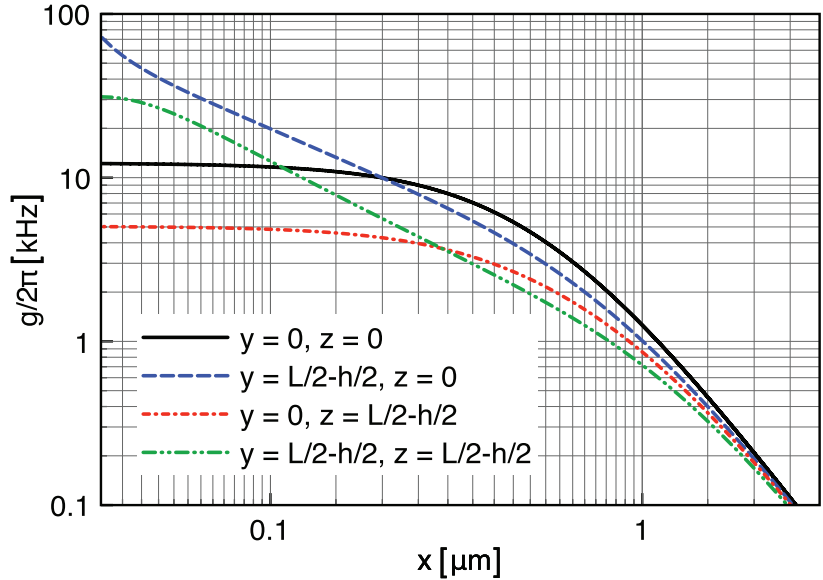}
 \caption[Coupling between a NV center and a flux qubit]{
Coupling $g$ between a FQ operated at the degeneracy point ($\varepsilon=0$) and a single NV center located at the position $(x,y,z)$ in a reference frame with axes as in Fig.~\ref{figure1paperDiamond} but centered in the middle of the square FQ.
 The NV crystal axis is assumed to be parallel to one of the wires  forming the flux qubit.
 The FQ has size $L=1\,\mu\rm{m}$, thickness $h=60\,\rm{nm}$, and  a critical current of $0.5\,\mu \rm{A}$.
 \label{figure2paperDiamond}
 }
 \end{center}
\end{figure}

The coupling constant $g \equiv \sin \theta W^{\rm FQ}_{\perp}/\sqrt{2}$ depends on the field perpendicular to the NV axis, and it is maximal at the degeneracy point $\varepsilon=0$, where $\theta = \pi/2$. Furthermore, this point has a `sweet-spot' property: 
the energetics of the system is insensitive to small fluctuations of $\varepsilon$,
as the eigenvalues have zero derivative with respect to $\varepsilon$
at this point. This reduces the dephasing of the FQ due to stray magnetic fields and, e.g., paramagnetic spins in the diamond crystal (see section \ref{FQdecSec}), and ensures that there will be no differential shifts of the resonance frequency of the NVs due to an inhomogeneous static field from the FQ. For the remainder of this chapter we will assume that we are working at the degeneracy point. Figure \ref{figure2paperDiamond} shows the coupling $g$ 
along four different vertical lines. We evaluate this using the magnetic field generated by a finite width square loop as given by the Biot-Savart law. 
For a FQ of size $L=1\,\mu\rm{m}$ and critical current $0.5\,\mu\rm{A}$, the
coupling reaches $g = 2 \pi \times 12\,\rm{kHz}$ for a single NV
center located at the center of the loop, which is about a factor
of 1000 larger than the coupling achieved with stripline resonators~\cite{Imamoglu09, Molmer09}.  This coupling $g$ is however too small
to achieve coherent transfer,
since current $T_{2}$ times in FQs are at best a few microseconds~\cite{Bertet05, Yoshihara06, Kakuyanagi07, Steffen10}.

\section{Remote spin coupling} \label{remotecouplingSec}

We now show that the FQ can be used as a virtual intermediary, allowing to couple coherently two or more NV centers with the same orientation
and detuned $\delta$ from the FQ.
For large enough detunings, $\delta \gg 1/T_{2}^{{\rm FQ}}, g$, we can
adiabatically eliminate the excited state of the FQ, and the two
coupled NVs have the states $\ket{11}$ and $(\ket{10} +
\ket{01})/\sqrt{2}$ shifted by an energy $2 g^2/\delta$.
In contrast, the states $(\ket{10} -
\ket{01})/\sqrt{2}$ and $\ket{00}$ are not shifted by the FQ. As a result, for a fixed time $t_X = \pi \delta / (4 g^2)$ an operation
resembling $\sqrt{\textrm{SWAP}}$ -- an entangling operation -- between two NV centers can  be implemented.

To analyze the gate fidelity, we notice that the coherence of FQs and NVs are subject to low-frequency noise.
Larger coherence times are typically obtained by removing this noise during single-qubit evolution, using spin-echo sequences.  Such sequences can be incorporated into the gate operation by following the prescription for composite pulses of the
CORPSE family~\cite{Cummins03}. ÊIn particular, turning on and off the
interaction with the FQ, e.g., by changing from a $m_S = +1,0$
to $m_S=-1,0$ superposition, local $\pi$ pulses (described by $\hat{\sigma}_x$)
allow to realize a sequence $\textrm{Ex}_{\pi/4- \phi/2} \hat{\sigma}_x^1
\hat{\sigma}_x^2 \textrm{w}\textrm{Ex}_{\phi} \textrm{w}
\hat{\sigma}_x^1 \hat{\sigma}_x^2 \textrm{Ex}_{\pi/4-\phi/2}$, where
$\phi = 2 \sin^{-1} (1/\sqrt{8}) \approx 41.4^{\circ}$,
$\textrm{Ex}_{\theta}$ is an exchange-type interaction for a rotation
$\theta$, with $\theta = \pi$ a `SWAP', and $\textrm{w}$ is a wait
for a time $t_W = t_X (1/2 - 2 \phi/\pi)$, which makes the time
between the $\pi$ pulses equal to twice the time on either end of the
sequence. ÊThis approach is only sensitive to detuning errors in
fourth order for both collective and individual noise on the two
spins, effectively integrating a Carr-Purcell-type spin echo with the
$\sqrt{\textrm{SWAP}}$ operation. Furthermore, the total time
$t_X$ spent interacting with the FQ during the sequence is
equal to the prior case, thus inducing no additional overhead in the virtual coupling through the FQ.

With a spin echo the coherence is often limited by energy relaxation, resulting in an exponential decay $\exp(-t/T_2^{{\rm FQ}})$ for the FQ \cite{Bertet05, Yoshihara06, Kakuyanagi07, Steffen10}, whereas the NVs decay as $\exp(-(t/T_2^{{\rm NV}})^3)$ \cite{Taylor08, Maze08}. During the operation, the finite chance of exciting the detuned FQ leads to an
induced decoherence rate $\gamma = 2 g^2 / (
\delta^2 T_{2}^{{\rm FQ}})$, resulting in a gate error  $\sim  ((t_X+2t_W) / T_{2}^{{\rmÊNV}})^3 + \gamma t_X$. 
Minimizing this expression we find an optimal detuning, which gives the
optimized error probability $\sim 2.2/(g ^2T_2^{{\rm FQ}}
T_2^{{\rm NV}})^{3/4}$ (see subsection \ref{swapSupSec}). ÊFor isotopically purified $^{12}$C diamond \cite{Balasubramanian09} with an optimistic
$T_{2}^{{\rm NV}} \approx 20$~ms, $g \approx 2 \pi \times 12$~kHz, and
$T_{2}^{{\rm FQ}} \approx 5\ \mu$s, the maximum achievable fidelity of
the $\sim \sqrt{\textrm{SWAP}}$ operation is $\gtrsim 0.98$ with an operating time $t_X+2t_W\approx 3.3$ ms. It is thus possible to achieve a high-fidelity coherent operation between two NVs separated by micrometer distances. This coupling may even be extended to NVs separated by large distances: If two FQs are strongly coupled (directly or through resonators) with a coupling exceeding the detuning $\delta$, two NVs residing in different FQs may be coupled through the dressed states of the two FQs. This results in a long-distance coupling between the NVs of roughly the same magnitude as derived above. 

\section{Collective coupling} \label{collectivecouplingSec}

Even though the FQ coherence times are much shorter than the coupling to a single NV center, it is possible to coherently transfer the quantum-state  from the FQ to an ensemble of many ($N$) NVs by benefiting from a $\sqrt{N}$ enhancement in the coupling constant~\cite{LukinRMP, Hammerer10}.
Consider the diamond crystal depicted in Fig.~\ref{figure1paperDiamond}a, with
density $n$ of NV centers, each of them with a fixed quantization axis
pointing along one of four possible crystallographic directions. If the orientations
are equally distributed among the four possibilities, and the external
field is homogeneous, a quarter of the
centers can be made resonant with the FQ. 
The total coupling will then be given by the FQ - single NV interaction summed over the resonant subensemble, whose state is conveniently expressed in terms of the collective angular momentum operator
$\hat{J}_{+} \equiv (1/G) \sum_{j} g_j
\hat{\sigma}_{+}^{(j)}$, where the sum runs over the resonant NVs,
and $G \equiv (\sum_{j} \vert g_j \vert^2)^{1/2}$
is the collective coupling constant. At the operating temperature of the FQ (tens of mK) the NVs are near full polarization, 
$\hat{\sigma}_z\approx\langle{\hat{\sigma}_{z}^{(j)}}\rangle \simeq -1$, and 
the collective spin operators fulfill
harmonic-oscillator commutation relations, thus having 
bosonic excitations in this limit.  
Tuning the system into resonance, the interaction between 
FQ and ensemble of NV centers is
 \beq
\hat{H}_{\rm{int}} = G \; \hat{\tau}_{-}\hat{J}_{+} +
{\rm H.c.} \label{Hcollective}, \eeq
whose dynamics can be complicated in general~\cite{Solano07, Christ07},
but close to full polarization takes place between the states
$|{1}\rangle_{\rm{FQ}}|0^N\rangle_{\rm{NV}}$ and
$|{0}\rangle_{\rm{FQ}}\hat{J}_+|0^N\rangle_{\rm{NV}}$, where $|0^N\rangle_{\rm{NV}}$ corresponds to all NVs being in the ground state~\cite{Taylor03}. One can thus reversibly transfer the quantum state between the FQ and the collective excitations of the NVs. 

\begin{figure}[t]
 \begin{center}
 \includegraphics[width=0.75\textwidth]{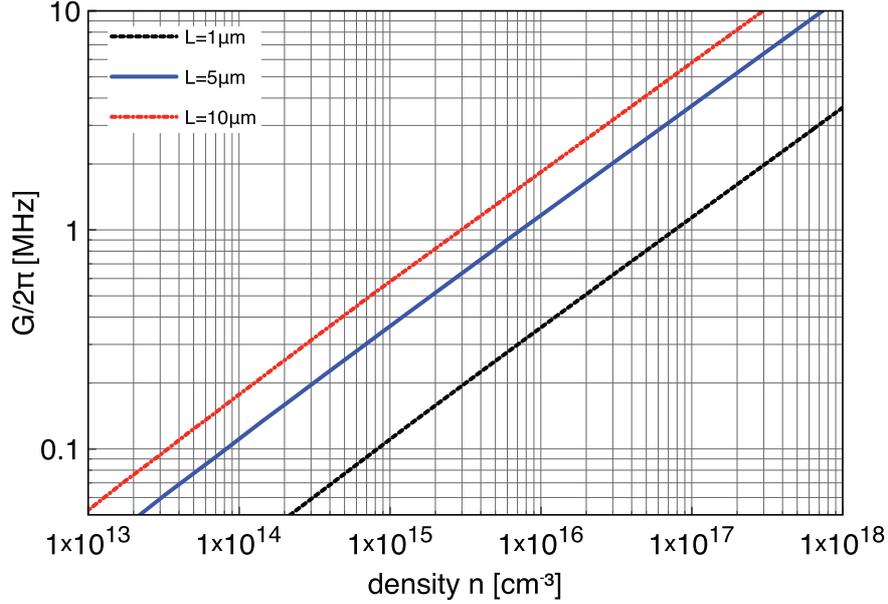}
 \caption[Coupling between an ensemble of NV centers and a flux qubit]{Coupling between a FQ and an ensemble of NV centers, as a function of the NV density $n$. Results are shown for three sizes $L$ of the square flux qubit, and are obtained by summing the inhomogeneous coupling constant, calculated as in Fig.~\ref{figure2paperDiamond}, over an ensemble of NVs located in a cubic diamond crystal of size $L_C=2 L$ placed at a height $h/2$ below the FQ. In all cases, the NV crystal axis is assumed to be parallel to one of the wires forming the FQ, the width of the FQ is $h=60\,\rm{nm}$, and its critical current is  $0.5\,\mu\rm{A}$.
   \label{figure3paperDiamond}
}
 \end{center}
\end{figure}

The collective coupling $G$ as a function of the density $n$ of NV centers, 
is shown in Fig.~\ref{figure3paperDiamond} for three different sized FQs. The coupling is obtained by summing the inhomogeneous coupling over NVs distributed in a crystal of size $(2L)^3$.
Taking the FQ coherence time to be $ \sim 5\,\mu\rm{s}$, 
we find that for a FQ with $L =5\,\mu\rm{m}$, coherent
transfer becomes possible at densities $n \gtrsim
10^{16}\,\rm{cm}^{-3}$, which have already been achieved in recent experiments~\cite{Waldermann07}. 
We note that it might be possible to increase the interaction strength with the FQ by designing circuits with higher critical currents.
Furthermore, an important feature of the present approach is that strong coupling to the FQ is achieved with ensembles containing a relatively small number of spins in a few micrometers, which makes it easier to achieve fast and identical manipulation of all spins, e.g., in spin-echo approaches.

The performance of the transfer of information between FQ and collective ensemble will be limited by paramagnetic impurities present in the diamond crystal, which will interact via a dipolar coupling with the NVs encoding the collective quantum state. In subsection \ref{DecNVensembleSec} we give a detailed discussion of decoherence induced by these impurities, which, due to the typically low Nitrogen to NV conversion efficiency, is dominated by unpaired Nitrogen electrons in the sample. For a FQ of size $L=5\ \mu$m and a density of $n=10^{17}$ cm$^{-3}$ we estimate a coherence time of $T_2^*=1.8\ \mu$s. This is sufficiently long for the infidelity induced by the paramagnetic impurities to be on the percent level, both for the transfer from the FQ to the spin wave and the nuclear storage discussed below. Furthermore the decoherence of the FQ induced by the impurities is negligible if we work close to the degeneracy point of the FQ, $\cos\theta\approx 0$.

\section{Nuclear spin memory and optical interface}

So far we have ignored the influence of the nuclear spin.
The strong hyperfine interaction with lattice $^{13}$C will be detrimental to the presented schemes, but this can be overcome using isotopically purified $^{12}$C diamond \cite{Balasubramanian09}. For the Nitrogen atoms forming the centers there are no stable isotopes without nuclear spin. This spin can, however, be polarized by transferring the nuclear state to the electron spin,
using a combination of radio-frequency and microwave pulses,
followed by polarization of the electron spin \cite{Childress06, Dutt07}, 
or directly through optical pumping using excited-state couplings \cite{Childress09}. For $^{14}$N, the  large quadrapolar field ($\sim$ 5 MHz) from the NV center leads to quantization of the spin-1 nucleus  along the NV axis. This suppresses spin-flip terms such that the hyperfine interaction just acts as an
additional parallel component to the external field.
Due to the long ($\gg 1$ s) relaxation
time of the $^{14}$N nuclear spin, this can be accounted for by an
appropriate detuning of the magnetic field so that the nuclear spin does not lead to decoherence.  
  
The nuclear spin can actually be turned into a valuable resource for long-term storage of quantum information. The electronic spin state ($m_S$) can be  transferred into the nuclear spin state ($m_I$) through a sequence of radio and microwave frequency pulses performing the evolution $\alpha\ket{00}+\beta\ket{10}\rightarrow \alpha \ket{00}+\beta\ket{11} \rightarrow \alpha \ket{00} + \beta\ket{01}$, where the states label the magnetic quantum numbers $\ket{m_S m_I}$.
This allows for a long-term memory in the system when other operations are performed, e.g., while the NVs interact with light for quantum communication.

\section{Decoherence of the NV centers}

\subsection{$\sqrt{{\rm SWAP}}$ operation between individual NV centers} \label{swapSupSec}

In section \ref{remotecouplingSec} we have described a method to perform an entangling
operation between two NV centers near the same FQ using this as a quantum bus. 
The interaction between different NVs is mediated off-resonantly by the FQ and can be described starting from the Hamiltonian (\ref{Hbare}), extended to the case of $2$ NVs. 
Taking the mixing angle $\theta
\rightarrow \pi/2$, such that the FQ is optimally biased with respect
to low-frequency flux noise, and a rotating frame such that the FQ is detuned $-\delta$, rather
than the NVs being detuned $\delta$, this Hamiltonian reads
\begin{equation}
\hat{H} = -\frac{\delta}{2}( \hat{\tau}_3 +1)+ \frac{1}{\sqrt{2}} \hat{\tau}_- \left(
W_{\perp,1}^{FQ}  \hat{\sigma}_+^{(1)} + W_{\perp,2}^{FQ} \hat{\sigma}_+^{(2)} \right) + {\rm H.c.}
\end{equation}
In this picture, it becomes clear that there are two zero-energy
eigenstates: $\hat{H} \ket{00}\ket{0}_{FQ} = \hat{H} \ket{D} \ket{0}_{FQ} = 0$, with the
dark state
$\ket{D} \equiv \frac{1}{\sqrt{2} G_2} \left(
  W_{\perp,2}^{FQ} \ket{10} - W_{\perp,1}^{FQ} \ket{01} \right)$ and the collective coupling constant for the two NVs $G_2 = \sqrt{(W_{\perp,1}^{FQ})^2 + (W_{\perp,2}^{FQ})^2}/\sqrt{2}=\sqrt{g_1^2+g_2^2}$.  There is a corresponding bright state $\ket{B} = \frac{1}{\sqrt{2} G_2} \left(  W_{\perp,1}^{FQ} \ket{10} + W_{\perp,2}^{FQ} \ket{01} \right)$, and the state $\ket{B} \ket{0}_{FQ}$ is coupled to
$\ket{00}\ket{1}_{FQ}$ with coupling strength $G_2$. Similarly the state
$\ket{11}\ket{0}_{FQ}$ is coupled to $\ket{B} \ket{1}_{FQ}$ with coupling
strength $G_2$.

Therefore, in second order perturbation theory, projecting onto the lower
energy state of the flux qubit, we have 
\beq 
\hat{H}_{\rm eff} =
\frac{G_2^2}{\delta} (\op{B}{B} + \op{11}{11}). 
\eeq 
Here we will assume $W_{\perp,1}^{FQ}\approx W_{\perp,2}^{FQ}$ and thus $G_2\approx \sqrt{2} g$. The
probability of finding the system in the excited state is then zero for the
two dark states, and $2 g^2/\delta^2$ for the two bright states of the
system.  Thus, FQ relaxation and Markovian dephasing, which occur
directly to the FQ at a rate $1/T_2^{\rm FQ}$, will lead to errors at a rate
$\sim {2 g^2}/{\delta^2 T_2^{\rm FQ}}$ for the two NV spins. Low-
frequency noise enters in small variations of $\delta$, which also
contribute at order $2 g^2/\delta^2$. Since such low-frequency noise is non-Markovian, it will only give a quadratic contribution  $\sim( g^2 t/(\delta^2 T_2^{\rm FQ}))^2 $, which will be less severe than the Markovian dephasing. We therefore take a worst-case scenario and assume the decoherence to be produced by Markovian dephasing.

To optimize the gate operation, we note that the overall time when the FQ is coupled is $t_{X} = \frac{\pi}{2} \frac{\delta}{2 g^2}$.  There is an additional wait time for the NV centers of $2t_W = t_X (1 - 4 \phi/\pi)$, where $\phi =2 \sin^{-1} (1/\sqrt{8})$, as explained in section \ref{remotecouplingSec}.  For NV spins in the states $\ket{00}, \ket{D}$, the error is entirely from the total time $t_X + 2t_W$, as the FQ is not included, while for NV spins in the states $\ket{11},\ket{B}$, both FQ and NV dephasing errors enter.  We take the latter case to overestimate the error induced in the operation, and find an effective dephasing of the NV center $\exp[- (\frac{t_X + 2 t_W}{T_2^{\rm NV}})^3]$, as expected from a dipole-dipole bath decorrelation, and an additional dephasing due to the admixture of the FQ as $\exp[-\frac{2 t_X g^2}{\delta^2T_2^{\rm FQ}}]$.  The only free parameter we can optimize over is the detuning $\delta$.  Rewriting the total fidelity as 
\beq
F = \exp[- \frac{\delta^3}{\alpha^3} - \frac{\beta}{\delta}],
\eeq
with $\alpha = \frac{2g^2 T_2^{\rm NV}}{\pi(1 - 2 \phi/\pi)}$ and $\beta = \frac{\pi}{2 T_2^{\rm FQ}}$, the optimum occurs for $\delta_* = ( \alpha^3\beta/3)^{1/4}$, giving $F_* = \exp[ - (\beta/\alpha)^{3/4} (3^{1/4} + 3^{-3/4})]$.  Substituting $\alpha$ and $\beta$ in $\delta_*$ and the numbers given in section \ref{remotecouplingSec}, we find $F_* = \exp[ - 2.18 (g^2 T_2^{\rm FQ} T_2^{\rm NV})^{-3/4}] \approx 0.98$, and $\delta_* = 0.96 [ g^6 (T_2^{\rm NV})^3/T_2^{\rm FQ}]^{1/4} \approx 2 \pi \times 3.6$ MHz.

 An additional error arises when $W_{\perp,1}^{FQ} \neq W_{\perp,2}^{FQ}$, since then the unitary operation is not performing exactly the desired evolution; this leads to a reduced fidelity of the entangling operation by a factor $W_{\perp,1}^{FQ} W_{\perp,2}^{FQ}/g^2$. It is important to note, however, that this imperfection is not a decoherence effect, and it may be possible to exploit the resulting unitary evolution even despite this imperfection, i.e., the resulting operation still resembles a $\sqrt{{\rm SWAP}}$.  

\subsection{Decoherence of the ensemble of NV centers} \label{DecNVensembleSec}

In section \ref{collectivecouplingSec} we have described the transfer of an excitation from the FQ to the collective state of an ensemble of NV centers. 
Here we estimate the dephasing of this collective state of the ensemble due to magnetic dipole-dipole interactions with other spins in the diamond crystal. The paramagnetic impurities in the crystal consist of both other NV centers, with either the same or different orientation, as well as unpaired electronic spins on substitutional Nitrogens. In most experiments the ratio of Nitrogen to NV centers is typically quite low, e.g. on the order of 1\% to 10\% \cite{Aharonovich09}. Below we refer to this ratio as the conversion efficiency. 

The dipole-dipole interaction is described by the Hamiltonian \cite{AbragamBook}
\begin{equation}
H_{\rm dip}=\sum_{j\neq k} \frac{1}{2} \frac{\mu_0}{4\pi r_{jk}^3}{\left[(\vec{m}^{(j)}\cdot \vec{m}^{(k)}-3(\vec{m}^{(j)}\cdot\vec{e}_{jk})( \vec{m}^{(k)}\cdot\vec{e}_{jk})\right]},
\end{equation}
where $r_{jk}$ is the distance between the dipoles $j$ and $k$, $\vec{e}_{jk}$ is a unit vector between them, and  $\vec{m}^{(j)}$ and $\vec{m}^{(k)}$ are their magnetic moments. These can be expressed in terms of the spin operators of the impurities, but the quantization axis for the different spins will be different: NV centers will be aligned along the symmetry axis of the center whereas electron spins on Nitrogens will be quantized along the axis of the applied static field. To describe this situation we introduce spin operators $\hat s_z^{(j)}$, $\hat s_+^{(j)}$ and $\hat s_-^{(j)}$, defined relative to the quantization axis of each particular spin, and take the different orientations into account in the coupling constant describing  the interaction between different spins. 

For the NV centers we furthermore ignore the $m_S=-1$ state which is assumed to be shifted out of resonance by a magnetic field. All the spins are therefore two-level systems which can be described by Pauli matrices $\hat{\sigma}_z^{(j)}$, $\hat{\sigma}_+^{(j)}$ and $\hat{\sigma}_-^{(j)}$. The spin operators  can then be expressed as 
\begin{equation}
\begin{split}
\hat s_z^{(j)}=\frac{1}{2}(\hat{\sigma}_z^{(j)}+l_j),
\qquad \hat s_+^{(j)}=\sqrt{1+l_j}\hat{\sigma}_+^{(j)},
\end{split}
\end{equation}
where the quantity $l_j$, which is unity for NV centers and vanishes for Nitrogen spin, accounts for the fact that the Nitrogen has spin 1/2, whereas the two-level NV system is made from the $m_S=0$ and $m_S=1$ states of a spin-1 particle. The magnetic dipole Hamilonian is then given by 
\begin{equation}
\begin{split}
\hat{H}_{{\rm dip}}=& \sum_{j,k\neq j}
  a_{jk}\hat{\sigma}^{(j)}_{+}\hat{\sigma}^{(k)}_{-} +
  b_{jk} (\hat{\sigma}^{(j)}_{z}+l_j)(\hat{\sigma}^{(k)}_{z}+l_k),
\end{split}
\label{eq:Hd}
\end{equation}
where we have used the rotating-wave approximation to ignore terms which do not conserve energy. By this approximation, the first term vanishes between spins with different resonance frequencies, i.e., with different orientations. Notice, that if the quantization axes for spins $j$ and $k$ are different, then the angle between these axes can be contained in the coupling constants $a_{jk}$ and $b_{jk}$. This is the case for the interaction between the NV and Nitrogen spins, whose quantization axes, determined by the crystal axis and the external magnetic field respectively, form  an angle $\beta$ between them. For the sake of simplicity, we here present the case $\beta=0$. This is a worst-case scenario for the $T_2^*$ derived below, and the general situation $\beta\neq 0$ only leads to minor modifications. 
With  parallel  quantization axes, we have
\begin{equation}\label{couplingconstants}
\begin{split}
a_{jk}&= \mu_{jk}^{(a)} \left[ 1- 6 e_{-}^{jk} e_{+}^{jk} \right] = \mu_{jk}^{(a)} \left[ 1 - \frac{3}{2} \sin^2\Theta_{jk} \right],\\
b_{jk}&= \mu_{jk}^{(b)} \left[ 1- 3 (e_z^{jk})^2 \right] = \mu_{jk}^{(b)} \left[ 1 - 3 \cos^2\Theta_{jk} \right],
\end{split}
\end{equation}
where $\Theta_{jk}$ is the angle between the vector $\vec{e}_{jk}$ and the crystal axis, $e_\pm^{jk}:=(e_x^{jk}\pm i e_y^{jk})/2$, and
\begin{equation}
\mu_{jk}^{(a)}:=\frac{1}{2}\frac{\mu_0}{4\pi} \frac{\mu_B^2 g_e^{(j)}g_e^{(k)} \sqrt{1+l_j}\sqrt{1+l_k}}{r_{jk}^3},\qquad \mu_{jk}^{(b)}:=\frac{1}{8}\frac{\mu_0}{4\pi} \frac{\mu_B^2 g_e^{(j)}g_e^{(k)} }{r_{jk}^3},
\end{equation}
with $\mu_B$ the Bohr magneton and $g_e^{(j)}$ the electron $g$-factor of the $j$th spin.

To evaluate the dephasing of the spin wave we will assume that we start out in a state $|00...0\rangle_{{\rm NV}}$, where all the NVs are initially prepared in their ground states, and that a superposition state $c_0|0\rangle_{{\rm FQ}}+c_1|1\rangle_{{\rm FQ}}$ is transferred into the collective spin wave, resulting in a state $(c_0+c_1\hat{J}_+)|00...0\rangle_{{\rm NV}}$, with $\hat{J}_+=(1/G)\sum_j g_j\hat\sigma_+^{(j)}$. We then evaluate the time evolution of the coherence
\begin{equation}
\langle \hat{J}_-(t)\rangle={\rm Tr} {\left[ e^{i\hat{H}t}\hat{J}_{-}e^{-i\hat{H}t} (c_0+c_1\hat{J}_+)|00...0\rangle_{{\rm NV}}\langle 0 ...00|(c_0^*+c_1^*\hat{J}_-) \hat{\rho}_{{\rm B}}
\right]},
\end{equation}
where $\hat{\rho}_{{\rm B}}$ is the initial density operator of the bath. Since the Hamiltonian (\ref{eq:Hd}) conserves the number of excitations in the NVs,  this expression can be simplified to 
\begin{equation}
\begin{split}
\langle \hat{J}_-(t)\rangle=c_0^*c_1
{\rm Tr} {\left[ e^{i\hat{H}t}\hat{J}_-e^{-i\hat{H}t} \hat{J}_+|00...0\rangle_{{\rm NV}}\langle 0 ...00| \hat{\rho}_{{\rm B}}
\right]}.
\end{split}
\end{equation}

Writing out the above equation in terms of single-spin operators we see that the dephasing is determined by the time evolution of the two-point correlation function $\langle \hat{\sigma}_-^{(j)}(t) \hat{\sigma}_+^{(k)}(t=0)\rangle$ in the state $|00...0\rangle_{{\rm NV}}\langle 0 ...00| \hat{\rho}_{{\rm B}}$. Since the density of Nitrogen is much higher than the density of NVs, the dephasing of the spin wave will predominantly be due to the interaction with the electronic spins of Nitrogen atoms, and we shall therefore ignore the interaction among the NVs. In this approximation there is no longer a mechanism in the Hamiltonian which can transfer the excitation from one NV to another and the correlation function $\langle \hat{\sigma}_-^{(j)}(t) \hat{\sigma}_+^{(k)}(t=0)\rangle$ vanishes exactly for $j\neq k$. The dephasing of the spin waves thus reduces to the calculation of the single-spin dephasing averaged over the spin wave:
\begin{equation}
\langle \hat{J}_-(t)\rangle=c_0^*c_1 \sum_j \frac{|g_j|^2}{G^2}
{\rm Tr} {\left[ e^{i\hat{H}t}\hat{\sigma}_-^{(j)}e^{-i\hat{H}t} \hat{\sigma}_+^{(j)}|00...0\rangle_{{\rm NV}}\langle 0 ...00| \hat{\rho}_{{\rm B}}
\right]},
\end{equation}
where the sum runs over all NVs in the spin wave. Assuming all NVs to be equivalent, this can be simplified to the calculation of the dephasing of a single-spin in a bath
\begin{equation}
\langle \hat{J}_-(t)\rangle=\langle \hat{J}_-(t=0)\rangle {\rm Tr} {\left[ e^{i\hat{H}t}\hat{\sigma}_-^{(j)}e^{-i\hat{H}t} \hat{\sigma}_+^{(j)}|0\rangle_{{\rm NV}}\langle 0| \hat{\rho}_{{\rm B}}
\right]}.
\label{eq:CohSingle}
\end{equation}
When considering a single NV, the action of the spin bath described by (\ref{eq:Hd}) essentially corresponds to a random magnetic field generated from the spin ensemble along the crystal axis. This effective field will be fluctuating in time because the dipole-dipole interaction among the Nitrogen spins introduces flip-flop processes.  When the Nitrogen concentration is much higher than that of NVs, the distance between Nitrogens is comparable to the distance between NVs and the nearest Nitrogen. These flip-flop processes will therefore take place on a time scale which is comparable or only slightly faster than the NV dephasing time. For simplicity we here  ignore the flip-flop processes and consider a static environment. This represents a worst-case scenario, since this approximation removes the time averaging of the field from these processes. The actual dephasing time $T_2^*$ will therefore be slightly larger than what we predict here. Within this approximation the expression in Eq. (\ref{eq:CohSingle}) can be reduced considerably:
\begin{equation}
\langle \hat{J}_-(t)\rangle=\langle \hat{J}_-(t=0)\rangle {\rm Tr} {\left[\prod_k e^{-i4b_{jk}\hat{\sigma}_z^{(k)}t}  \hat{\rho}_{{\rm B}}
\right]},
\label{eq:CohSingleSimp}
\end{equation}
where the product is over the Nitrogen spins. 

With expression (\ref{eq:CohSingle}) we can evaluate the dephasing for a given bath. If the NV density is $n$, the typical strength of the spin-spin interaction is $\mu_0\mu_B^2 g_e^2 n/4\pi$, of the order of $\sim 25\;\mu$K for high Nitrogen densities $n_{{\rm N}}\sim 10^{19}$ cm$^{-3}$. This is much smaller than the  typical operating temperature of the FQ (tens of mK), and we can therefore neglect the interaction among the spins for determining the initial density matrix of the bath $\hat{\rho}_{{\rm B}}$. Furthermore, any non-vanishing mean value $\langle \hat{\sigma}_k^{(j)}\rangle\neq 0$ only results in a mean shift of the resonance frequency of the ensemble. The dephasing will thus be determined by spin fluctuations. A worst-case scenario can be obtained by assuming that the mean value vanishes $\langle \hat{\sigma}_k^{(j)}\rangle=0$, in which case the variance is maximal. The evolution of the coherence is then given by 
\begin{equation}
\langle \hat{J}_-(t)\rangle=\langle \hat{J}_-(t=0)\rangle \prod_k \cos(4b_{jk}t).
\end{equation}
To simplify this expression we expand it in time and find
\begin{equation}
\langle \hat{J}_-(t)\rangle=\langle \hat{J}_-(t=0)\rangle {\left(1- 2 t^2 n_{{\rm N}}\int d^3\vec r_k b_{jk}^2\right)},
\end{equation}
where we have replaced the sum by an integral and introduced the Nitrogen density $n_{{\rm N}}$. Using Eq. (\ref{couplingconstants}) this expression reduces to 
\begin{equation}
\begin{split}
\langle \hat{J}_-(t)\rangle=\langle \hat{J}_-(t=0)\rangle {\left[1- t^2
{\left(\frac{\mu_0}{4\pi} \frac{\mu_B^2g_e^{{\rm N}} g_e^{{\rm NV}}}{8}\right)}^2 4\pi n_{{\rm N}} \int_0^{\pi} \sin\Theta (1-3\cos^2\Theta)^2 d\Theta \int \frac{1}{r^4} dr \right]}.
\end{split}
\end{equation}
The angular integral gives $8/5$ but the radial integral has a strong divergence at $r\to 0$. Since the integral is over the distance between the NV and the Nitrogen impurity, this divergency represents the very fast dephasing of NV centers which happen to have a nearby Nitrogen spin. Such NV centers with a nearby Nitrogen spin will, however, also be far out of resonance  during the interaction with the Flux qubit.   
This interaction takes a time $\sim 1/G$, and therefore any NV with a dipole interaction stronger than $G$ will effectively not participate in the spin wave. We exclude these NVs by truncating the integral at a distance $r_{\rm min}$ when the interaction strength reaches the value of the coupling constant $G$, i.e., 
\beq
G=\frac{1}{8}\frac{\mu_0}{4\pi} \frac{\mu_B^2 g_e^{N}g_e^{NV} }{r_{\rm min}^3}.
\eeq
This truncation of the integral gives
\begin{equation} \label{Jdecay}
\langle \hat{J}_-(t)\rangle=\langle \hat{J}_-(t=0)\rangle {\left[1- t^2 
{{\frac{\mu_0\mu_B^2g_e^{{\rm N}} g_e^{{\rm NV}}}{15\eta} n_{{\rm NV}} G}} \right ]},
\end{equation}
where $\eta$ is the Nitrogen-to-$NV$ conversion efficiency. The result (\ref{Jdecay}) corresponds to the first term in an expansion of an exponential decay $\exp(-(t/T_2^*)^2)$, with coherence time
\begin{equation}
T_2^*=\frac{1}{\sqrt{\frac{\mu_0\mu_B^2g_e^{{\rm N}} g_e^{{\rm NV}}}{15\eta} n_{{\rm NV}} G}}.
\end{equation}
Notice that in order to be applicable, our regularization procedure requires $r_{{\rm min}}$ to be much smaller than the typical distance between spins, $1/\sqrt[3]{n_{{\rm N}}}$, since we require that only a small fraction of the NVs are excluded. This condition is fulfilled in the interesting regime $G T_2^*\gg1$. Taking a conversion efficiency $\eta=0.05$, we find $T_2^*\approx 0.3 \ \mu$s for a FQ with $L=5\ \mu$m and a density $n_{{\rm NV}}=10^{18}$ cm$^{-3}$, corresponding to $G\approx 2\pi\times 15$ MHz. For a full transfer of the 
state from the FQ to the spin wave and back we need a transfer time $t=\pi/G$, corresponding to an infidelity of the order of $1-F\sim(t/2T_2^*)^2 < 0.5\%$ (the factor of 2 accounting for the fact that the excitation only spends half of the transfer time in the spin wave). For different densities the infidelity scales as $1-F\propto \sqrt{n_{{\rm NV}}}$, becoming smaller at lower densities due to the reduced dipole-dipole interaction. For example, at $n_{{\rm NV}}=10^{16}$ cm$^{-3}$, the error is reduced by an order of magnitude but then we are approaching the limit where the FQ decoherence becomes important. The transfer of excitations from the FQ to the spin wave is thus feasible in the regime $10^{16}$ cm$^{-3} \lesssim n_{{\rm NV}} \lesssim 10^{18}$ cm$^{-3}$. In order for the spin system to be useful as a long-term memory the coherence time should, however, also be sufficiently long to allow for the transfer  to the nuclear spin for long-term storage. Since this can at best be achieved on a time scale set by the hyperfine interaction ($\sim 5$ MHz), it would exclude working at the highest densities in this interval. Working at $n_{{\rm NV}}\sim 10^{17}$ cm$^{-3}$ leads to $T_2^*\approx  1.8 \mu$s, which is sufficient to allow a transfer of the excitation from the electron spin to the nuclear spin with an infidelity $1-F$ at the percent level.  

The estimates above indicate that it is realistic to achieve a transfer of excitations from the FQ to the spin wave and back even without extending the coherence time by spin-echo techniques. These may, however, be desirable in order to achieve even longer coherence times. In particular, $T_2$ may be extended if the NVs can be flipped by an external AC field on a time scale faster than the Nitrogen flip-flop processes. One should, however, be careful about applying spin-echo to the collective NV spin since, e.g. errors in the pulse area may give rise to collective decoherence processes. These can be more detrimental to quantum states stored in collective degrees of freedom than in individual spins (for instance flipping the $|0\rangle-|1\rangle$ transition would require control of the pulse area to an accuracy better than $1/\sqrt{N}$ in order to preserve the collective state). Since the dephasing of the NV spin wave is dominated by Nitrogen spins, a more desirable solution could be to apply the external driving field to the electron spins on the Nitrogen atoms. If these spins are flipped on a time scale much faster than the flip-flop processes it would lead to an increased coherence time of the NV spin wave.  Since Nitrogen spins have a considerably different resonance frequency, this could be achieved with little influence on the NV spin wave where the quantum information is stored.

There are different methods to estimate the decoherence of the spin ensemble other than the one presented in this subsection. For example, the effect of dipolar interactions described by the Hamiltonian (\ref{eq:Hd}) can be also studied using short-time expansions \cite{Lowe57}. Furthermore, the fidelity of the transfer of information between flux qubit and spin wave can be calculated treating $\hat{H}_\mathrm{dip}$ as a perturbation (see appendix \ref{FidelityAppendix}).

\section{Decoherence of the flux qubit due the spin bath} \label{FQdecSec}

A different concern is the extent to which the diamond crystal with a high density of spins may cause dephasing of the flux qubit, of particular importance for the coupling with an ensemble of NVs. As discussed above, the paramagnetic impurities in the diamond crystal consist primarily of NV centers and unpaired electron spins on Nitrogen atoms which were not converted into NVs centers during the annealing process. Due to the typically low conversion efficiency from Nitrogen to NV centers, we shall first consider the effect of Nitrogen impurity spins and then discuss the role of the NV centers. The coupling of the FQ to paramagnetic impurities can be described by a Hamiltonian similar to Eq.~(\ref{Hbare}):
\begin{equation}  \hat{H} = \hat{H}_{{\rm spin}}+\varepsilon \hat{\tau}_3/2 + \lambda \hat{\tau}_1+ 
  \hat{\tau}_3 \sum_j \vec{W}^{\rm FQ}(\vec{r}_j) \cdot \vec{S}^{(j)},
  \label{eq:Hbare}
\end{equation}
where the spin Hamiltonian $\hat{H}_{{\rm spin}}=\sum_j \Delta_j S_z^{(j)}/2+\hat H_{{\rm int}}$ describes an energy splitting $\Delta_j$ of the individual spins, and the interaction between them is encapsulated in $\hat H_{{\rm int}}$. Changing to the dressed-state picture of the flux qubit, this Hamiltonian is transformed into
\begin{equation}  \hat{H} = \hat{H}_{{\rm spin}}+\omega \hat{\tau}_3/2 + 
(\cos\theta  \hat{\tau}_3-\sin\theta\hat{\tau}_1) \sum_j \vec{W}^{\rm FQ} (\vec{r}_j)\cdot \vec{S}^{(j)}.
  \label{eq:Hdressed}
\end{equation}
In the rotating frame with respect to $\omega\hat{\tau}_3/2$, the operator $\hat{\tau}_3$ remains stationary whereas $\hat{\tau}_1$ oscillates at a frequency $\omega$ -- much higher than any time scale of the bath, provided that there are no near resonant impurities in the diamond sample. This is the case for the Nitrogen spins, whose splitting is determined by the applied field. The NVs are near resonance and will be dealt with below. The slowly varying and rapidly oscillating terms have different qualitative behavior, so we will consider them separately. 

Let us first consider the slowly-varying contribution. As we will show now this contribution vanishes if we work close to the degeneracy point of the FQ, $\cos\theta\approx 0$, where the left- and right-circulating current states are degenerate. The slowly-varying contribution is described by
\beq
\hat{H}_\mathrm{slow} = \hat{\tau}_3 W_\mathrm{eff}^{FQ},
\eeq
with
\beq
W_\mathrm{eff}^{FQ} \equiv \cos(\theta) \sum_j \left[ W_z^{FQ}(\vec{r}_j) \cos(\beta_j) - W_{\perp}^{FQ}(\vec{r}_j) \sin(\beta_j) \right] S_3^{(j)},
\eeq
and  $\beta_j$ being the angle between the quantization axis of the $j$th spin and the NV axis (recall that the $z$-axis is defined by the crystal axis). For simplicity, we define
\beq
\kappa_j := W_z^{FQ}(\vec{r}_j) \cos(\beta_j) - W_{\perp}^{FQ}(\vec{r}_j) \sin(\beta_j).
\eeq
Since the interaction of the flux qubit with an individual spin is of the order of $g$, the former has little influence on the state of an individual spin for the duration of the interaction $\sim 1/G\sim 1/g\sqrt{N}$. Furthermore the coupling of the FQ to collective degrees of freedom in the spin bath, will only have a limited influence  on the state of an unpolarized bath, since the FQ can flip at most a single-spin. We can therefore ignore the influence of the FQ on the spin bath and consider $W_\mathrm{eff}^{FQ}$ as a fluctuating external field. Any mean value of this field will merely give rise to a shift of the resonance frequency which can be compensated and we thus need to consider the fluctuations. The root mean square of the fluctuating field is given by 
\beq
\delta W_\mathrm{eff}^{FQ} &=& \cos(\theta) \Big[ \sum_{j,k} \kappa_j\kappa_k \langle\!\langle S_3^{(j)} S_3^{(k)} \rangle\!\rangle \Big]^{1/2},
\eeq
where we have used the cumulant notation $\langle\!\langle a b \rangle\!\rangle := \langle (a - \langle a \rangle)(b - \langle b \rangle) \rangle$.
As discussed in the previous section, the temperature is typically high compared to the dipole-dipole interaction energy. We can therefore ignore correlations of different spins and consider only the $j=k$ contribution, giving
\beq
\delta W_\mathrm{eff}^{FQ} &\approx& \cos(\theta) \Big[ \sum_j \kappa_j^2 \langle\!\langle (S_3^{(j)})^2 \rangle\!\rangle \Big]^{1/2}. 
\eeq
Considering a Nitrogen-to-NV conversion of $\eta\sim 5\%$, the sum over the impurities in the ensemble will be dominated by the Nitrogen impurities. An estimate of this expression can be obtained by noting that the coupling to the Nitrogen spin $\kappa$ is comparable to the coupling $g$ to the NVs. This sum and the sum leading to the collective coupling constant $G$ only differ by the number of terms in the sum and we obtain 
\beq \label{deltaWeff}
\delta W_\mathrm{eff}^{FQ} \sim \frac{\cos(\theta)}{\sqrt{\eta}} G.
\eeq
A more accurate treatment taking into account the full spatial distribution of the field over a sample of the dimensions considered in the text only changes the estimate by a factor of less than 2. Eq. (\ref{deltaWeff}) quantifies how close to the degeneracy point we need to be so that the dephasing of the FQ induced by the bath of spins is negligible. In particular, from (\ref{deltaWeff}) we derive that we can safely neglect this dephasing on a time scale $1/G$ for a conversion efficiency $\eta=0.05$ if $\cos(\theta)\lesssim 0.01 $, for which $\delta W_\mathrm{eff}^{FQ}/G \lesssim 0.05$.

Next we turn to the rapidly oscillating part of the coupling, described by the term containing $\hat{\tau}_1$, and assume that we are near the degeneracy point, that is $\sin \theta\approx1$. We will now argue that since $\hat{\tau}_1$ is oscillating rapidly, the slowly-varying dynamics of the spin bath will only have a very weak influence on the FQ. Specifically, the spin bath can influence the FQ either through direct transitions to the spin bath or through a slow dephasing. The direct coupling can be excluded by noting that the coupling constant to collective excitations of the bath is limited by $G/\sqrt{\eta}$, but for reasonably low applied magnetic fields, $B^{{\rm ext}}\lesssim 10$ mT, the detuning $\omega-\Delta_j$ will be of the order of the NV zero-field splitting (a few GHz). The probability to transfer the excitation is $\sim G^2/(\omega-\Delta_j)^2\eta$, and can thus safely be neglected since $G\sim$ MHz. The dephasing of the FQ caused by the bath can be estimated by calculating the energy shift of the FQ in second order perturbation theory. Assuming that apart from the free precession, the spins change slowly on a time scale set by 
$\omega-\Delta_j$, the effective interaction is given by 
\begin{equation}
\hat{H}_{{\rm eff}}=\hat \tau_3\sum_l\frac{{W}^{\rm FQ}_z (\vec{r}_l)^2}{4\omega}+\frac{{W}^{\rm FQ}_\perp (\vec{r}_l)^2\omega}{4(\omega^2-\Delta_j^2)}-\frac{{W}^{\rm FQ}_\perp (\vec{r}_l)^2\Delta_j}{2(\omega^2-\Delta_j^2)} \hat{S}_z^{(j)}(t).
\end{equation}
Here the first two terms are independent of the state of the impurities, and therefore merely lead to a mean shift of the energy, which can be compensated by a magnetic field.  It is thus only the last term which leads to dephasing. Again, we can estimate the root mean square value of this term by neglecting the correlations among the impurities and we find a typical energy shift 
\begin{equation}
\Delta E \sim \frac{G}{\sqrt{\eta}} \frac{g\Delta_j}{\omega^2-\Delta_j^2},
\end{equation}
which we can safely neglect, since the single-spin coupling $g\sim $ kHz is much smaller than any other quantities in the system. 

A remaining problem is the influence of the NVs with different orientations. The interaction with these centers is similar to that with the resonant subensemble and of order $\sim G$. We assume that we select a single orientation by applying a magnetic field with a component along the axis of the NV center. This will shift all other centers out of resonance by an amount $\delta\omega\sim g_e\mu_B B^{{\rm ext}}$, such that the resulting error can be estimated to be $G^2/\delta\omega^2$. For a reasonable applied field, e.g., $B^{{\rm ext}}\sim  10$ mT, this error is negligible.

\section{Conclusions}

We have shown how to magnetically couple a superconducting FQ to NV centers in diamond. This may be used to achieve strong coupling of distant centers or to transfer the state of a FQ to an ensemble of NVs. The latter opens the possibility of long-term storage of the information in the nuclear spins. Using the strong optical transitions of the NVs, the system may enable an interface between superconducting qubits and light. 


\chapter{Ultra-strong coupling effects in cavity quantum electrodynamics\footnote{The results presented in this chapter have been published in \cite{Forn10}.}} 
\label{Chapter6}
\lhead{Chapter 6. \emph{Ultra-strong coupling effects in cavity QED}} 

\begin{flushright}

\textit{``The understanding can intuit nothing, the senses can think nothing.\\ Only through their union can knowledge arise.''}

I. Kant

\end{flushright}

\begin{small}

We discuss the physics that arise beyond the rotating wave approximation in atom-cavity systems. We contrast our theory with the measurement of the dispersive energy-level shift of an $LC$ resonator magnetically coupled to a superconducting qubit, which clearly shows that the system operates in the ultrastrong coupling regime. The large mutual kinetic inductance provides a coupling energy of $\sim0.82$~GHz, requiring the addition of counter-rotating-wave terms in the description of the Jaynes-Cummings model. We find a $\sim$ 50~MHz Bloch-Siegert shift when the qubit is in its symmetry point, fully consistent with the experimental result. 

\end{small}

\newpage

\section{Introduction}

The study of driven quantum two-level systems has been at the heart of important discoveries over the last century. 
A generic example is the field of nuclear magnetic resonance, where the dynamics of nuclear spins is controlled by the application of radio-frequency pulses, 
resulting in coherent Rabi oscillations of the spin moments \cite{CohenQMbook}. In the usual description, the applied harmonic field is decomposed into co-rotating and counter-rotating components with respect to the spin (Larmor) precession. At resonance, and in the weak-driving limit, only the co-rotating component interacts constructively with the spins, leading to a Rabi frequency that scales linearly with the driving strength. In this regime, the rotating-wave approximation (RWA) is known to hold. However, if the driving is so strong that the Rabi frequency approaches the Larmor frequency, the counter-rotating terms (CRTs) need to be included in the description, and the RWA is no longer valid. This leads to an energy shift in the level transition, the so-called Bloch-Siegert shift \cite{Bloch-Siegert40, KlimovBook}. In atomic systems, the qubit-cavity coupling $g$ is typically low, $g/\omega_q\sim g/\omega_r\sim 10^{-4}$ (with $\omega_q$ the qubit frequency and $\omega_r$ the cavity frequency) and therefore non-RWA effects are negligible. In this situation, one can nonetheless use a strong driving field to observe the Bloch-Siegert (BS) shift \cite{Shirley65}. Without driving, however, a much larger qubit-cavity coupling is required to observe the quantum BS shift. This situation will be demonstrated here in a flux qubit -- resonator system.


In circuit quantum electrodynamics (QED) \cite{Blais04}, superconducting qubits play the role of artificial atoms. With energy-level transitions in the microwave regime, they can be easily cooled to the ground state at standard cryogenic temperatures. These `atoms' can interact strongly with on-chip resonant circuits and reproduce many of the physical phenomena that had been previously observed in cavities with natural atoms \cite{Raimond01, Mooij03, Johansson06, Baur09, Hofheinz08}. The large dipolar coupling achievable in superconducting circuits enables to explore the strong-dispersive limit \cite{Schuster07}. It is only very recently that the ultra-strong coupling (USC) regime $g/\omega_r\sim1$ \cite{Feranchuk96, Irish07, Nataf10, Ashhab10} is being reached \cite{Niemczyk10}. In this chapter we present theory and experiment giving rise to the BS shift in a system consisting of a flux qubit coupled to a LC resonator in the USC regime, $g/\omega_q\sim g/\omega_r\gtrsim 0.1$.

\section{Description of the system}

\begin{figure}[t]
\center
\includegraphics[width=0.7\textwidth]{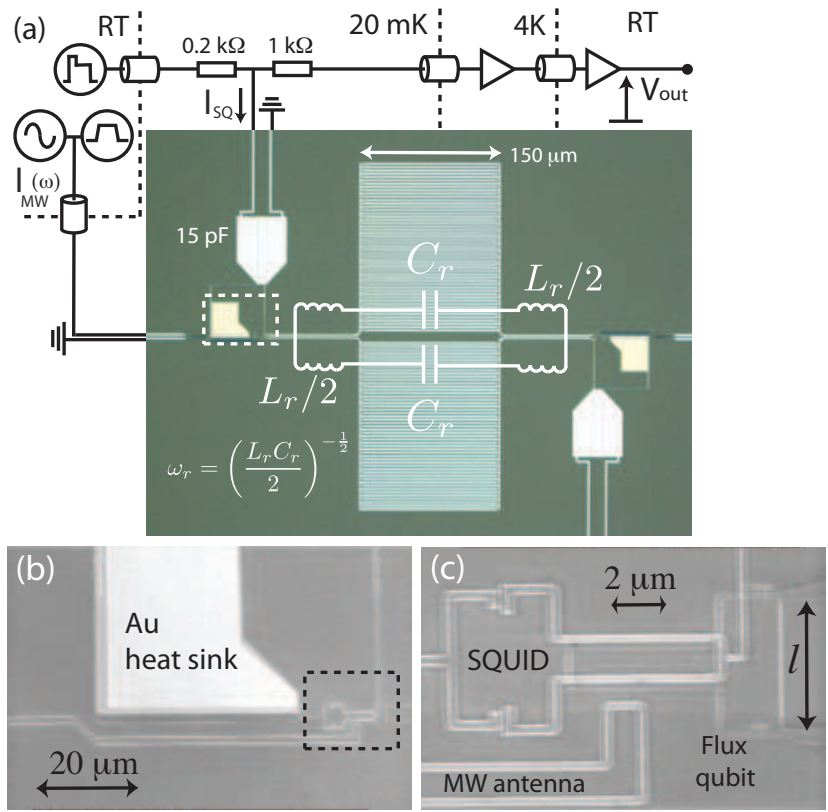}
\caption[Measurement setup of two flux qubits coupled to a $LC$ resonator]{Circuit layout and images of the device. a) Schematics of the measurement setup. 
The interdigitated capacitor of the $LC$ resonator can be seen in the center of the optical image, with the circuitry of the two SQUIDs next to it (top left and bottom right); $C_r/2\simeq0.25$~pF and $L_r\simeq1.5$~nH. b) Scanning electron micrograph (SEM) picture of the SQUID circuit (zoom in of dashed region in (a)). The readout line is made to overlap with a big volume of AuPd and Au to thermalize the quasiparticles when the SQUID switches. c) SEM picture of the qubit with the SQUID on top (zoom in of dashed region in (b)). On the right of the picture the coupling wire to the resonator of length $l$ can be seen.
\label{fig1paperBS}
}
\end{figure}

Our system consists of a four-Josephson-junction flux qubit, in which one junction is made smaller than the other three by a factor of approximately 0.5. The qubit is galvanically connected to a lumped-element $LC$ resonator (see Fig.~\ref{fig1paperBS}). In previous works the employed $LC$ resonators were strongly coupled to the flux qubit \cite{Chiorescu04, Johansson06, Fedorov10}, but since they were loaded by the impedance of the external circuit their quality factor was low. Flux qubits have also been successfully coupled to high-quality transmission line resonators \cite{Abdumalikov08, Steffen10}. In the present experiment we use an interdigitated finger capacitor in series with a long superconducting wire, following the ideas from lumped-element kinetic inductance detectors \cite{Doyle08}. In order to read out the qubit state, a dc-switching SQUID magnetometer is placed on top of the qubit. The detection procedure can be found in~\cite{Bertet05}.

The qubit and the resonator were fabricated in the same layer of evaporated aluminum using standard lithography techniques \cite{Bertet05}. A second aluminum layer galvanically isolated from the first one contains the SQUID and its circuitry, together with a microwave antenna to control the local frustration and to produce flux and microwave pulses in the qubit (see Fig.~\ref{fig1paperBS}). An external coil is used to generate a magnetic field in the qubit and SQUID in order to bias them at their operating points. A second qubit with its own circuitry was also coupled to the resonator, but during the experiment it was always flux biased such that it did not affect the measurements.

The resonator is made of two capacitors, each containing 50 fingers of 150~$\mu$m length and 1.5~$\mu$m width, separated by 2~$\mu$m (see Fig.~\ref{fig1paperBS}a). The two capacitors are linked by two 500~$\mu$m long superconducting wires of 1~$\mu$m width. With these parameters we estimate a capacitance $C_r\simeq0.5$~pF and an inductance $L_r\simeq1.5$~nH, corresponding to a resonance frequency $\omega_r/(2\pi)=1/(2\pi\sqrt{L_rC_r/2})\simeq8.2$~GHz. At temperatures $\sim30$~mK the resonator will be mostly in its ground state, with zero-point current fluctuations $\displaystyle I_{\text{rms}}=\sqrt{\hbar\omega_r/2L_r}\simeq40$~nA.

The flux qubit, with an externally applied magnetic flux $\Phi\approx\Phi_0/2$, being $\Phi_0=h/2e$ the flux quantum, behaves effectively as a two-level system. Since the second excited state is at a much higher energy (typically $\sim$ 30~GHz), the effective Hamiltonian can be written as\footnote{In this chapter we obviate the hat in the notation for operators.} $H_q=-(\varepsilon\sigma_z+\Delta\sigma_x)/2$ using the Pauli matrix notation in the basis of the persistent current states $\lbrace|\circlearrowright\,\rangle$, $|\circlearrowleft\,\rangle\rbrace$. Here $\varepsilon=2I_p(\Phi-\Phi_0/2)$, with $I_p$ the persistent current in the qubit loop. $\Delta$ is the tunnel coupling between the two persistent current states. The qubit is inductively coupled to a dc-SQUID detector with a mutual inductance of $M_{\text{SQ}}\simeq5$~pH.

The qubit is galvanically attached to the resonator (see Fig.~\ref{fig1paperBS}c) with a coupling wire of length $l=5~\mu$m, width $w=100$~nm and thickness $t=50$~nm. To achieve a high coupling we use the kinetic inductance $L_K$ of the wire that can easily be made larger than the geometric contribution. The kinetic inductance for our narrow dirty wire is found from its normal state resistance \cite{TinkhamBook} $L_K=0.14\hbar R_n/k_BT_c\simeq(25\pm2)$~pH. The strength of the coupling can be approximated by $ g=I_pI_{\mathrm{rms}}L_K$ \cite{Lindstrom07, Bourassa09}. Since our $\sim500$ $\mu$m $LC$ resonator is much smaller than the wavelength at the resonance frequency ($\lambda_r\approx20$ mm), the current is uniform along the superconducting wires connecting the capacitor plates. Therefore the position of the qubit along the inductor will not affect the coupling strength. 

\section{Theory}

Let us consider the Hamiltonian of two qubits coupled to a common resonator. The qubit-cavity coupling is of dipolar nature, and thus in the basis $\lbrace|\circlearrowright\, , n \rangle,|\circlearrowleft\, , n \rangle\rbrace$, where $\{\ket{n}\}$ refers to a Fock-state basis of the resonator, we can write
\beq \label{Hamiltonian2FQsCavity}
H = \sum_{i=1}^2 \left( \frac{\varepsilon_i}{2} \sigma_z^{(i)} + \frac{\Delta_i}{2} \sigma_x^{(i)} \right) + \omega_r \left( a^{\dagger}a + 1/2 \right) + \sum_{i=1}^2 g_i \sigma_z^{(i)} a^{\dagger} + \mathrm{H.c.},
\eeq
where $g$ is the intensity of qubit-cavity coupling and $a^{\dagger}$ ($a$) creates (annihilates) a photon. Switching to the qubit eigenbasis $\{\ket{g,n},\ket{e,n}\}$ this Hamiltonian reads
\beq \label{Hamiltonian2FQsCavityDiagonal}
H = \sum_{i=1}^2 \frac{\omega_q^{(i)}}{2} \sigma_z^{(i)} + \omega_r \left( a^{\dagger}a + 1/2 \right) + \sum_{i=1}^2 g_i \left( \cos(\theta_i) \sigma_z^{(i)} - \sin(\theta_i) \sigma_x^{(i)} \right) \left( a^{\dagger} + a \right),
\eeq
where $\omega_q^{(i)} := \sqrt{\varepsilon_i^2 + \Delta_i^2}$ is the eigenfrequency of qubit $i$, $\tan(\theta_i):=\Delta_i/\varepsilon_i$, and we have assumed $g_i$ to be real.
In terms of rising and lowering operators $\sigma_{\pm} := (\sigma_x \pm i \sigma_y)/2$ we have
\beq \label{H2FQsCavityCRTs}
H &=& \sum_{i=1}^2 \frac{\omega_q^{(i)}}{2} \sigma_z^{(i)} + \omega_r \left( a^{\dagger}a + 1/2 \right) + \sum_{i=1}^2 g_i \cos(\theta_i) \sigma_z^{(i)} (a^{\dagger} + a) \nonumber \\&& - \sum_{i=1}^2 g_i \sin(\theta_i) \left( \sigma_{+}^{(i)} a + \sigma_{-}^{(i)} a^{\dagger} \right) -  \sum_{i=1}^2 g_i \sin(\theta_i) \left( \sigma_{+}^{(i)} a^{\dagger} + \sigma_{-}^{(i)} a \right).
\eeq
Here, the third component corresponds to the co-rotating terms, while the last component corresponds to the counter-rotating terms. For $g\ll \omega_r, \omega_q$ the CRTs are typically neglected. In the USC regime, however, these play an important role in the dynamics, as we will see in the following. The relevance of the CRTs can already be seen from the Rabi formula (\ref{RabiFormula}). There, $\delta\to \omega_r-\omega_q$ for the co-rotating terms, and $\delta\to \omega_r+\omega_q$ for the CRTs. This gives a vanishing transition probability for $\delta\gg g$, which close to resonance, $\omega_r\approx \omega_q$, and in the weak coupling regime, $g\ll \omega_r, \omega_q$, occurs only for the counter-rotating component. Alternatively, the neglect of the CRTs (RWA) can be reasoned from Eq.~(\ref{nonSecME}). For the co-rotating component one obtains a similar equation with $\omega_{ab}\to \omega_r$ and $\omega_{cd}\to \omega_q$, while one has $\omega_{ab}\to \omega_r$ and $\omega_{cd}\to -\omega_q$ for the CRTs. Since $\Delta t \sim 1/g$, this makes that for large $g$ the integral is not negligible for neither of the contributions.

The Hamiltonian (\ref{H2FQsCavityCRTs}) cannot be diagonalized analytically without further approximation. We will perturbatively eliminate the interaction concerning the CRTs. The idea is that choosing a proper perturbative parameter, the RWA will be appropriate from certain order. First, we transform the Hamiltonian according to
\beq
H \to e^S H e^{-S} = H + [S,H] + \frac{1}{2!} [S,[S,H]] + \ldots
\eeq
with
\beq
S = \sum_j \gamma_j \left( \sigma_{+}^{(j)} a^{\dagger} - \sigma_{-}^{(j)} a \right),
\eeq
where we have made implicit that the sum runs for $j=1,2$. This transformation gives
\begin{equation} \label{termsBStransformation}
\begin{array}{l}
\Big[ S,\sum_i \frac{\omega_q^{(i)}}{2} \sigma_z^{(i)} \Big] = - \sum_i \gamma_i\omega_q^{(i)}  \left( \sigma_{+}^{(i)} a^{\dagger} + \sigma_{-}^{(i)} a \right). \\ \\
\frac{1}{2!} \Big[ S, \Big[ S, \sum_i \frac{\omega_q^{(i)}}{2} \sigma_z^{(i)} \Big] \Big] =  - \sum_i \frac{\gamma_i^{2}}{2} \omega_q^{(i)} \sigma_z^{(i)} (1+2\hat{n}). \\ \\
\Big[ S, \omega_r \left( a^{\dagger}a +1/2 \right) \Big] = - \sum_i \gamma_i \omega_r \left( \sigma_{+}^{(i)} a^{\dagger} + \sigma_{-}^{(i)} a \right). \\ \\
\frac{1}{2!}\Big[ S, \Big[ S, \omega_r \left( a^{\dagger}a +1/2 \right) \Big]\Big] = - \sum_i \frac{\gamma_i^2}{2} \omega_r \sigma_z^{(i)} (1+2\hat{n}) +  \sum_{ij} \frac{\gamma_i\gamma_j}{2} \omega_r \left( \sigma_{+}^{(i)}\sigma_{-}^{(j)} + \sigma_{-}^{(i)}\sigma_{+}^{(j)} \right). \\ \\
\Big[ S, -\sum_i g_i \sin(\theta_i) \left( \sigma_{+}^{(i)} a + \sigma_{-}^{(i)} a^{\dagger} \right) \Big] \\ = -\sum_i \gamma_i g_i \sin(\theta_i) \sigma_z^{(i)} \left( (a^{\dagger})^2 + a^2 \right) + \sum_{ij} \gamma_i g_i \sin(\theta_i) \left( \sigma_{+}^{(i)}\sigma_{+}^{(j)} + \sigma_{-}^{(i)}\sigma_{-}^{(j)} \right). \\ \\
\frac{1}{2!}\Big[ S, \Big[ S, -\sum_i g_i \sin(\theta_i) \left( \sigma_{+}^{(i)} a + \sigma_{-}^{(i)} a^{\dagger} \right) \Big]\Big] \\ = \sum_i \gamma_i^2 g_i \sin(\theta_i) \left( \sigma_{+}^{(i)} (a^{\dagger})^3 + \sigma_{-}^{(i)} a^3 \right) + \sum_i \gamma_i^2 g_i \sin(\theta_i) \left( \sigma_{+}^{(i)} a \hat{n} + \sigma_{-}^{(i)} \hat{n} a^{\dagger} \right) \\ + \sum_{ij} \frac{\gamma_i^2}{2} g_i \sin(\theta_i) \left( \sigma_{+}^{(i)} a + \sigma_{-}^{(i)} a^{\dagger} \right) \sigma_z^{(j)} + \sum_{ij} \gamma_i\gamma_j g_i \sin(\theta_i) \left( \sigma_{+}^{(i)} a + \sigma_{-}^{(i)} a^{\dagger} \right) \sigma_z^{(j)}. \\ \\
\Big[ S, -\sum_i g_i \sin(\theta_i) \left( \sigma_{+}^{(i)} a^{\dagger} + \sigma_{-}^{(i)} a \right) \Big] \\ = -\sum_i \gamma_i g_i \sin(\theta_i) \sigma_z^{(i)} (1+2\hat{n}) + \sum_{ij} \gamma_i g_i \sin(\theta_i) \left( \sigma_{+}^{(i)}\sigma_{-}^{(j)} + \sigma_{-}^{(i)}\sigma_{+}^{(j)} \right). \\ \\
\frac{1}{2!}\Big[ S, \Big[ S, -\sum_i g_i \sin(\theta_i) \left( \sigma_{+}^{(i)} a^{\dagger} + \sigma_{-}^{(i)} a \right) \Big]\Big] \\ = \sum_i 2\gamma_i^2 g_i \sin(\theta_i) \left( \sigma_{+}^{(i)} \hat{n} a^{\dagger} + \sigma_{-}^{(i)} a \hat{n} \right) + \sum_{ij} \frac{\gamma_i^2}{2} g_i \sin(\theta_i) \left( \sigma_{+}^{(i)} a^{\dagger} + \sigma_{-}^{(i)} a \right) \sigma_z^{(j)} \\ + \sum_{ij} \gamma_i\gamma_j g_i \sin(\theta_i) \left( \sigma_{+}^{(i)} a^{\dagger} + \sigma_{-}^{(i)} a \right) \sigma_z^{(j)}. \\ \\
\Big[ S, \sum_i g_i \cos(\theta_i) \sigma_z^{(i)} (a^{\dagger} + a) \Big] \\ = -\sum_i 2\gamma_i g_i \cos(\theta_i) \left( \sigma_{+}^{(i)} a^{\dagger} + \sigma_{-}^{(i)} a \right) (a^{\dagger} + a) - \sum_{ij} \gamma_i g_i \cos(\theta_i) \left( \sigma_{+}^{(i)} + \sigma_{-}^{(i)} \right) \sigma_z^{(j)}. \\ \\
\frac{1}{2!}\Big[ S, \Big[ S, \sum_i g_i \cos(\theta_i) \sigma_z^{(i)} (a^{\dagger} + a) \Big]\Big] = \mbox{(second-order off-resonant terms)}.
\end{array}
\end{equation}
Here $\hat{n}\equiv a^{\dagger}a$. Now, choosing
\beq \label{gamma}
\gamma_i = \frac{-g_i \sin(\theta_i)}{\omega_q^{(i)} + \omega_r},
\eeq
the counter-rotating term $-  \sum_{i=1}^2 g_i \sin(\theta_i) \left( \sigma_{+}^{(i)} a^{\dagger} + \sigma_{-}^{(i)} a \right)$ is eliminated under the transformation. Furthermore, close to resonance, $\omega_r\approx\omega_q$, the result can be simplified neglecting oscillating terms of order $\gamma_i^2$. This can now be done as the qubit-cavity coupling term has acquired a prefactor $\gamma^2 = g^2\sin^2(\theta)/(\omega_q + \omega_r)^2$ under the transformation, which for the experimental parameters, $g/(2\pi)\simeq 0.81$~GHz, $\omega_q/(2\pi)\simeq 4.20$~GHz and $\omega_r/(2\pi)\simeq 8.13$~GHz as we will see, takes the value $\gamma^2 \simeq 0.004$ at the degeneracy point, $\sin(\theta)=1$. Accordingly, we can safely neglect second-order off-resonant terms after the transformation, that is, 
$\sum_i \gamma_i^2 g_i \sin(\theta_i) \left( \sigma_{+}^{(i)} (a^{\dagger})^3 + \sigma_{-}^{(i)} a^3 \right)$
in the sixth term of (\ref{termsBStransformation}) and the eighth and tenth terms in (\ref{termsBStransformation}). Higher-order terms can also be neglected. 
Doing this we obtain the effective Hamiltonian
\beq \label{HamiltonianBSafterUtransf}
H&=&\sum_i \frac{\omega_q^{(i)}}{2} \sigma_z^{(i)} + \omega_r \left( a^{\dagger}a + 1/2 \right) + \sum_i \frac{g_i^2 \sin^2 (\theta_i)}{2( \omega_q^{(i)} + \omega_r )} \sigma_z^{(i)} (1+2\hat{n}) \nonumber \\&-& \sum_i g_i \sin(\theta_i) \left( \sigma_{+}^{(i)} a + \sigma_{-}^{(i)} a^{\dagger} \right) + \sum_i \frac{g_i^2 \sin^2(\theta_i)}{\omega_q^{(i)}+\omega_r} \sigma_z^{(i)} \left( (a^{\dagger})^2 + a^2 \right) \nonumber \\ &+& \sum_i \frac{g_i^3\sin^3(\theta_i)}{(\omega_q^{(i)}+\omega_r)^2} \left( \sigma_{+}^{(i)} a \hat{n} + \sigma_{-}^{(i)} \hat{n} a^{\dagger} \right) + \sum_i \frac{2g_i^2 \sin(\theta_i)\cos(\theta_i)}{\omega_q^{(i)} + \omega_r} \left( \sigma_{+}^{(i)} a^{\dagger} + \sigma_{-}^{(i)} a \right) (a^{\dagger} + a) \nonumber \\ &+& \sum_{ij} \left[ \frac{g_i \sin(\theta_i)}{2(\omega_q^{(i)}+\omega_r)} + \frac{g_j \sin(\theta_j)}{\omega_q^{(j)}+\omega_r} \right] \frac{g_i^2 \sin^2(\theta_i)}{\omega_q^{(i)}+\omega_r} \left( \sigma_{+}^{(i)} a + \sigma_{-}^{(i)} a^{\dagger} \right) \sigma_z^{(j)} \nonumber\\ &+& \sum_{ij} \left[ \frac{\omega_r g_j \sin(\theta_j)}{2(\omega_q^{(j)}+\omega_r)} - g_i \sin(\theta_i) \right] \frac{g_i \sin(\theta_i)}{\omega_q^{(i)} + \omega_r} \left( \sigma_{+}^{(i)} \sigma_{-}^{(j)} + \sigma_{-}^{(i)}\sigma_{+}^{(j)}  \right) \nonumber\\ &-& \sum_{ij} \frac{g_i^2 \sin^2(\theta_i)}{\omega_q^{(i)} + \omega_r} \left( \sigma_{+}^{(i)}\sigma_{+}^{(j)} + \sigma_{-}^{(i)}\sigma_{-}^{(j)} \right) + \sum_{ij} \frac{g_i^2 \sin(\theta_i) \cos(\theta_i)}{\omega_q^{(i)} + \omega_r} \left( \sigma_{+}^{(i)} + \sigma_{-}^{(i)} \right) \sigma_z^{(j)}.
\eeq
Let us analyze the different terms appearing in this Hamiltonian:\\
$\bullet$ The third term corresponds to the so-called Bloch-Siegert shift. It is a shift in the spectrum of the qubit proportional to the number of photons in the cavity. Alternatively, it can be seen as a shift of the cavity frequency depending on the state of the qubit. \\
$\bullet$ The fifth term corresponds to two-photon processes. This can be eliminated by a new canonical transformation, giving rise to an effective coupling between the two qubits proportional to the number of photons in the cavity (which we can name BS interaction). \\
$\bullet$ The sixth term is a correction to the Jaynes-Cummings (JC) coupling (fourth term). It gives a small renormalization of the system-cavity coupling constant. \\
$\bullet$ The seventh term represents a shift of the JC frequency proportional to the position of the oscillator. This term is only relevant when we are in a regime not too close to the degeneracy point ($\cos(\theta)\approx 0$) and not too far from it ($\sin(\theta)\approx 0$). \\
$\bullet$ The eight, ninth, tenth and eleventh terms involve interaction between qubits. This can occur via the cavity (eighth term), or through a direct coupling (ninth, tenth and eleventh terms). Notice that in the summations, the condition $i\neq j$ is not implied. For $i=j$, the tenth term vanishes when we take matrix elements, while the other terms contribute as a small energy shift. Below we will focus on the single-qubit case, thereby not having a contribution from these terms in the energy spectrum.

The two-photon first-order process (fifth term in (\ref{HamiltonianBSafterUtransf}) can be eliminated through the transformation
\beq
H \to e^{V} H e^{-V}, \;\;\mbox{with}\;\; V = \sum_j \eta_j \sigma_z^{(j)} \left( (a^{\dagger})^2 - a^2 \right).
\eeq
This gives
\begin{equation} \label{termsBStransformation2}
\begin{array}{l}
\Big[ V, \sum_i \frac{\omega_q^{(i)}}{2} \sigma_z^{(i)} \Big] = 0. \\ \\
\Big[ V, \omega_r (a^{\dagger}a+1/2) \Big] = -\sum_i 2\eta_i \omega_r \sigma_z^{(i)} \left( (a^{\dagger})^2 + a^2 \right). \\ \\
\frac{1}{2!}\Big[ V, \Big[ V, \omega_r (a^{\dagger}a+1/2) \Big]\Big] = -\sum_{ij} 4 \eta_i\eta_j \omega_r \sigma_z^{(i)}\sigma_z^{(j)} (1+2\hat{n}). \\ \\
\Big[ V, -\sum_i g_i \sin(\theta_i) \left( \sigma_{+}^{(i)} a + \sigma_{-}^{(i)} a^{\dagger} \right) \Big] \\ = - \sum_i 2 \eta_i g_i \sin(\theta_i) \left( (a^{\dagger})^2 - a^2 \right) \left( \sigma_{+}^{(i)} a - \sigma_{-}^{(i)} a^{\dagger} \right) + \sum_{ij} \eta_i g_i \sin(\theta_i) \sigma_z^{(i)} \left( \sigma_{+}^{(j)} a^{\dagger} + \sigma_{-}^{(j)} a \right).
\end{array}
\end{equation}
Taking
\beq
\eta_i = \frac{g_i^2 \sin^2(\theta_i)}{2\omega_r(\omega_q^{(i)}+\omega_r)},
\eeq
the two-photon process is eliminated. Also notice that the last term in (\ref{termsBStransformation2}) will be neglected since close to resonance, $\omega_r\approx\omega_q$, it corresponds to a second order ($\gamma_i^2$) term oscillating rapidly. 
The rest of the summands transformed under $V$ give also either second-order off-resonant contributions or $\gamma_i^3$ terms, so they can be neglected. Under this second transformation, the Hamiltonian reads then
\beq
H&=&\sum_i \frac{\omega_q^{(i)}}{2} \sigma_z^{(i)} + \omega_r \left( a^{\dagger}a + 1/2 \right) + \sum_i \frac{g_i^2 \sin^2 (\theta_i)}{2( \omega_q^{(i)} + \omega_r )} \sigma_z^{(i)} (1+2\hat{n}) - \sum_i g_i \sin(\theta_i) \left( \sigma_{+}^{(i)} a + \sigma_{-}^{(i)} a^{\dagger} \right) \nonumber \\&+& \sum_i \frac{g_i^3\sin^3(\theta_i)}{(\omega_q^{(i)}+\omega_r)^2} \left( \sigma_{+}^{(i)} a \hat{n} + \sigma_{-}^{(i)} \hat{n} a^{\dagger} \right) + \sum_i \frac{2g_i^2 \sin(\theta_i)\cos(\theta_i)}{\omega_q^{(i)} + \omega_r} \left( \sigma_{+}^{(i)} a^{\dagger} + \sigma_{-}^{(i)} a \right) (a^{\dagger} + a) \nonumber \\ &-& \sum_{ij} \frac{g_i^2 \sin^2(\theta_i) g_j^2 \sin^2(\theta_j)}{\omega_r(\omega_q^{(i)} + \omega_r)(\omega_q^{(j)} + \omega_r)} \sigma_z^{(i)}\sigma_z^{(j)} (1+2\hat{n}) \nonumber \\ &+& \sum_{ij} \left[ \frac{g_i \sin(\theta_i)}{2(\omega_q^{(i)}+\omega_r)} + \frac{g_j \sin(\theta_j)}{\omega_q^{(j)}+\omega_r} \right] \frac{g_i^2 \sin^2(\theta_i)}{\omega_q^{(i)}+\omega_r} \left( \sigma_{+}^{(i)} a + \sigma_{-}^{(i)} a^{\dagger} \right) \sigma_z^{(j)} \nonumber\\ &+& \sum_{ij} \left[ \frac{\omega_r g_j \sin(\theta_j)}{2(\omega_q^{(j)}+\omega_r)} - g_i \sin(\theta_i) \right] \frac{g_i \sin(\theta_i)}{\omega_q^{(i)} + \omega_r} \left( \sigma_{+}^{(i)} \sigma_{-}^{(j)} + \sigma_{-}^{(i)}\sigma_{+}^{(j)}  \right) \nonumber\\ &-& \sum_{ij} \frac{g_i^2 \sin^2(\theta_i)}{\omega_q^{(i)} + \omega_r} \left( \sigma_{+}^{(i)}\sigma_{+}^{(j)} + \sigma_{-}^{(i)}\sigma_{-}^{(j)} \right) + \sum_{ij} \frac{g_i^2 \sin(\theta_i) \cos(\theta_i)}{\omega_q^{(i)} + \omega_r} \left( \sigma_{+}^{(i)} + \sigma_{-}^{(i)} \right) \sigma_z^{(j)}.
\eeq

Finally, we can eliminate the sixth term of the previous Hamiltonian by applying the transformation
\beq
A = \sum_j \zeta_j \left( \sigma_{+}^{(j)} a^{\dagger} - \sigma_{-}^{(j)} a \right) \left( a^{\dagger} + a \right),
\eeq
which gives
\beq
\Big[ A, \sum_i \frac{\omega_q^{(i)}}{2} \sigma_z^{(i)} \Big] = -\sum_i \zeta_i \omega_q^{(i)} \left( \sigma_{+}^{(i)} a^{\dagger} + \sigma_{-}^{(i)} a \right) \left( a^{\dagger} + a \right),
\eeq
with
\beq
\zeta_i = \frac{2 g_i^2 \sin(\theta_i)\cos(\theta_i)}{\omega_q^{(i)}(\omega_q^{(i)} + \omega_r)},
\eeq
to eliminate the cited term. The remaining summands give $\sim\gamma_i^2$ off-resonant contributions.
In total, we have the Hamiltonian
\beq
H&=&\sum_i \frac{\omega_q^{(i)}}{2} \sigma_z^{(i)} + \omega_r \left( a^{\dagger}a + 1/2 \right) + \sum_i \frac{g_i^2 \sin^2 (\theta_i)}{2( \omega_q^{(i)} + \omega_r )} \sigma_z^{(i)} (1+2\hat{n}) - \sum_i g_i \sin(\theta_i) \left( \sigma_{+}^{(i)} a + \sigma_{-}^{(i)} a^{\dagger} \right) \nonumber \\&+& \sum_i \frac{g_i^3\sin^3(\theta_i)}{(\omega_q^{(i)}+\omega_r)^2} \left( \sigma_{+}^{(i)} a \hat{n} + \sigma_{-}^{(i)} \hat{n} a^{\dagger} \right) - \sum_{ij} \frac{g_i^2 \sin^2(\theta_i) g_j^2 \sin^2(\theta_j)}{\omega_r(\omega_q^{(i)} + \omega_r)(\omega_q^{(j)} + \omega_r)} \sigma_z^{(i)}\sigma_z^{(j)} (1+2\hat{n}) \nonumber \\ &+& \sum_{ij} \left[ \frac{g_i \sin(\theta_i)}{2(\omega_q^{(i)}+\omega_r)} + \frac{g_j \sin(\theta_j)}{\omega_q^{(j)}+\omega_r} \right] \frac{g_i^2 \sin^2(\theta_i)}{\omega_q^{(i)}+\omega_r} \left( \sigma_{+}^{(i)} a + \sigma_{-}^{(i)} a^{\dagger} \right) \sigma_z^{(j)} \nonumber\\ &+& \sum_{ij} \left[ \frac{\omega_r g_j \sin(\theta_j)}{2(\omega_q^{(j)}+\omega_r)} - g_i \sin(\theta_i) \right] \frac{g_i \sin(\theta_i)}{\omega_q^{(i)} + \omega_r} \left( \sigma_{+}^{(i)} \sigma_{-}^{(j)} + \sigma_{-}^{(i)}\sigma_{+}^{(j)}  \right) \nonumber\\ &-& \sum_{ij} \frac{g_i^2 \sin^2(\theta_i)}{\omega_q^{(i)} + \omega_r} \left( \sigma_{+}^{(i)}\sigma_{+}^{(j)} + \sigma_{-}^{(i)}\sigma_{-}^{(j)} \right) + \sum_{ij} \frac{g_i^2 \sin(\theta_i) \cos(\theta_i)}{\omega_q^{(i)} + \omega_r} \left( \sigma_{+}^{(i)} + \sigma_{-}^{(i)} \right) \sigma_z^{(j)}.
\eeq

Let us consider the single-qubit case, in which the second qubit is far detuned from resonance. In this situation, the previous Hamiltonian reduces to
\beq \label{1qubit-cavity}
H = \frac{\omega_q}{2} \sigma_z + \omega_r \left( \hat{n} +1/2 \right) + \omega_{BS} \sigma_z (\hat{n} + 1/2) + \sigma_{+} a g_n + g_n \sigma_{-} a^{\dagger},
\eeq
where
\beq
\omega_{BS} &\equiv& \frac{g^2\sin^2(\theta)}{\omega_q+\omega_r}. \\
g_n &\equiv& - g \sin(\theta) \left( 1 - \hat{n} \frac{\omega_{BS}}{\omega_q+\omega_r} \right).
\eeq
The Hamiltonian (\ref{1qubit-cavity}) is of a JC-type, and projected onto the basis $\lbrace |g,n+1\rangle,|e,n\rangle\rbrace$ is box-diagonal. Every box can be diagonalized to give the eigenvalues
\beq
\lambda_{n,\pm} = (n+1)\omega_r - \omega_{BS}/2 \pm \sqrt{\left[\delta/2+(n+1)\omega_{BS}\right]^2 + (n+1)g_{n+1}^2},
\eeq
with $\delta\equiv\omega_q-\omega_r$. Notice that JC limit (c.f. appendix \ref{JCappendix}) is recovered, taking $\omega_{BS}\to0$:
\beq
\lambda_{n,\pm}^{JC} = (n+1)\omega_r \pm \sqrt{(\delta/2)^2 + (n+1)\tilde{g}^2},
\eeq
being $\tilde{g}:=-g\sin(\theta)$. To fit the spectrum we need to consider the difference with the ground state
\beq \label{EnergySpectrumEq}
\Lambda_{n\pm} := \lambda_{n,\pm} - \lambda_{0,g}, \; \mbox{with} \; \lambda_{0,g} \equiv - \left( \frac{\delta}{2} + \frac{\omega_{BS}}{2} \right)
\eeq
\begin{figure}[t]
\center
\includegraphics[width=0.8\textwidth]{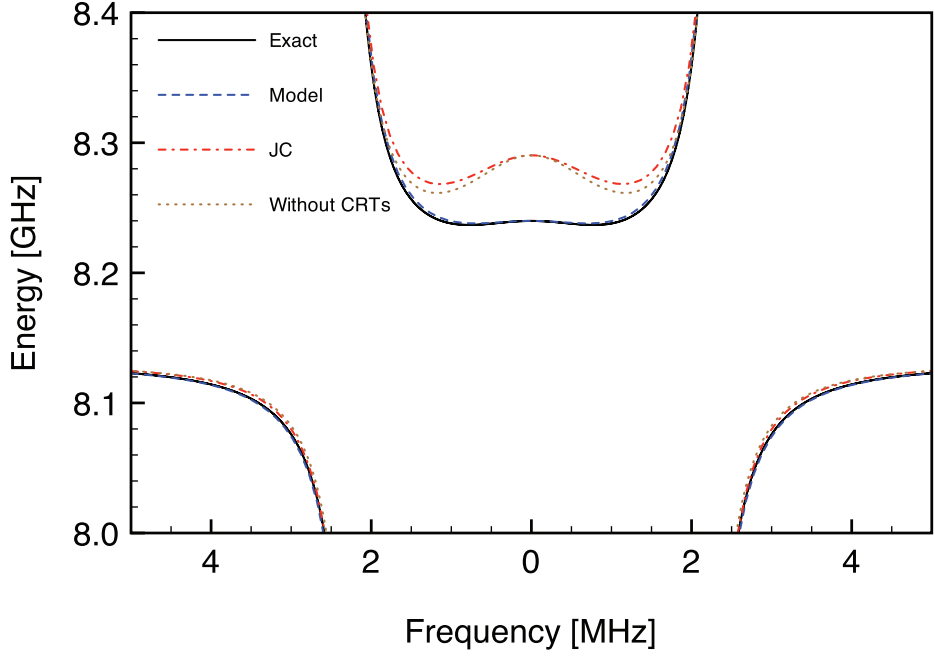}
\caption[Predicted energy spectrum of a flux qubit coupled to a $LC$ resonator]{Energy spectrum corresponding to a single flux qubit coupled to a $LC$ resonator. We show the exact solution, the model presented in the text ($\Lambda_{1-}$ given by Eq.~(\ref{EnergySpectrumEq}), the Jaynes-Cummings model spectrum (JC), and exact solution without counter-rotating terms in the Hamiltonian (Without CRTs). The values of the parameters are $g/(2\pi)=0.81$~GHz, $\omega_q/(2\pi)=4.20$~GHz and $\omega_r/(2\pi)=8.13$~GHz. The model approximates very well to the exact solution. Also, we can see that the difference with the typically employed JC model is due to the CRTs. The frequency on the horizontal axis is proportional to the bias $\varepsilon$.}
\label{spectrumBSpaper}
\end{figure}
The observed energy shift of the first exited state with respect to the JC model will be given by
\beq \label{xi}
\xi = \sqrt{(\delta/2+\omega_{BS})^2 + g_1^2} - \sqrt{(\delta/2)^2 + \tilde{g}^2}.
\eeq
Substituting $g/(2\pi)=0.81$~GHz, $\omega_q/(2\pi)=4.20$~GHz and $\omega_r/(2\pi)=8.13$~GHz, the shift at the degeneracy point is
\beq
\xi/(2\pi) \simeq -50.4 \; \mbox{MHz}.
\eeq

In figure \ref{spectrumBSpaper} we show the energy spectrum close to the first transition, $|e,0\rangle \leftrightarrow |g,1\rangle$. For the parameters given above we compare the exact solution with that given by the model presented here (Eq.~\ref{xi}). The Jaynes-Cummings result and the solution corresponding to the exact model without the counter-rotating terms are also shown. These two differ by a term $g \cos(\theta) \sigma_z (a^{\dagger} + a)$ in the Hamiltonian, which in the weak coupling regime can be neglected due to its oscillating character. We therefore observe an energy shift with respect to the standard approximations and experiments with weak coupling. Notice that this shift is encapsulated in the Bloch-Siegert frequency $\omega_{BS}$, that is, it vanishes as $\omega_{BS}\to 0$.

\begin{figure}[t]
\center
\includegraphics[width=0.7\textwidth]{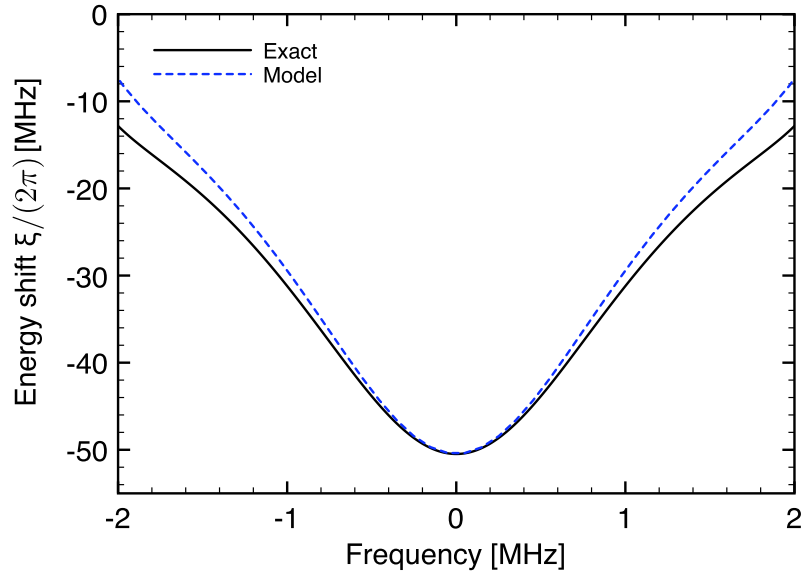}
\caption[Predicted Bloch-Siegert shift in a qubit-oscillator system]{Energy shift given by equation (\ref{xi}), and by the exact solution referred to the JC model. At the sweet-spot both curves coincide, but they start to disagree as $\varepsilon\neq 0$. The frequency on the horizontal axis is proportional to the bias $\varepsilon$.}
\label{EnergyShiftBSpaper}
\end{figure}

In figure \ref{EnergyShiftBSpaper} we show the energy shift $\xi$ as a function of frequency as given by the exact solution referred to the JC model and by Eq.~(\ref{xi})). The shift is maximum at the degeneracy point, where both curves coincide. As we turn on the bias, the energy shift becomes smaller and both curves start to differ. As we will see, the agreement with the experimental data is remarkably good.

\section{Experiment}

\begin{figure}[b]
\center
\includegraphics[width=0.75\textwidth]{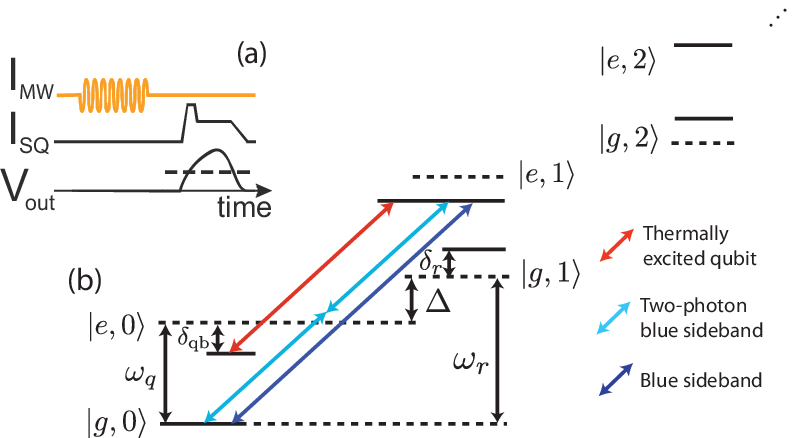}
\caption[Schematics of the Jaynes-Cummings energy levels]{\label{fig2paperBS} Measurement scheme and energy-level diagram. a)~Schematics of the measurement protocol to perform qubit spectroscopy. b)~JC ladder depicting the energy-level structure of a flux qubit coupled to a $LC$ resonator. The levels are drawn for the case 
$\delta\equiv\omega_q-\omega_r<0$. The arrows represent the level-transitions that are visible in the spectrum. The dashed lines represent the uncoupled qubit and resonator states. $\delta_q$ and $\delta_r$ are the dispersive shifts that the qubit and the resonator induce to each other.}
\end{figure}

\begin{figure}[t]
\center
\includegraphics[width=0.95\textwidth]{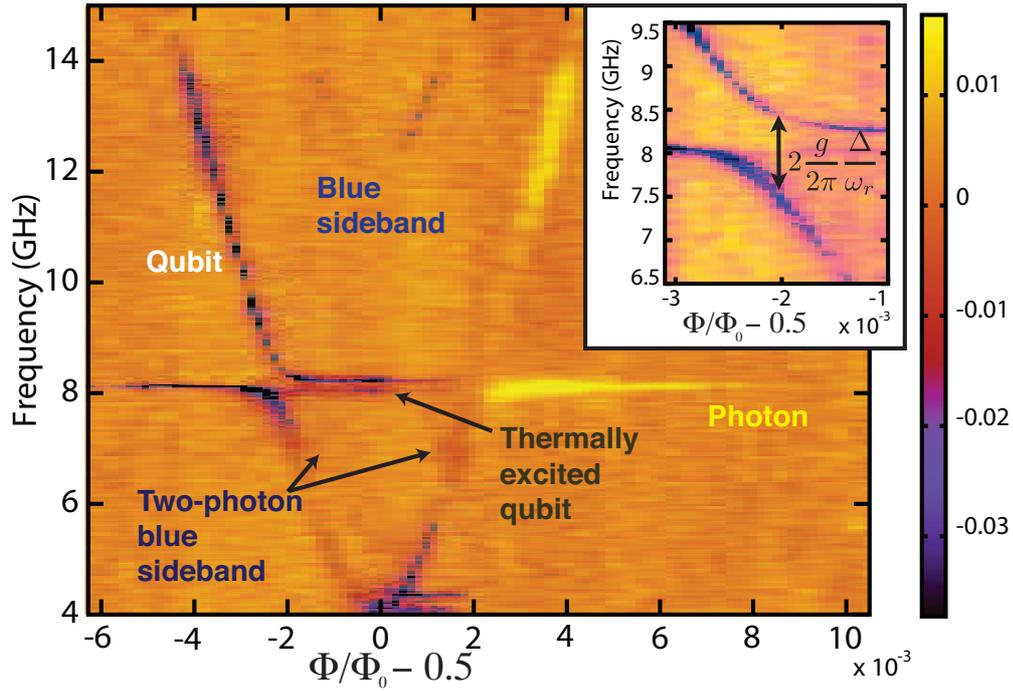}
\caption[Energy spectrum of a flux qubit coupled to a $LC$ resonator]{\label{fig3paperBS} Spectrum of the flux qubit coupled to the $LC$ resonator. An avoided-level crossing is observed at a frequency of $8.13$~GHz. 
The weak transition near $8$~GHz is associated with excited photons due to thermal population of the qubit excited state ($T_{\rm eff}\sim100~$mK at $\sim4-5~$GHz energy splitting). In the inset we show a zoom in around the resonance between qubit and oscillator. The splitting on resonance is $2g\sin(\theta)/(2\pi)\simeq0.9$~GHz.} 
\end{figure}

\begin{figure}[t]
\center
\includegraphics[width=0.7\textwidth]{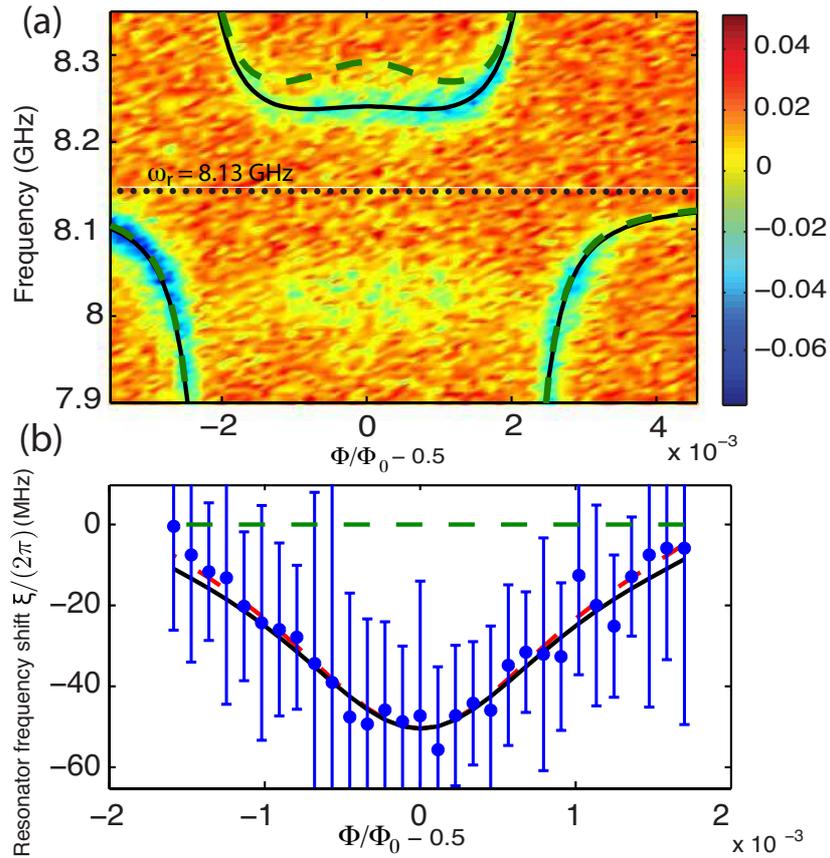}
\caption[Bloch-Siegert shift in a qubit-oscillator system]{\label{fig4paperBS} Bloch-Siegert shift. a)~Spectrum in proximity to the resonator frequency obtained using lower driving power than in Fig.~\ref{fig3paperBS} and flux pulses. The solid black line corresponds to a fit of Eq.~(\ref{Hamiltonian2FQsCavityDiagonal}), while the dashed green line is a plot of the JC model. The dashed straight line indicates the bare resonator frequency $\omega_r$. A clear deviation between the JC model and the data can be observed around the symmetry point of the qubit. b)~Difference between measurement (blue dots) and the solution given by the JC model. The solid black curve is the same as the solid black curve in (a) and the dashed red curve represents $\Lambda_{1-}$. All the curves are subtracted from the JC model. The blue dots are peak values extracted from Lorentzian fits to frequency scans at fixed flux, with the error bars representing the full width at half maximum of each Lorentzian.}
\end{figure}

We prepare the qubit in the ground state by cooling the sample to 20~mK in a dilution refrigerator. Using the protocol shown in Fig.~\ref{fig2paperBS}a, we measure the spectrum of the qubit-resonator system (Fig.~\ref{fig3paperBS}). To obtain a higher resolution in the relevant region around $8.15$~GHz, we repeated the spectroscopy using lower driving power in combination with the application of flux pulses in order to equalize the qubit signal by reading out far from its degeneracy point (Fig.~\ref{fig4paperBS}). We can identify the energy-level transitions on the basis of the JC ladder shown in Fig.~\ref{fig2paperBS}b. A large avoided crossing between states $|g,1\rangle$ and $|e,0\rangle$ is observed around a frequency of $\sim8$~GHz. This is very close to the estimated resonance frequency of the oscillator. The energy splitting $2g_\mathrm{eff}\equiv2g(\Delta/\omega_r)$ (inset of Fig.~\ref{fig3paperBS}) is approximately $0.9$~GHz. A least-squares fit of the data in Fig.~\ref{fig3paperBS} to the full Hamiltonian (Eq.~\ref{Hamiltonian2FQsCavityDiagonal}) gives the parameter values $\Delta/(2\pi)=(4.20\pm0.02)$~GHz, $I_p=(500\pm10)$~nA, 
$\omega_r/(2\pi)=(8.13\pm0.01)$~GHz 
and $g/(2\pi)=(0.82\pm0.03)$~GHz. 
The value of $g$ obtained is in good agreement with $I_pI_{\text{rms}}L_K/h=(0.83\pm0.08)$~GHz. We therefore have the ratio $g/\omega_r\approx0.1$. This large value brings us into the USC regime, and below we will demonstrate that the system indeed shows the characteristics shown in the previous section.

The spectral line of the resonator can be resolved when it is detuned several GHz away from the qubit (see Fig.~\ref{fig3paperBS}). This could be caused by the external driving when it is resonant with the oscillator. By loading photons in it, the oscillator can drive the qubit off-resonantly because of their large coupling. Another possibility is an adiabatic shift during state readout through the anticrossing of the qubit and resonator energies. The qubit readout pulse produces a negative shift of -2 m$\Phi_0$ in magnetic flux, making the spectral amplitude asymmetric with respect to the qubit symmetry point. For our parameters, this shift is coincidental with the avoided level crossing with the oscillator. Then, a state containing one photon in the resonator (e.g. $\Phi/\Phi_0-0.5=4$~m$\Phi_0$ in Fig.~\ref{fig3paperBS}) can be converted into an excited state of the qubit with very high probability, as the Landau-Zener tunneling rate is very low. Both effects, off-resonant driving and adiabatic shifting, would explain that the sign of the spectral line of the resonator coincides with the one of the qubit on both sides of the symmetry point. Irrespective of the mechanism, the spectral features of Fig.~\ref{fig3paperBS} allow us to give a lower bound for the quality factor of the resonator $Q>10^3$.

In Fig.~\ref{fig4paperBS}a a marked difference in the resonator frequency between the prediction given by our model (solid black line) and the JC solution can be clearly resolved. The difference is largest ($\sim50$~MHz as predicted in the previous section) at the symmetry point of the qubit. This energy difference is the Bloch-Siegert shift $\omega_{BS}$ associated with the counter-rotating terms. The maximum difference occurs at the symmetry point as the effective coupling $g\sin(\theta)$ decreases with increasing $\varepsilon$. Figure~\ref{fig4paperBS}b shows Eq.~(\ref{xi}) corresponding to the model presented in previous section (red dahsed line) together with the exact solution (black solid line) and the experimental measurements (blue dots). The agreement between theory and experiment is very good. The qubit should experience the same shift $\omega_{BS}$ that the resonator, but with opposite sign. Since the qubit line-width at the symmetry point around $4$~GHz is very large ($\sim80$~MHz), the Bloch-Siegert shift cannot be clearly resolved in this case. 

\section{Discussion}

In the theory presented above, we gave a result to second order in the perturbative parameter $\gamma_i=\frac{-g_i \sin(\theta_i)}{\omega_q^{(i)}+\omega_r}$. For larger couplings, namely $g_i\gtrsim \omega_q+\omega_r$, this result becomes no longer appropriate. It is however possible to rewrite the Hamiltonian (\ref{Hamiltonian2FQsCavityDiagonal}) using a transformation that allows us to extract a solution to the problem for arbitrary large coupling $g_i$. Such transformation is:
\beq
H \to e^{S} H e^{-S}, \;\;\mbox{with}\;\; S = \sum_j \lambda_j \left( \cos(\theta_j) \sigma_z^{(j)} - \sin(\theta_j)\sigma_x^{(j)}\right) a^{\dagger} - \mathrm{H.c.}
\eeq
Now, instead of transforming the full Hamiltonian (\ref{Hamiltonian2FQsCavityDiagonal}) according to $H \to H + [S,H] + \frac{1}{2!}[S,[S,H]] + \ldots$, we apply this expansion to each of the operators in the Hamiltonian. Doing this we can sum the full series, and thus find a solution to arbitrary large coupling. Under the transformation $S$, the operators change to all orders in $\lambda_j$ as
\beq
\sigma_z^{(i)} \to \sigma_z^{(i)} &-& \sin(\theta_i) \left( \cos(\theta_i) \sigma_x^{(i)} + \sin(\theta_i) \sigma_z^{(i)} \right) \left( 1- \cosh\{2\lambda_i(a^{\dagger} - a)\} \right) \nonumber\\ &+& i \sin(\theta_i) \sigma_y^{(i)} \sinh\{2\lambda_i(a^{\dagger}-a)\}. \\
\sigma_x^{(i)} \to \sigma_x^{(i)} &-& \cos(\theta_i) \left( \cos(\theta_i) \sigma_x^{(i)} + \sin(\theta_i) \sigma_z^{(i)} \right) \left( 1- \cosh\{2\lambda_i(a^{\dagger} - a)\} \right) \nonumber\\ &+& i \cos(\theta_i) \sigma_y^{(i)} \sinh\{2\lambda_i(a^{\dagger}-a)\}. \\
a \to a &-& \sum_i \lambda_i \left( \cos(\theta_i) \sigma_z^{(i)} - \sin(\theta_i) \sigma_x^{(i)} \right). \\
a^{\dagger} \to a^{\dagger} &-& \sum_i \lambda_i \left( \cos(\theta_i) \sigma_z^{(i)} - \sin(\theta_i) \sigma_x^{(i)} \right). \\
a^{\dagger}a \to a^{\dagger}a &-& \sum_i \lambda_i \left( \cos(\theta_i) \sigma_z^{(i)} - \sin(\theta_i) \sigma_x^{(i)} \right) \left(a^{\dagger} + a \right) \nonumber\\ &-& \sum_{ij} \lambda_i \lambda_j \cos(\theta_i)\cos(\theta_j) \sigma_z^{(i)} \sigma_z^{(j)} - \sum_{ij} \lambda_i \lambda_j \sin(\theta_i)\sin(\theta_j) \sigma_x^{(i)} \sigma_x^{(j)} \nonumber\\ &+& \sum_{ij} \lambda_i \lambda_j \sin(\theta_i)\cos(\theta_j) \sigma_x^{(i)} \sigma_z^{(j)} + \sum_{ij} \lambda_i \lambda_j \cos(\theta_i)\sin(\theta_j) \sigma_z^{(i)} \sigma_x^{(j)}.
\eeq
Taking $\lambda_i=g_i/\omega_r$, this gives the Hamiltonian
\beq
H &=& \sum_i \frac{\omega_q^{(i)}}{2} \sigma_z^{(i)} + \omega_r \left( a^{\dagger}a +1/2 \right) + \nonumber\\
&+& \frac{\omega_q^{(i)}}{2}\sin(\theta_i) \left[ \left( \cos(\theta_i) \sigma_x^{(i)} + \sin(\theta_i) \sigma_z^{(i)} \right) \cosh\left\{ \frac{2g_i}{\omega_r} (a^{\dagger}-a) \right\} + i\sigma_y^{(i)} \sinh\left\{ \frac{2g_i}{\omega_r} (a^{\dagger}-a) \right\} \right] \nonumber\\ &-& \sum_{ij} \frac{g_i^2}{\omega_r} \cos(\theta_i)\cos(\theta_j) \sigma_z^{(i)} \sigma_z^{(j)} - \sum_{ij} \frac{g_i^2}{\omega_r} \sin(\theta_i)\sin(\theta_j) \sigma_x^{(i)} \sigma_x^{(j)} \nonumber\\ &+& \sum_{ij} \frac{g_i^2}{\omega_r} \sin(\theta_i)\cos(\theta_j) \sigma_x^{(i)} \sigma_z^{(j)} + \sum_{ij} \frac{g_i^2}{\omega_r} \cos(\theta_i)\sin(\theta_j) \sigma_z^{(i)} \sigma_x^{(j)}.
\eeq
An analytical solution to this Hamiltonian can be found using, for example, a generalized RWA \cite{Irish07}, which will allow us to explore the physics of the system in the regime $g^2\gtrsim \omega_q\omega_r$, achievable with present circuit QED systems \cite{Devoret07}. In this regime, we expect a superradiant phase transition, thereby having a ground state with a finite number of photons \cite{KlimovBook}.

\section{Conclusions}

We have presented theory and experiment giving rise to the Bloch-Siegert shift in a system composed of a $LC$ resonator strongly coupled to a flux qubit. This demonstrates the failure of the rotating-wave approximation in this ultra-strong coupling regime of circuit QED. The large coupling $g/\omega_r\approx 0.1$ is achieved using the kinetic inductance of the wire shared by the two systems. The coupling could easily be further enhanced by increasing the kinetic inductance or by inclusion of a Josephson junction \cite{Niemczyk10, Bourassa09}. This will allow us to explore new physics that arise when $g^2\gtrsim \omega_q\omega_r$.



\chapter{Conclusions} 
\label{Chapter7}
\lhead{Chapter 7. \emph{Conclusions}} 

\begin{flushright}

\textit{``The reasonable man adapts himself to the world; the unreasonable one persists in trying\\ to adapt the world to himself. Therefore all progress depends on the unreasonable man.''}

George Bernard Shaw

\end{flushright}

In this thesis we have investigated quantum-mechanical effects that arise in different nanoscopic systems. In particular, we have centered our attention on artificial atomic systems, such as quantum dots, NV centers, or superconducting qubits. 

In chapter~\ref{Chapter1} we have presented an overview of these systems, and discussed important effects, such as the Coulomb blockade (see subsection~\ref{CBsection}). 
The central results of this thesis concern quantum transport systems, and their coupling to other systems, such as cavity resonators, optical atomic systems, and to dissipative baths. Special emphasis is made in the study of the current noise spectrum through quantum transport systems. We have learned that depending of the parameter regime (see Fig.~\ref{FDTlimitsFig}), this noise reveals different physics. Interestingly, a Markovian theory of electron transport (chapter~\ref{Chapter3}) does not capture part of this physics, namely that of quantum fluctuations, and a non-Markovian transport theory (chapter~\ref{Chapter4}) is needed to that end. 

In chapter~\ref{Chapter2} we have studied the widely used mathematical methods in quantum transport. Special emphasis has been made on the techniques to calculate correlation functions of a quantum-mechanical variable. In this chapter, different models have been studied to illustrate the distinct theorethical approaches to quantum transport. 

After this introduction, the main results of the thesis have been presented, and can be summarized as follows:

\begin{itemize}

\item In section~\ref{CNDQDsSection} we have presented an experiment showing Pauli spin blockade in carbon-nanotube double quantum dots. This, has been used to characterize the hyperfine interaction in carbon devices, contrasting both $^{12}$C and $^{13}$C samples. The hyperfine constant for a $^{13}$C host has been measured, obtaining a value two orders of magnitude larger than expected.

\item In section~\ref{FCSwPaper1Sec} we have given an introduction to our theory of frequency-dependent counting statistics of electron transport. The theory presented in this section is based on the Markovian density operator approach, and allows us to calculate arbitrary-order current correlation functions at finite frequencies (c.f. Eqs. (\ref{genF}) and (\ref{corr})). We have applied the formalism to present novel results for the skewness of the current distribution in prototypical models of quantum transport.

\item In chapter~\ref{Chapter3} we have presented in detail our theory of Markovian counting statistics of electron transport. Analytic expressions for the frequency-dependent noise and skewness have been given (Eqs. (\ref{noiseformula}) and (\ref{skewnessformula})). The results predicted by our theory for a single resonant level model have been compared with the exact solution and the fluctuation-dissipation theorem. From this comparison, we have extracted that the Markovian approximation hinders the physics of vaccum fluctuations (see Fig.~\ref{Fig5FCSpaper2}).

\item In chapter~\ref{Chapter4} we have extended our theory of counting statistics to include the case in which the system-reservoir coupling is non-Markovian. In this situation, it is essential to include initial system-bath correlations, and these contribute to the moment generating function in the form of an inhomogeneous term (see Eqs. (\ref{inhomo2}) and (\ref{Gamma})). A general non-Markovian noise formula has been derived in this chapter (see Eqs.~(\ref{Sz})-(\ref{J1nonMarkovian})), and it has been applied to obtain the spectrum of quantum noise of double quantum dot (see Fig.~\ref{Fig3paperNM}).

\item In chapter~\ref{Chapter5} we have proposed a novel hybrid quantum system, consisting of a superconducting flux qubit (FQ) coupled to an ensemble of NV centers in diamond (see Fig.~\ref{figure1paperDiamond}). The coupling between FQ and a single NV center reaches tens of kHz, but due to the low FQ coherence times is not sufficient to coherently transfer the quantum information between both systems. However, the superconducting qubit can act as a quantum bus to mediate a coherent interaction between distant NV centers. Moreover, an ensemble of NV centers can be used to achieve coupling of tens of GHz (see Fig.~\ref{figure2paperDiamond}), thereby opening the possibility of coherent transfer of the quantum information between FQ and NV ensemble.

\item In chapter~\ref{Chapter6} we have presented an experiment, along with the related theory, measuring the Bloch-Siegert shift in a qubit-oscillator system. A superconducting flux qubit has been coupled to a $LC$ resonator in the ultra-strong coupling (USC) regime, where the coupling energy is comparable to the qubit and cavity frequencies. We have discussed that in this regime, new physics is expected, such as a superradiant phase transition.

\end{itemize}

We expect that the work presented in this thesis opens new lines of research. For example, the noise theory presented in chapters \ref{Chapter3} and \ref{Chapter4} can be applied to a variety of systems, such as nano-mechanical resonators, superconductors, and quantum-optical systems. In the field of hybrid quantum systems, the coupling between solid-state qubits and light is still a challenging problem, and in this direction it will be interesting to study the coupling with plasmonic waveguides. Finally, the field of USC promises to undercover new physical phenomena, such as the mentioned Dicke phase transition with a small number of qubits.

~

~

~

~

~

~

~

~

~

~

~

~

~

~

~

~

~

~

~

~

~

~


\chapter*{Conclusiones} 
\label{Chapter7spanish}
\lhead{\emph{Conclusiones}} 





En esta tesis hemos investigado distintos efectos mecano-cu\'anticos en sistemas nanosc\'opicos. En particular, hemos centrado nuestra atenci\'on en sistemas at\'omicos artificiales, como son los puntos cu\'anticos, los centros NV en diamante o los qubits superconductores.

En el cap\'itulo~\ref{Chapter1} hemos presentado un resumen de estos sistemas, discutiendo efectos importantes, tales como el bloqueo de Coulomb (ver subsecci\'on~\ref{CBsection}).
Los resultados centrales de la tesis est\'an en relaci\'on con sistemas de transporte cu\'antico, y su acoplo con otros sistemas, como por ejemplo cavidades resonantes, sistemas \'opticos at\'omicos y ba\~nos disipativos. Especial inter\'es ha sido prestado al estudio del ruido de corriente en sistemas de transporte electr\'onico. Hemos concluido que dependiendo del r\'egimen de par\'ametros (ver Fig.~\ref{FDTlimitsFig}), este ruido captura una f\'isica diferente. Una conclusi\'on interesante es que una teor\'ia Markoviana del transporte (cap\'itulo~\ref{Chapter3}) no captura la f\'isica de las fluctuaciones de vac\'io, y una generalizaci\'on no Markoviana de la teor\'ia resulta necesaria para capturar dicha f\'isica (cap\'itulo~\ref{Chapter4}).

En el cap\'itulo~\ref{Chapter2} hemos estudiado los principales m\'etodos te\'oricos de inter\'es en el transporte cu\'antico. Especial \'enfasis ha sido hecho en las t\'ecnicas para calcular funciones de correlaci\'on de una variable cu\'antica. En este cap\'itulo, hemos estudiado distintos modelos de transporte para ilustrar las t\'ecnicas te\'oricas.

Tras esta introducci\'on, los principales resultados de la tesis se recogen en los cap\'itulos y secciones subsiguientes, y pueden resumirse como sigue:

\begin{itemize}

\item En la secci\'on~\ref{CNDQDsSection}, hemos presentado un experimento en el que se demuestra el bloqueo de esp\'in en dobles puntos cu\'anticos formados en nanotubos de carbono. Usando dicho bloqueo de esp\'in, hemos caracterizado la interacci\'on hiperfina en dispositivos de carbono, comparando muestras de $^{12}$C y $^{13}$C. La constante hiperfina para las muestras de $^{13}$C es dos \'ordenes de magnitude mayor que lo esperado por distintas teor\'ias.

\item En la secci\'on~\ref{FCSwPaper1Sec} hemos presentado una introducci\'on a nuestra teor\'ia de estad\'istica de contaje dependiente de la frecuencia. La teor\'ia desarrollada en esta parte est\'a basada en el formalismo del operador densidad y en su aproximaci\'on Markoviana; como hemos visto, esta permite el c\'alculo de funciones de correlaci\'on a frecuencias finitas y orden arbitrario (ver ecuaciones (\ref{genF}) y (\ref{corr})). El formalismo ha sido utilizado para derivar resultados para la funci\'on de correlaci\'on de corriente de tercer orden en ejemplos protot\'ipicos del transporte cu\'antico.

\item En el cap\'itulo~\ref{Chapter3}, hemos presentado en detalle nuestra teor\'ia Markoviana de estad\'istica de contaje en sistemas de transporte. Expresiones anal\'iticas para el ruido y funci\'on de correlaci\'on de tercer orden a frecuencias finitas han sido derivadas con el formalismo (ecuaciones (\ref{noiseformula}) y (\ref{skewnessformula})). Los resultados predichos por la teor\'ia para el modelo de un nivel resonante han sido comparados con la soluci\'on exacta y el teorema de fluctuaci\'on-disipaci\'on. De esta comparaci\'on, hemos extraido que la aproximaci\'on Markoviana no captura la f\'isica de las fluctuaciones de vac\'io en estos modelos (ver Fig.~\ref{Fig5FCSpaper2}).

\item En el cap\'itulo~\ref{Chapter4}, hemos generalizado nuestra teor\'ia de estad\'istica de contaje para incluir el caso en que el acoplo entre sistema cu\'antico y ba\~no es no Markoviano. En esta situaci\'on, resulta esencial incluir las correlaciones iniciales entre sistema y ba\~no; estas a\~naden un t\'ermino inomog\'eneo en la ecuaci\'on para la funci\'on generatriz (ver ecuaciones (\ref{inhomo2}) y (\ref{Gamma})). Una f\'ormula no Markoviana general ha sido derivada en este cap\'itulo (ver ecuaciones (\ref{Sz})-(\ref{J1nonMarkovian})), y ha sido utilizada para obtener el espectro de ruido de un doble punto cu\'antico (ver Fig.~\ref{Fig3paperNM}).

\item En el cap\'itulo~\ref{Chapter5}, hemos propuesto un sistema h\'ibrido cu\'antico, compuesto de un qubit superconductor de flujo acoplado a un conjunto de centros NV en diamante (ver Fig.~\ref{figure1paperDiamond}). El acoplo entre qubit de flujo y un solo centro NV alcanza unos cuantos kHz, el cual, debido a los bajos tiempos de coherencia de los qubits de flujo, no es suficiente para acoplar ambos sistemas coherentemente. En cambio, el qubit superconductor puede actuar como mediador de una interacci\'on coherente entre varios centros NV. Adem\'as, hemos demostrado que un conjunto de centros NV puede ser usado para lograr un acoplo coherente entre el qubit de flujo y el conjunto de centros NV.

\item En el cap\'itulo~\ref{Chapter6}, hemos presentado un experimento, junto con la teor\'ia subyacente, donde se realiza la medida de la variaci\'on energ\'etica ``Bloch-Siegert'' en un sistema qubit-oscillador. Un qubit de flujo ha sido acoplado a un resonador $LC$ en el r\'egimen de acoplo ultra-fuerte, en el que la energ\'ia de acoplo entre ambos sistemas es comparable a las frecuencias del qubit y la cavidad. Hemos discutido c\'omo en este r\'egimen, esperamos una f\'isica nueva, teniendo por ejemplo una transici\'on de fase superadiante.

\end{itemize}

Esperamos que el trabajo presentado en esta tesis abra nuevas l\'ineas de investigaci\'on. Por ejemplo, la teor\'ia de ruido presentada en los cap\'itulos \ref{Chapter3} y \ref{Chapter4} puede ser aplicada a una variedad de sistemas, tales como resonadores nano-mec\'anicos, superconductores y sistemas optico-cu\'anticos. En el plano de los qubits h\'ibridos, el acoplo entre sistemas de estado s\'olido y la luz es a\'un un desaf\'io experimental y te\'orico, y en esta direcci\'on ser\'a interesante estudiar por ejemplo el acoplo de estos sistemas con guias de onda plasm\'onicas. Finalmente, el campo del acoplo ultra-fuerte en \'optica cu\'antica promete revelar nuevos fen\'omenos f\'isicos, tales como la mencionada transici\'on de Dicke en sistemas con un n\'umero de qubits peque\~no.

~

~

~

~

~

~

~

~

~

~

~

~

~

~

~

~

~

~

~

~

~

~

~

~

~

~

~

~

~

~

~

~

~

~

~

~

~

~

~

~

~ 


\addtocontents{toc}{\vspace{1em}} 

\appendix 


\chapter{The quantum two-level system}
\label{2LSappendix}
\lhead{Appendix B. \emph{The quantum two-level system}}

The quantum two-level system (2LS) is one of the most recursive examples in quantum mechanics. In the basis $\{\ket{1},\ket{0}\}$, the Hamiltonian is described by the matrix
\beq \label{HamiltonianMatrix2LS}
\cal{\hat{H}} =
\begin{pmatrix}
\varepsilon & T_c \\ T_c & -\varepsilon
\end{pmatrix}.
\eeq
Defining $\tan(\theta):=T_c/\varepsilon$, the eigenvalues and associated eigenvectors read
\beq
\mu_+=\sec(\theta) \to \ket{+} = \begin{pmatrix} \cos(\theta/2) \\ \sin(\theta/2) \end{pmatrix}; &
\mu_-=-\sec(\theta) \to \ket{-} = \begin{pmatrix} -\sin(\theta/2) \\ \cos(\theta/2) \end{pmatrix};
\eeq
which means that the change of basis can be written as
\beq
\begin{pmatrix}
\ket{+} \\ \ket{-}
\end{pmatrix}
=
\begin{pmatrix}
\cos(\theta/2) & \sin(\theta/2) \\ -\sin(\theta/2) & \cos(\theta/2)
\end{pmatrix}
\begin{pmatrix}
\ket{1} \\ \ket{0}
\end{pmatrix}.
\eeq
The level splitting is given by $\Delta\equiv 2\sqrt{\varepsilon^2+T_c^2}$, which is the fundamental energy scale of the 2LS. With $\ket{1}\to\ket{L}$ and $\ket{0}\to\ket{R}$, the Hamiltonian (\ref{HamiltonianMatrix2LS}) describes the double quantum dot studied throughout the text. 
The definition of the mixing angle is of course not unique, some authors prefer to take $\sin(\theta/2)=\frac{T_c}{\sqrt{(\varepsilon+\Delta/2)^2+T_c^2}}$ and $\cos(\theta/2)=\frac{\varepsilon+\Delta/2}{\sqrt{(\varepsilon+\Delta/2)^2+T_c^2}}$ in the notation above.

~

~

~

~

~

~

~

~

~

~

~

~

~

~

~

~

~

~

~

~

~

~


\chapter{The Jaynes-Cummings model}
\label{JCappendix}
\lhead{Appendix B. \emph{The Jaynes-Cummings model}}

The Jaynes-Cummings model \cite{Jaynes-Cummings63} is a basic model in quantum optics. It captures the interaction between a two-level system (2LS) and the quantized radiation field. The Hamiltonian is
\beq
{\cal \hat{H}} = \frac{\omega_q}{2} \hat{\sigma}_z + \omega_c \left( \hat{a}^{\dagger}\hat{a} + \frac{1}{2} \right) + g \left( \hat{a}^{\dagger}\hat{\sigma}_{-} + \mathrm{H.c.} \right).
\eeq
This Hamiltonian can be written as
\begin{equation} \label{JCmatrix}
{\cal H}=
\left(
\begin{array}{cc|cc|cc}
\frac{\omega_q}{2}+\frac{\omega_c}{2} & g &  &  &  &  \\
g & -\frac{\omega_q}{2}+\frac{3\omega_c}{2} &  &  &  &  \\ \cline{1-6}
 &  & \frac{\omega_q}{2}+\frac{3\omega_c}{2} & g\sqrt{2} &  &  \\
 &  & g\sqrt{2} & -\frac{\omega_q}{2}+\frac{5\omega_c}{2} &  &  \\ \cline{1-6}
 &  &  &  & \ddots & \\
 &  &  &  &  & 
\end{array}
\right),
\end{equation}
in the basis $\{\ket{e,n},\ket{g,n+1}$, where $\ket{g}$ and $\ket{e}$ denote the ground and excited states of the 2LS respectively, and $\{\ket{n}\}$ with $n=0,1,\ldots$ is a Fock-state basis capturing the number of photons in the system. The infinite matrix (\ref{JCmatrix}) is box-diagonal, with blocks of the form
\begin{equation}
\begin{pmatrix}
\frac{\omega_q}{2}+\left(n+\frac{1}{2}\right)\omega_c & g\sqrt{n+1} \\
g\sqrt{n+1} & -\frac{\omega_q}{2}+\left(n+\frac{3}{2}\right)\omega_c 
 \end{pmatrix},
 \end{equation}
which can be readily diagonalized to give the eigenvalues
\beq
\mu_{\pm} = (n+1) \omega_c \pm \sqrt{(\delta/2)^2+(n+1)g^2},
\eeq
and respective eigenvectors
\beq
\ket{+}_n &=& \cos(\theta_n/2) \ket{e,n} + \sin(\theta_n/2) \ket{g,n+1}. \\
\ket{-}_n &=& -\sin(\theta_n/2) \ket{e,n} + \cos(\theta_n/2) \ket{g,n+1}.
\eeq
Here, we have defined the quantities $\delta\equiv\omega_q-\omega_c$, $\sin(\theta_n/2)\equiv \frac{R_n -\omega_q}{\sqrt{(R_n-\omega_q)^2+4(n+1)\omega_q^2}}$, $\cos(\theta_n/2)\equiv \frac{2g\sqrt{n+1}}{\sqrt{(R_n-\omega_q)^2+4(n+1)\omega_q^2}}$, and $R_n\equiv\sqrt{\omega_q^2+4(n+1)g^2}$.


\chapter{Supplement to section \ref{FCSwPaper1Sec}}
\label{SuppToFCSwPaper1Sec}
\lhead{Appendix C. \emph{Supplement to section \ref{FCSwPaper1Sec}}}

In this appendix we show explicitly how the result (\ref{F3wwFCSpaper1}) for the skewness arises. It is also shown how cross correlations and `total' cumulants can be calculated with the formalism presented in section \ref{FCSwPaper1Sec}.

\begin{figure}
  \begin{center}
    \includegraphics[width=\textwidth]{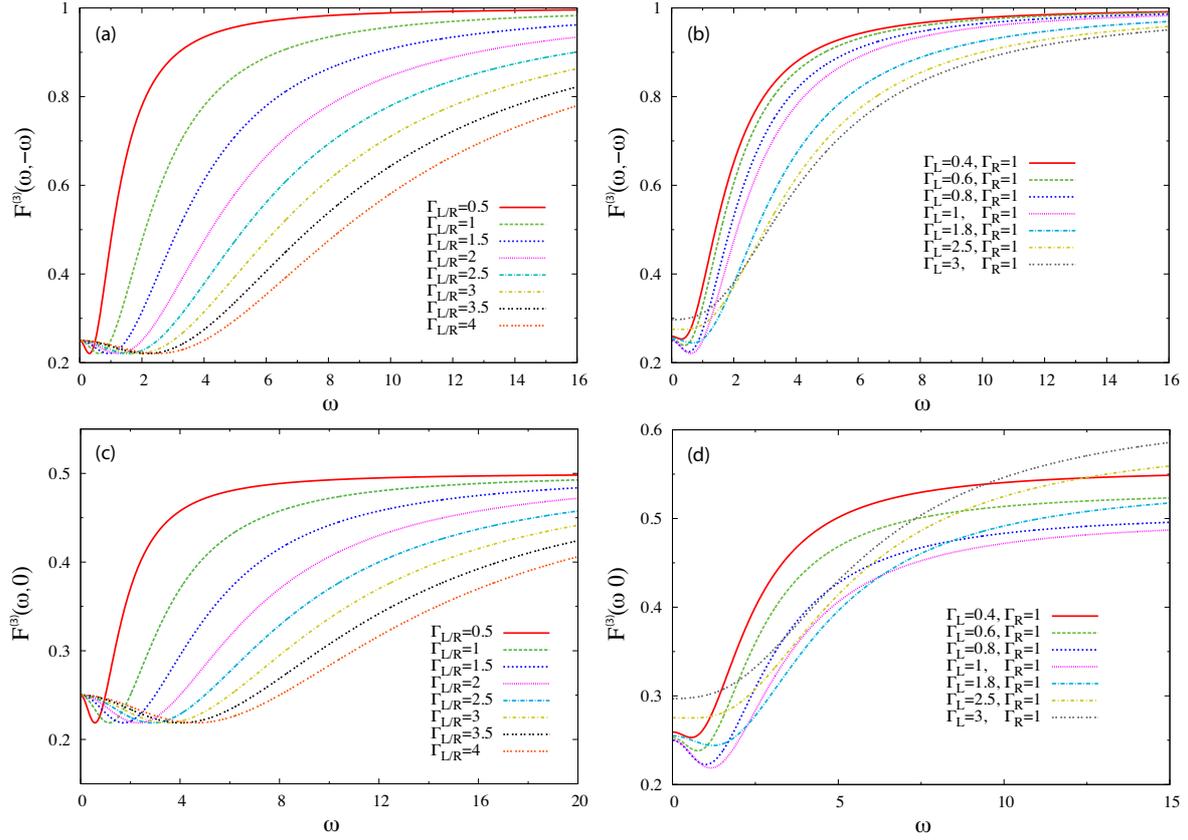}
  \end{center}  
  \caption[Behaviour of the frequency-dependent skewness]{Behaviour of the frequency-dependent third-order Fano factor. a)~$F^{(3)}(\omega,-\omega)$ for $\Gamma_L=\Gamma_R$, as this rate varies. b)~$F^{(3)}(\omega,-\omega)$ for $\Gamma_L\neq\Gamma_R$ and different rate ratios. c)~$F^{(3)}(\omega,0)$ for $\Gamma_L=\Gamma_R$, as this rate varies. d)~$F^{(3)}(\omega,0)$ for $\Gamma_L\neq\Gamma_R$ and different rate ratios.
}
  \label{F3wwCutsFig}
\end{figure}

Our starting point is the multi-time cumulant generating function (\ref{genF}) for the case $N=3$, and with the kernel (\ref{kernelSRLrightFCS}) corresponding to the single resonant level (SRL) model. This gives the correlation functions
\begin{eqnarray}\label{f_{321}}
f_{321}^{(3)} := \langle n(t_1)n(t_2)n(t_3) \rangle_c^{t_3\geq t_2\geq t_1} &=& (-i)^3\partial_{\chi_1}\partial_{\chi_2}\partial_{\chi_3} {\cal F}\;\Big|_{\chi_1=\chi_2=\chi_3=0} \nonumber\\ &=& \frac{\Gamma_L^2\Gamma_R^2\left(\Gamma_L-\Gamma_R\right)^2}{\left(\Gamma_L+\Gamma_R\right)^6}\left[ 3 + 3e^{-\Gamma\tau_1} + e^{-\Gamma\tau_2} \right] \nonumber \\ &+& \frac{\Gamma_L\Gamma_R\left(\Gamma_L^5 -\Gamma_L^4\Gamma_R + 4\Gamma_L^3\Gamma_R^2 + 4\Gamma_L^2\Gamma_R^3 - \Gamma_L\Gamma_R^4 + \Gamma_R^5 \right)}{\left(\Gamma_L+\Gamma_R\right)^6}t_1 \nonumber \\ &-& \frac{2\Gamma_L^3\Gamma_R^3}{\left(\Gamma_L+\Gamma_R\right)^5}\tau_1\left[ 2e^{-\Gamma\tau_1} + e^{-\Gamma\tau_2} \right].
\end{eqnarray}
\begin{eqnarray}\label{f_{21}}
f_{21}^{(3)}:= \langle n(t_1)n(t_2)n(t_2) \rangle_c^{t_2\geq t_1} &=& (-i)^3\partial_{\chi_2}^2\partial_{\chi_1} {\cal F}(\chi_1,\chi_2,\chi_3)\;\Big|_{\chi_1=\chi_2=\chi_3=0} \nonumber\\ &=& \frac{3\Gamma_L^2\Gamma_R^2\left(\Gamma_L-\Gamma_R\right)^2}{\left(\Gamma_L+\Gamma_R\right)^6}\left[ 1 + e^{-\Gamma\tau_1} \right] \nonumber \\ &+& \frac{\Gamma_L\Gamma_R\left(\Gamma_L^5 -\Gamma_L^4\Gamma_R + 4\Gamma_L^3\Gamma_R^2 + 4\Gamma_L^2\Gamma_R^3 - \Gamma_L\Gamma_R^4 + \Gamma_R^5 \right)}{\left(\Gamma_L+\Gamma_R\right)^6}t_1 \nonumber \\ &-& \frac{4\Gamma_L^3\Gamma_R^3}{\left(\Gamma_L+\Gamma_R\right)^5}\tau_1 e^{-\Gamma\tau_1}.
\end{eqnarray}
\begin{eqnarray}\label{f_{1}}
f_{1}^{(3)}:= \langle n(t)n(t)n(t) \rangle_c &=& (-i)^3\partial_{\chi_1}^3 {\cal F}(\chi_1,\chi_2,\chi_3)\;\Big|_{\chi_1=\chi_2=\chi_3=0} \nonumber\\ &=& \frac{6\Gamma_L^2\Gamma_R^2\left(\Gamma_L-\Gamma_R\right)^2}{\left(\Gamma_L+\Gamma_R\right)^6} \nonumber \\ &+& \frac{\Gamma_L\Gamma_R\left(\Gamma_L^5 -\Gamma_L^4\Gamma_R + 4\Gamma_L^3\Gamma_R^2 + 4\Gamma_L^2\Gamma_R^3 - \Gamma_L\Gamma_R^4 + \Gamma_R^5 \right)}{\left(\Gamma_L+\Gamma_R\right)^6}t. \; \; \; \; \; \;
\end{eqnarray}
Here, we have defined $\tau_1\equiv t_2-t_1$ and $\tau_2\equiv t_3-t_2$. Of course expressions (\ref{f_{21}}) and (\ref{f_{1}}) can be obtained as limiting cases of (\ref{f_{321}}) in which various times are equal. The symmetrized time-correlation function of the number of tunneled particles through the right contact can be then obtained as
\begin{eqnarray}
\langle n(t_1)n(t_2)n(t_3)\rangle_c &=& f_{321}^{(3)}\theta_{32}\theta_{21} + f_{312}^{(3)}\theta_{31}\theta_{12} + f_{231}^{(3)}\theta_{23}\theta_{31} \nonumber\\ &+&  f_{213}^{(3)}\theta_{21}\theta_{13} +  f_{132}^{(3)}\theta_{13}\theta_{32} +  f_{123}^{(3)}\theta_{12}\theta_{23},
\end{eqnarray}
where $\theta(t)$ is the unit step function. We are nevertheless interested in the current cumulant, which can be simply obtained as 
\beq \label{number-current-relation}
\langle I(t_1) I(t_2) I(t_3) \rangle_c = \partial_{t_1}\partial_{t_2}\partial_{t_3} \langle n(t_1)n(t_2)n(t_3) \rangle_c.
\eeq
The frequency-dependent skewness, $S^{(3)}(\omega_1,\omega_2,\omega_3)$, follows from the Fourier transform of the resulting expression. Due to time translational invariance of the stationary state, $S^{(3)}$ is proportional to $\delta(\omega_1+\omega_2+\omega_3)$, so it only depends on two frequencies. Notice that to obtain the frequency dependent skewness, it is not necessary to perform the time derivatives in expression (\ref{number-current-relation}), since the Fourier transform of a derivative gives simply multiplication by the associated frequency. Following this procedure, expression (\ref{F3wwFCSpaper1}) is obtained. This has the interesting limits
\beq
F^{(3)}(\omega,-\omega) &=& 1 - \frac{6\Gamma_L\Gamma_R\left( \Gamma_L^2 + \Gamma_R^2 +3\omega^2 \right)}{\Gamma^4+5\Gamma^2\omega^2+4\omega^4}. \\
F^{(3)}(\omega,0) &=& 1 - \frac{2\Gamma_L\Gamma_R \left(\Gamma_L^2+\Gamma_R^2+\omega^2\right)\left(3\Gamma^2+\omega^2\right)}{\Gamma^2\left[\Gamma^2+\omega^2\right]^2}. \\
F^{(3)}(\omega,\infty) &=& 1 - \frac{2\Gamma_L\Gamma_R}{\Gamma^2+\omega^2} = F^{(2)}(\omega). \\
F^{(3)}(0,0) &=& \frac{\Gamma_L^4-2\Gamma_L^3\Gamma_R+6\Gamma_L^2\Gamma_R^2-2\Gamma_L\Gamma_R^3+\Gamma_R^4}{\Gamma^4}. \\
F^{(3)}(0,\infty) &=& \frac{\Gamma_L^2+\Gamma_R^2}{\Gamma^2} = F^{(2)}(0). \\
F^{(3)}(\infty,\infty) &=& 1.
\eeq
The behaviour of $F^{(3)}$ in a SRL model for different system parameters is shown in Fig.~\ref{F3wwCutsFig}.

It is important to notice that the step function used above includes $t=0$ in its definition, such that $\theta(t)+\theta(-t)=1$ for all $t$. If $t=0$ is excluded from the definition of the step function, we can proceed adding the self-correlation terms `by hand' \cite{Hershfield93}. That is, we would need to consider also the terms $f_{21}^{(2)} :=\langle n(t_1)n(t_2)\rangle_c^{t_2\geq t_1}$ and $f_{1}^{(1)} := \langle n(t) \rangle_c = \langle n(t) \rangle$, which for the SRL model read respectively $ \frac{\Gamma_L^2\Gamma_R^2}{\left(\Gamma_L+\Gamma_R\right)^4} + \frac{\Gamma_L\Gamma_R\left(\Gamma_L^3+\Gamma_L^2\Gamma_R+\Gamma_L\Gamma_R^2+\Gamma_R^3\right)}{\left(\Gamma_L+\Gamma_R\right)^4}t_1 + \frac{\Gamma_L^2\Gamma_R^2}{\left(\Gamma_L+\Gamma_R\right)^4} e^{-(\Gamma_L+\Gamma_R)\tau_1}$ and $\frac{\Gamma_L\Gamma_R}{\left(\Gamma_L+\Gamma_R\right)}t$, where again $\tau_1\equiv t_2-t_1$. Then, the symmetric third-order correlation function is derived as
\beq
\langle n(t_1)n(t_2)n(t_3)\rangle_c &=&f_{321}^{(3)}\theta_{32}\theta_{21} + f_{312}^{(3)}\theta_{31}\theta_{12} + f_{231}^{(3)}\theta_{23}\theta_{31} +  f_{213}^{(3)}\theta_{21}\theta_{13} \nonumber \\ &+&  f_{132}^{(3)}\theta_{13}\theta_{32} +  f_{123}^{(3)}\theta_{12}\theta_{23} + f_{32}^{(2)}\theta_{32}\delta_{21} + f_{23}^{(2)}\theta_{23}\delta_{21} \nonumber\\  &+& f_{31}^{(2)}\theta_{31}\delta_{32} + f_{13}^{(2)}\theta_{13}\delta_{32} + f_{21}^{(2)}\theta_{21}\delta_{31} + f_{12}^{(2)}\theta_{12}\delta_{31} +  f_{1}^{(1)}\delta_{32}\delta_{21}. \;\;\;\;\;\;\;\;\;\;
\eeq

Cross correlations of the form $\langle I_L(t_1)I_R(t_2)\rangle_c$, concerning currents at different contacts, can be also obtained with the formalism presented in section \ref{FCSwPaper1Sec}. Similarly to what was done in this section, we use the quantum jump approach, and resolve the density operator in two indexes, $n$ and $m$, referring to the number of particles tunneled through the left and right junctions respectively:
\beq
\dot{\bm \rho}_S^{(n,m)}(t) = \mathcal{W}_0 {\bm \rho}_S^{(n,m)}(t) + \mathcal{W}_L {\bm \rho}_S^{(n-1,m)}(t) + \mathcal{W}_R {\bm \rho}_S^{(n,m-1)}(t).
\eeq
Here, ${\cal W}_{L/R}$ is the part of the kernel containing the `jump' processes at the left/right barrier, ${\cal W}_0$ containing no `jump' processes, and we have assumed unidirectional tunneling. 
Multipliying by $e^{-in\chi^L}e^{im\chi^R}$ and summing over $n$ and $m$ we get
\beq
\dot{\bm \rho}_S(\chi^L,\chi^R,t) = \mathcal{W}_0 {\bm \rho}_S(\chi^L,\chi^R,t) + \mathcal{W}_L  e^{-i\chi^L} {\bm \rho}_S(\chi^L,\chi^R,t) + \mathcal{W}_R e^{i\chi^R} {\bm \rho}_S(\chi^L,\chi^R,t),
\eeq
where we have defined $\rho_S(\chi^L,\chi^R,t) := \sum_{n,m} e^{-in\chi^L}e^{im\chi^R} \rho_S^{(n,m)}(t)$. The probability distribution of having counted $n$ particles at the left contact and $m$ particles at the right contact after time $t$ is simply $P(n,m,t)=\mathrm{Tr} \{\rho_S^{(n,m)}(t)\}$, and the cross correlations can be calculated from the derivatives of the associated cumulant generating function (CGF):
\beq
\langle n_L(t_1)n_R(t_2)\rangle_c = (i)(-i) \partial_{\chi_1^L}\partial_{\chi_2^R} {\cal F} \Big|_{\bm \chi=0}.
\eeq
This CGF is given by
\beq
&&e^{{\cal F}(\chi_1^L,\chi_1^R,\chi_2^L,\chi_2^R)} = \sum_{n_1,m_1,n_2,m_2} e^{-in_1\chi_1^L}e^{im_1\chi_1^R}e^{-in_2\chi_2^L}e^{im_1\chi_1^R} P(n_1,m_1,t_1;n_2,m_2,t_2) \nonumber \\ &&= \sum_{n_1,m_1} e^{-in_1\chi_1^L}e^{im_1\chi_1^R} P(n_1,m_1,t_1) \sum_{n_2,m_2} e^{-in_2\chi_2^L}e^{im_2\chi_2^R} P(n_2,m_2,t_2|n_1,m_1,t_1) \nonumber \\ &&= \sum_{n_1,m_1} e^{-in_1\chi_1^L}e^{im_1\chi_1^R} \mathrm{Tr} \lbrace \rho_S^{(n_1,m_1)}(t_1) \rbrace \underbrace{\sum_{n_2,m_2} e^{-in_2\chi_2^L}e^{im_2\chi_2^R} \mathrm{Tr} \lbrace \rho_S^{(n_2,m_2|n_1,m_1)}(t_1) \rbrace}_{\frac{\mathrm{Tr}\lbrace \Omega(\chi_2^L,\chi_2^R,t_2-t_1) \rho_S(\chi_2^L,\chi_2^R,t_1) \rbrace}{\mathrm{Tr} \lbrace \rho_S^{(n_1,m_1)}(t_1) \rbrace}} \nonumber \\ &&= \mathrm{Tr} \Big\lbrace \sum_{n_1,m_1} \Omega(\chi_2^L,\chi_2^R,t_2-t_1) e^{-in_1(\chi_1^L+\chi_2^L)}e^{im_1(\chi_1^R+\chi_2^R)} \rho_S^{(n_1,m_1)}(t_1) \Big\rbrace \nonumber \\ &&= \mathrm{Tr} \lbrace \Omega(\chi_2^L,\chi_2^R,t_2-t_1) \rho_S(\chi_1^L+\chi_2^L,\chi_1^R+\chi_2^R,t_1) \rbrace \nonumber \\ &&= \mathrm{Tr} \lbrace \Omega(\chi_2^L,\chi_2^R,t_2-t_1) \Omega(\chi_1^L+\chi_2^L,\chi_1^R+\chi_2^R,t_1) \rho_S(0) \rbrace.
\eeq
Using the kernel (\ref{kernelinfty}) we obtain the cross correlation for the SRL model
\beq \label{correlator}
\langle I_L(t_1)I_R(t_2)\rangle_c = \langle I \rangle \Bigg[\frac{\Gamma_R^2}{\Gamma} e^{-\Gamma\tau} \theta(\tau) + \frac{\Gamma_L^2}{\Gamma} e^{\Gamma\tau} \theta(-\tau) \Bigg],
\eeq
where $\Gamma\equiv\Gamma_L+\Gamma_R$. In the Fourier space, Eq.~(\ref{correlator}) gives
\beq
S^{(2)}_{LR}(\omega) +S^{(2)}_{RL}(\omega) = 2 \langle I \rangle \frac{ \Gamma_L^2 + \Gamma_R^2 }{\Gamma^2 + \omega^2}.
\eeq
This result, together with the one for the particle noise
\beq
S^{(2)}_{L}(\omega) = S^{(2)}_{R}(\omega) =  \langle I \rangle \frac{ \Gamma_L^2 + \Gamma_R^2 + \omega^2 }{\Gamma^2 + \omega^2}
\eeq
reproduces Eq.~(\ref{totalFanonoiseSRL}) using the total-noise formula
\beq \label{totalnoiseLR}
S^{(2)}_{tot}(\omega) = \alpha^2 S^{(2)}_{L}(\omega) + \beta^2 S^{(2)}_{R}(\omega) + \alpha\beta \left( S^{(2)}_{LR}(\omega) +S^{(2)}_{RL}(\omega) \right),
\eeq
where $\alpha$ and $\beta$ account for the current partitioning, as discussed in subsection \ref{FCStotaccum}. Finally, we notice that, similarly to the total noise, which can be obtained using the formalism explained in subsection \ref{FCStotaccum}, or equations (\ref{totalnoiseLR})/(\ref{totalnoiseEq}), the total skewness can be also derived using these different methods, with the equivalent formula to (\ref{totalnoiseEq}), being
\beq \label{S3tot}
S^{(3)}_{tot}(\omega,\omega') &=& \alpha^2 (\alpha-\beta) S^{(3)}_{L}(\omega,\omega') + \beta(\beta^2-\alpha^2) S^{(3)}_{R}(\omega,\omega') + \nonumber \\ &+& \frac{\alpha\beta}{2} \Big[ S^{(3)}_{\mathcal{J}}(\omega,\omega') + (\alpha-\beta) S^{(3)}_{\mathcal{I}}(\omega,\omega') \Big].
\eeq
Here, ${\cal J}\equiv I_L + I_R$ and ${\cal I}\equiv I_L + I_R$, whose skewness contributions can be calculated using the jump operators ${\cal J}_{\chi,L}^{(n)}+{\cal J}_{\chi,R}^{(n)}$ and ${\cal J}_{\chi,L}^{(n)}-{\cal J}_{\chi,R}^{(n)}$ respectively (c.f. subsection \ref{FCStotaccum}).

~

~

~

~

~

~

~

~

~

~

~

~

~

~

~

~

~

~

~

~

~

~


\chapter{Derivation of frequency-dependent cumulants}
\label{appDiagrams}
\lhead{Appendix D. \emph{Derivation of frequency-dependent cumulants}}

The expressions (\ref{currentformula})-(\ref{skewnessformula}) follow from derivatives of moment generating functions. Performing derivatives of (\ref{GN}) we find
\beq 
\langle I(z) \rangle^{>} &=& z\partial_{\chi} \Big\langle \Omega(\chi, z) \Big\rangle\Big\vert_0= z^{-1}\langle {\cal J}_0^{(1)} \rangle, \label{Igreater} 
  \\
S_m^{(2)>}(z_1,z_2) &=& \left(z_1z_2\right) \left.
   \partial_{\chi_1} \partial_{\chi_2} \Big\langle
   \Omega(\chi_2,z_2)\Omega(\chi_{12},z_{12})
   \Big\rangle
   \right|_0 \nonumber \\
    &=&\left(z_1z_2\right) z_2^{-1}z_{12}^{-2} \langle 2 {\cal J}_0^{(1)} \Omega_0(z_{12}) {\cal J}_0^{(2)} + z_{12} {\cal J}_0^{(2)} \Omega_0(z_2) \Omega_0(z_{12}) {\cal J}_0^{(1)} + {\cal J}_0^{(2)} \rangle, \;\;\;\;\;\;\;\;\;\;    \label{S2greater} 
   \\
S_m^{(3)>}(z_1,z_2,z_3) &=&  \left(z_1z_2z_3\right) \left.
   \partial_{\chi_1}\partial_{\chi_2}\partial_{\chi_3}
    \Big\langle
    \Omega(\chi_3,z_3)
    \Omega(\chi_{23},z_{23})
    \Omega(\chi_{123},z_{123})
	  \Big\rangle
   \right|_0 \nonumber \\
	  &=& \left(z_1z_2z_3\right) z_3^{-1} z_{123}^{-1} \nonumber \\  
	&&\times \langle
		  {\cal J}_0^{(1)} \Omega_0(z_{3}) \Omega_0(z_{23})
		  {\cal J}_0^{(1)} \Omega_0(z_{23})\Omega_0(z_{123}){\cal J}_0^{(1)} \nonumber \\
		  &&+ 2{\cal J}_0^{(1)} \Omega_0(z_{3}) \Omega_0(z_{23})\Omega_0(z_{123})
		  {\cal J}_0^{(1)} \Omega_0(z_{123}){\cal J}_0^{(1)}
	  \nonumber\\ 
		  &&+4 z_{23}^{-1} {\cal J}_0^{(1)} \Omega_0(z_{23}) \Omega_0(z_{123})
		  {\cal J}_0^{(1)}\Omega_0(z_{123}){\cal J}_0^{(1)}
	  \nonumber\\ 
		&&
		  + 2 z_{23}^{-1} {\cal J}_0^{(1)} \Omega_0(z_{23})
		  {\cal J}_0^{(1)} \Omega_0(z_{23})\Omega_0(z_{123}){\cal J}_0^{(1)} \nonumber \\	
		&&
		  +6 z_{23}^{-1} z_{123}^{-1} {\cal J}_0^{(1)}  \Omega_0(z_{123})
		  {\cal J}_0^{(1)} \Omega_0(z_{123}){\cal J}_0^{(1)} \nonumber \\		    	  
	    &&+ {\cal J}_0^{(1)}  \Omega_0(z_{3})\Omega_0(z_{23})\Omega_0(z_{123}){\cal J}_0^{(2)} \nonumber \\
	    &&+2 z_{23}^{-1} {\cal J}_0^{(1)}  \Omega_0(z_{23})\Omega_0(z_{123}){\cal J}_0^{(2)}
	    	  \nonumber\\ 
		  &&+ z_{23}^{-1} {\cal J}_0^{(2)}  \Omega_0(z_{23})\Omega_0(z_{123}){\cal J}_0^{(1)} \nonumber\\	
		  &&+3 z_{23}^{-1} z_{123}^{-1} {\cal J}_0^{(2)}  \Omega_0(z_{123}){\cal J}_0^{(1)}
	  \nonumber\\ 		  	  
		&&
	    +3 z_{23}^{-1} z_{123}^{-1} {\cal J}_0^{(1)}  \Omega_0(z_{123}){\cal J}_0^{(2)} \nonumber\\
		  &&+z_{23}^{-1} z_{123}^{-1}{\cal J}_0^{(3)}
	  \rangle
	  , \label{S3greater}
\eeq
where $z_{ij}:=z_i+z_j$, $z_{ijk}:=z_i+z_j+z_k$. Next we use
\beq \label{simplificationOmegas}
\Omega_0(z_2)\Omega_0(z_1+z_2) = \frac{1}{z_1} \Big[ \Omega_0(z_2) - \Omega_0(z_1+z_2) \Big],
\eeq
and add the `lesser' part ($<$) corresponding to negative Laplace frequencies. At this point we change to `physical' frequencies $\omega:=\omega_2+\ldots+\omega_N$, $\omega':=\omega_3+\ldots+\omega_N$, etc., and symmetrize the result. This means adding the expressions corresponding to all the possible frequency switchings. In this step we take into account that 
\beq \label{deltaIdentity}
\lim_{\eta\to 0} \left(\frac{1}{i\omega + \eta} + \frac{1}{-i\omega+\eta} \right)= \lim_{\eta\to 0} \frac{2\eta}{\omega^2+\eta^2} = 2\pi\delta(\omega),
\eeq 
where $\eta\rightarrow 0$ is a small parameter coming from the `greater' ($>$) or `lesser' ($<$) parts. Finally, we make use of this energy conservation inherited from the time-translational symmetry of the cumulants. We then arrive to the equations (\ref{currentformula})-(\ref{skewnessformula}).
Importantly, after frequency symmetrization, one can realize that the first three cumulant formulae are equal to their moment counterparts.

\subsection*{A.1. Diagrams}

\begin{figure}[h!]
 \begin{center}
 \includegraphics[width=0.7\textwidth]{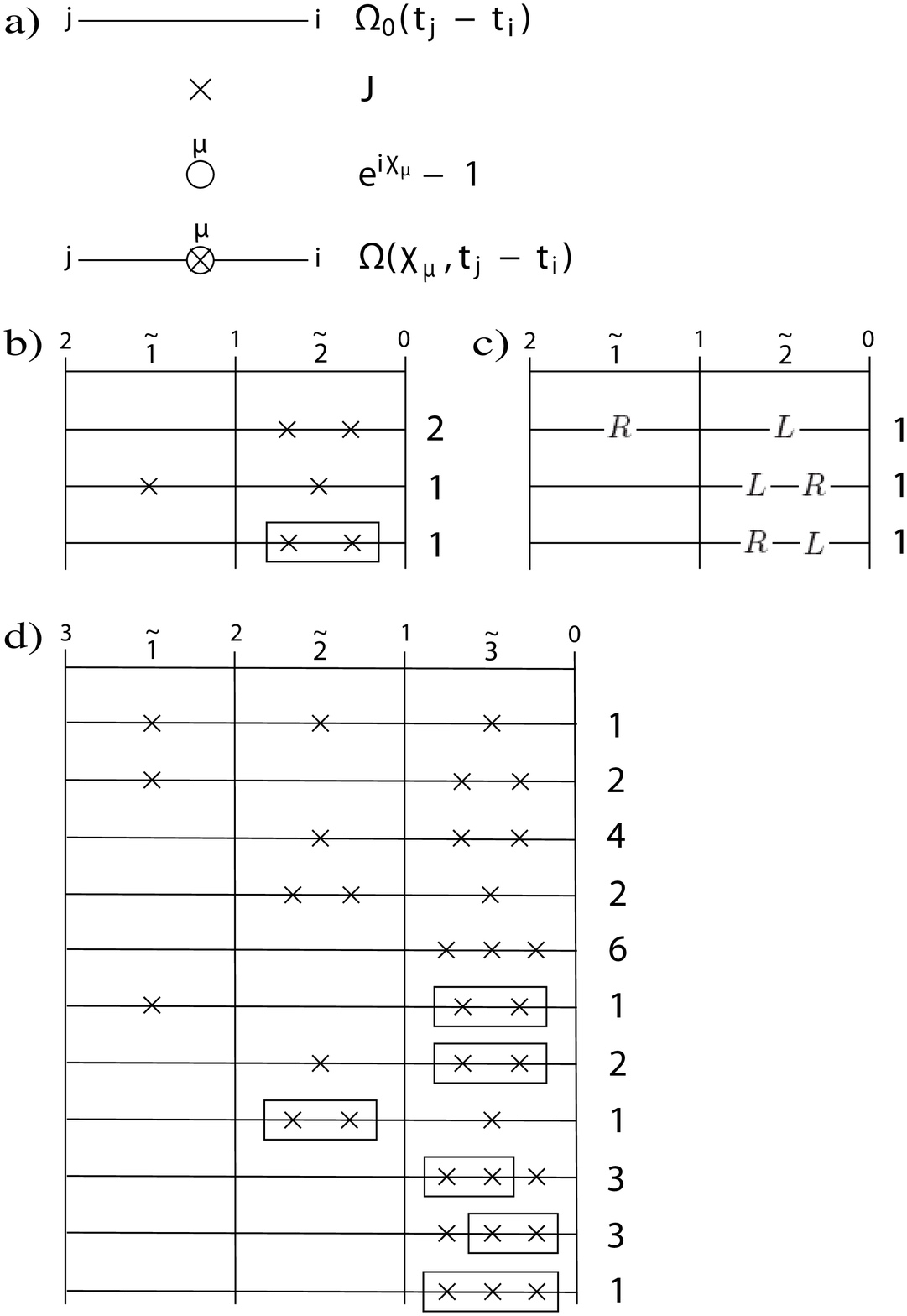}
 \end{center}
  \caption[Diagrammatics to obtain frequency-dependent cumulants]{Diagrammatics to obtain frequency-dependent cumulants.
     a)~Building pieces for the diagrammatic technique. A line is associated to a bare propagator, a cross with a jump operator, and a circle with the time dependence of ${\cal J}$ (term $(e^{i\chi} - 1)$ in the single-particle unidirectional tunneling case). Derivatives with respect to counting fields eliminate circles correspondingly. Diagrams can be simplified using rules like (\ref{simpDiagrams}). b)~Diagrams for the noise. Reading this table we find the expression $2\Omega_0(\tilde{z}_1)\Omega_0(\tilde{z}_2){\cal J}^{(1)}_0\Omega_0(\tilde{z}_2){\cal J}^{(1)}_0\Omega_0(\tilde{z}_2) + \Omega_0(\tilde{z}_1){\cal J}^{(1)}_0\Omega_0(\tilde{z}_1)\Omega_0(\tilde{z}_2){\cal J}^{(1)}_0\Omega_0(\tilde{z}_2) + \Omega_0(\tilde{z}_1)\Omega_0(\tilde{z}_2){\cal J}^{(2)}_0\Omega_0(\tilde{z}_2)$. c)~Diagrams for the second order cross correlation $S^{(2)}_{LR}$. Reading this table we find the expression $\Omega_0(\tilde{z}_1){\cal J}^{(1)}_{0,R}\Omega_0(\tilde{z}_1)\Omega_0(\tilde{z}_2){\cal J}^{(1)}_{0,L}\Omega_0(\tilde{z}_2) + \Omega_0(\tilde{z}_1)\Omega_0(\tilde{z}_2){\cal J}^{(1)}_{0,L}\Omega_0(\tilde{z}_2){\cal J}^{(1)}_{0,R}\Omega_0(\tilde{z}_2) + \Omega_0(\tilde{z}_1)\Omega_0(\tilde{z}_2){\cal J}^{(1)}_{0,R}\Omega_0(\tilde{z}_2){\cal J}^{(1)}_{0,L}\Omega_0(\tilde{z}_2)$. d)~Diagrams to derive the frequency-dependent skewness formula.
}
  \label{diagrams}
\end{figure}

Interestingly, the results (\ref{Igreater})-(\ref{S3greater}) given above can be derived following a diagrammatic technique, similarly to how this is done with Feynman diagrams in the expansion of the partition function or the $S$-matrix. This can be done if the CGF is written as a series expansion, either in the time domain or in the frequency space. To that end we expand each of the $\chi$-dependent propagators in the CGF as a Dyson series:
\beq\Omega(\chi,z) =
 \frac{1}{z-\mathcal{W}(\chi)}
=\Omega_0(z)\sum_{n=0}^{\infty} \left[
\mathcal{J}_{\chi} \Omega_0(z) \right]^n.
\eeq
This suggests the use of diagrams of the form given in Fig.~\ref{diagrams}a.
In the frequency domain\footnote{If we work in the time domain, it is enough to label the propagating lines with the corresponding times at the beginning and end of each line (see Fig.~\ref{diagrams}a).} these rules are:
\begin{itemize}
\item To each bare propagator $\Omega_0(\tilde{z}_k)$ in the expansion we associate a line with a superscript $\tilde{k}\equiv \sum_{i=N+1-k}^N i$, where $N$ is the order of the cumulant we want to obtain.
\item To each jump operator ${\cal J}_{\tilde{\chi}_k}$ in the expansion we associate an encircled cross with superscript $\tilde{k}$.
\end{itemize}
The formula for the generating function to a given order can therefore be written diagrammatically. For example, to second order we have 
\beq
{\cal G}(\bm{\chi},\bm{z}) &=& {\cal T}_S \mathrm{Tr}\{\Omega(\tilde{\chi}_1,\tilde{z}_1)\Omega(\tilde{\chi}_2,\tilde{z}_2)\rho_\mathrm{S}^{stat}\} \nonumber\\ &=& {\cal T}_S \mathrm{Tr}\{ \left( \Omega_0(z_2) + \Omega_0(z_2){\cal J}_{\chi_2} \Omega_0(z_2) + \ldots \right) \nonumber\\ &&\times \left( \Omega_0(z_1+z_2) + \Omega_0(z_1+z_2){\cal J}_{\chi_1+\chi_2} \Omega_0(z_1+z_2) + \ldots \right) \rho_\mathrm{S}^{stat} \}.
\eeq
We can then multiply the different terms using diagrams as described above. The multiplication of propagators implies joining them together. The result can be simplified using the property ${\cal J}_{\chi_1+\chi_2}={\cal J}_{\chi_1}{\cal J}_{\chi_2}+{\cal J}_{\chi_1}+{\cal J}_{\chi_2}$, which diagrammatically is denoted as
\beq \label{simpDiagrams}
\bigotimes^{12}=\bigotimes^1\bigotimes^2+\bigotimes^1+\bigotimes^2.
\eeq
Here, the super-index $12$ denotes an associated frequency $z_1+z_2$ and counting field $\chi_1+\chi_2$. 
Next, to arrive to the frequency-dependent \textit{moment}, we take derivatives with respect to counting fields. Diagrammatically, the derivative $\partial_{\chi_k}$ means removing a circle with index $k$. From here we can rewrite the expression analytically. The outcome can be simplified using (\ref{simplificationOmegas}), and needs to be multiplied by $z_1z_2$ (case $N=2$), coming from the Fourier transform of the time derivatives in the frequency domain. We finally need to take the average in the stationary state and symmetrize the result as dictated by ${\cal T}_S$. 

With the diagrammatic approach we realize that the diagrams contributing to the final result can be arranged in tables (see Fig.~\ref{diagrams}b to Fig.~\ref{diagrams}d). These reproduce the results given in (\ref{Igreater})-(\ref{S3greater}). To construct these tables one must proceed according to the following rules:
\begin{itemize}
\item To arrive to an expression for the cumulant of order $N$, write a table with $N$ time (frequency) intervals and corresponding superscripts $\tilde{k}$. The propagation of time will be taken from right to left.
\item Write all the possible diagrams having $N$ crosses (jumps) distributed in the different intervals, with the constraint that the maximum number of crosses in each is set by the corresponding index $\tilde{k}$. Diagrams with $n$ jumps occurring at the same time have to be included as well. These crosses are enclosed together with a box, and contribute with the jump operator $\mathcal{J}_0^{(n)}:=\partial_{\chi}^n\mathcal{J}_{\chi}|_{\chi=0}$.
\item Taking into account that jumps occurring in the same interval are indistinguishable, and that each cross can be associated to one of the possible counting fields $\chi_1,\ldots,\chi_k$ present in that interval, write the multiplicity of each diagram on the right. 
\item Write the mathematical expression corresponding to each diagram (see Fig.~\ref{diagrams}a) and sum the different terms evaluated at $z=i\omega$. 
\item Take the average in the stationary state, and multiply by $(-i)^N$. The resulting expression corresponds to the {\it unsymmetrized} (`greater', $>$) {\it moment of the number of particles}. 
\item Multiply by $(i\omega_1)\ldots (i\omega_N)$. This gives the {\it unsymmetrized moment of the current distribution}.
\item Add the `lesser' ($<$) part, that is, the expression corresponding to negative frequency. 
\item Finally, symmetrize the result, adding all the possible switchings of frequencies. This gives the {\it symmetrized moment of the current distribution}.
The result can be simplified using (\ref{simplificationOmegas}) and (\ref{deltaIdentity}).
\end{itemize}

As mentioned above, explicit derivation gives the same result for the expressions of \textit{cumulants} of the current distribution as those derived for the moments up to $N=3$. To higher orders it is unknown for us if this property still holds or not.
Expressions for the cross correlations, e.g. $S^{(2)}_{LR}:= \langle I(t_1)I(t_2) \rangle_c$, between two (or more) stochastic processes, e.g.
$L$ and $R$, can also be derived with this technique. To this end we simply need to label each of the jumps occurring at $L$ or $R$ accordingly (see Fig.~\ref{diagrams}c), having two types of jump operators, ${\cal J}_L$ and ${\cal J}_R$. Also, expressions for the total current ($\alpha I_L +\beta I_R$) and accumulated current ($I_L-I_R$) can be derived using the jump operators (\ref{Jtot}) and (\ref{Jaccum}), respectively, in the diagrams.

\subsection*{A.2. Equivalent form}

We can write down an equivalent form to expressions (\ref{currentformula})-(\ref{skewnessformula}). This will allow us to obtain an analytical expression for their zero-frequency limit, which is not well defined in the form given above. To this end we make use of the projectors $P:=\vert 0 \rangle\!\rangle \langle\!\langle  \widetilde{0} \vert$ and $Q:=\mathds{1}-P$, where $P$ projects onto the subspace spanned by the stationary state\footnote{The state $\langle\!\langle \widetilde{0}\vert$ denotes the left eigenvector of the Liouvillian. The tilde indicates that it is not the adjoint to $\vert 0 \rangle\!\rangle$, since the Liouvillian is not Hermitian.} $\vert 0 \rangle\!\rangle \equiv \bm{\rho}_\mathrm{S}^{stat}$; and we define the pseudo-inverse $R_0(z):= Q \Omega_0(z) Q$, such that $\Omega_0(z) = R_0(z) + P/z$.
Making this change in (\ref{Igreater})-(\ref{S3greater}), and symmetrizing the expression (including positive and negative frequencies), we get
\beq
 i \langle I (z) \rangle = \delta(z) \langle {\cal J}_0^{(1)} \rangle,
\eeq
\beq
 i^2 S^{(2)}(z_1,z_2)= \delta(z_1+z_2)
  \langle {\cal J}_0^{(2)} + {\cal J}_0^{(1)}R_0(z_1){\cal J}_0^{(1)} + {\cal J}_0^{(1)}R_0(z_2) {\cal J}_0^{(1)} \rangle,
\eeq
\beq
  i^3 S^{(3)}(z_1,z_2,z_3) &=& \delta(z_1+z_2+z_3) \nonumber\\
  &&\times\langle
	  {\cal J}_0^{(3)}
	  +
	  {\cal J}_0^{(1)} \left[R_0(z_1) + R_0(z_2)+R_0(z_3) \right]{\cal J}_0^{(2)} \nonumber \\
	  &&+
	  {\cal J}_0^{(2)}  \left[R_0(z_{12}) + R_0(z_{23})+R_0(z_{13})\right]{\cal J}_0^{(1)}
	\nonumber\\
	&&
		+ 
		{\cal J}_0^{(1)} R_0(z_1) {\cal J}_0^{(1)} \left[R_0(z_{12}) + R_0(z_{13})\right] {\cal J}_0^{(1)} \nonumber \\
	&&	+ 
		{\cal J}_0^{(1)} R_0(z_2) {\cal J}_0^{(1)} \left[R_0(z_{12}) + R_0(z_{23})\right] {\cal J}_0^{(1)}
	\nonumber\\
	&&
		+ 
		{\cal J}_0^{(1)} R_0(z_3) {\cal J}_0^{(1)} \left[R_0(z_{13}) + R_0(z_{23})\right] {\cal J}_0^{(1)}
	\nonumber\\
	&&
	  +
	  z_1^{-1} \ew{{\cal J}_0^{(1)}} 
	  {\cal J}_0^{(1)} \left[ R_0(z_{12}) \right. \nonumber\\ 
	  &&\left. -R_0(z_2)+ R_0(z_{13}) -R_0(z_3) \right] {\cal J}_0^{(1)}
	\nonumber\\
	&&
	  +
	  z_2^{-1} \ew{{\cal J}_0^{(1)}} 
	  {\cal J}_0^{(1)} \left[ R_0(z_{12}) \right. \nonumber\\
	  && \left. -R_0(z_1)+ R_0(z_{23}) -R_0(z_3) \right] {\cal J}_0^{(1)}
	\nonumber\\
	&&
	  +
	  z_3^{-1} \ew{{\cal J}_0^{(1)}} 
	  {\cal J}_0^{(1)} \left[ R_0(z_{13}) \right. \nonumber\\
	  && \left. -R_0(z_1)+ R_0(z_{23}) -R_0(z_2) \right] {\cal J}_0^{(1)}
	\rangle.
\eeq
Now we make use of the delta function to write $z_2=-z_1$ in the noise expression and $z_3=-z_1-z_2$ in the skewness result. Performing the change of variables $z_1 \to -i \omega$, $z_2 \to i \omega$ in the noise and $z_1\to -i\omega$, $z_2\to i\omega-i\omega'$, $z_3\to i\omega'$ in the skewness, we obtain
\beq \label{IRw}
 i I_{stat} = \langle {\cal J}_0^{(1)} \rangle,
\eeq
\beq \label{S2Rw}
 i^2 S^{(2)}(\omega)=
  \langle {\cal J}_0^{(2)} + {\cal J}_0^{(1)}R_0(i\omega){\cal J}_0^{(1)} + {\cal J}_0^{(1)}R_0(-i\omega') {\cal J}_0^{(1)} \rangle,
\eeq
\beq \label{S3Rw}
  i^3 S^{(3)}(\omega,\omega') &=& 
  \langle
	  {\cal J}_0^{(3)}
	  +
	  {\cal J}_0^{(1)} \left[R_0(-i\omega) + R_0(i\omega-i\omega')+R_0(i\omega')\right]{\cal J}_0^{(2)} \nonumber \\
	  &&+
	  {\cal J}_0^{(2)}  \left[R_0(-i\omega') + R_0(i\omega)+R_0(i\omega'-i\omega)\right]{\cal J}_0^{(1)}
	\nonumber\\
	&&
		+ 
		{\cal J}_0^{(1)} R_0(-i\omega) {\cal J}_0^{(1)} \left[R_0(-i\omega') + R_0(i\omega'-i\omega)\right] {\cal J}_0^{(1)} \nonumber \\
	&&	+ 
		{\cal J}_0^{(1)} R_0(i\omega-i\omega') {\cal J}_0^{(1)} \left[R_0(-i\omega') + R_0(i\omega)\right] {\cal J}_0^{(1)}
	\nonumber\\
	&&
		+ 
		{\cal J}_0^{(1)} R_0(i\omega') {\cal J}_0^{(1)} \left[R_0(i\omega'-i\omega) + R_0(i\omega)\right] {\cal J}_0^{(1)}
	\nonumber\\
	&&
	  +
	  (-i\omega)^{-1} \ew{{\cal J}_0^{(1)}} 
	  {\cal J}_0^{(1)} \left[ R_0(-i\omega') \right. \nonumber\\ 
	  &&\left. -R_0(i\omega-i\omega')+ R_0(i\omega'-i\omega) -R_0(i\omega') \right] {\cal J}_0^{(1)}
	\nonumber\\
	&&
	  +
	  (i\omega-i\omega')^{-1} \ew{{\cal J}_0^{(1)}} 
	  {\cal J}_0^{(1)} \left[ R_0(-i\omega') \right. \nonumber\\
	  && \left. -R_0(-i\omega)+ R_0(i\omega) -R_0(i\omega') \right] {\cal J}_0^{(1)}
	\nonumber\\
	&&
	  +
	  (i\omega')^{-1} \ew{{\cal J}_0^{(1)}} 
	  {\cal J}_0^{(1)} \left[ R_0(i\omega'-i\omega) \right. \nonumber\\
	  && \left. -R_0(-i\omega)+ R_0(i\omega) -R_0(i\omega-i\omega') \right] {\cal J}_0^{(1)}
	\rangle.
\eeq
The limit $\omega\to 0$ of these expressions is well defined, and they can therefore be used to check that the proper result is recovered in that limit.

\subsection*{A.3. Zero-frequency limit}

As mentioned, expressions (\ref{IRw})-(\ref{S3Rw}) are well behaved when $\omega\to 0$. The zero-frequency noise comes straightforwardly from (\ref{S2Rw}) setting $\omega=0$. For the skewness, this limit requires nevertheless noticing that
\beq
\lim_{\omega\to 0} \left[R_0(i\omega) - R_0(-i\omega)\right] = 2i \omega \partial_\omega R_0(i\omega)\vert_{\omega=0}.
\eeq
So the zero-frequency skewness can be written as
\beq
iS^{(3)}(0,0) &=&  \ew{
	  {\cal J}_0^{(3)}
	  +
	  3 {\cal J}_0^{(1)} R_0(0){\cal J}_0^{(2)}
	  +
	  3 {\cal J}_0^{(2)} R_0(0) {\cal J}_0^{(1)} \nonumber\\
		&&+ 6 {\cal J}_0^{(1)} R_0(0) {\cal J}_0^{(1)} R_0(0) {\cal J}_0^{(1)}} \nonumber\\
		&&+   6 \ew{{\cal J}_0^{(1)}}\ew{{\cal J}_0^{(1)} \partial_\omega R_0(0) {\cal J}_0^{(1)}}.
\eeq
Now, since $\partial_\omega R_0(0) = R(0) R(0)$, we have 
\beq
iS^{(3)}(0,0) &=&  \ew{
	  {\cal J}_0^{(3)}
	  +
	  3 {\cal J}_0^{(1)} R_0(0){\cal J}_0^{(2)}
	  +
	  3 {\cal J}_0^{(2)} R_0(0) {\cal J}_0^{(1)} \nonumber\\
		&&+ 6 {\cal J}_0^{(1)} R_0(0) {\cal J}_0^{(1)} R_0(0) {\cal J}_0^{(1)}} \nonumber\\
		&&+   6 \ew{{\cal J}_0^{(1)}}\ew{{\cal J}_0^{(1)} R_0(0) R_0(0) {\cal J}_0^{(1)}}.
\eeq
Which is the zero-frequency limit found in \cite{Flindt05b}.

~

~

~

~

~

~

~

~

~

~

~

~

~

~

~

~

~

~

~

~

~

~


\chapter{Derivation of the self-energy}
\label{appKernel}
\lhead{Appendix E. \emph{Derivation of the self-energy}}

In this appendix we show the form of the kernel (\ref{EOMchi}) and the self-energies (\ref{1pointSE})-(\ref{2pointSE}) at finite bias voltage. We follow the
perturbative treatment by Schoeller and coworkers
\cite{Schoeller09, Leijnse08}. Let $\mathcal{L}_S$, $\mathcal{L}_R$
and $\mathcal{L}_V$ be the corresponding Liouvillians to
(\ref{Hs}), (\ref{Hr}) and (\ref{Hv}), respectively. The last can be written in the form \beq
\mathcal{L}_V = -i \sum_{\eta,\alpha,m,\xi,p} {\cal V}_{\eta\alpha m}
G_{m}^{\xi p} J_{\eta\alpha}^{\xi p}(\chi), \eeq where $\xi = +$
$(-)$ refers to the creation (annihilation) of particles in the
leads, and $G_{m}^{\xi p}$ and $J_{\eta\alpha}^{\xi p}(\chi)$ are system
and reservoir super-operators respectively. These act on an operator
$A$ as 
\begin{equation}
\begin{array}{c}
G_{m}^{\xi p} A = \sigma^p \times \left\{ \begin{array}{c} \;\; g_m ^{\xi} A \;\; \mbox{if} \;\; p=+ \\
                                -A g_m ^{\xi} \;\; \mbox{if} \;\; p=- \end{array} \right. .
\\ \\
J_{\eta\alpha}^{\xi p}(\chi) A = \left\{ \begin{array}{c} \;\; j_{\eta\alpha}^{\xi}(\chi) A \;\; \mbox{if} \;\; p=+ \\
                                              A j_{\eta\alpha}^{\xi}(\chi) \;\; \mbox{if} \;\; p=- \end{array} \right. .
\end{array}
\end{equation}
Here, $\sigma^p$ is a super-operator with matrix elements $\left[ \sigma^p \right]_{s,s';-s,-s'} = \delta_{s,-s}\delta_{s,-s'}p^{N_s-N_{s'}}$, being $N_s$ the number of electrons in state $s$. The operators $g_{\eta\alpha} ^{\xi} $ and $j_m ^{\xi} $ are defined as
\beq
&&g_m^{+} = \sum_{aa'} \langle a \vert d_m \vert a' \rangle \vert a \rangle \langle a' \vert, \\
&&j_{\eta\alpha}^{+}(\chi) = c_{\eta \alpha}^{\dagger} e^{i
s_{\alpha}\chi_{\alpha}/2}, \eeq and $g_m^{-} =
(g_m^{+})^{\dagger}$, $j_{\eta\alpha}^{-}(\chi) = \left(
j_{\eta\alpha}^{+}(\chi) \right)^{\dagger}$. The index $s_{\alpha}=\pm 1$ is taken according to the sign convention for the current flow in lead $\alpha$.

With this notation, the self-energy to order $|{\cal V}|^2$
reads \cite{Schoeller09,Leijnse08,Emary09}
\beq \label{selfenergy}
&&\Sigma(\chi,z) \rho_\mathrm{S} (t_0) = \sum \mathrm{Tr_R} \Big\lbrace {\cal V}_{\eta\alpha m} G_{m}^{\xi p} J_{\eta \alpha}^{\xi p}(\chi) \frac{-1}{z-\mathcal{L}_S-\mathcal{L}_R} {\cal V}_{\eta' \alpha' m'} G_{m'}^{\xi' p'} J_{\eta' \alpha'}^{\xi' p'}(\chi) \rho(t_0) \Big\rbrace \nonumber \\
&&= \frac{1}{2\pi} \sum  \int_{-D}^{D} \Gamma_{\alpha m m'}^{\xi p p'} (\varepsilon,\chi) G_m^{\xi p} \frac{-p'}{z -i\xi (\varepsilon + \mu_{\alpha}) -i\lambda_a} \vert a \rangle\!\rangle \langle\!\langle \widetilde{a} \vert G_{m'}^{-\xi p'} f(-\xi p \varepsilon/kT) d\varepsilon \rho_\mathrm{S}(t_0) \nonumber \\
&&= \sum -pp' \Gamma_{\alpha m m'}^{\xi p p'}(\varepsilon,\chi) G_m^{\xi p} \vert a \rangle\!\rangle \langle\!\langle \widetilde{a} \vert G_{m'}^{-\xi p'} \nonumber\\ &&~~\times \Big[ \frac{1}{2} f\left( p (\lambda_a +\xi\mu_{\alpha} -iz) \right) + \frac{ip}{2\pi} \phi\left( p (\lambda_a +\xi\mu_{\alpha} -iz) \right) \Big] \rho_\mathrm{S}(t_0), 
\eeq 
where the summations run over all scripts, and $D$ is a high-energy cutoff set by the bandwidth of the Fermi leads -- larger than the rest of energy scales in the problem. 
In this expression, we have
introduced a complete set of eigenstates of the system
Liouvillian, $\mathcal{L}_S \vert a \rangle\!\rangle = i \lambda_a \vert a
\rangle\!\rangle$, and the definitions $f(x) := (e^{x/kT} +1)^{-1}$ and
\beq \label{Gammachi}
&&\Gamma_{\alpha m m'}^{\xi p p'}(\varepsilon,\chi) := \Gamma_{\alpha m m'} (\varepsilon) e^{i s_{\alpha} \xi \left(\frac{p'-p}{2}\right) \chi_{\alpha}}, \\
&&\phi(x) := \mbox{Re} \; \Psi \left( \frac{1}{2} + i \frac{x}{2\pi kT}
\right) - \mbox{ln} \; \frac{D}{2\pi kT}. \eeq 
Here, $\Gamma_{\alpha m m'} (\varepsilon) \equiv \frac{2\pi}{\hbar} \sum_{\eta} {\cal V}_{\eta\alpha m} {\cal V}_{\eta \alpha m'} \delta(\varepsilon - \varepsilon_{\eta\alpha})$ (which we take to be independent of the energy $\Gamma_{\alpha m m'} (\varepsilon)\approx \Gamma_{\alpha m m'}$), and $\Psi$ is the digamma function. The self-energy (\ref{selfenergy}) is important as it allows us to explore correctly the low bias limit ($eV\lesssim kT$) to sequential tunneling order. This self-energy is non-Markovian as the Markovian approximation has not been made up to this point. This can be made (together with the secular approximation) by taking the limit $z\to 0$ of (\ref{selfenergy}), and this is precisely what was used in chapter \ref{Chapter3}. However, the $z$-dependent expression was used in chapter \ref{Chapter4}.
The two-point self-energy (\ref{2pointSE}) can be similarly derived. It reads
\beq \label{2pointselfenergy}
&&\Pi(\chi,\chi',z) \rho_\mathrm{S} (t_0) = \sum \mathrm{Tr_R} \Big\lbrace {\cal V}_{\eta\alpha m} G_{m}^{\xi p} J_{\eta \alpha}^{\xi p}(\chi) \frac{-1}{z-\mathcal{L}_S-\mathcal{L}_R} {\cal V}_{\eta' \alpha' m'} G_{m'}^{\xi' p'} J_{\eta' \alpha'}^{\xi' p'}(\chi') \rho(t_0) \Big\rbrace \nonumber \\
&&= \frac{1}{2\pi} \sum  \int_{-D}^{D} \Gamma_{\alpha m m'}^{\xi p p'} (\varepsilon,\chi,\chi') G_m^{\xi p} \frac{-p'}{z -i\xi (\varepsilon + \mu_{\alpha}) -i\lambda_a} \vert a \rangle\!\rangle \langle\!\langle \widetilde{a} \vert G_{m'}^{-\xi p'} f(-\xi p \varepsilon/kT) d\varepsilon \rho_\mathrm{S}(t_0) \nonumber \\
&&= \sum -pp' \Gamma_{\alpha m m'}^{\xi p p'}(\varepsilon,\chi,\chi') G_m^{\xi p} \vert a \rangle\!\rangle \langle\!\langle \widetilde{a} \vert G_{m'}^{-\xi p'} \nonumber\\ &&~~\times \Big[ \frac{1}{2} f\left( p (\lambda_a +\xi\mu_{\alpha} -iz) \right) + \frac{ip}{2\pi} \phi\left( p (\lambda_a +\xi\mu_{\alpha} -iz) \right) \Big] \rho_\mathrm{S}(t_0)
\eeq 
where now
\beq
\Gamma_{\alpha m m'}^{\xi p p'}(\varepsilon,\chi,\chi') := \Gamma_{\alpha m m'} (\varepsilon) e^{i s_{\alpha} \xi \left(\frac{p'\chi'_{\alpha}-p\chi_{\alpha}}{2}\right)}.
\eeq

To derive (\ref{selfenergy}) and (\ref{2pointselfenergy}) the following integral has been used:
\beq
&&\int_{-D/kT}^{D/kT} \frac{1}{x-q_{\alpha}} \frac{1}{e^x+1} dx = \oint_{C} \frac{1}{s - q_{\alpha}} \frac{1}{e^s+1} ds - \oint_{\curvearrowleft} \frac{1}{s - q_{\alpha}} \frac{1}{e^s+1} ds \nonumber\\ &&=2\pi i \frac{1}{e^{q_{\alpha}}+1} + 2\pi i \sum_{n=0}^{n_D^{\alpha}} \frac{1}{(2n+1)\pi i - q_{\alpha}} - \int_0^{\pi} \frac{1}{s - q_{\alpha}} \frac{1}{e^s +1} i s d\varphi \Big\vert_{s= \frac{De^{i\varphi}}{kT}} \nonumber \\ &&= 2\pi i f(q_{\alpha}) + \sum_{n=0}^{n_D^{\alpha}} \frac{1}{n + \left( \frac{1}{2} + i\frac{q_{\alpha}}{2\pi} \right)} -\int_{\pi/2}^{\pi} i d\varphi \nonumber\\ &&= 2\pi i f(q_{\alpha}) + \ln \left( n_D^{\alpha} \right) - \Psi \left(\frac{1}{2} +i\frac{q_{\alpha}}{2\pi}\right) - i \pi/2 \nonumber\\ &&= i\pi f(q_{\alpha}) - \mathrm{Re} \left\lbrace \Psi\left(\frac{1}{2}+i\frac{q_{\alpha}}{2\pi}\right) \right\rbrace + \ln \left( n_D^{\alpha} \right),
\eeq
where we have defined $n_D^{\alpha} := \frac{D}{2\pi kT}$, and used the property $i \mathrm{Im} \left\lbrace \Psi\left( 1/2 + ix \right) \right\rbrace = i\pi f(2\pi x) -i\pi/2$, being $\Psi$ the digamma function. $C$ is the closed circuit lying along the real axis and encircling the upper complex plane with radius $D$, while $\curvearrowleft$ denotes the upper part of the circuit.

~

~

~

~

~

~

~

~

~

~

~

~

~

~

~

~

~

~

~

~

~

~


\chapter{Calculation of the fidelity}
\label{FidelityAppendix}
\lhead{Appendix F. \emph{Calculation of the fidelity}}

In this appendix we give a useful expression for the fidelity of a qubit operation in the presence of a dissipative bath. 
We start with the definition of the fidelity, which using (\ref{rhoPauli}), can be written as
\beq
F&=&\bra{\psi_f} \mathrm{Tr}_B \{ \hat{\rho} \} \ket{\psi_f} = \mathrm{Tr} \left\{ \ket{\psi_f}\bra{\psi_f} \hat{\rho} \right\} \nonumber\\ &=& \mathrm{Tr} \left\{ \ket{\psi_f}\bra{\psi_f} \frac{1}{2} U(t_f) \left( \mathds{1} + \vec{p}(0)\hat{\vec{\sigma}}(0) \right) \otimes \hat{\rho}_B(0) U^{\dagger}(t_f) \right\} \nonumber\\ &=& \frac{1}{4} \mathrm{Tr} \left\{ U^{\dagger}(t_f) \left( \mathds{1} + \vec{p}(t_f)\hat{\vec{\sigma}}(t_f) \right) U(t_f) \left( \mathds{1} + \vec{p}(0)\hat{\vec{\sigma}}(0) \right) \otimes \hat{\rho}_B(0) \right\}.
\eeq
Here, $\hat{\vec{\sigma}}\equiv (\hat{\sigma}_x,\hat{\sigma}_y,\hat{\sigma}_z)_T$ is a vector of Pauli operators, $\vec{p}\equiv \ew{\hat{\vec{\sigma}}}$, $\hat{\rho}_B$ the bath density operator, and $U(t)$ is the evolution operator for a time $t$. The state $\ket{\psi_f}$ is that of the system at time $t_f$, in which we are interested. Here, we choose $t_f$ such that $\hat{\vec{\sigma}}(t_f)=\hat{\vec{\sigma}}(0)$ (one complete cycle), so in this case the fidelity reads
\beq \label{FidelityInitialEq}
\frac{1}{4} \mathrm{Tr} \left\{ U^{\dagger}(t_f) \left( \mathds{1} + \vec{p}(0)\hat{\vec{\sigma}}(0) \right) U(t_f) \left( \mathds{1} + \vec{p}(0)\hat{\vec{\sigma}}(0) \right) \otimes \hat{\rho}_B(0) \right\}.
\eeq
We assume that the Hamiltonian describing the whole system is of the form (\ref{Hform}). Switching to the interaction picture, we can write evolution operator as (c.f. Eq.~(\ref{EvolutionOperatorEq}))
\beq \label{Uexpansion}
U(t) = \mathds{1} + U_1(t) + U_2(t) +\ldots
\eeq
with $U_1(t)\equiv -i \int_0^t \hat{\tilde{\cal H}}_\mathrm{V}$$(t') dt'$, $U_2(t)\equiv - \int_0^t \int_0^{t'} \hat{\tilde{\cal H}}_\mathrm{V}$$(t') \hat{\tilde{\cal H}}_\mathrm{V}$$(t'') dt'' dt'$, etc., and $\hat{\tilde{\cal H}}_\mathrm{V}$ defined according to (\ref{InteractionPictureDef}). Inserting the expansion (\ref{Uexpansion}) into Eq.~(\ref{FidelityInitialEq}), and using $U_1^{\dagger}(t)=-U_1(t)$, along with $U_2^{\dagger}(t)+U_2(t)=U_1(t)U_1(t)$ and the cyclic property of the trace, to second order in the coupling Hamiltonian $\hat{\cal H}_\mathrm{V}$ we find
\beq
F=1- \frac{1}{4}{\rm Tr}{\left\{{\left[U_1(t_f),{\left(\mathds{1}+\vec{p}(0)\hat{\vec{\sigma}}(0)\right)}\right]} U_1(t_f) {\left(\mathds{1}+\vec{p}(0)\hat{\vec{\sigma}}(0)\right)} \hat{\rho}_B\right\}}.
\eeq
This expression can be simplified using $\left[ U_1(t), \mathds{1} \right]=0$ and averaging over the Bloch sphere, which using $\ew{\sigma_x}_\mathrm{Bloch}=\ew{\sigma_y}_\mathrm{Bloch}=\ew{\sigma_z}_\mathrm{Bloch}=0$ and $\ew{\sigma_x^2}_\mathrm{Bloch}=\ew{\sigma_y^2}_\mathrm{Bloch}=\ew{\sigma_z^2}_\mathrm{Bloch}=1/3$, gives
\beq \label{FidelityFinal}
 F&=&1-\frac{1}{12}{\rm Tr}{\left\{[U_1(t_f),\hat{\sigma}_z(0)]U_1(t_f)\hat{\sigma}_z(0) \hat{\rho}_B \right\} } \nonumber \\ &&~ -\frac{1}{6}{\rm Tr}{\left\{[U_1(t_f), \hat{\sigma}_{-}(0)]U_1(t_f)\hat{\sigma}_{+}(0) \hat{\rho}_B\right\}} \nonumber\\ &&~ -\frac{1}{6}{\rm Tr}{\left\{[U_1(t_f), \hat{\sigma}_{+}(0)]U_1(t_f)\hat{\sigma}_{-}(0) \hat{\rho}_B\right\}},
\eeq
where $\hat{\sigma}_{\pm}\equiv (\hat{\sigma}_x\pm \hat{\sigma}_y)/2$.

Equation (\ref{FidelityFinal}) can be applied for instance to calculate the fidelity of a qubit operation of duration $t_f$ in the presence of a Wigner-Weisskopf bath. This is defined by the interaction Hamiltonian (\ref{Wigner-Weisskopf-Eq}), and gives
\beq
F=1-\frac{\Gamma t_f}{3},
\eeq
where $\Gamma\approx\Gamma(\varepsilon)\equiv 2\pi \sum_k |g_k|^2 \delta(\varepsilon-\varepsilon_k)$ is the qubit decay rate due to the bath. More complicated examples can be studied using Eq.~(\ref{FidelityFinal}). For example, the collective spin model explained in section \ref{collectivecouplingSec} gives $F=1-\frac{\Gamma t_f}{6}$, assuming a Wigner-Weisskopf bath, and although through a rather lengthy calculation, Eq.~(\ref{FidelityFinal}) can be used to estimate the fidelity of a flux-qubit operation in the presence of the dipolar bath.

\addtocontents{toc}{\vspace{1em}} 

\clearpage  





\addtotoc{List of publications}  
\publications{
\addtocontents{toc}{\vspace{1em}}  

\begin{itemize}

\item Non-Markovian Effects in the Quantum Noise of Interacting Nanostructures.\\
\textit{D. Marcos, C. Emary, T. Brandes, and R. Aguado.}\\
Phys. Rev. B {\bf 83}, 125426 (2011).
\item Observation of the Bloch-Siegert Shift in a Qubit-Oscillator System in the Ultrastrong Coupling Regime.\\
\textit{P. Forn-D\'iaz, J. Lisenfeld, D. Marcos, J. J. Garc\'ia-Ripoll, E. Solano, C. J. P. M. Harmans, and J. E. Mooij.}\\
Phys Rev. Lett. {\bf 105}, 237001 (2010).
\item Finite-frequency Counting Statistics of Electron Transport: Markovian Theory.\\
\textit{D. Marcos, C. Emary, T. Brandes, and R. Aguado.}\\
New J. Phys. {\bf 12}, 123009 (2010).
\item Coupling Nitrogen-Vacancy Centers in Diamond to Superconducting Flux Qubits.\\
\textit{D. Marcos, M. Wubs, J. M. Taylor, R. Aguado, M. D. Lukin, and A. S. S\o rensen.}\\
Phys. Rev. Lett. {\bf 105}, 210501 (2010).
\item Electron-Nuclear Interaction in ${}^{13}\textrm{C}$ Nanotube Double Quantum Dots.\\
\textit{H. O. H. Churchill, A. J. Bestwick, J. W. Harlow, F. Kuemmeth, D. Marcos, C. H. Stwertka, S. K. Watson, and C. M. Marcus}.\\
Nature Physics {\bf 5}, 321 (2009).
\item Frequency-dependent Counting Statistics in Interacting Nanoscale Conductors.\\
\textit{C. Emary, D. Marcos, R. Aguado, and T. Brandes.}\\
Phys. Rev. B {\bf 76}, 161404(R) (2007).

\end{itemize}
}
\clearpage  

\addtotoc{Curriculum Vitae}  
\curriculum{
\addtocontents{toc}{\vspace{1em}}  

\begin{tabular}{ll}

May 20th. 1982 & Born in Madrid, Spain. \\
Sept. 1998 -- June 2000 & Baccalaureate in Science. \\ & Instituto Marqu\'es de Suanzes, Madrid, Spain. \\
Sept. 2000 -- Sept. 2005 & Graduate in Physics. \\ & Universidad Aut\'onoma de Madrid. Spain. \\
Sept. 2005 -- June 2006 & Master in Condensed Matter Physics. \\ & Universidad Aut\'onoma de Madrid. Spain. \\
June 2006 -- December 2010 & PhD thesis. Group of Professor Ram\'on Aguado. \\ & Instituto de Ciencia de Materiales de Madrid (CSIC), Spain. \\
Sept. 2006 -- Dec. 2006 & Visiting Scholar. Group of Professor Tobias Brandes. \\ & Technische Universit\"at Berlin, Germany. \\
Sept. 2007 -- Jan. 2008 & Visiting Scholar. Group of Professor Charles Marcus. \\ & Harvard University, Cambridge, USA. \\
Sept.  2008 -- Jan. 2009 & Visiting Scholar. Group of Professor Anders S\o rensen. \\ & Niels Bohr Institutet, Copenhagen, Denmark. \\
Sept. 2009 -- Jan. 2010 & Visiting Scholar. Group of Professor Leo Kouwenhoven. \\ & Technische Universiteit Delft, The Netherlands.

\end{tabular}

\clearpage  


\addtocontents{toc}{\vspace{0em}}  
\backmatter
\label{Bibliography}
\lhead{\emph{Bibliography}}  
\bibliographystyle{alpha}  
\bibliography{Bibliography}  

\end{document}